\documentclass[twocolumn,showpacs,preprintnumbers,amsmath,amssymb]{revtex4-1}
\usepackage{graphicx,color}
\usepackage{dcolumn}
\usepackage{bm}
\usepackage{longtable}
\usepackage{amsmath}
\usepackage{mathrsfs}
\usepackage{amsfonts}
\usepackage{textcomp}
\usepackage{esvect}
\usepackage{float}

\def\gsim {\mbox{\hbox{ \lower-.6ex\hbox{$>$}
\kern-1.12em \lower.5ex\hbox{$\sim$}\kern+.35em}}}
\def\lsim {\mbox{\hbox{ \lower-.6ex\hbox{$<$}
\kern-1.12em \lower.5ex\hbox{$\sim$}\kern+.35em}}}

\begin{document}


\title{Analytic formulas for the rapid evaluation of the orbit 
       response matrix and chromatic functions from lattice 
       parameters in circular accelerators\vspace{-0.0 cm}}

\author{Andrea Franchi\footnote[*]{\email{andrea.franchi@esrf.fr}} and 
        Simone Maria Liuzzo}
\affiliation{ESRF, CS 40220, 38043 Grenoble Cedex 9, France}
\author{Zeus Mart\'i}
\affiliation{CELLS, 08193 Bellaterra, Spain}

\date{\today}

\begin{abstract}\vspace{0.0 cm}
Measurements and analysis of orbit response matrix have been 
providing for decades a formidable tool in the detection 
of linear lattice imperfections and their correction. Basically 
all storage-ring-based synchrotron light sources across the world 
make routinely use of this technique in their daily operation, 
reaching in some cases a correction of linear optics down to $1\%$ 
beta beating and 1\textperthousand \  coupling. During the design phase 
of a new storage ring it is also applied in simulations for the 
evaluation of magnetic and mechanical tolerances. 
However, this technique is known for its intrinsic slowness 
compared to other methods based on turn-by-turn beam position 
data, both in the measurement and in the data analysis. 
In this paper analytic formulas are derived and discussed that 
shall greatly speed up this second part. The mathematical formalism 
based on the Lie algebra and the resonance driving terms is extended 
to the off-momentum regime and explicit analytic formulas for 
the evaluation of chromatic functions from lattice parameters 
are also derived. The robustness of these formulas, which are 
linear in the magnet strengths, is tested with different lattice 
configurations. 
\end{abstract}



\maketitle 

\section{Introduction and motivation}
Measurement and correction of focusing errors in circular 
accelerators is one of the top priorities in colliders and 
storage ring-based light sources to provide users with beam 
sizes and divergences as close as possible to the design 
values and to limit the possible detrimental effects on the 
beam lifetime caused by the integer and half-integer resonances. 
To this end, so many different techniques have been developed 
and successfully tested since decades that they already occupy 
entire chapters in textbooks~\cite{Zimmermann-book}. 
A more recent historical overview highlighting the great 
advancements on this domain can be found in~\cite{Ro-Review}. 

The ever increasing BPM resolution and computing power made 
the analysis and correction of linear optics (focusing error 
and betatron coupling) via measurements of the orbit response 
matrix (ORM) a routine task in basically all light sources 
worldwide~\cite{LOCO1,LOCO2}. Simulated ORM analysis is 
also carried out during the design phase of new 
storage-ring-based light sources for the evaluation of 
magnetic and mechanical tolerances~\cite{SimoneThesis}. 
Since comprehensive analytic formulas for 
its evaluation have not yet been found (they exist for 
the ideal case with no betatron coupling), the ORM response 
to a lattice error is computed numerically by optics codes 
evaluating at least one ORM for each source of error 
(typically quadrupole and dipole). Unless it is parallelized 
over several processor units, this computation becomes 
time-consuming in large rings and in new lattices design with 
even larger number of magnets. This paper aims at speeding 
up this computation by presenting and testing new analytic 
formulas for a rapid evaluation of the ORM response to 
linear lattice errors, with no need of orbit distortion 
computation. 

Another known drawback of the ORM analysis is its lengthy 
procedure for a single measured, which typically foresees 
a sequence of current changes in orbit correctors and the 
retrieval of the corresponding orbit data. In the old (1994-2018) ESRF 
storage ring, this phase takes about 10 minutes for a 
partial ORM (32 out of 192 steerers), or 1 hour for a 
complete one. In larger machines such as the Large 
Hadron Collider (LHC) of CERN the time needed to scan 
the entire magnetic cycle makes this approach unsuitable 
for operational purposes. However, a new approach making use 
of alternating-current steerers, fast BPM acquisition system 
(at 10 kHz) and harmonic analysis of orbit data was proved to 
obtain the same measurement with simultaneous magnet excitations at 
different frequencies, hence reducing dramatically the measurement 
time~\cite{AC-ORM,AC-ORM2}. Still, superconducting machines like the LHC 
may not benefit from this ploy. These experimental 
aspects are not discussed in this paper. 

The ORM is the main observable, though not the 
only ingredient for a complete analysis of linear magnetic 
errors. The latter do indeed modulate and generate 
dispersion in the horizontal and vertical 
planes, respectively. Analytic formulas establishing 
the correlations between lattice errors and linear 
dispersion are also inferred. The mathematical formalism 
developed for their derivation provides handy formulas 
for the computation of other chromatic functions, such as 
the chromatic beating (i.e. the dependence of the beta 
functions upon the energy deviation), chromatic coupling 
(i.e. how betatron coupling varies when particles go off 
energy) and the derivative of the dispersion function. These 
three quantities scale linearly with sextupole fields 
(normal and skew), providing a tool for the evaluation 
of the sextupolar model of a circular accelerators and for 
a fast correction of their deviations from design values.

The paper is structured as follows. The principles of the 
ORM analysis are presented in Sec.~\ref{ORM} for a mere 
sake of nomenclature. In Sec.~\ref{closed-orbit} a new 
expression for the closed-orbit condition in the presence 
of lattice errors including betatron coupling is 
reported. The analytic formulas for the 
evaluation of the ORM and linear dispersion 
from linear lattice errors are presented 
and discussed in Sec.~\ref{ORM-Formulas}, whereas 
Sec.~\ref{CHROM-Formulas} contains the expressions 
for the chromatic functions. Two schemes for the 
analysis of sextupolar errors based on the measurement 
of off-energy ORMs are eventually discussed in 
Sec.~\ref{ORMdelta}. All mathematical derivations are 
put in separated appendices: Appendix~\ref{app:1} for the 
ORM formulas and Appendix~\ref{app:2} for the chromatic 
functions, Appendix~\ref{app:3} for the corrections to 
the previous formulas accounting for the  variation 
along magnets of the optical parameters (thick-magnet 
corrections).

\section{Quick review of the linear optics from closed orbit (LOCO) }
\label{ORM}
After introducing an orbit distortion via horizontal 
and vertical deflections, represented by two vectors 
$\vec{\Theta}_x=(\Theta_{x,1},\ \Theta_{x,2},\ ..., \Theta_{x,N_S})^T$ and 
$\vec{\Theta}_y=(\Theta_{y,1},\ \Theta_{y,2},\ ..., \Theta_{y,N_S})^T$, 
where $T$ denotes the transpose and 
$N_s$ is the number of available magnets, the horizontal and vertical 
orbits recorded at $N_B$ BPMs $\vec{O}_x=(O_{x,1},\ O_{x,2},\ ...,
 O_{x,N_B})^T$ and $\vec{O}_y=(O_{y,1},\ O_{y,2},\ ..., O_{y,N_B})^T$ can 
be recorded and written as
\begin{eqnarray}\hskip-0.0cm
\left(\begin{array}{c}\vec{O}_{x}\\ \vec{O}_{y}\end{array}\right)&=&\mathbf{ORM}
\left(\begin{array}{c}\vec{\Theta}_{x}\\ \vec{\Theta}_{y}\end{array}\right),\quad 
\mathbf{ORM}=
\left(\begin{array}{c c}\mathbf{O^{(xx)}} &\mathbf{O^{(xy)}} \\
                        \mathbf{O^{(yx)}} &\mathbf{O^{(yy)}} \end{array}
\right),\nonumber \\ && \label{eq:ORM_01}
O^{(xx)}_{wj}=\frac{\partial O_{x,j}}{\partial \Theta_{x,w}}\ ,\quad
O^{(xy)}_{wj}=\frac{\partial O_{x,j}}{\partial \Theta_{y,w}}\ , \\ &&
O^{(yx)}_{wj}=\frac{\partial O_{y,j}}{\partial \Theta_{x,w}}\ ,\quad 
O^{(yy)}_{wj}=\frac{\partial O_{y,j}}{\partial \Theta_{y,w}}\ . \nonumber
\end{eqnarray}
Optics codes such as MADX~\cite{madx} or AT~\cite{AT} can easily compute 
$\mathbf{ORM}$ for the ideal (or initial model) lattice model and the 
difference between the measured and expected matrix may be written as 
\begin{eqnarray}
\mathbf{\delta ORM}&=&\mathbf{ORM^{(meas)}-ORM^{(mod)}}\ .\label{eq:ORM_02}
\end{eqnarray}
The dispersion function at the BPMs (both horizontal and 
vertical) is also measured and its deviation from the 
ideal model may be computed as 
\begin{eqnarray}
\vec{\delta D}_{x,y}=\vec{D}_{x,y}^{(meas)}-\vec{D}_{x,y}^{(mod)}\ .
\end{eqnarray}
Both $\mathbf{\delta  ORM}$ and $\vec{\delta D}_{x,y}$ depend linearly 
on the linear lattice errors (i.e. from bending and quadrupole magnets). 
By sorting the elements of each ORM block sequentially in a vector, the 
dependence reads
\begin{eqnarray}
\left(\begin{array}{c}\delta\vec{O}^{(xx)}\\ 
      \delta\vec{O}^{(yy)}\\\delta\vec{D}_x\end{array}
\right)&=&\mathbf{N}
\left(\begin{array}{c}\delta\vec{K}_{1}\\ \delta\vec{K}_{0}
\end{array}\right)\ , \label{eq:ORM_04}\\\label{eq:ORM_05}
\left(\begin{array}{c}\delta\vec{O}^{(xy)}\\ 
      \delta\vec{O}^{(yx)}\\ \delta\vec{D}_y\end{array}
\right)&=&\mathbf{S}
\left(\begin{array}{c}\vec{J}_1\\ \vec{J}_0
\end{array}\right)\ .
\end{eqnarray}
$\delta\vec{K}_{1}$ and $\delta\vec{K}_{0}$ are the vectors containing 
the quadrupole and dipole errors, respectively, whereas 
$\vec{J}_1$ and $\vec{J}_0$ denote the skew quadrupole fields 
and the vertical dipole strengths. The latter may be replaced 
in Eq.~\eqref{eq:ORM_05} by the corresponding tilt angles $\theta$, 
since 
\begin{eqnarray}
J_{1}=-K_{1}\sin{(2\theta^{(quad)})}\ ,\quad
J_{0}=-K_{0}\sin{(\theta^{(bend)})}\ .\quad
\end{eqnarray}
Throughout the paper, the MADX nomenclature for the multipolar expansion 
of magnetic fields is adopted,
\begin{equation}\label{eq:MADX}
-\Re\left[\sum_{n}{(K_{w,n-1}+iJ_{w,n-1})
            \frac{(x_w+iy_w)^n}{n!}}\right]\ ,
\end{equation}
with $K$ and $J$ referring to the integrated normal and 
skew magnetic strengths (normalized to the magnetic 
rigidity). Multipole coefficients in AT and MADX are 
defined differently and scaling factors depending on the 
multipole order need to be taken into account when converting 
them between the two codes. By pseudo-inverting 
the two systems of Eqs.~\eqref{eq:ORM_04}-\eqref{eq:ORM_05}, 
for instance via singular value 
decomposition (SVD), effective models that best fit 
the measured ORM and dispersion can be built. An 
unique model may not be extracted, since a trade-off 
between accuracy (i.e. large number of eigen-values 
in the decomposition) and reasonableness of the 
errors (i.e. low number of eigen-values to prevent 
numerical instabilities) shall be fixed on a subjective 
basis. Moreover, the systems of 
Eqs.~\eqref{eq:ORM_04}-\eqref{eq:ORM_05} ignore 
contributions from the feed-down effects of quadrupoles 
and sextupoles induced by their misalignments 
and/or off-axis orbit at their locations. The closed 
orbit distortion resulting from this modelling renders 
the analysis more complex without adding values to 
the physical observables (betatron phase and amplitude 
at the the BPMs) and are usually {\sl absorbed} by 
additional dipole errors (accounting for quadrupole 
misalignments) and quadrupole errors (representing 
the quadrupolar feed-down in sextupoles). In optics 
codes dipole errors induce a distortion of the 
reference orbit, though not of the closed one. 
Eqs.~\eqref{eq:ORM_04}-\eqref{eq:ORM_05} are the core 
of the {\em Linear Optics from Closed Orbit} (LOCO) 
analysis~\cite{LOCO1,LOCO2}. Additional fit parameters 
may be included in the r.h.s. of the two equations, 
such as calibration factors and rolls of steerers 
and BPMs. Once the errors ($\delta\vec{K}_{1},\ \delta\vec{K}_{0}$,  
$\vec{J}_1$ and $\vec{J}_0$) are included into the 
lattice model, the optical parameters (such as $\beta$,  
$\phi$, and $D$) can be computed by the optics 
codes and compared to the expected ones. 
Eqs.~\eqref{eq:ORM_04}-\eqref{eq:ORM_05} are usually 
modified by inserting weights and imposing fixed tunes 
to obtain an effective model.

The pseudo-inversion of Eqs.~\eqref{eq:ORM_04}-\eqref{eq:ORM_05}
is a quick task. However, the overall analysis 
is quite time consuming, since the responses  
$\mathbf{M}$ and $\mathbf{S}$ of the ORM on the lattice 
errors ($\delta\vec{K}$ and $\vec{J}$) is usually 
computed by simulating an ORM for each error: A 
heavy computation (a few minutes) already for the 
old ESRF storage ring with 256 quadrupoles and 64 
dipoles, which can only become more lengthy in 
larger machines and future light sources. If this 
computational time may still be tolerated when 
periodically correcting the linear lattice of an 
existing machine, it becomes the main computational 
overhead is simulation studies of new lattice designs, 
where tens of thousands of scans (including errors and 
corrections) are required to determine the best 
magnet arrangements and working point, as well as to 
specify (magnetic and mechanical) tolerances. Large 
computing farms came to the help of lattice designers 
in the last decade to reduce the time needed for such 
scan (and to increase the revenues of IT companies). 
The analytic formulas derived in this paper aim at 
further reducing the calculation time with no need of 
upgrading the computing farm.

\section{Closed orbit condition in the presence of lattice errors}
\label{closed-orbit}
Textbook formulas for the evaluation of the 
closed-orbit distortion induced by a dipolar 
perturbation are reported in 
Eqs.~\eqref{eq:CO_standard0}-~\eqref{eq:orm_classic}. 
Even though they still hold in the presence of 
focusing errors, provided that the modified 
Courant-Snyder (C-S) parameters are used, they 
do not account for betatron coupling, which 
transfers part of the orbit in one transverse 
plane in the other one. In the first part of 
Appendix~\ref{app:1} a condition including 
betatron coupling is derived. This requires 
an analysis in the complex domain and  
the introduction of some (complex) quantities. 
First, the complex C-S coordinates need to 
be introduced, $h_{z,\pm}=\tilde{z}\pm i\tilde{p}_z$, 
where $z$ stands for either $x$ or $y$ The 
orbit is retrieved from $h_{\pm}$ according 
to $z=\sqrt{\beta_z}\Re\{h_{z,\pm}\}$. In the 
decoupled complex C-S space, the linear one-turn 
map is represented by a diagonal matrix, 
$e^{i\mathbf{Q}}=\hbox{diag}(e^{2\pi iQ_x},e^{-2\pi iQ_x},
e^{2\pi iQ_y},e^{-2\pi iQ_y})$. The 
linear transport between two elements $w$ and $j$ 
is represented by another phase space rotation
$e^{i\mathbf{\Delta\phi}_{wj}}=\hbox{diag}(e^{i\Delta\phi_{x,wj}},
e^{-i\Delta\phi_{x,wj}},e^{i\Delta\phi_{y,wj}},e^{-i\Delta\phi_{y,wj}})$, 
where the phase advance between $\Delta\phi_{wj}$, 
must be a positive quantity. However, if it 
is computed from the ideal betatron phases 
$\phi_{j}$ and $\phi_{w}$ with a fixed origin, 
it becomes negative whenever the position $w$ 
is downstream $j$: In this case the tune (i.e. 
the total phase advance over one turn) needs 
to be added, namely
\begin{eqnarray}\label{eq:Text-deltaphisign}
\left\{\begin{aligned}
&\Delta\phi_{x,wj}=(\phi_{x,j}-\phi_{x,w})
                    \hspace{1.3cm}\ ,\ \hbox{if\ }\phi_{x,j}>\phi_{x,w}\\
&\Delta\phi_{x,wj}=(\phi_{x,j}-\phi_{x,w})+2\pi Q_x\ ,\ 
                    \hbox{if\ }\phi_{x,j}<\phi_{x,w}
\end{aligned}\right. \ . \qquad
\end{eqnarray}
See Eq.~\eqref{eq:deltaphisign} for more 
details. The effect at a generic 
position $j$ of focusing errors can 
be represented by two resonance driving 
terms (RDTs)~\cite{prstab_esr_coupling} , 
one for each plane: 
\begin{equation}\label{eq:Text-RDT-beat0}
\begin{aligned}
f_{2000,j} =-\frac{\sum\limits_{m=1}^M \beta_{m,x}^{(mod)}\delta K_{m,1}
            e^{2i\Delta\phi_{x,mj}^{(mod)}}}{1-e^{4\pi iQ_x^{(mod)}}}
            \ +O(\delta K_{1}^2)\\
f_{0020,j} =\frac{\sum\limits_{m=1}^M \beta_{m,y}^{(mod)}\delta K_{m,1}
            e^{2i\Delta\phi_{y,mj}^{(mod)}}}{1-e^{4\pi iQ_y^{(mod)}}}
            \ +O(\delta K_{1}^2)
\end{aligned}\quad ,
\end{equation}
where $\delta K_{1}$ denotes the quadrupolar 
errors, the sum extends over all sources of  
error, and the C-S parameters $\beta^{(mod)}$ and 
$\Delta\phi^{(mod)}$ refer to the ideal lattice, i.e. 
not including the above focusing errors. The 
remainder is proportional to $\delta K_{1}^2$.
Betatron coupling can also be described by 
two RDTs,
\begin{eqnarray}\label{eq:f1001}
\begin{aligned}
f_{1001,j}=\displaystyle\frac{\sum\limits_{m=1}^M  J_{m,1}\sqrt{\beta_{m,x}\beta_{m,y}} 
                    e^{i(\Delta\phi_{x,mj} - \Delta\phi_{y,mj})}}
               {4\left[1-e^{2\pi i(Q_x-Q_y)}\right]} \ +O(J_{1}^2)\\
f_{1010,j}=\displaystyle\frac{\sum\limits_{m=1}^M  J_{m,1}\sqrt{\beta_{m,x}\beta_{m,y}} 
                    e^{i(\Delta\phi_{x,mj} + \Delta\phi_{y,mj})}}
               {4\left[1-e^{2\pi i(Q_x+Q_y)}\right]} \ +O(J_{1}^2)
\end{aligned}\ \ ,\qquad
\end{eqnarray}
where $J_1$ is the skew quadrupole strength 
and the remainder scales with its square. 
The linear tunes in the above denominators 
shall be replaced by the eigen-tune if either 
resonance condition is approached. The C-S 
parameters  $\beta$ and $\Delta\phi$ refer in 
this case to the lattice with focusing errors 
already included in the model. A complex 
matrix $\mathbf{B}$ containing the above four 
RDTs can be constructed to describe the evolution 
of the complex C-S coordinate vector 
$\vec{h}=(h_{x,-},h_{x,+},h_{y,-},h_{y,+})^T$:
\begin{eqnarray}\label{eq:Text-Bmatrix}
\begin{aligned}
&\mathbf{B}_w\simeq\hskip-1mm \left(
\begin{array}{c c c c}
\hskip-1.5mm    1             & 4if_{2000,w}    &\ \ 2if_{1001,w}&\ \ 2if_{1010,w} \\
\hskip-1.5mm -4if_{2000,w}^*  &        1        &-2if_{1010,w}^* &-2if_{1001,w}^*  \\
\hskip-1.5mm\ \ 2if_{1001,w}^*&\ \  2if_{1010,w}&       1        &\ \  4if_{0020,w}\\
\hskip-1.5mm-2if_{1010,w}^*   &-2if_{1001,w}    &-4if_{0020,w}^* &       1
\end{array}
\hskip-1mm\right) , \\ & \\
&\mathbf{B}_j^{-1}\simeq\hskip-1mm \left(
\begin{array}{c c c c}
\hskip-1.5mm    1             & -4if_{2000,j}  &-2if_{1001,j}     &-2if_{1010,j}     \\
\hskip-1.5mm\ \ 4if_{2000,j}^*&        1       &\ \ 2if_{1010,j}^*&\ \ 2if_{1001,j}^*\\
\hskip-1.5mm-2if_{1001,j}^*   &-2if_{1010,j}   &       1          &-4if_{0020,j}     \\
\hskip-1.5mm\ \ 2if_{1010,j}^*&\ \ 2if_{1001,j}&\ \ 4if_{0020,j}^*&       1 
\end{array}
\hskip-1mm\right) ,
\end{aligned}
\end{eqnarray}
where the remainder in the above definitions 
is proportional to the square of the RDTs, whereas 
$w$ and $j$ refer to two generic positions along 
the ring. The two matrices at the same location 
are each other's inverse to first order in 
the RDTs.

The equation for closed orbit distortion induced 
by $W$ horizontal and vertical deflections $\Theta_w$ 
in the complex C-S 
coordinates then reads 
\begin{eqnarray}\label{eq:Text-co1B}
\vec{h}_j&=&\mathbf{B}_j^{-1}\sum_{w=1}^W\left\{
\frac{e^{i\mathbf{\Delta\phi}_{wj}}}{\mathbf{1}-e^{i\mathbf{Q}}}
\mathbf{B}_w\ \delta\vec{h}_w\right\}\ ,
\end{eqnarray}
where $\mathbf{1}$ is a $4\times4$ identity matrix, 
and $\delta\vec{h}_{w}=(-\sqrt{\beta_{w,x}}\Theta_{w,x},
\sqrt{\beta_{w,x}}\Theta_{w,x},-\sqrt{\beta_{w,y}}\Theta_{w,y},
\sqrt{\beta_{w,y}}\Theta_{w,y})^T$. Since 
$O_j=\sqrt{\beta_j}\Re\{h_j\}$, the ORM blocks 
of Eq.~\eqref{eq:ORM_01} eventually read 
\begin{eqnarray}\label{eq:Text-co2}
\begin{aligned}
&O^{(xx)}_{wj}=\sqrt{\beta_{j,x}\beta_{w,x}}\Re\left\{i
                  \mathbf{B}_j^{-1}
\frac{e^{i\mathbf{\Delta\phi}_{wj}}}{\mathbf{1}-e^{i\mathbf{Q}}}
\mathbf{B}_w\right\}^{(1,1\rightarrow2)}\ , \\
&O^{(xy)}_{wj}=\sqrt{\beta_{j,x}\beta_{w,y}}\Re\left\{i
                  \mathbf{B}_j^{-1}
\frac{e^{i\mathbf{\Delta\phi}_{wj}}}{\mathbf{1}-e^{i\mathbf{Q}}}
\mathbf{B}_w\right\}^{(1,3\rightarrow4)}\ , \\
&O^{(yx)}_{wj}=\sqrt{\beta_{j,y}\beta_{w,x}}\Re\left\{i
                  \mathbf{B}_j^{-1}
\frac{e^{i\mathbf{\Delta\phi}_{wj}}}{\mathbf{1}-e^{i\mathbf{Q}}}
\mathbf{B}_w\right\}^{(3,1\rightarrow2)}\ , \\
&O^{(yy)}_{wj}=\sqrt{\beta_{j,y}\beta_{w,y}}\Re\left\{i
                  \mathbf{B}_j^{-1}
\frac{e^{i\mathbf{\Delta\phi}_{wj}}}{\mathbf{1}-e^{i\mathbf{Q}}}
\mathbf{B}_w\right\}^{(3,3\rightarrow4)}\ .
\end{aligned}\qquad
\end{eqnarray}
In the above notation, given a $4\times4$ matrix 
$\mathbf{A}$, $\mathbf{A}^{(a,b\rightarrow c)}=A_{ac}-A_{ab}$. 

If focusing errors are included in the 
model, $f_{2000}=f_{0020}=0$ anywhere along the ring 
and the more explicit expressions for the four 
ORM blocks of Eq.~\eqref{eq:co3} can be derived.

\section{Analytic formulas for the evaluation of ORM and 
         linear dispersion from lattice parameters}
\label{ORM-Formulas}
Equation~\eqref{eq:Text-co2} is further expanded 
in Appendix~\ref{app:1} to derive the ORM 
response to a focusing error and to a skew 
quadrupole field, i.e. to infer the betatronic 
blocks of the matrices $\mathbf{N}$ and $\mathbf{S}$ 
of Eqs.~\eqref{eq:ORM_04}-\eqref{eq:ORM_05}. As far 
as the former is concerned, the expressions truncated 
to first order in $\delta K_1$ for the two diagonal 
blocks read
\begin{widetext}
\begin{eqnarray}\nonumber 
N^{(xx)}_{wj,m}&\simeq&-
  \frac{\sqrt{\beta_{j,x}^{(mod)}\beta_{w,x}^{(mod)}}\beta_{m,x}^{(mod)}}
       {2\sin{(\pi Q_x^{(mod)})}}
 \Bigg\{\frac{\cos{(\tau_{x,wj}^{(mod)})}}
              {4\sin{(2\pi Q_x^{(mod)})}}
              \Big[\cos{(2\tau_{x,mj}^{(mod)})}
                  +\cos{(2\tau_{x,mw}^{(mod)})}\Big]
              \\ \nonumber &&\hspace{3.6cm}
        +\frac{\ \sin{(\tau_{x,wj}^{(mod)})}}
              {4\sin{(2\pi Q_x^{(mod)})}}
             \Big[\sin{(2\tau_{x,mj}^{(mod)})}
               -\ \sin{(2\tau_{x,mw}^{(mod)})}\Big]
              \\ \nonumber &&\hspace{3.6cm}
       +\frac{1}{2}\sin{(\tau_{x,wj}^{(mod)})}
        \left[\Pi(m,j)-\Pi(m,w)+\Pi(j,w)\right]
       +\frac{\cos{(\Delta\phi_{x,wj}^{(mod)})}}
            {4\sin{(\pi Q_x^{(mod)})}}\Bigg\}\ ,
\\ &&\hspace{0.4cm} \label{eq:Nmatrix1} \\ \nonumber
N^{(yy)}_{wj,m}&\simeq&+
  \frac{\sqrt{\beta_{j,y}^{(mod)}\beta_{w,y}^{(mod)}}\beta_{m,y}^{(mod)}}
       {2\sin{(\pi Q_y^{(mod)})}}
 \Bigg\{\frac{\cos{(\tau_{y,wj}^{(mod)})}}
              {4\sin{(2\pi Q_y^{(mod)})}}
              \Big[\cos{(2\tau_{y,mj}^{(mod)})}
                  +\cos{(2\tau_{y,mw}^{(mod)})}\Big]
              \\ \nonumber &&\hspace{3.6cm}
        +\frac{\ \sin{(\tau_{y,wj}^{(mod)})}}
              {4\sin{(2\pi Q_y^{(mod)})}}
             \Big[\sin{(2\tau_{y,mj}^{(mod)})}
               -\ \sin{(2\tau_{y,mw}^{(mod)})}\Big]
              \\ \nonumber &&\hspace{3.6cm}
       +\frac{1}{2}\sin{(\tau_{y,wj}^{(mod)})}
        \left[\Pi(m,j)-\Pi(m,w)+\Pi(j,w)\right]
       +\frac{\cos{(\Delta\phi_{y,wj}^{(mod)})}}
            {4\sin{(\pi Q_y^{(mod)})}}\Bigg\}\ , 
\end{eqnarray}
\end{widetext}
where the function $\Pi$ is defined as 
\begin{eqnarray}\label{eq:def_Pi}
\Pi(a,b)=1\quad\hbox{if\ }s_a<s_b\ ,\quad
\Pi(a,b)=0\quad\hbox{if\ }s_a\ge s_b\ .\qquad
\end{eqnarray}
The quantity $\tau_{ab}$ is a mere shifted phase advance 
between two locations $a$ and $b$,  
\begin{eqnarray}\label{eq:def_tau}
\tau_{z,ab}=\Delta\phi_{z,ab}-\pi Q_z\ ,\quad z=x,y\quad ,
\end{eqnarray}
where the phase advance $\Delta\phi_{wj}$ 
is evaluated as usual according to 
Eq.~\eqref{eq:Text-deltaphisign}.

\begin{figure}[!t]
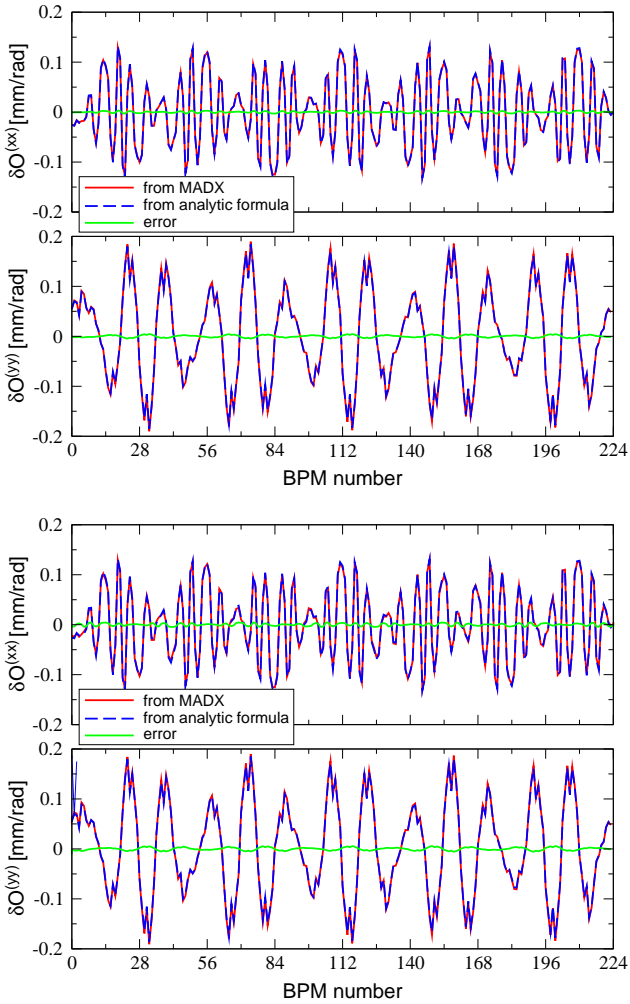

\rule{0mm}{0mm}
\centerline{\includegraphics[width=8.3cm]{ORM_Nxxyy_test1.eps}}\\
\rule{0mm}{1mm}
\centerline{\includegraphics[width=8.3cm]{ORM_Nxxyy_test2.eps}}
  \caption{\label{fig_Oxxyy1} (Color) $\delta\vec{O}^{(xx)}$ 
    and $\delta\vec{O}^{(yy)}$ induced by a steerer (horizontal 
    and vertical, respectively) in the presence of a single quadrupole 
    error inducing an rms beta beating of $6.9\%$ and $2.2\%$ in the 
    two planes (top 2 plots) and with an additional source of betatron 
    coupling generating an emittance ratio $\mathscr{E}_y/\mathscr{E}_x\simeq1\%$ 
    (bottom 2 plots). The red curves result from the computation of 
    the orbit distortion by MADX, whereas the blue dashed lines are 
    derived from Eqs.~\eqref{eq:Nmatrix1} and \eqref{eq:ORM-formula1}
    which do not require any orbit calculation. The agreement 
    between the two evaluations is within $3\%$ rms. The lattice 
    of the old ESRF storage ring has been used.}
\rule{0mm}{3mm}
\end{figure}
\begin{figure}
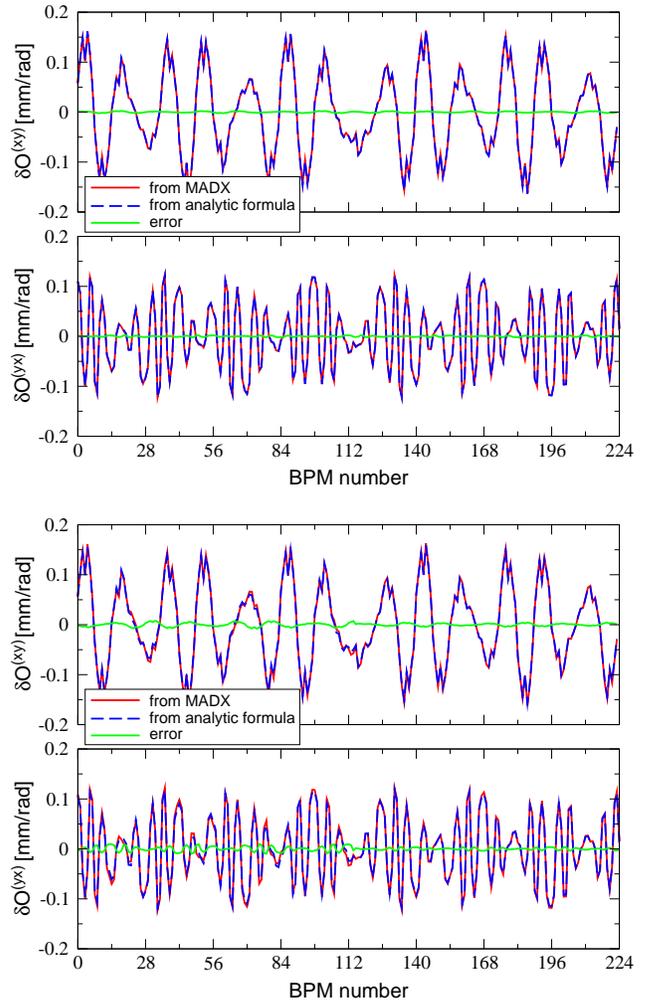

\rule{0mm}{0mm}
\centerline{\includegraphics[width=8.3cm]{ORM_Sxyyx_test1.eps}}\\
\rule{0mm}{1mm}
\centerline{\includegraphics[width=8.3cm]{ORM_Sxyyx_test2.eps}}
  \caption{\label{fig_Oxyyx1} (Color) $\delta\vec{O}^{(xy)}$ 
    and $\delta\vec{O}^{(yx)}$ induced by a steerer (vertical 
    and horizontal, respectively) in the presence of an rms beta beating 
    of $6.9\%$ and $2.2\%$ in the two planes and a single skew quadrupole 
    error generating an emittance ratio $\mathscr{E}_y/\mathscr{E}_x\simeq1\%$. 
    The red curves result from the computation of 
    the orbit distortion by MADX, whereas the blue dashed lines are 
    derived from Eqs.~\eqref{eq:ORM-formula2}-~\eqref{eq:Smatrix1}
    which do not require any orbit calculation. The agreement 
    between the two evaluations is within $2\%$ rms if the C-S 
    parameters including focusing errors are used in 
    Eq.~\eqref{eq:Smatrix1} (top 2 plots), whereas it increases up to
    $7\%$ rms if the ideal lattice parameters (of the old ESRF storage ring 
    in this case) are used (bottom 2 plots).}
\rule{0mm}{3mm}
\end{figure}

Equation~\eqref{eq:Nmatrix1} describes the response of a
$\mathbf{\delta}${\bf ORM} diagonal block element 
$wj$ (where $w$ refers to the steerer and $j$ 
to the BPM) to a quadrupole error $m$, namely
\begin{eqnarray}
N^{(xx)}_{wj,m}=\frac{\delta O^{(xx)}_{wj}}{\delta K_{m,1}}
\quad ,\quad
N^{(yy)}_{wj,m}=\frac{\delta O^{(yy)}_{wj}}{\delta K_{m,1}}
\quad . 
\end{eqnarray}
The deviation of the ORM diagonal blocks from the 
ideal values are then computed according to
\begin{eqnarray} 
\delta O^{(xx)}_{wj}\simeq\sum_{m=1}^M  N^{(xx)}_{wj,m}\delta K_{m,1}\ ,\ 
\delta O^{(yy)}_{wj}\simeq\sum_{m=1}^M  N^{(yy)}_{wj,m}\delta K_{m,1}\ .
\nonumber \\\label{eq:ORM-formula1} 
\end{eqnarray}
Note that all C-S parameters $\beta^{(mod)}$ and 
$\Delta\phi_{wj}^{(mod)}$ refer to the ideal or 
initial lattice model, implying that the responses 
$N^{(xx)}_{wj,m}$ and $N^{(yy)}_{wj,m}$ can be 
computed post-processing a single output file 
or table from any optics code, with no need of 
launching it to compute the ORM for each 
quadrupolar error. The phase advance $\Delta\phi_{wj}$ 
is evaluated according to Eq.~\eqref{eq:Text-deltaphisign}.

In Fig.~\ref{fig_Oxxyy1} two examples are reported showing 
the deviation of one column of the ORM diagonal blocks from the 
ideal model, $\delta\vec{O}^{(xx)}$ and $\delta\vec{O}^{(yy)}$, 
at the 224 BPMs of the old ESRF storage ring. The two blocks 
are computed from the direct evaluation by MADX of the 
orbit distortion induced by two steerers in the presence 
of an error in one thick quadrupole, 
$\delta K_1=5\times10^{-3}$ m$^{-1}$ (red curves), as 
well as from Eqs.~\eqref{eq:Nmatrix1} and 
\eqref{eq:ORM-formula1} (blue dashed curves). In the former case, 
two complete ORMs need to be computed (with and without the 
quadrupole error), whereas the two formulas require a single 
evaluation of the ideal C-S parameters and a few lines of 
post-processing code: a computation by far much faster than 
the direct calculation of the ORM. The rms error of the 
ORM blocks computed via Eqs.~\eqref{eq:Nmatrix1} and
\eqref{eq:ORM-formula1} with respect to the direct 
computation of the matrices is within $3\%$. The 
quadrupole error induces an rms beta beating of $6.9\%$ 
and $2.2\%$ in the two planes (top 2 plots). The addition 
of a skew quadrupole inducing a ratio between the two 
transverse equilibrium emittances 
$\mathscr{E}_y/\mathscr{E}_x\simeq1\%$ (bottom 2 plots of 
Fig.~\ref{fig_Oxxyy1}) does not deteriorate the level of 
accuracy.

The expressions in Eq.~\eqref{eq:Nmatrix1} have been 
derived assuming a constant value of the beta 
function ($\beta_m$) across a generic quadrupole 
$m$, usually computed at its center. 
The phase advance $\Delta\phi_{mj}$ between 
the magnet and a generic location $j$ refers to 
its center too. This approximation may not be sufficiently 
accurate in general and in particular for lattices 
comprising combined-function 
magnets, along which the beta function varies 
considerably. In Appendix~\ref{app:3}
corrections accounting for that variation are derived 
assuming hard-edged quadrupoles (i.e. ignoring fringe 
fields). The terms to be replaced in Eq.~\eqref{eq:Nmatrix1}
are
\begin{eqnarray}
\begin{aligned}
\beta_{m}\hskip.5mm &\longrightarrow\hskip.5mm I_{\beta,m}\\
\beta_{m}\sin{(2\tau_{mj})}\hskip.5mm &\longrightarrow\hskip.5mm
         I_{S,mj}\ \ , \quad 
\beta_{m}\cos{(2\tau_{mj})}\hskip.5mm \longrightarrow\hskip.5mm
         I_{C,mj}\ \ , \\
\beta_{m}\sin{(2\tau_{mw})}\hskip.5mm &\longrightarrow\hskip.5mm 
         I_{S,mw}\ , \quad 
\beta_{m}\cos{(2\tau_{mw})}\hskip.5mm \longrightarrow\hskip.5mm
         I_{C,mw}\ ,
\end{aligned} \nonumber \\ 
\end{eqnarray}
where $I_{\beta,m}$, $I_{C,m}$ and $I_{S,m}$ 
are computed from the quadrupole coefficients (length 
and non-integrated strength) and C-S parameters at the 
magnet entrance ($s_{m}$) according to 
Eqs.~\eqref{eq:Ib2}-\eqref{eq:IS3}. 
An even more general case is considered where the quadrupolar 
field error is sought in other type of magnets (such as steerers and 
nonlinear elements): The corresponding expressions for the above 
integrals are given in Eqs.~\eqref{eq:Ibeta_drift}-\eqref{eq:Ic_drift}. 
As far as the old ESRF 
storage ring is concerned, which does not include 
combined-function magnets, the above corrections reduce 
the rms error of Eqs.~\eqref{eq:Nmatrix1} and
\eqref{eq:ORM-formula1} by about a factor 2.
Steerers $w$ are also assumed to be of zero length 
in Eq.~\eqref{eq:Nmatrix1}. A further generalization 
accounting for thick deflectors is also presented 
in Sec.~\ref{app:thicksteerers} at the end of 
Appendix~\ref{app:3}.

The response of a $\mathbf{\delta}${\bf ORM} 
off-diagonal block element $wj$ to a 
skew quadrupole $m$ can be written as 
\begin{eqnarray}
S^{(xy)}_{wj,m}=\frac{\partial O^{(xy)}_{wj}}{\partial J_{m,1}}
\quad ,\quad
S^{(yx)}_{wj,m}=\frac{\partial O^{(yx)}_{wj}}{\partial J_{m,1}}
\quad .
\end{eqnarray}
Assuming an uncoupled ideal (or initial) lattice model,  
the deviation of the ORM off-diagonal blocks from the 
ideal values (which are zeros) corresponds to the block 
themselves and can be evaluated according to
\begin{eqnarray} 
\delta O^{(xy)}_{wj}\simeq\sum_{m=1}^M S^{(xy)}_{wj,m}J_{m,1}\ ,\ 
\delta O^{(yx)}_{wj}\simeq\sum_{m=1}^M S^{(yx)}_{wj,m}J_{m,1}\ ,
\nonumber \\\label{eq:ORM-formula2} 
\end{eqnarray}
where the remainders scales with $J_1^2$ and 
the matrix elements $S^{(xy)}_{wj,m}$ and 
$S^{(yx)}_{wj,m}$ read
\begin{widetext}
\begin{eqnarray}\nonumber 
S^{(xy)}_{wj,m}&\simeq&
      \frac{1}{8}\sqrt{\beta_{j,x}\beta_{w,y}\beta_{m,x}\beta_{m,y}}
      \left\{\frac{1}{\sin{[\pi(Q_x-Q_y)]}}\left[
             \frac{\cos{(\tau_{x,mj}-\tau_{y,mj}+\tau_{y,wj})}}{\sin{\pi Q_y}}
            -\frac{\cos{(\tau_{x,mw}-\tau_{y,mw}+\tau_{x,wj})}}{\sin{\pi Q_x}}
      \right]\right.\nonumber \\ 
     &&\hskip 3.25cm\left.
            +\frac{1}{\sin{[\pi(Q_x+Q_y)]}}\left[
             \frac{\cos{(\tau_{x,mj}+\tau_{y,mj}-\tau_{y,wj})}}{\sin{\pi Q_y}}
            +\frac{\cos{(\tau_{x,mw}+\tau_{y,mw}+\tau_{x,wj})}}{\sin{\pi Q_x}}
      \right]\right\}\ ,
\nonumber\\ &&\hspace{0.4cm} \label{eq:Smatrix1} \\ \nonumber
S^{(yx)}_{wj,m}&\simeq&
      \frac{1}{8}\sqrt{\beta_{j,y}\beta_{w,x}\beta_{m,x}\beta_{m,y}}
      \left\{\frac{1}{\sin{[\pi(Q_x-Q_y)]}}\left[
            -\frac{\cos{(\tau_{x,mj}-\tau_{y,mj}-\tau_{x,wj})}}{\sin{\pi Q_x}}
            +\frac{\cos{(\tau_{x,mw}-\tau_{y,mw}-\tau_{y,wj})}}{\sin{\pi Q_y}}
      \right]\right.\nonumber \\ 
     &&\hskip 3.25cm\left.
            +\frac{1}{\sin{[\pi(Q_x+Q_y)]}}\left[
             \frac{\cos{(\tau_{x,mj}+\tau_{y,mj}-\tau_{x,wj})}}{\sin{\pi Q_x}}
            +\frac{\cos{(\tau_{x,mw}+\tau_{y,mw}+\tau_{y,wj})}}{\sin{\pi Q_y}}
      \right]\right\}\ .\nonumber
\end{eqnarray}
\end{widetext}
Note that all C-S parameters $\beta$ and 
$\Delta\phi_{wj}$ refer this time to the lattice 
model including the focusing errors,
quadrupolar error. This requires that the 
analysis of Eq.~\eqref{eq:ORM_04} is 
carried out before matching the measured 
ORM off-diagonal blocks of Eq.~\eqref{eq:ORM_05}. 
If the ideal (or initial) C-S parameters are 
used, the accuracy of Eq.~\eqref{eq:Smatrix1} 
is deteriorated. 

As expected, when either tune approaches the integer 
or half-integer resonance, both 
Eqs.~\eqref{eq:Nmatrix1}-\eqref{eq:Smatrix1}
diverge. In the presence of betatron coupling the 
same is true for the off-diagonal ORM blocks 
$\delta\vec{O}^{(xy)}$ and $\delta\vec{O}^{(yx)}$ 
of Eq.~\eqref{eq:Smatrix1}, when the sum resonance 
is approached, i.e. $Q_x+Q_y\simeq N$, where N in an 
integer. On the other hand, the denominators 
dependent on $(Q_x-Q_y)$ do not diverge when 
the difference resonance is approached, since the 
eigen-tunes remain separated by $\Delta Q_{min}$.

In Fig.~\ref{fig_Oxyyx1} an example of deviation of 
one column from the ORM diagonal off-blocks $\delta\vec{O}^{(xy)}$ 
and $\delta\vec{O}^{(yx)}$ evaluated by MADX and 
Eqs.~\eqref{eq:ORM-formula2}-\eqref{eq:Smatrix1}
is displayed, along with the errors of the 
analytic formulas. When the C-S parameters including 
focusing errors are used in Eq.~\eqref{eq:Smatrix1} 
the relative rms error is 
of about $2\%$ (top 2 plots), whereas it increases to 
$7\%$ if the ideal C-S parameters are used. This 
confirms the need of evaluating the focusing error 
model (from the ORM diagonal blocks) before fitting  
the off-diagonal blocks. 

If the variation of $\beta$ and $\tau$ along 
the skew (or tilted) quadrupole $m$ is to be taken 
into account, the same procedure described in  
Appendix~\ref{app:3} can be followed. The cosine 
terms of Eq.~\eqref{eq:Smatrix1} 
can be manipulated so to factorize the ones 
dependent on the magnet $m$ only, and replace 
them with their integrals, namely 
\begin{eqnarray}
\begin{aligned}
\sqrt{\beta_{m}}\sin{\tau_{mj}}&\longrightarrow&J_{S,mj}\ ,  \\
\sqrt{\beta_{m}}\cos{\tau_{mj}}&\longrightarrow&J_{C,mj}\ .
\end{aligned}\label{eq:Smatrix1-thick}
\end{eqnarray}
These integrals can be computed analytically 
via Eq.~\eqref{eq:JCmj2}-\eqref{eq:JCmj3}
and inserted in Eq.~\eqref{eq:Smatrix1}. 
If the variation of the C-S parameters across the 
steerer $w$ is to be taken into account, the same 
procedure carried out in Sec.~\ref{app:thicksteerers} 
at the end of Appendix~\ref{app:3} can be followed 
(not reported here).

\begin{figure}
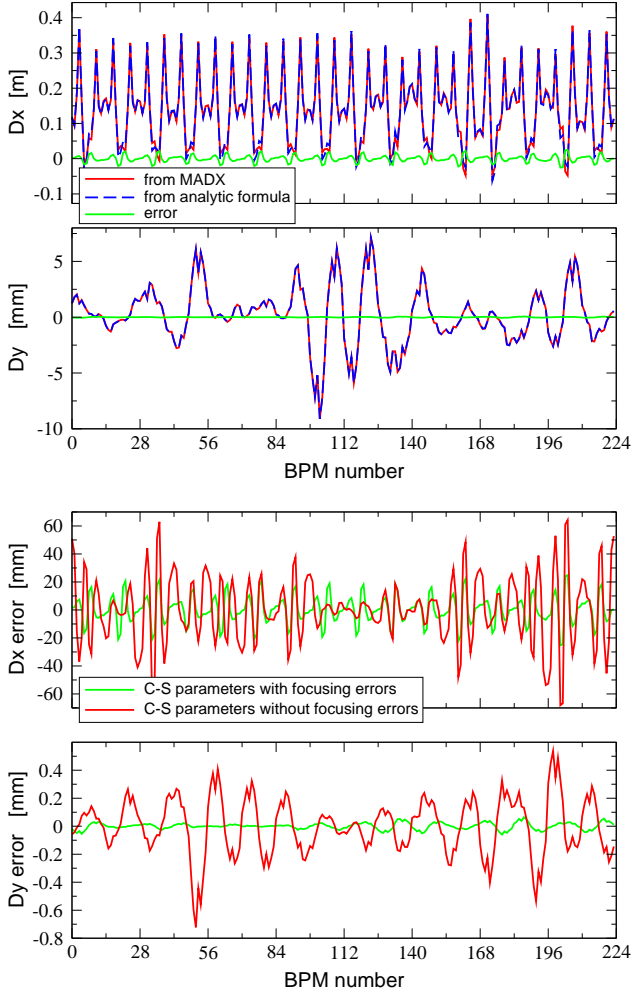

\rule{0mm}{0mm}
\centerline{\includegraphics[width=8.3cm]{Linear_Disp_formulas_1.eps}}\\
\rule{0mm}{1.5mm}
\centerline{\includegraphics[width=8.3cm]{Linear_Disp_formulas_2.eps}}
  \caption{\label{fig_Disp1} (Color) Top 2 plots: Dispersion function computed
           by MADX (PTC module) and by Eq.~\eqref{eq:Text_Disp1} for the 
           lattice of the old ESRF storage ring comprising a model of 
           focusing errors and betatron coupling inferred from an ORM 
           measurement. The bottom two plots show the error of 
           Eq.~\eqref{eq:Text_Disp1} with different lattice parameters:
           from the model including the focusing errors (black), from 
           the ideal model (red), and from the model with edge focusing 
           from dipoles on top of the previous set of focusing errors (green). 
           In the first and latter cases the relative error is within 
           $10\%$ and $1.5\%$ for $D_x$ and $D_y$, respectively.}
\rule{0mm}{3mm}
\end{figure}

In Appendix~\ref{app:2} an analytic expression 
to evaluate the linear dispersion at a generic 
location $j$ in the presence of betatron coupling 
is derived:
\begin{eqnarray}
\begin{aligned}
D_x(j)&\simeq+\frac{\sqrt{\beta_{j,x}}}{2\sin{(\pi Q_x)}}\sum_{m=1}^M 
\left(K_{m,0}+J_{m,1}D_{m,y}\right)\sqrt{\beta_{m,x}} \times \\
 &\hspace{3.2cm}\cos(\tau_{x,mj})\ , \\
D_y(j)&\simeq-\frac{\sqrt{\beta_{j,y}}}{2\sin{(\pi Q_y)}}\sum_{m=1}^M 
\left(J_{m,0}-J_{m,1}D_{m,x}\right)\sqrt{\beta_{m,y}} \times \\
 &\hspace{3.2cm}\cos(\tau_{y,mj})\ ,
\end{aligned} \label{eq:Text_Disp1}
\end{eqnarray}
where $\tau$ is the same shifted phase advance of 
Eq.~\eqref{eq:def_tau} and the dispersion function at the 
magnets $D_m$ refers to the uncoupled lattice, i.e. that 
generated by horizontal and vertical bending 
magnets ($K_0$ and $J_0$ for $D_{m,x}$ and $D_{m,y}$, 
respectively). $D_m$ shall then be computed from 
the above equations putting $J_{m,1}=0$ and 
then inserted in the complete formulas to 
obtain the final dispersion $D(j)$. 
Equation~\eqref{eq:Text_Disp1} indeed describes the 
entanglement between the horizontal and vertical 
dispersion functions due to skew quadrupole fields. 
In the presence of focusing errors the above 
equations are still valid, provided that the 
corresponding C-S parameters $\beta$, $\phi$ and 
dispersion are used. If the ideal lattice parameters 
are inserted, a larger error is to be expected. In 
Fig.~\ref{fig_Disp1} an example is shown with 
the dispersion function computed by MADX (PTC 
module) and by Eq.~\eqref{eq:Text_Disp1} for the lattice 
of the old ESRF storage ring comprising a model of focusing 
errors and betatron coupling inferred from an ORM 
measurement. If the C-S parameters including 
focusing errors are used, the rms error is within 
$10\%$ and $1\%$ for $D_x$ and $D_y$, respectively, 
whereas it increases to $25\%$ and $10\%$ if the 
ideal C-S parameters are used in Eq.~\eqref{eq:Text_Disp1}. 

\begin{figure}
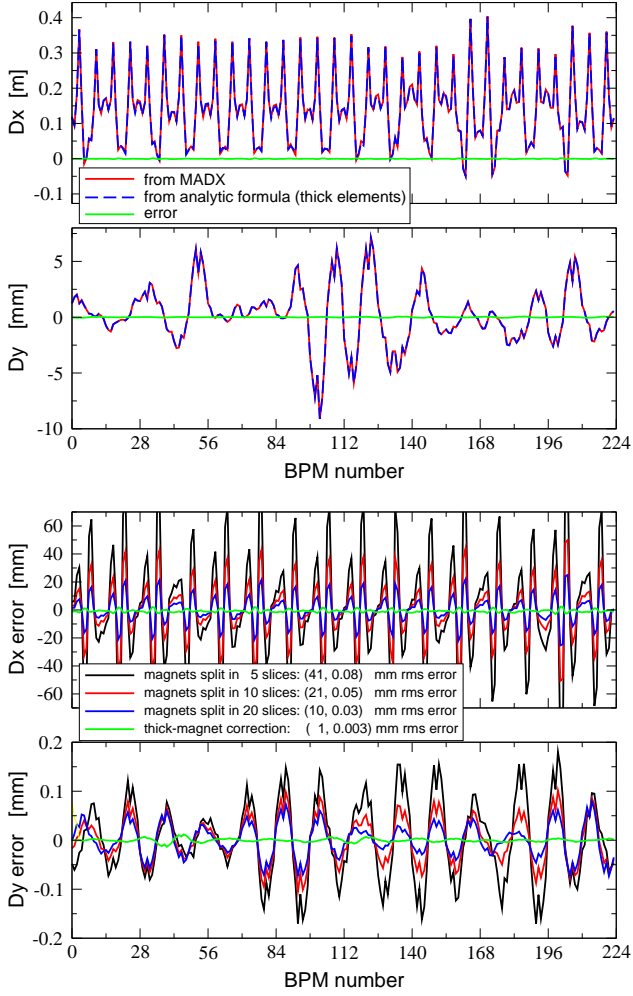

\rule{0mm}{0mm}
\centerline{\includegraphics[width=8.3cm]{Linear_Disp_formulas_1_tck.eps}}\\
\rule{0mm}{1.5mm}
\centerline{\includegraphics[width=8.3cm]{Linear_Disp_formulas_2_tck.eps}}
  \caption{\label{fig_Disp2} (Color) Top 2 plots: The same dispersion 
           function of Fig.~\ref{fig_Disp1} computed by MADX (PTC module)
           and by Eq.~\eqref{eq:Text_Disp1} with the thick-magnet 
           correction of Eq.~\eqref{eq:text_disp_tck}. The bottom two plots 
           show the dependence of the error on the number of magnet slices
           for Eq.~\eqref{eq:Text_Disp1} compared to (much lower) 
           discrepancy of Eq.~\eqref{eq:text_disp_tck}. In the legend, rms 
           errors are given for $D_x$ and $D_y$, respectively.}
\rule{0mm}{3mm}
\end{figure}

In Eq.~\eqref{eq:Text_Disp1} constant C-S parameters 
and dispersion across the magnet $m$ are assumed. 
In order to account for their variation, the latter 
can be divided in several sub-elements to better 
retrieve the correct profile of those functions 
(Fig.~\ref{fig_Disp1} is obtained after slicing 
the magnets in twenty elements). 
Once again, analytic expressions exist to overcome 
this inconvenience and are derived in  
Appendix~\ref{app:3}. 
The terms to be replaced in Eq.~\eqref{eq:Text_Disp1} 
are
\begin{eqnarray}
\begin{aligned}
\sqrt{\beta_{m}}\cos{(\tau_{mj})}&
         \hskip.5mm\longrightarrow\hskip.5mm J_{C,mj}\ , \\
\sqrt{\beta_{m,x}}D_{m,y}\cos{(\tau_{x,mj})}&
         \hskip.5mm\longrightarrow\hskip.5mm J_{C,mj}^{(D_y)}\ , \\
\sqrt{\beta_{m,y}}D_{m,x}\cos{(\tau_{y,mj})}&
         \hskip.5mm\longrightarrow\hskip.5mm J_{C,mj}^{(D_x)}\ , 
\end{aligned}\qquad \label{eq:text_disp_tck}
\end{eqnarray}
where $J_{C,mj}$ is computed via 
Eqs.~\eqref{eq:JCmj0}-\eqref{eq:JCmj1} for pure 
(sector) bending magnets and via 
Eqs.~\eqref{eq:JCmj2}-\eqref{eq:JCmj3} for 
combined-functions magnets.  $J_{C,mj}^{(D_y)}$ 
and $J_{C,mj}^{(D_x)}$ depend instead on the (skew or tilted) 
quadrupole parameters and can be evaluated from 
Eq.~\eqref{eq:JCDxy}.
As shown by Eqs.~\eqref{eq:JCmj0}-\eqref{eq:JCmj1} 
of Appendix~\ref{app:3}, $J_{C,mj}$ exhibits  
a dependence on the bending angle $K_{m,0}$, which 
can be ignored as long as $K_{m,0}\ll 1$, i.e. for 
large rings. On the other hand, in small rings 
with strong bending angles, the dependence of the 
dispersion function on $K_{m,0}$ becomes nonlinear. 
The effectiveness of 
the thick-magnet correction of Eq.~\eqref{eq:text_disp_tck} 
can be appreciated in Fig.~\ref{fig_Disp2}: In 
this example, the rms error turns out to be one 
order of magnitude lower than the one obtained by 
using Eq.~\eqref{eq:Text_Disp1} after slicing all 
magnets in twenty parts. 

From Eq.~\eqref{eq:Text_Disp1} the $D_x$
response to a dipole error $\delta K_0$, and the 
one of $D_y$ on vertical dipole fields $J_0$ and 
skew quadrupole strength $J_1$ can be easily 
inferred. 
\begin{eqnarray}
\begin{aligned}
N_{j,m}^{(\delta K_0\rightarrow D_x)}&\simeq
       +\frac{\sqrt{\beta_{j,x}}}{2\sin{(\pi Q_x)}}
        \left\{\sqrt{\beta_{m,x}}\cos(\tau_{x,mj})\right\}\ , \\
S_{j,m}^{(J_0\rightarrow D_y)}&\simeq
       -\frac{\sqrt{\beta_{j,y}}}{2\sin{(\pi Q_y)}}
        \left\{\sqrt{\beta_{m,y}}\cos(\tau_{y,mj})\right\}\ , \\
S_{j,m}^{(J_1\rightarrow D_y)}&\simeq
       +\frac{\sqrt{\beta_{j,y}}}{2\sin{(\pi Q_y)}}
        \left\{\sqrt{\beta_{m,y}}D_{m,x}\cos(\tau_{y,mj})\right\}\ .
\end{aligned} \label{eq:Text_Disp1B}
\end{eqnarray}
If needed, the terms in the above curly brackets 
can be replaced by the thick-magnet corrections 
of Eq.~\eqref{eq:text_disp_tck}. The dispersive 
parts of Eqs.~\eqref{eq:ORM_04}-\eqref{eq:ORM_05}
then read
\begin{eqnarray}
\begin{aligned}
\delta D_{j,x}&=\sum_{m=1}^M N_{j,m}^{(\delta K_0\rightarrow D_x)}\delta K_{m,0}\ ,\\
\delta D_{j,y}&=\sum_{m=1}^M \left(S_{j,m}^{(J_0\rightarrow D_y)}J_{m,0}+
                         S_{j,m}^{(J_1\rightarrow D_y)}J_{m,1}\right)\ .
\end{aligned} \label{eq:Text_Disp1C}
\end{eqnarray}
The contribution to $\delta D_x$ stemming from 
the product $J_{m,1}D_{m,y}$ in 
Eq.~\eqref{eq:Text_Disp1} has been ignored 
as it is of a perturbation of second 
order. 

In conclusion, Eqs.~\eqref{eq:Nmatrix1}, \eqref{eq:Smatrix1} 
and \eqref{eq:Text_Disp1B} provide explicit 
expressions to evaluate the ORM and dispersion 
response matrices $\mathbf{N}$ and $\mathbf{S}$ of 
Eqs.~\eqref{eq:ORM_04}-\eqref{eq:ORM_05} from 
lattice parameters with no need of evaluating 
any ORM.

The impact of sextupoles in the measurement of the 
ORM is discussed in Sec.~\ref{sec:app_sext} of 
Appendix~\ref{app:1}. The orbit distortion induced 
by steerer magnets generates normal and skew 
quadrupole feed-down fields, 
$\delta K_{1}=-K_{2}x_{\rm c.o.}$ and 
$J_{1}=K_{2}y_{\rm c.o.}$, where $K_{2}$ 
denotes the integrated strength of a generic sextupole 
and $(x_{\rm c.o.},\ y_{\rm c.o.})$ is the 
corresponding closed orbit. Dipolar feed-down 
fields proportional to $(x_{\rm c.o.}^2,\ y_{\rm c.o.}^2)$ 
are also generated. It is demonstrated that if 
the ORM is measured via 
a double symmetric distortion $\pm\theta_w$, 
the quadrupolar feed-down generated by sextupoles 
is canceled out, leaving a residual error proportional 
to $\theta_w^2$ (a few permil rms for the old ESRF storage ring).

\section{Analytic formulas for the phase advance shifts
                 induced by quadrupole errors}
\label{PhAdShift-Formulas}
An alternative to the ORM for the linear analysis 
of lattice errors is the measurement and fit 
of the BPM phase advances obtained from turn-by-turn 
data. In Ref.~\cite{Andrea-Linear-arxiv}
analytic formulas relating the actual 
betatron phase advance to the ideal one (from the 
model), detuning terms and the RDTs were derived. 
In Appendix~\ref{app:2} those formulas have been 
further manipulated so to make the dependence of 
the phase advance on the quadrupole errors 
$\delta K_{1}$ explicit, yielding
\begin{eqnarray}
\begin{aligned}
\Delta\phi_{x,wj}&\simeq\Delta\phi_{x,wj}^{(mod)}
   +\sum_{m=1}^M \delta K_{m,1}\frac{\beta_{m,x}^{(mod)}}{4} \\ 
   &\hspace{5mm}\times 
    \Bigg\{2\Big[\Pi(m,j)-\hspace{-.5mm}\Pi(m,w)+\hspace{-.5mm}\Pi(j,w)\Big] \\
   &\hspace{10mm}+\frac{\sin{(2\tau_{x,mj}^{(mod)})}-\hspace{-.5mm}\sin{(2\tau_{x,mw}^{(mod)})}}
         {\sin{(2\pi Q_x^{(mod)})}}\Bigg\}\quad ,\hspace{8mm}\\
\Delta\phi_{y,wj}'&\simeq\Delta\phi_{y,wj}^{(mod)}
   -\sum_{m=1}^M \delta K_{m,1}\frac{\beta_{m,y}^{(mod)}}{4} \\
   &\hspace{5mm}\times 
    \Bigg\{2\Big[\Pi(m,j)-\hspace{-.5mm}\Pi(m,w)+\hspace{-.5mm}\Pi(j,w)\Big] \\
   &\hspace{10mm}+\frac{\sin{(2\tau_{y,mj}^{(mod)})}-\hspace{-.5mm}\sin{(2\tau_{y,mw}^{(mod)})}}
         {\sin{(2\pi Q_y^{(mod)})}}\Bigg\}\quad ,\hspace{8mm}
\end{aligned} \label{eq:Text_PhAd_1} 
\end{eqnarray}
where $(mod)$ refers to the lattice model not including 
the quadrupole errors $\delta K_{1}$, whereas the 
functions $\Pi$ and $\tau$ are the same of 
Eqs.~\eqref{eq:def_Pi} and \eqref{eq:def_tau}, 
respectively. In the above expressions, the remainder  
is proportional to $\delta K_{1}^2$. A response matrix 
$\mathbf{P}$ can be computed from the ideal C-S 
parameters with no need of going through the harmonic 
analysis of single-particle tracking data for each 
quadrupole error, since Eq.~\eqref{eq:Text_PhAd_1} 
can be rewritten as
\begin{eqnarray}
\left(\begin{array}{c}\delta\vec{\Delta\phi}_x\\ 
      \delta\vec{\Delta\phi}_y\end{array}\right)
&=&\mathbf{P}\cdot\delta\vec{K}_{1} +O(\delta K_1^2)
\ . \label{eq:Text_PhAd_2}
\end{eqnarray}
The effect of sextupoles and other higher-order 
multipoles can be neglected in the above system 
only if the amplitude of the turn-by-turn data 
is kept sufficiently low. If this is not the 
case, the more realistic harmonic analysis of 
simulated data is to be applied for a  
numerical evaluation of $\mathbf{P}$~\cite{Andrea-Linear-arxiv}.

\begin{figure}
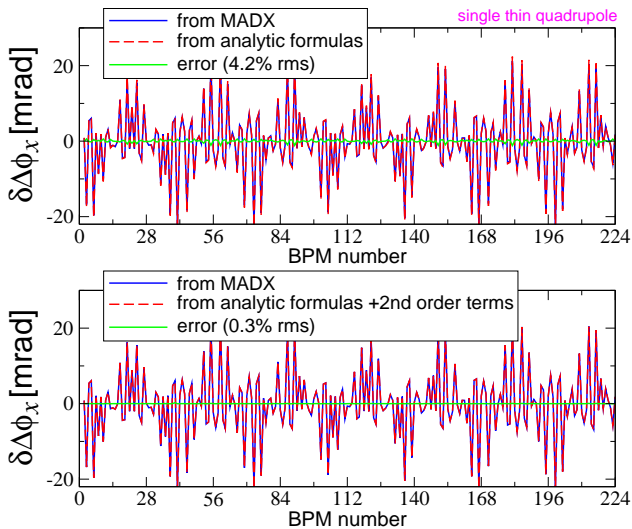

\rule{0mm}{0mm}
\centerline{\includegraphics[width=8.3cm]
           {madx_test_BPMPhAdv_xOnly_SingleThinQuad1.eps}}\\
\vskip .3mm
\centerline{\includegraphics[width=8.3cm]
           {madx_test_BPMPhAdv_xOnly_SingleThinQuad2.eps}}
  \caption{\label{fig_PhAd1} (Color) Top: Simulated horizontal BPM phase 
   advance shift induced by a single thin quadrupole of the old ESRF 
   storage ring as computed by MADX and via 
   Eqs.~\eqref{eq:Text_PhAd_1}-\eqref{eq:Text_PhAd_2}. The rms error 
   is of about 0.4 mrad ($4.2\%$ in relative terms). Bottom: When
   second-order terms are added to the above formulas the errors reduces 
   to 0.02 mrad (i.e. $0.3\%$).}
\rule{0mm}{3mm}
\end{figure}
\begin{figure}
\rule{0mm}{0mm}
\centerline{\includegraphics[width=8.3cm]
           {madx_test_BPMPhAdv_xOnly_SeveralThinQuad_DetuningWeak1.eps}}\\
\vskip .3mm
\centerline{\includegraphics[width=8.3cm]
           {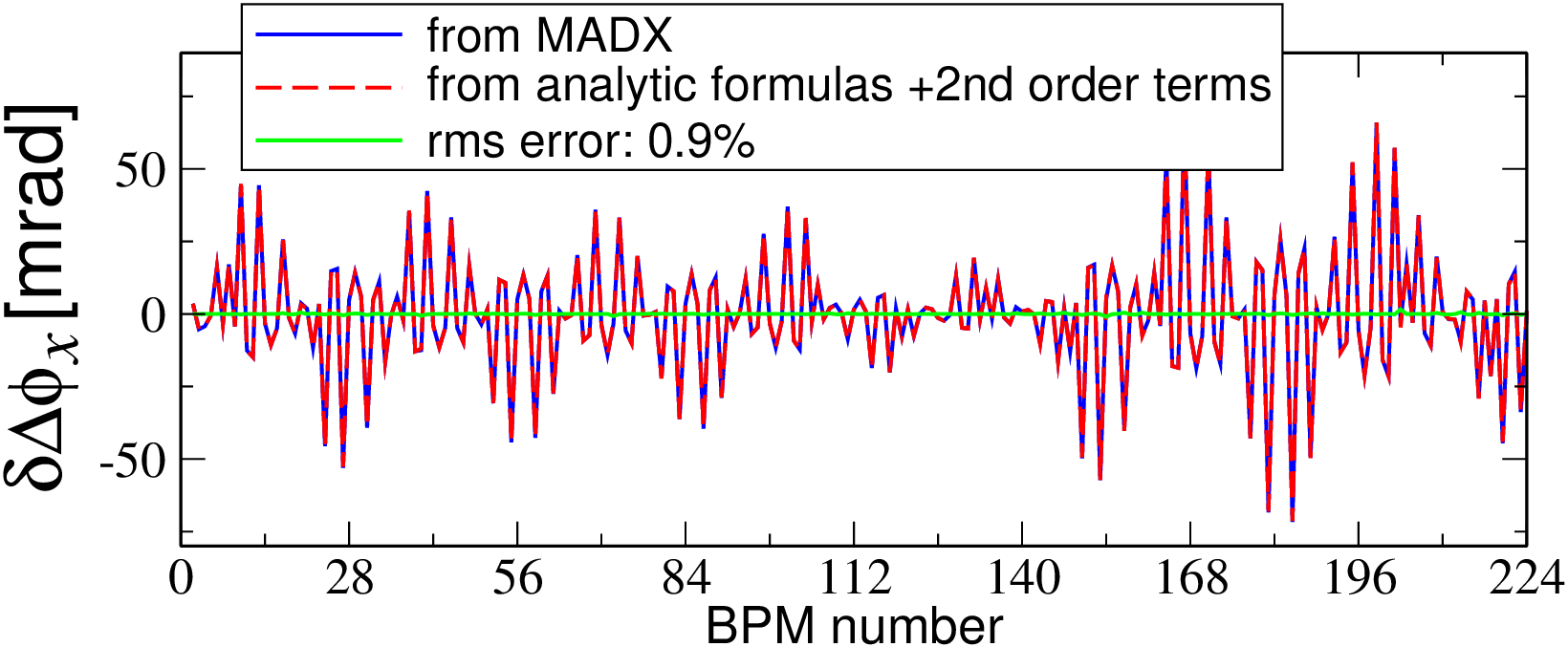}}
  \caption{\label{fig_PhAd2} (Color) Top: Simulated horizontal BPM phase 
   advance shift induced by six thin quadrupoles of the old ESRF 
   storage ring generating a weak detuning $\Delta Q_x\simeq1.5\times10^{-4}$ 
   as computed by MADX and via 
   Eqs.~\eqref{eq:Text_PhAd_1}-\eqref{eq:Text_PhAd_2}. The rms error 
   is of about 1.7 mrad ($7.9\%$ in relative terms). Bottom: When
   second-order terms are added to the above formulas the errors reduces 
   to 0.2 mrad (i.e. $0.9\%$).}
\rule{0mm}{3mm}
\end{figure}
\begin{figure}
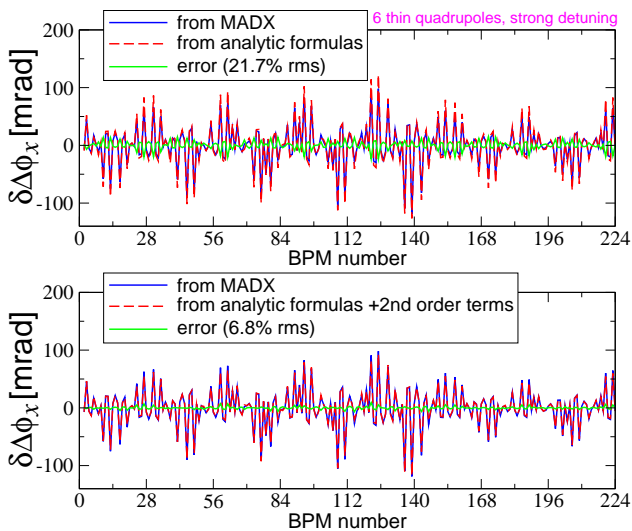

\rule{0mm}{0mm}
\centerline{\includegraphics[width=8.3cm]
           {madx_test_BPMPhAdv_xOnly_SeveralThinQuad_DetuningStrong1.eps}}\\
\vskip .3mm
\centerline{\includegraphics[width=8.3cm]
           {madx_test_BPMPhAdv_xOnly_SeveralThinQuad_DetuningStrong2.eps}}
  \caption{\label{fig_PhAd3} (Color) Top: The same six thin quadrupoles of 
   Fig.~\ref{fig_PhAd2} are modified so to induce a strong detuning 
   $\Delta Q_x\simeq-1.4\times10^{-2}$ and the agreement between MADX and 
   Eqs.~\eqref{eq:Text_PhAd_1}-\eqref{eq:Text_PhAd_2} worsens to 7.3 mrad 
   rms (or $21.7\%$ in relative terms). Bottom: When
   second-order terms are added to the above formulas the rms error reduces 
   to 2.3 mrad (i.e. $6.8\%$).}
\rule{0mm}{3mm}
\end{figure}
\begin{figure}
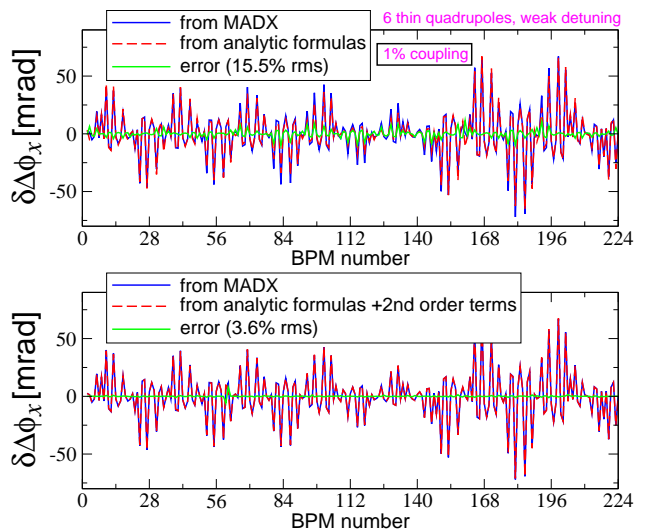

\rule{0mm}{0mm}
\centerline{\includegraphics[width=8.3cm]
           {madx_test_BPMPhAdv_xOnly_SeveralThinQuad_DetuningWeak_Coupling41pm1.eps}}\\
\vskip .3mm
\centerline{\includegraphics[width=8.3cm]
           {madx_test_BPMPhAdv_xOnly_SeveralThinQuad_DetuningWeak_Coupling41pm2.eps}}
  \caption{\label{fig_PhAd4} (Color) Top: Nine skew quadrupoles generating 
    an emittance ratio $\mathscr{E}_y/\mathscr{E}_x=1\%$ are added to the same 
    six thin quadrupoles of Fig.~\ref{fig_PhAd2} 
    and the agreement between MADX and 
   Eqs.~\eqref{eq:Text_PhAd_1}-\eqref{eq:Text_PhAd_2} deteriorates to 3.2 mrad 
   rms (or $15.5\%$ in relative terms). Bottom: second-order terms reduce 
   the rms error to 0.7 mrad (i.e. $3.6\%$).}
\rule{0mm}{3mm}
\end{figure}

Eqs.~\eqref{eq:Text_PhAd_1}-\eqref{eq:Text_PhAd_2} 
and more generally the linear response of the 
phase advance shift against the integrated quadrupole
strength $\delta K_{1}$ have been tested against the 
actual values computed at the BPMs by MADX for the lattice 
of the old ESRF storage ring. \\
First, the simplest case with a single, thin quadrupole 
error has been analyzed. Results for the horizontal BPM 
phase advance shifts are shown in the top plot of Fig.~\ref{fig_PhAd1} 
(similar plots and results are obtained in the vertical 
plane): The rms relative error of 
Eqs.~\eqref{eq:Text_PhAd_1}-\eqref{eq:Text_PhAd_2} 
is of about $4.2\%$. The sizable tune shift induced by 
this quadrupole, $\Delta Q_x\simeq-2.4\times10^{-3}$ and 
the non-negligible rms error suggest to seek for 
second-order terms $\propto\delta K_{1}^2$: This corresponds 
to keep all terms proportional to $f_{2000}^2$ in the 
various truncations and approximation made to derive 
Eqs.~\eqref{eq:Text_PhAd_1} and to include second-order 
RDTs following the procedure described in Ref.~\cite{Andrea-arxiv}. 
Handy formulas cannot be provided in this case and this 
correction needs to be computed numerically from the 
C-S parameters and first-order RDTs and Hamiltonian 
terms. The bottom plot of Fig.~\ref{fig_PhAd1}  shows 
indeed how second-order terms efficiently account for 
most of the initial error, the latter dropping to $0.3\%$. \\
A second numerical test was carried out by introducing 
six thin quadrupole errors generating a weak tune 
shift of $\Delta Q_x\simeq1.5\times10^{-4}$. The linear 
response of Eqs.~\eqref{eq:Text_PhAd_1}-\eqref{eq:Text_PhAd_2} 
can predict the BPM phase advance shift up to $7.9\%$ rms 
only (see top plot of Fig.~\ref{fig_PhAd2}). Most of 
this error stems from second-order terms, the error 
going below $1\%$ when these are included (bottom plot 
of Fig.~\ref{fig_PhAd2}). Second and higher order terms 
are generated by non-zero detuning terms (negligible in 
this example) and by cross-terms between the several 
quadrupole errors. \\
To confirm this point, a third simulation was run with 
the same six quadrupole errors where one quadrupole only 
was changed to generate a large tune shift 
$\Delta Q_x\simeq-1.4\times10^{-2}$. As expected, the linear 
dependence of the phase advance shift on the quadrupole 
errors of Eq.\eqref{eq:Text_PhAd_2} is by far less accurate, 
as shown in the top plot of Fig.~\ref{fig_PhAd3}: The rms 
error reaches almost $22\%$. Second-order terms help reduce 
the discrepancy to less than $7\%$ (bottom plot in the same 
figure), though suggesting that even higher-order terms 
play a role in this (unrealistic) example. \\
An additional source of second and higher-order terms that may 
spoil the linear analysis of Eq.\eqref{eq:Text_PhAd_2} is 
represented by betatron coupling. A fourth simulation was 
launched with the same 6 thin quadrupole errors of 
Fig.~\ref{fig_PhAd2} (with negligible tune shift) and 
additional nine thin skew quadrupoles generating a 
large ratio between the two transverse equilibrium 
emittances of $\mathscr{E}_y/\mathscr{E}_x=1\%$ (The old ESRF 
storage ring usually operated at a ratio close to $0.1\%$). 
Betatron coupling decreases the accuracy of
Fig.~\ref{fig_PhAd2} from $7.9\%$ to $15.5\%$ rms, as 
illustrated by the top plot of Fig.~\ref{fig_PhAd4}. When 
second-order terms are taken into account the rms error 
lessens to $3.6\%$. \\
Simulations with errors in thick quadrupoles (not 
shown here) revealed a general decrease of 
the predictive and correcting power of 
Eqs.~\eqref{eq:Text_PhAd_1}-\eqref{eq:Text_PhAd_2}. 
In order to account for the variation of the C-S parameters 
across the quadrupoles, the following substitutions can 
be made in Eq.~\eqref{eq:Text_PhAd_1}
\begin{eqnarray}
\begin{aligned}
\beta_{m}\sin{(2\tau_{mj})}\hskip.5mm &\longrightarrow\hskip.5mm 
         I_{S,mj} \\
\beta_{m}\sin{(2\tau_{mw})}\hskip.5mm &\longrightarrow\hskip.5mm 
         I_{S,mw} \\
\end{aligned}\quad , 
\end{eqnarray}
where the integrals $I_{S,mj}$ and $I_{S,mw}$ are 
evaluated via Eq.~\eqref{eq:IS2}. The labels $j$ and $w$ 
denote here BPMs which are assumed to be of zero length. 

The above numerical studies suggest some precautions 
need to taken when using the BPM phase advance 
errors and the linear system of Eq.~\eqref{eq:Text_PhAd_2} 
(regardless the way the linear response $\mathbf{P}$ 
is computed) to infer focusing lattice errors. 
As far as the old ESRF storage ring is concerned,
when pseudo-inverting Eq.~\eqref{eq:Text_PhAd_2}, 
an intrinsic accuracy as large as $4\%$ is to be 
expected even in the most ideal and simple case. 
Any fit of quadrupole errors leading to a fit error  
below this value is to be considered as unreliable. 
The accuracy deteriorates in the presence of large 
betatron coupling and detuning. Preliminary simulations 
shall then be run with the expected lattice configuration 
and errors, in order to estimate the level of 
accuracy expected when fitting quadrupole errors 
via Eq.~\eqref{eq:Text_PhAd_2}. 


\section{Analytic formulas for the evaluation of chromatic functions 
                 from lattice parameters}
\label{CHROM-Formulas}
In order to derive Eq.~\eqref{eq:Text_Disp1}, an 
off-momentum Hamiltonian formalism is used in 
Appendix~\ref{app:2}. With the same algebra other 
chromatic functions have been derived. They 
represent an extension of existing formulas 
for an ideal lattice of Ref.~\cite{Bengtsson} to a more 
general case including magnet errors and tilts 
from dipoles up to sextupoles. Possible use 
of these equations is discussed in Sec.~\ref{ORMdelta}. 

As for the linear dispersion, the edge 
focusing provided by nonzero dipole pole-face 
angles is not included in the lattice representation, 
the magnetic modelling being based on the 
multipolar expansion of Eq.~\eqref{eq:MADX}. 

\subsection{Linear chromaticity}
Off-energy particles experience the nominal focusing 
forces provided by quadrupoles and an additional one 
induced by the quadrupolar feed-down generated by the 
non-zero dispersive orbit at the sextupoles. The main 
consequence for such particles is a shift of their 
betatron tune, $Q(\delta)=Q+Q'\delta$, where $Q'$ is the 
linear chromaticity. The latter reads
\begin{eqnarray}
\begin{array}{l}\displaystyle
Q_x'=-\displaystyle\frac{1}{4\pi}\sum\limits_{m=1}^M \left(K_{m,1}-
      K_{m,2}D_{m,x}+ J_{m,2}D_{m,y} \right)\beta_{m,x}\ ,
      \vspace{ 1.5mm}\\ \displaystyle
Q_y'=+\displaystyle\frac{1}{4\pi}\sum\limits_{m=1}^M \left(K_{m,1}-
      K_{m,2}D_{m,x}+ J_{m,2}D_{m,y} \right)\beta_{m,y}\ .
\end{array}\label{eq:Text_Chrom1}
\end{eqnarray}
As expected, both quantities do not depend 
on the longitudinal position (or the betatron 
phase) and differ 
only by the sign and the beta functions, the 
argument within the parenthesis being the same 
in both planes. This indeed represent the effective 
quadrupole forces experienced by off-energy 
particles. The above relations 
require some comments. First, textbook 
formulas are retrieved when removing either 
vertical dispersion $D_{m,y}$ or the skew sextupole 
strengths $J_{2}$. Second, skew quadrupole 
fields $J_{1}$ do not influence explicitly 
linear chromaticity, at least to first 
order. Betatron coupling enters only indirectly 
in Eq.~\eqref{eq:Text_Chrom1} through vertical 
dispersion $D_y$. 
Beta and dispersion functions in 
Eq.~\eqref{eq:Text_Chrom1} refer to the lattice 
model including focusing errors, if any. 

In order to account for the variation of the 
beta function across quadrupoles, the following 
substitution can be made
\begin{eqnarray}
\beta_{m}\hskip.5mm &\longrightarrow\hskip.5mm I_{\beta,m} \quad ,\hskip 1cm
\beta_{m}D_m\hskip.5mm &\longrightarrow\hskip.5mm L_{\beta,D,m} \quad ,
\end{eqnarray}
where $I_{\beta,m}$ is defined in Eq.~\eqref{eq:Ib2} 
and $L_{\beta,D,m}$ is evaluated in Eq.~\eqref{eq:LBD2} 
(sextupoles are modelled as drifts).

\subsection{Chromatic beating}
\label{sec:ChromBeat}
Another consequence of the additional focusing 
experienced by off-momentum particles is a 
modulation of beta functions. Even an ideal lattice 
with no focusing error (i.e. no on-momentum 
{\sl geometric} beta-beating) is unavoidably 
subjected to an energy-dependent modulation of the 
betas  and hence to the corresponding 
half-integer resonance. This chromatic beating 
can be simply defined as the derivative of the 
beta function with respect to $\delta$, since 
\begin{eqnarray}\label{eq:Text_ChromBeat1}
\beta(\delta)=\beta+
     \frac{\partial\beta}{\partial\delta}\bigg|_{\delta=0}\delta
     +(\delta^2)\ .
\end{eqnarray}
In practice it is more convenient to express 
the beating as the normalized derivative
\begin{eqnarray}\label{eq:Text_ChromBeat2}
\tilde{\beta}'=\frac{1}{\beta}
     \frac{\partial\beta}{\partial\delta}\bigg|_{\delta=0}\ .
\end{eqnarray}
This quantity has the great advantage of being 
a dimensionless observable which is not affected 
by BPM calibration errors. In Appendix~\ref{app:2} 
the following expressions are derived for the 
chromatic beating in the two transverse planes:
     \begin{eqnarray}\label{eq:Text_ChromBeat3}
     \begin{array}{l}\displaystyle
     \tilde{\beta}_x'(j)\simeq 
     +\Big\{
     \sum_{m=1}^M \left(K_{m,1}-K_{m,2}D_{m,x}+ J_{m,2}D_{m,y}\right)
                    \times \\ \displaystyle
      \hspace{2.6cm}\beta_{m,x}\cos{(2\tau_{x,mj})}\Big\}
                    \frac{1}{2\sin{(2\pi Q_x)}}-1
     \\ \ \\ \displaystyle
     \tilde{\beta}_y'(j)\simeq 
     -\Big\{
     \sum_{m=1}^M \left(K_{m,1}-K_{m,2}D_{m,x}+ J_{m,2}D_{m,y}\right)
                    \times \\ \displaystyle
      \hspace{2.6cm}\beta_{m,y}\cos{(2\tau_{y,mj})}\Big\}
                    \frac{1}{2\sin{(2\pi Q_y)}}-1
     \end{array}\quad ,
     \end{eqnarray}
where the shifted phase advance 
$\tau_{mj}=\Delta\phi_{mj}-\pi Q$ is the same 
of Eq.~\eqref{eq:def_tau} and the phase 
advance $\Delta\phi_{mj}$ is to be computed 
according to Eq.~\eqref{eq:Text-deltaphisign}.
Note that the argument within the above 
parentheses is the same in both planes and equal 
to the one of Eq.~\eqref{eq:Text_Chrom1}, as it 
represents the effective quadrupole strengths 
experienced by off-energy particles. The structure 
of the above summations, which is responsible for the 
modulation of the beating along the ring, is also 
identical to the one of the formulas for the 
geometric beta beating induced by focusing 
errors~\cite{Andrea-Linear-arxiv}.
It is worthwhile noticing that the above 
expressions differ from the ones 
found in the literature~\cite{Luo-chrom,Tevatron}
for the presence of the $-1$ (or $-\beta$ if 
the un-normalized derivative is used) in the r.h.s., 
which stems from the invariant. This term does not 
affect the construction of a response matrix to 
correct the chromatic beating with sextupoles, 
since it cancels out. Similarly, it does not affect 
the evaluation of the difference between the model 
and the measured chromatic beating, provided that 
the former is computed by an optics code, such as 
MADX or PTC, which includes automatically this term.

\begin{figure}
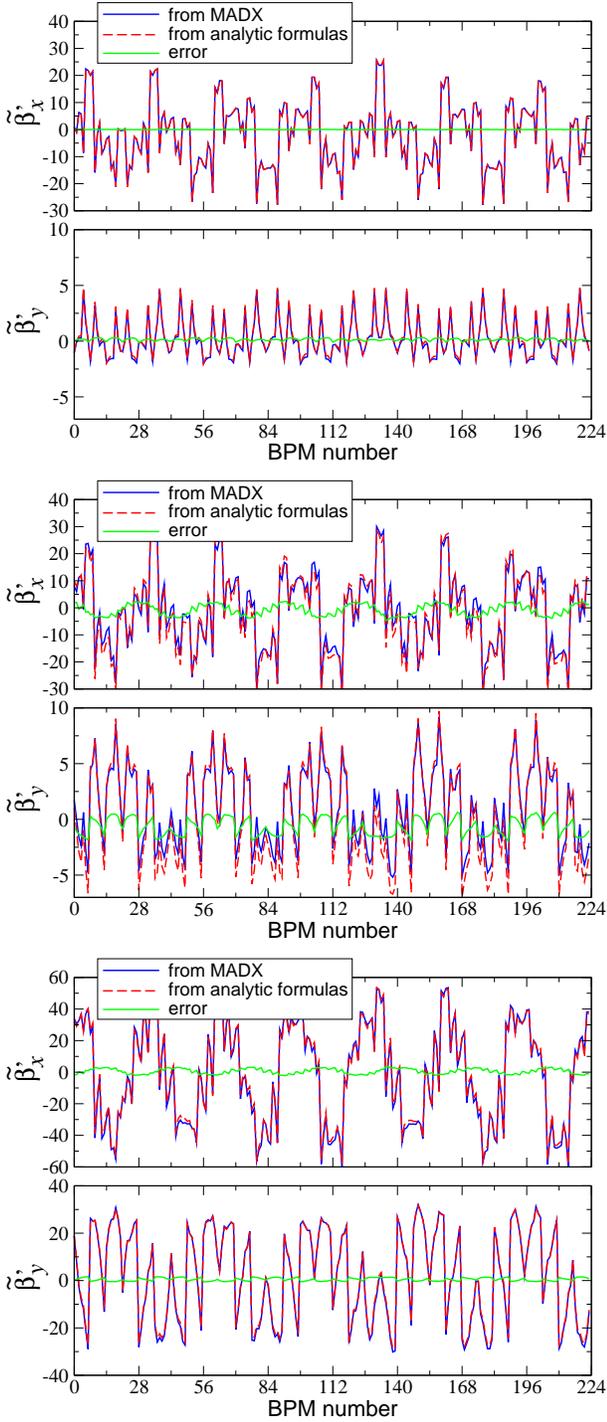

\rule{0mm}{0mm}
\centerline{\includegraphics[width=8.cm]{madx_test_ChromBeat_ideal.eps}}
\vskip 2mm
\centerline{\includegraphics[width=8.cm]{madx_test_ChromBeat_SkewSext.eps}}
\vskip 2mm
\centerline{\includegraphics[width=8.cm]{madx_test_ChromBeat_SkewSext_NoEdgeFoc.eps}}
  \caption{\label{fig_ChromBeat1} (Color) Examples of chromatic beating 
    computed by MADX-PTC and by Eq.~\eqref{eq:Text_ChromBeat3}. Top: 
    The ideal lattice of the old ESRF storage ring with no skew sextupole 
    is used and the rms error of the analytic formulas is of about 
    0.10 ($6\%$), mostly in the vertical plane. Center: A strong skew 
    sextupole and a tilted dipole are introduced in the lattice model 
    to enhance the contribution of the $J_{m,2}D_{m,y}$ term in 
    Eq.~\eqref{eq:Text_ChromBeat3}, leading to larger chromatic beating 
    and greater error of the analytic formula ($22\%$ rms). Bottom: 
    The latter lattice model is further modified by removing the edge focusing 
    in the dipoles, which increases even more the vertical chromatic 
    beating while reducing the error of Eq.~\eqref{eq:Text_ChromBeat3} 
    to below $5\%$.}
\rule{0mm}{0mm}
\end{figure}

The robustness of Eq.~\eqref{eq:Text_ChromBeat3} was 
tested numerically against the values computed by MADX 
via the \verb PTC_twiss  module for several configurations.
The ideal lattice of the old ESRF storage ring  including 
the edge focusing in the bending magnets (not included 
explicitly in the analytic formulas) was used for a 
first test, whose results are reported in the top two plots 
of Fig.~\ref{fig_ChromBeat1}: The agreement is of about $6\%$ 
rms, mostly in the vertical plane (it is of $0.3\%$
horizontally). The chromatic beating is not periodic 
because of one insertion optics with a non-standard 
quadrupole and sextupole layout (around the BPM number 135). 
In order to asses the validity of the 
$J_{m,2}D_{m,y}$ term, a strong skew sextupole and a large 
vertical deflection were then introduced into the model 
so to generate a sizable vertical dispersion and alter 
significantly the chromatic beating compared to the 
nominal lattice. The result of this test is shown in the 
central two plots of Fig.~\ref{fig_ChromBeat1}: The beating is 
indeed rather different, especially in the vertical plane, 
and Eq.~\eqref{eq:Text_ChromBeat3}  could reproduce this 
change quite well, even though the relative error increases 
to about $22\%$ rms in this example. This test and the 
fact the this contribution is of second order 
(in $J_{m,2}D_{m,y}$ both vertical dispersion and skew 
sextupole field components are orders of magnitude lower than 
the horizontal dispersion and normal sextupole strengths of 
$K_{m,2}D_{m,x}$) suggest 
that Eq.~\eqref{eq:Text_ChromBeat3} is not suitable for the 
evaluation of skew sextupole field components in real 
machines. In the attempt of better understanding the source 
of such discrepancy, a third test was carried out with the 
same two strong magnets, though removing the edge focusing 
in the dipoles (without retuning the baseline lattice). The 
chromatic beating of this unrealistic model changed 
completely, as demonstrated by 
the bottom two plots of Fig.~\ref{fig_ChromBeat1} and 
the accuracy of Eq.~\eqref{eq:Text_ChromBeat3} improved 
greatly, reaching an rms error of $5\%$, this time mainly 
in the horizontal plane (it is of $3.6\%$ for $\tilde{\beta}_y'$).

As for the previous formulas, the accuracy of 
Eq.~\eqref{eq:Text_ChromBeat3} can be improved by 
accounting for the variation of the C-S parameters 
and dispersion across the magnets, i.e. replacing
\begin{eqnarray}
\begin{aligned}
\beta_{m}\cos{(2\tau_{mj})}\hskip.5mm &\longrightarrow\hskip.5mm 
         I_{C,mj}\quad , \\
D_{m,q}\beta_{m,p}\cos{(2\tau_{p,mj})}\hskip.5mm &\longrightarrow\hskip.5mm 
         L_{C_p,D_q,mj}\ ,\ (p,q=x,y)\quad ,\quad
\end{aligned}
\end{eqnarray}
where the integral $I_{C,mj}$ is evaluated via Eq.~\eqref{eq:IS2}
and $L_{C_p,D_q,mj}$ is computed in Eq.~\eqref{eq:LBDCS}. 
In both case, the transport over a thick sextupole is 
modelled as a drift space. 

\subsection{chromatic phase advance shift}
Quadrupole errors induce a betatron phase shift to 
particles with nominal energy. When going off 
momentum the additional focusing provided by 
the off-axis closed orbit across sextupoles generate 
a similar {\sl chromatic} phase shift. 
In Appendix~\ref{app:2} the following expressions 
are derived for the derivative of the phase advance 
shift with respect to $\delta$
\begin{eqnarray}
\Delta\phi_{x,wj}'&=&\frac{\partial\Delta\phi_{x,wj}}{\partial\delta}\Bigg|_{\delta=0}
 \label{eq:Text-ChromPhAdx} \\
    &\simeq&
   -\frac{1}{4}\sum_{m=1}^M\left(K_{m,1}-K_{m,2}D_{m,x}+J_{m,2}D_{m,y}
   \right)\beta_{m,x}\nonumber\\
   &&\hspace{12mm}\times \Big\{2\Big[\Pi(m,j)-\Pi(m,w)+\Pi(j,w)\Big]
      \nonumber\\ 
   &&\hspace{17mm}+\frac{\sin{(2\tau_{x,mj})}-\sin{(2\tau_{x,mw})}}
         {\sin{(2\pi Q_x)}}\Big\} , \nonumber
\end{eqnarray}
where the functions $\Pi$ and $\tau$ are the same of 
Eqs.~\eqref{eq:def_Pi} and \eqref{eq:def_tau}, 
respectively. The chromatic shift in the 
vertical plane reads
\begin{eqnarray}
\Delta\phi_{y,wj}'&=&\frac{\partial\Delta\phi_{y,wj}}{\partial\delta}\Bigg|_{\delta=0}
 \label{eq:Text-ChromPhAdy} \\
    &\simeq&
   +\frac{1}{4}\sum_{m=1}^M\left(K_{m,1}-K_{m,2}D_{m,x}+J_{m,2}D_{m,y}
   \right)\beta_{m,y}\nonumber\\
   &&\hspace{12mm}\times \Big\{2\Big[\Pi(m,j)-\Pi(m,w)+\Pi(j,w)\Big]
      \nonumber\\ 
   &&\hspace{17mm}+\frac{\sin{(2\tau_{y,mj})}-\sin{(2\tau_{y,mw})}}
         {\sin{(2\pi Q_x)}}\Big\} . \nonumber
\end{eqnarray}
\begin{figure}
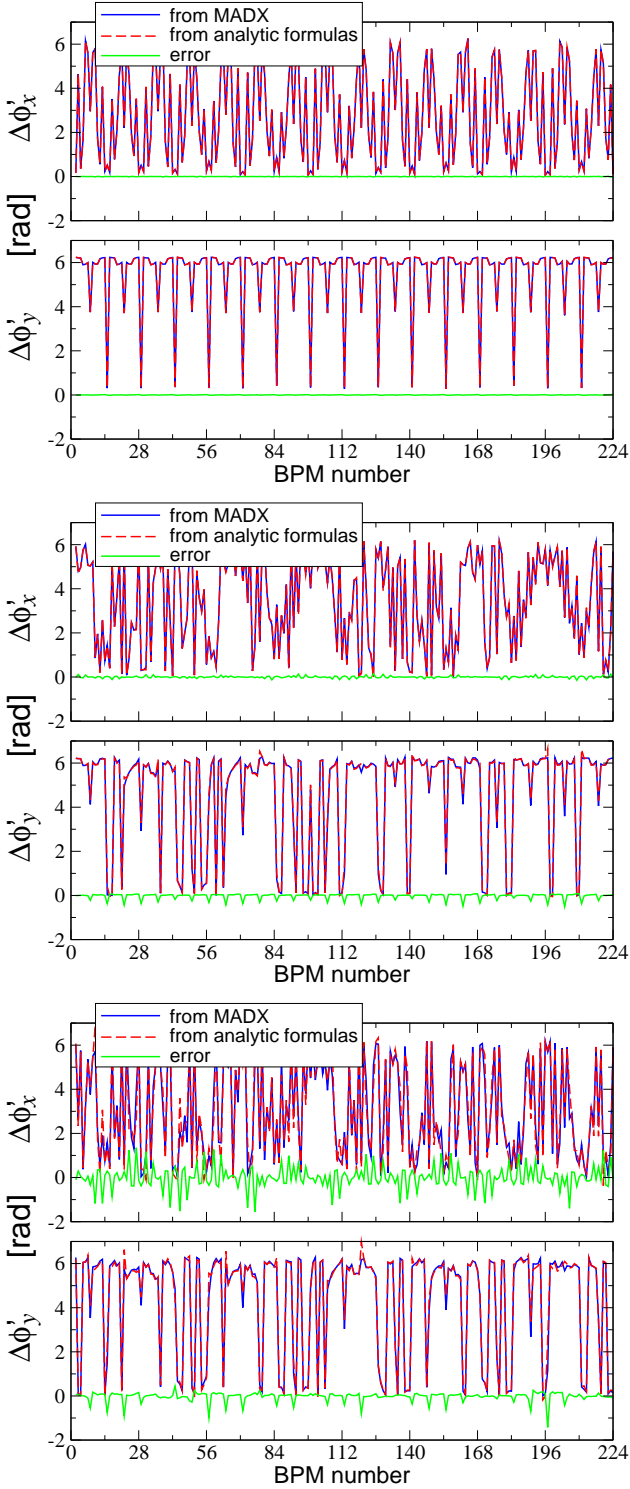

\rule{0mm}{0mm}
\centerline{\includegraphics[width=8.3cm]{madx_test_ChromPhAdv_ideal2012_NoEdge.eps}}
\vskip 2mm
\centerline{\includegraphics[width=8.3cm]{madx_test_ChromPhAdv_ideal2016_SkewSext.eps}}
\vskip 2mm
\centerline{\includegraphics[width=8.3cm]{madx_test_ChromPhAdv_mdt2016_SkewSext.eps}}
  \caption{\label{fig_ChromPhAd_1} (Color) Examples of comparison 
    between the chromatic phase advance shift computed by 
    Eqs.~\eqref{eq:Text-ChromPhAdx}-\eqref{eq:Text-ChromPhAdy} and MADX-PTC.
    Top two plots: ideal lattice of the old ESRF storage ring without the 
    dipole edge focusing. Center two plots: same lattice, after reintroducing 
    the nominal edge focusing in the bending magnets and adding 4 strong skew 
    sextupoles and a tilted dipole. Bottom two plots: A typical set of 
    linear errors (focusing and coupling) inferred from ORM measurement is 
    added to the second lattice model.} 
\rule{0mm}{0mm}
\end{figure}

As for the chromatic beating of Sec.~\ref{sec:ChromBeat}
the accuracy of the above formulas was tested 
numerically against the same quantities  
computed by MADX-PTC for several configurations. 
In Fig.~\ref{fig_ChromPhAd_1} the comparison with 
three different lattice models is reported. In the 
first two plots, $\Delta\phi_{wj}'$ is evaluated 
for the ideal lattice of the old ESRF storage ring 
without the dipole edge focusing, resulting in 
an excellent agreement within $0.5\%$ rms. In the 
second pair of plots, the phase advance shift is 
calculated from the same lattice, after reintroducing 
the nominal edge focusing in the bending magnets and 
including 4 strong skew sextupoles and a 100 mrad 
tilt in a dipole so to enhance the term $J_{m,2}D_{m,y}$ 
in Eqs.~\eqref{eq:Text-ChromPhAdx}-\eqref{eq:Text-ChromPhAdy}: 
The agreement is worse, at about $2\%$ and $5\%$ rms 
in the horizontal and vertical planes, respectively. 
The last two graphs correspond to the later lattice 
model with a typical set of linear errors (focusing 
and coupling) inferred from ORM measurement. The 
presence of betatron coupling which excites higher-order 
terms not included in the above formulas (of which more 
in Sec.~\ref{sec:SecOrdHam}) worsen the predictive power 
of the analytic formulas, with rms errors of about $20\%$ 
and $9\%$ in the two planes.

Once again, the accuracy of 
Eqs.~\eqref{eq:Text-ChromPhAdx}-\eqref{eq:Text-ChromPhAdy}
can be improved by accounting for the variation of the C-S parameters 
and dispersion across the magnets with the following 
substitutions
\begin{eqnarray}
\begin{aligned}
\beta_{m}\hskip.5mm &\longrightarrow\hskip.5mm I_{\beta,m} \quad , \\
\beta_{m}\sin{(2\tau_{mj})}\hskip.5mm &\longrightarrow\hskip.5mm 
         I_{S,mj}\quad , \\
D_{m,q}\beta_{m,p}\sin{(2\tau_{p,mj})}\hskip.5mm &\longrightarrow\hskip.5mm 
         L_{S_p,D_q,mj}\ ,\ (p,q=x,y)\quad ,\quad
\end{aligned}
\end{eqnarray}
where the integrals $I_{\beta,m}$, $I_{S,mj}$ and 
$L_{S_p,D_q,mj}$ are evaluated via 
Eqs.~\eqref{eq:Ib2},~\eqref{eq:IS2} and~Eq.~\eqref{eq:LBDCS}, 
respectively.

\subsection{ Second-order dispersion}
The linear dependence of the closed orbit on 
the energy (i.e. the dispersion function) is 
a function of mainly the bending magnets and the 
on-momentum linear optics, as demonstrated by 
Eq.~\eqref{eq:Text_Disp1}. At larger energy 
deviation the quadratic dependence of the 
orbit on $\delta$ needs to be taken into account. 
This corresponds to the derivative of the 
dispersion function with respect to $\delta$, 
namely 
\begin{eqnarray}\label{eq:def_disp2}
D_x'=\frac{\partial^2 x_{co}}{\partial\delta^2}=
\frac{\partial D_x}{\partial\delta}\quad .
\end{eqnarray}
The same definition applies to the vertical 
plane. Some authors~\cite{Bengtsson} define 
the second-order dispersion from the Taylor 
expansion in $\delta$, hence introducing a 
factor $1/2$. In order to follow the MADX-PTC 
nomenclature, Eq.~\eqref{eq:def_disp2} is 
used in this paper to define $D'$. Conversely 
to the linear dispersion, $D'$ depends on the 
modified off-momentum optics as well as on the dipolar 
feed-down from quadrupoles and sextupoles. 
Like for the chromatic beating, it is of interest 
to evaluate the dispersion normalized to 
the square root of the beta function, in order 
to make this observable independent of any 
possible BPM calibration error. 
In Appendix~\ref{app:2} the one-turn map 
of Eq.~\eqref{eq:Text-co1B} is used along with 
the computation of the Hamiltonian terms 
proportional to $\delta^2$ to derive the 
following analytic relations
\begin{eqnarray}\label{eq:disp2_1}
\left\{
\begin{aligned}
\tilde{D}_{x}'(j)&=\frac{D_x'(j)}{\sqrt{\beta_x(j)}}
                  =\Re\left\{\tilde{d}_{x,-}'(j)\right\}\\
\tilde{D}_{y}'(j)&=\frac{D_y'(j)}{\sqrt{\beta_y(j)}}
                  =\Re\left\{\tilde{d}_{y,-}'(j)\right\}
\end{aligned}\right .\quad ,
\end{eqnarray}
where $\tilde{d}_{x,-}'=\tilde{D}_x - i\tilde{D}_x'$ and 
$\tilde{d}_{y,-}'=\tilde{D}_y - i\tilde{D}_y'$ are the first 
and third elements of the complex C-S 
dispersion vector $\vec{d}=(\tilde{d}_{x,-},\tilde{d}_{x,+},
\tilde{d}_{y,-},\tilde{d}_{y,+})^T$. 
The latter reads
\begin{eqnarray}\label{eq:disp2_2}
\vec{d}'(j)\simeq\mathbf{B}_j^{-1}\sum_{m=1}^M\left\{
     \frac{e^{i\mathbf{\Delta\phi}_{mj}}}{1-e^{i\mathbf{Q}}}
     \mathbf{B}_m\ 4i
     \left(\begin{array}{r} h_{m,10002}\\-h_{m,10002}\\h_{m,00102}\\-h_{m,00102}
     \end{array}\right)\right\}\ ,\ \qquad
\end{eqnarray}
where the sum extends over all (normal and skew) 
$M$ dipoles, quadrupoles and sextupoles along 
the ring, whereas the RDT matrices $\mathbf{B}_j^{-1}$ 
and $\mathbf{B}_m$ are the same of  
Eq.~\eqref{eq:Text-Bmatrix}, and the Hamiltonian 
coefficients are
\begin{eqnarray}
\hspace{-3mm}\left\{\begin{aligned}
h_{m,10002}&=\frac{\sqrt{\beta_{m,x}}}{2}
              \bigg[-K_{m,0}-J_{m,1}D_{m,y}+ K_{m,1}D_{m,x}
              \hspace{-4mm}\\ &\hspace{1mm}
             -\frac{1}{2}K_{m,2}\left(D_{m,x}^2-D_{m,y}^2\right) 
             +J_{m,2}D_{m,x}D_{m,y}\bigg] \hspace{-4mm}\\ 
h_{m,00102}&=\frac{\sqrt{\beta_{m,y}}}{2}
              \bigg[\ \ J_{m,0}-J_{m,1}D_{m,x}- K_{m,1}D_{m,y}
              \hspace{-4mm}\\ &\hspace{1mm}
             +\frac{1}{2}J_{m,2}\left(D_{m,x}^2-D_{m,y}^2\right) 
             +K_{m,2}D_{m,x}D_{m,y}\bigg]\hspace{-4mm}
\end{aligned}\right .\quad . \nonumber \\ \label{eq:disp2_3}
\end{eqnarray}
The calculation of the second-order dispersion 
requires hence the preliminary evaluation of 
the coupling RDTs in order to infer the  
$\mathbf{B}$ matrices. Focusing errors are to
be included into the linear model to evaluate 
the C-S parameters and the linear dispersion, 
so to have $f_{2000}=f_{0020}=0$ anywhere along 
the ring and to compute the above Hamiltonian 
coefficients more accurately. The calculation 
simplifies greatly in the absence of linear coupling  
and tilted magnets, i.e. with $\mathbf{B}_m=\mathbf{B}_j^{-1}
=\mathbf{I}$, $D_{m,y}=0$, $J_{m,0}=J_{m,1}=J_{m,2}=0$ 
and hence $h_{m,00102}=h_{m,00012}=0$: 
\begin{eqnarray}
\left\{
\begin{aligned}\displaystyle
\tilde{D}_{x}'(j)&=4
             \Re\left\{\frac{e^{-i\Delta\phi_{x,mj}}}{1-e^{i2\pi Q_x}}
             i\ h_{m,01002}\right\}\\
\tilde{D}_{y}'&=0
\end{aligned}
\right.\ .\qquad
\end{eqnarray}
The ideal second-order horizontal dispersion then reads
\begin{eqnarray}\nonumber
\tilde{D}_{x}'(j)&=&\frac{1}{\sin{(\pi Q_x)}}
\sum_{m=1}^M \Big[-K_{m,0}+ K_{m,1}D_{m,x}\\ && \hspace{5mm}
       -\frac{1}{2}K_{m,2}D_{m,x}^2\Big]
       \sqrt{\beta_{m,x}}\cos{(\Delta\phi_{x,mj}-\pi Q_x)}
\nonumber \\  \nonumber 
&=&\hspace{-1mm}-2\tilde{D}_x(j)+\frac{1}{\sin{(\pi Q_x)}}\hspace{-1mm}
\sum_{m=1}^M \left[K_{m,1}-\frac{1}{2}K_{m,2}D_{m,x}\right] \\
       && \hspace{5mm}\times \label{eq:2ndDx_Id}
       D_{m,x}\sqrt{\beta_{m,x}}\cos{(\Delta\phi_{x,mj}-\pi Q_x)}\quad ,
\end{eqnarray}
corresponding to Eq.(112) of Ref.~\cite{Bengtsson} 
multiplied by a factor two. In the above equation, 
the linear dispersion $D_x(j)$ of Eq.~\eqref{eq:Text_Disp1}
has been extracted from the summation. 
As usual, the phase advance $\Delta\phi_{mj}$ is 
to be computed as in 
Eq.~\eqref{eq:Text-deltaphisign}. If the mere 
difference between the two betatron phases 
at the positions $m$ and $j$ is used, the 
absolute value $|\Delta\phi_{x,mj}|$ 
shall then be used, as done in textbooks. 
$\tau_{x,mj}=\Delta\phi_{x,mj}-\pi Q_x$ has been 
omitted here to ease the comparison with 
the standard formula. 
For a lattice with errors the more general 
Eqs.~\eqref{eq:disp2_1}-\eqref{eq:disp2_3}
shall be used and numerically evaluated. 

\begin{figure}
\rule{0mm}{0mm}
\centerline{\includegraphics[width=8.3cm]{madx_test_Disp2_SkewSext_LatErrors.eps}}
\vskip 2mm
\centerline{\includegraphics[width=8.3cm]{madx_test_Disp2_LatErrors.eps}}
\vskip 2mm
\centerline{\includegraphics[trim={0 0 0 0},clip=true,width=8.3cm]
          {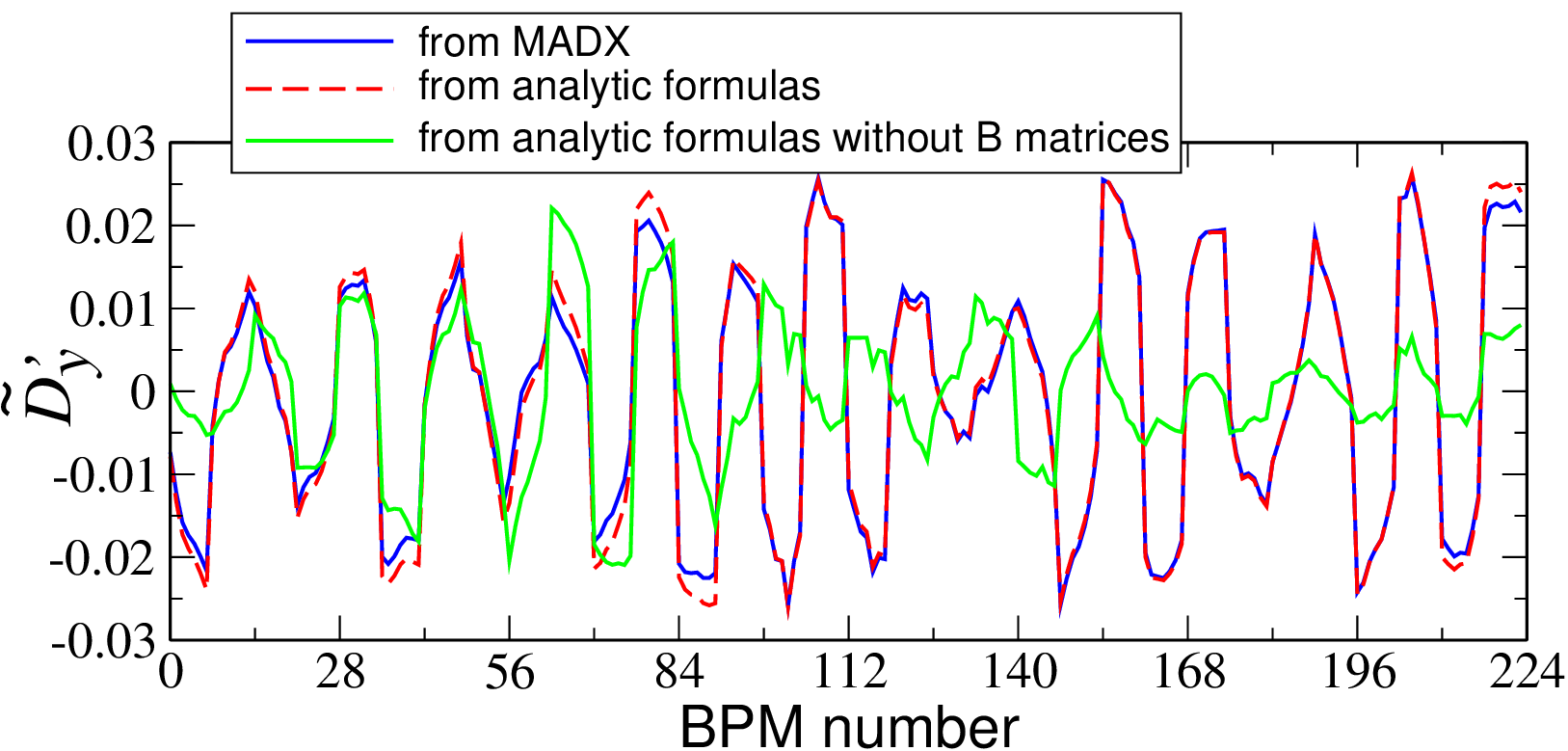}}
  \caption{\label{fig_Disp2_1} (Color) Examples of second-order dispersion 
  computed via Eqs.~\eqref{eq:disp2_1}-\eqref{eq:disp2_3} and by MADX-PTC. 
  Top plots: typical linear lattice errors inferred from ORM measurements 
  are included into the model of the old ESRF storage ring, along with two 
  strong skew magnets, one quadrupole (to enhance coupling and vertical 
  dispersion) and one sextupole (to have all contributions in 
  Eq.~\eqref{eq:disp2_2} 
  active). The analytic formulas predict $D_x'$ within $5\%$, whereas the 
  rms relative error for $D_y'$ is of about $2.5\%$. Center plots: The 
  previous two strong skew magnets are removed, providing the typical 
  second-order dispersion of the old ESRF storage ring. The smaller $D_y'$ 
  is accompanied by a larger relative error ($10\%$). $D_y'$ was also 
  computed by removing the RDT matrices  $\mathbf{B}$ in 
  Eq.~\eqref{eq:disp2_2}: The bottom plot shows how they cannot indeed 
  be ignored in the evaluation of $D_y'$.} 
\rule{0mm}{0mm}
\end{figure}

Once again, 
the accuracy of the above formulas was tested 
numerically against the second-order dispersion 
computed by MADX-PTC for several configurations, out of 
which two examples are reported here.
First, the lattice of the old ESRF storage ring  including 
the edge focusing in the bending magnets (not included 
explicitly in the analytic formulas), as well as typical linear 
lattice errors (beta beating and betatron coupling) inferred 
from ORM measurements, was used along with one strong 
skew quadrupole and one skew sextupole, so to have all 
terms in the square brackets of Eq.~\eqref{eq:disp2_3} 
active. The second-order dispersion predicted by 
Eqs.~\eqref{eq:disp2_1}-\eqref{eq:disp2_3} is compared 
to the one computed by MADX-PTC in the top plots of 
Fig.~\ref{fig_Disp2_1}: The agreement is of about $5\%$ 
rms for $D_x'$ and $2.5\%$ for $D_y'$. In the center plots 
of the same figure the comparison refers to the same lattice 
without the strong skew quadrupole and sextupole, hence 
representing a typical operational scenario for the old ESRF 
storage ring. While $D_x'$ is weakly altered (as it is 
dominated by the main bending magnets via $h_{m,10002}$ 
of Eq.~\eqref{eq:disp2_3} and the rms relative errors 
remains at the $5\%$ level, the derivative of the vertical 
dispersion is much smaller and the relative rms errors 
increases to about $10\%$. In order to asses the weight 
of the RDT matrices $\mathbf{B}$, $D_y'$ was also 
calculated by replacing them with the identity matrix 
$\mathbf{I}$. The bottom plot of Fig.~\ref{fig_Disp2_1} 
shows how they are indeed an essential ingredient in the 
correct evaluation of $D_y'$. 

The usual thick-magnet correction to account for the 
variation of C-S parameters and dispersion across magnets 
can be in principle carried out also here, though only 
for the ideal horizontal dispersion of 
Eq.~\eqref{eq:2ndDx_Id}, by following the same procedure 
described in Section~\ref{app:3}. For the more general 
formulas Eqs.~\eqref{eq:disp2_1}-\eqref{eq:disp2_3} 
a different approach needs to be defined.

\subsection{ Chromatic coupling}
Betatron coupling between the two transverse 
planes is generated by tilted quadrupoles, 
and non-zero vertical closed 
orbit inside sextupole magnets, whose feed-down 
field is of the skew-quadrupole type. Betatron 
coupling induces some vertical dispersion, on 
top of the one generated by any source of vertical 
deflection along the ring. When going off momentum, 
vertical dispersion adds an additional vertical 
beam displacement across the sextupoles, hence 
generating a new {\sl chromatic} coupling. If 
skew sextupole fields are also present, the 
horizontal displacements induced by the natural 
horizontal dispersion contribute also to 
coupling. Betatron coupling is completely 
described by the two RDTs $f_{1001}$ and 
$f_{1010}$. Hence, in order to describe the 
linear dependence of betatron coupling 
on the energy offset, i.e. chromatic coupling, it 
is natural to look for analytic formulas 
for the derivative of the two RDTs with respect 
to $\delta$, 
\begin{eqnarray}\label{eq:Text-chrCoupDef}
f_{10011}(j)=\hspace{-1mm}\frac{\partial f_{1001}(j)}{\partial\delta}\Bigg|_{\delta=0} 
              \hspace{-1mm} , \ 
f_{10101}(j)=\hspace{-1mm}\frac{\partial f_{1010}(j)}{\partial\delta}\Bigg|_{\delta=0}
\hspace{-1mm}.\qquad\quad\hspace{-4mm}
\end{eqnarray}
In Appendix~\ref{app:2} 
it is shown how the effective coupling terms
experienced by off-energy particles is 
represented by the Hamiltonian coefficients 
$h_{m,10011}=h_{m,10101}$ which depend 
linearly on skew quadrupoles, sextupoles (both 
normal and skew) and dispersion, according to 
\begin{eqnarray}
h_{m,10011}\hspace{-.7mm}=\hspace{-.7mm}
           -\frac{\sqrt{\beta_{m,x}\beta_{m,y}}}{4}\hspace{-.5mm}
           \left(J_{m,1}\hspace{-.5mm}-\hspace{-.6mm}K_{m,2}D_{m,y}
                 -\hspace{-.5mm}J_{m,2}D_{m,x}\right) .
\nonumber \\
\label{eq:Text-chrCoup1}
\end{eqnarray}
Chromatic coupling is then described by the 
following functions
\begin{eqnarray}
\begin{aligned}
f_{10011}(j)&
             \hspace{-.2mm}\simeq\hspace{-.4mm}F_{10011}(j,J_1) 
             \hspace{-.3mm}+\hspace{-.5mm}\frac{\sum\limits_{m=1}^M 
             \hspace{-.8mm}h_{m,10011}
              e^{i(\Delta\phi_{x,mj}-\Delta\phi_{y,mj})}}
              {1-e^{2\pi i(Q_x-Q_y)}}, \\
f_{10101}(j)&
             \hspace{-.2mm}\simeq\hspace{-.4mm}F_{10101}(j,J_1) 
             \hspace{-.3mm}+\hspace{-.5mm}\frac{\sum\limits_{m=1}^M
             \hspace{-.8mm}h_{m,10101}
              e^{i(\Delta\phi_{x,mj}+\Delta\phi_{y,mj})}}
             {1-e^{2\pi i(Q_x+Q_y)}} ,
\end{aligned} \nonumber  \\ \label{eq:Text-chrCoup2}
\end{eqnarray}
where $F_{10011}$ and $F_{10101}$ are defined in 
Eq.~\eqref{eq:chrCoup7} and depend mainly 
on skew quadrupole fields, and weakly on sextupole 
strengths, and the sum runs over all skew quadrupoles 
and sextupoles (both normal and skew) present in 
the machine. As usual, the phase advance 
$\Delta\phi_{mj}$ is to be computed as in 
Eq.~\eqref{eq:Text-deltaphisign}. The remainder in 
Eq.~\eqref{eq:Text-chrCoup2} is proportional to 
$J_1^2$. 
MADX-PTC does not compute directly the chromatic 
coupling RDTs. In order to test Eq.~\eqref{eq:Text-chrCoup2} 
they are then computed from simulated off-energy 
single-particle tracking data and the harmonic 
analysis, as done in Ref.~\cite{prstab_coup}. 
The derivative is then numerically computed from two 
sets of RDTs at $\delta=\pm10^{-3}$. In the first 
test, the ideal lattice of the old ESRF storage ring 
is used with one tilted bending magnet (to generate 
vertical dispersion) and one skew sextupole, though no 
skew quadrupole. By doing so, 
$f_{1001}=f_{1010}=F_{10011}=F_{10101}=0$ around the ring 
and the Hamiltonian coefficient of Eq.~\eqref{eq:Text-chrCoup1} 
contains only the coefficient $(K_{m,2}D_{m,y}-J_{m,2}D_{m,x})$ 
with no higher-order terms proportional to $J_1^2$ 
corrupting Eq.~\eqref{eq:Text-chrCoup2}. This test is 
important in assessing whether this equation can be 
effectively used to compute a response matrix to 
correct chromatic coupling with skew sextupoles. Results 
are shown in Fig.~\ref{fig_test_NoSkew2}, where the 
real and imaginary parts of the $f_{10011}$ and $f_{10101}$ 
are displayed from tracking and from  Eq.~\eqref{eq:Text-chrCoup2}: 
The agreement is well within $2\%$ rms. A second test is 
instead performed by removing the tilted bending magnet 
and the skew sextupole, after including a typical set 
of linear lattice errors (including skew quadrupole fields) 
obtained from ORM measurements. In this case, chromatic 
coupling is dominated by $F_{10011}$ and $F_{10101}$ along 
with the skew quadrupole term in $h_{m,10011}$ of 
Eq.~\eqref{eq:Text-chrCoup1} and the higher-order terms
are no longer zero. $F_{10011}$ and $F_{10101}$ are also 
computed with a further numerical approximation of 
Eq.~\eqref{eq:RDTd-5}. Results are reported in 
Fig.~\ref{fig_test_mdt2}, revealing a much worse agreement, 
of about $20\%$ rms, mostly for the sum RDT $f_{10101}$. 
Even if this test shows an intrinsic limitation in the 
capability of Eq.~\eqref{eq:Text-chrCoup2} in reproducing 
the real chromatic coupling, the first simulation shows how 
it can be effective in its correction by using skew sextupoles 
only, once skew quadrupoles are optimized to minimize 
the (on-momentum) betatron coupling.

\begin{widetext}
\ \\
\begin{figure}
\centerline
  {$\hspace{0.5cm}$\includegraphics[width=14.5cm,angle=0]
  {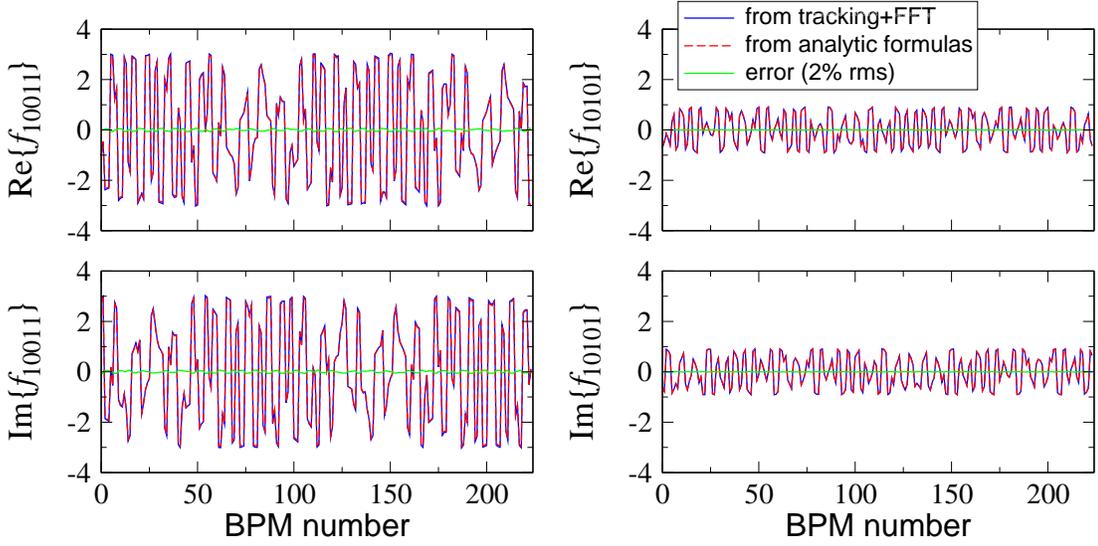}} 
  \caption{\label{fig_test_NoSkew2} (Color) Chromatic coupling, expressed 
          by the chromatic RDTs $f_{10011}$ and $f_{10101}$ for the ideal old 
          ESRF lattice with one vertical bending magnet and one skew sextupole.
          The functions computed from off-energy single-particle tracking 
          and the harmonic analysis of Ref.~\cite{prstab_coup} are in 
          blue, whereas the dashed red curves refer to those evaluated 
          from Eq.~\eqref{eq:Text-chrCoup2}. Even in the absence of geometric 
          betatron coupling ($f_{1001}=f_{1010}=F_{10011}=F_{10101}=0$ along the
          ring), the sextupoles (normal and 
          skew) couple with the dispersion functions (both horizontal and 
          vertical) to generate a linear coupling for all off-momentum 
          particles.}
\end{figure}
\begin{figure}
\centerline
  {$\hspace{0.5cm}$\includegraphics[width=14.5cm,angle=0]
  {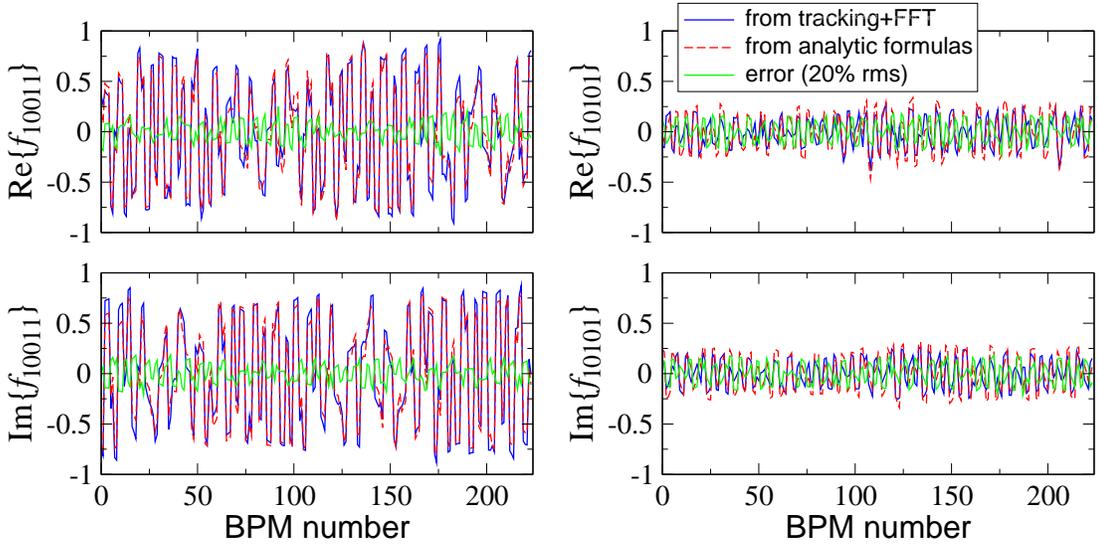}} 
  \caption{\label{fig_test_mdt2} (Color) ) Chromatic coupling, expressed 
          by the chromatic RDTs $f_{10011}$ and $f_{10101}$ for the
          ESRF lattice with a set of lattice errors (focusing and 
          betatron coupling) inferred from beam-based measurements.
          The functions computed from off-energy single-particle tracking 
          and the harmonic analysis of Ref.~\cite{prstab_coup} are in 
          blue, whereas the dashed red curves refer to those evaluated 
          from Eq.~\eqref{eq:Text-chrCoup2}. 
          Conversely to Fig.~\ref{fig_test_NoSkew2}, 
          chromatic coupling is generated here by the non-zero geometric 
          betatron coupling. The agreement between tracking and the 
          analytic predictions of Eq.~\eqref{eq:Text-chrCoup2} is 
          worse than in Fig.~\ref{fig_test_NoSkew2}, since there is 
          a non-zero second-order Hamiltonian contribution (see 
          Sec.~\ref{sec:SecOrdHam}) stemming from skew quadrupoles.}
\end{figure}
\end{widetext}

\subsection{Impact of higher-order Hamiltonian terms on the 
            chromatic functions}
\label{sec:SecOrdHam}
In evaluating the robustness of Eq.~\eqref{eq:Text_PhAd_1} 
it has been observed that second-order terms account for 
a large fraction of its error. This is true for all other 
observables. Nonlinear contributions from magnet strengths 
$K_n$ and $J_n$ to these observables originate from a 
series of truncations and approximations which remove 
terms proportional to powers higher than 1 of the RDTs. 
Moreover if focusing errors are not included 
in the model, betatron coupling is present in the lattice 
and linear chromaticity differs from zero, there is an 
additional contribution to the linear chromatic functions 
stemming from cross-product between Hamiltonian terms. 
The procedure for their (numerical) evaluation is 
presented in Sec.~\ref{sec:app-SecOrdHam} of 
Appendix~\ref{app:2}.

\section{Linear analysis of off-momentum ORM for the evaluation 
         of a sextupolar lattice model}
\label{ORMdelta}
Linear dispersion and on-momentum ORM are 
routinely measured and used to fit linear 
lattice errors by pseudo-inverting the two 
systems of Eqs.~\eqref{eq:ORM_04}-\eqref{eq:ORM_05}, 
where the two response matrices $\mathbf{N}$ 
and $\mathbf{S}$ can be either obtained by 
simulating the measurement (slower, but more 
accurate) or analytically computed from the 
equations presented in Sec.~\ref{ORM-Formulas} 
(quicker, but less precise).

The same approach can be extended to 
off-momentum ORM and second-order dispersion $D'$. 
Indeed, the off-axis orbit across sextupoles 
generated by the energy offset via the linear dispersion 
generates quadrupole feed-down field (which is 
linear in the sextupole strengths) altering 
both the linear optics (and hence the ORM) and 
dispersion. How strong, and hence observable, is 
this effect depends mainly on the dispersion 
function at the sextupoles: It is then suitable 
for chromatic sextupoles, less so for the harmonic 
ones. The systems of 
Eqs.~\eqref{eq:ORM_04}-\eqref{eq:ORM_05} can 
be extended to the case with $\delta\ne0$
according to
\begin{eqnarray}
\begin{array}{c}
\left(\begin{array}{c}\delta\vec{O}^{(xx)}\\ 
      \delta\vec{O}^{(yy)}\\\delta\vec{D}_x\end{array}
\right)_{\delta\ne0}=\mathbf{N}_{\delta}
\left(\begin{array}{c}\delta\vec{K}_{2}\\ \delta\vec{K}_{1}\\ \delta\vec{K}_{0}
\end{array}\right)\ , 
\vspace{2mm}\\
\left(\begin{array}{c}\delta\vec{O}^{(xy)}\\ 
      \delta\vec{O}^{(yx)}\\ \delta\vec{D}_y\end{array}
\right)_{\delta\ne0}=\mathbf{S}_{\delta}
\left(\begin{array}{c}\vec{J}_2\\ \vec{J}_1\\ \vec{J}_0
\end{array}\right)\ .
\end{array}\label{eq:ORM_D1}
\end{eqnarray}
$\delta\vec{K}_{2}$ and $\delta\vec{J}_{2}$ 
are the vectors containing the sextupole errors 
and tilts (represented by skew sextupole integrated 
strengths). $\delta{\mathbf O}_{\delta\ne0}$ and 
$\delta\vec{D}_{\delta\ne0}$ denote instead the 
deviation between the measured and the model off-energy 
ORM and dispersion, whereas $\mathbf{N}_{\delta}$ 
and $\mathbf{S}_{\delta}$ are the response matrices 
of Eqs.~\eqref{eq:ORM_04}-\eqref{eq:ORM_05} 
including sextupole magnets and computed at a 
given $\delta\ne0$. However, the linear lattice errors 
inferred from the on-momentum ORM and dispersion can 
be inserted in the model used to compute the corresponding 
off-momentum quantities. If the deviations 
$\delta{\mathbf O}_{\delta\ne0}$ and 
$\delta\vec{D}_{\delta\ne0}$ are then computed with 
respect to this modified model, the above systems 
simplifies to 
\begin{eqnarray}
\begin{array}{c}
\left(\begin{array}{c}\delta\vec{O}^{(xx)}\\ 
      \delta\vec{O}^{(yy)}\\\delta\vec{D}_x\end{array}
\right)_{\delta\ne0}=\mathbf{N'}_{\delta}\ \delta\vec{K}_{2}\ , 
\vspace{2mm}\\
\left(\begin{array}{c}\delta\vec{O}^{(xy)}\\ 
      \delta\vec{O}^{(yx)}\\ \delta\vec{D}_y\end{array}
\right)_{\delta\ne0}=\mathbf{S'}_{\delta}\ \vec{J}_2\ ,\hskip 4mm\ 
\end{array}\label{eq:ORM_D3}
\end{eqnarray}
where both $\mathbf{N'}_{\delta}$ and $\mathbf{S'}_{\delta}$ 
are now computed from the model including the linear 
errors. The pseudo-inversion of these two later systems 
can be then used to infer an error model for the 
(chromatic) sextupoles. Eq.~\eqref{eq:ORM_D3} may
be modified by inserting weights and fixing 
chromaticity to the measured value in order to 
obtain an effective model. \\

Alternatively, two measurements at $\pm\delta$ of 
both ORM and linear dispersion can be performed. 
From the linear analysis of Eqs.~\eqref{eq:ORM_04}-\eqref{eq:ORM_05}, 
which shall include the energy offset and sextupole magnets, 
the linear lattice parameters ($\beta,\ f_{1001},\ f_{1010}$ 
and $D_{x,y}$) at the BPMs can be computed at $\pm\delta$. 
Their derivative with respect to $\delta$, i.e. the 
chromatic functions of Sec.~\ref{CHROM-Formulas}, 
can be then evaluated:
\begin{eqnarray}\label{eq:ORM_D5}
\left(\begin{array}{c}
   \delta{\mathbf O}_{\pm\delta} \vspace{1mm}\\
   \vec{D}_{\pm\delta}
\end{array}\right)
\Rightarrow 
\begin{array}{c}
   \left(\begin{array}{c}\vec{\beta}\vspace{1mm}\\ \vec{f}_{1001} 
   \vspace{1mm}\\ \vec{f}_{1010}\vspace{1mm}\\\vec{D} 
\end{array}\right)_{\pm\delta} \end{array}
\Rightarrow 
\left(\begin{array}{c}
\vec{\beta}'\vspace{1mm}\\ \vec{f}_{10011} \vspace{1mm}\\ 
\vec{f}_{10101} \vspace{1mm}\\ \vec{D}'
\end{array}\right)\ .
\end{eqnarray}
The vector with the difference between measured  
and model chromatic 
functions can be expressed in terms of sextupole 
errors (strengths and tilts) according to
\begin{eqnarray}\label{eq:ORM_D6}
\left(\begin{array}{c}
\vec{\beta}'\\\hskip-1mm \vec{f}_{10011}\hskip-2mm\ \vspace{1mm}\\
              \hskip-1mm \vec{f}_{10101}\hskip-2mm\ \vspace{1mm}\\ \vec{D}'
\end{array}\right)^{\hskip-1mm(meas)}\hskip-4mm -\hskip 1mm
\left(\begin{array}{c}
\vec{\beta}'\\\hskip-1mm \vec{f}_{10011}\hskip-2mm\ \vspace{1mm}\\
              \hskip-1mm \vec{f}_{10101}\hskip-2mm\ \vspace{1mm}\\ \vec{D}'
\end{array}\right)^{\hskip-1mm(mod)}\hskip-4mm
=\hskip 1mm\mathbf{T}
\left(\begin{array}{c}\delta\vec{K}_{2} \vspace{1mm}\\ \vec{J}_2\end{array}\right)\ ,
\end{eqnarray}
where the betatronic block of the response matrix 
$\mathbf{T}$ is computed from Eq.~\eqref{eq:Text_ChromBeat3}, 
the part corresponding to the chromatic coupling 
is obtained from Eqs.~\eqref{eq:Text-chrCoup1}-\eqref{eq:Text-chrCoup2}, 
whereas the terms for the second-order dispersion 
are derived from Eqs.~\eqref{eq:disp2_1}-\eqref{eq:disp2_3}.
Once again, weights between the different parameters 
and constant chromaticity (see Eq.~\eqref{eq:Text_Chrom1}) 
shall be included to the above system to obtain 
a realistic model. 
The complex chromatic RDTs $f_{10011}$ and $f_{10101}$ 
may be split in real and imaginary parts to preserve 
the linearity of the system.
Interestingly, when evaluating $\mathbf{T}$ it is 
not necessary to include terms either constant,  
such as the -1 in the formulas for $\beta'$ of 
Eq.~\eqref{eq:Text_ChromBeat3}, or independent on 
sextupole strengths, e.g. the complicated functions 
$F_{10011}$ and $F_{10101}$ in Eq.~\eqref{eq:Text-chrCoup2}.
The BPM chromatic phase advance shift can be also 
included in Eq.~\eqref{eq:ORM_D6} or replace the 
chromatic beating. In this case the system to 
be pseudo-inverted would read 
\begin{eqnarray}\label{eq:ORM_D7}
\left(\begin{array}{c}
\Delta\vec{\phi}' \\\hskip-1mm \vec{f}_{10011}\hskip-2mm\ \vspace{1mm}\\
                    \hskip-1mm \vec{f}_{10101}\hskip-2mm\ \vspace{1mm}\\ \vec{D}'
\end{array}\right)^{\hskip-1mm(meas)}\hskip-4mm -\hskip 1mm
\left(\begin{array}{c}
\Delta\vec{\phi}' \\\hskip-1mm \vec{f}_{10011}\hskip-2mm\ \vspace{1mm}\\
                    \hskip-1mm \vec{f}_{10101}\hskip-2mm\ \vspace{1mm}\\ \vec{D}'
\end{array}\right)^{\hskip-1mm(mod)}\hskip-4mm
=\hskip1mm\mathbf{T'}
\left(\begin{array}{c}\delta\vec{K}_{2} \vspace{1mm}\\ \vec{J}_2\end{array}\right)\ ,
\end{eqnarray}
where the $\mathbf{T'}$ differs from $\mathbf{T}$ 
for the block corresponding to the chromatic 
phase shift, which is computed from 
Eqs.~\eqref{eq:Text-ChromPhAdx}-\eqref{eq:Text-ChromPhAdy}.

The advantage of using the chromatic functions 
instead of the off-momentum ORM is that 
the same systems of Eqs.~\eqref{eq:ORM_D6}-\eqref{eq:ORM_D7} 
can be defined irrespective of the measurement 
technique. For example, chromatic functions can be 
measured from the harmonic analysis of BPM 
turn-by-turn off-momentum data. 

\section{Conclusion}
Analytic formulas for the computation of the distortion of 
orbit response matrix (ORM) induced by quadrupole errors and 
rotations have been derived and tested by using the 
lattice model of the ESRF storage rings (old and new). An accuracy at 
the level of a few percent (rms) has been demonstrated. 
Explicit formulas for the evaluation of chromatic 
functions (beta beating, phase shift, coupling and 
second-order dispersion) were also derived. Their 
robustness depends largely on the suppression of 
higher-order terms that can be minimized by including 
focusing errors in the model and correcting coupling. 
By doing so, a chromatic sextupole error model can be 
inferred from the analysis of either the off-momentum 
ORM or the chromatic functions, the correlation being 
linear with the sextupole strengths and tilts. 
\\




\begin{widetext}
\appendix
\section{Derivation of the ORM response due to quadrupole errors and 
         tilts}
\label{app:1}

The standard procedure to evaluate the closed orbit distortion 
induced by a dipole horizontal perturbation $\Theta_w$ is based 
on the closed-orbit condition 
\begin{eqnarray}\label{eq:CO_standard0}
\mathbf{M}{X}_+={X}_-\ , \qquad 
{X}_-=\left(\begin{array}{c} x_w \\ p_{x,w}-\theta_w\end{array}\right)\ ,\quad
{X}_+=\left(\begin{array}{c} x_w \\ p_{x,w}\end{array}\right)\ ,\quad
\mathbf{M}\ \hbox{is the ideal one-turn matrix}
\end{eqnarray}
In the absence of lattice errors, the two planes are decoupled and an 
equivalent relation applies to the vertical plane. In the 
Courant-Snyder (C-S) coordinates the above system reads
\begin{eqnarray}\label{eq:CO_standard}
\mathbf{R}\tilde{X}_+=\tilde{X}_-\ , \qquad 
\tilde{X}_-=\left(\begin{array}{c} \tilde{x}_w \\ \tilde{p}_{x,w}-
            \sqrt{\beta_{w,x}}\Theta_w\end{array}\right)\ ,\quad
\tilde{X}_+=\left(\begin{array}{c} \tilde{x}_w \\ \tilde{p}_{x,w}
            \end{array}\right)\ ,\quad
\mathbf{R}=\left(\begin{array}{c c} \cos{(2\pi Q_x)} & \sin{(2\pi Q_x)} \\
              -\sin{(2\pi Q_x)} & \cos{(2\pi Q_x)}\end{array}\right)\ ,\quad
\end{eqnarray}
whose solution reads
\begin{eqnarray}
\left\{
\begin{aligned}\displaystyle
\tilde{x}_w&=\frac{\sqrt{\beta_{w,x}}\Theta_w\cos{(\pi Q_x)}}
                 {2\sin{(\pi Q_x)}}\\
\tilde{p}_x&=\frac{\sqrt{\beta_{w,x}}\Theta_w}{2}
\end{aligned}
\right.  \qquad\Rightarrow\qquad
\left\{
\begin{aligned}\displaystyle
x_w&=\frac{\beta_{w,x}\Theta_w\cos{(\pi Q_x)}}
                 {2\sin{(\pi Q_x)}}\\
p_x&=\frac{\Theta_w}{2\sin{(\pi Q_x)}}
       (\sin{\pi} Q_x-\alpha_x\cos{(\pi Q_x)})
\end{aligned}
\right. \quad .
\end{eqnarray}
The closed orbit at a generic location $j$ is obtained again first in 
the C-S coordinates, where the transport between the position $w$ and 
$s$ is a mere rotation by the corresponding phase advance, and then 
in the Cartesian ones:
\begin{eqnarray}
\mathbf{R}\tilde{X}_+=\tilde{X}_-\ , \quad 
\left(\begin{array}{c} \tilde{x}_j \\ \tilde{p}_{x,j}
            \end{array}\right)=\mathbf{R}\left(\Delta\phi_{x,wj}\right)
\left(\begin{array}{c} \tilde{x}_w \\ \tilde{p}_{x,w}\end{array}\right) 
\quad\Rightarrow\quad
\left\{
\begin{aligned}\displaystyle
\tilde{x}_j&=\frac{\sqrt{\beta_{w,x}}\Theta_w}
                 {2\sin{(\pi Q_x)}}\cos{(\Delta\phi_{x,wj}-\pi Q_x)}\\
x_j&=\frac{\sqrt{\beta_{w,x}\beta_{j,x}}\Theta_w}
                 {2\sin{(\pi Q_x)}}\cos{(\Delta\phi_{x,wj}-\pi Q_x)}
\end{aligned}
\right. \quad .\label{eq:orm_classic}
\end{eqnarray}
The phase advance between the BPM $j$ and the magnet $w$, 
$\Delta\phi_{x,wj}$, must be a positive quantity. However, 
if it is computed from the ideal betatron phases $\phi_{x,j}$ 
and $\phi_{x,w}$ with a fixed origin, it becomes negative 
whenever the magnet is downstream the BPM: In this case the 
total phase advance (i.e. over one turn) needs to be added, namely
\begin{eqnarray}\label{eq:deltaphisign}
\left\{\begin{aligned}
&\Delta\phi_{x,wj}=(\phi_{x,j}-\phi_{x,w})
                    \hspace{1.3cm}\ ,\ \hbox{if\ }\phi_{x,j}>\phi_{x,w}\\
&\Delta\phi_{x,wj}=(\phi_{x,j}-\phi_{x,w})+2\pi Q_x\ ,\ 
                    \hbox{if\ }\phi_{x,j}<\phi_{x,w}
\end{aligned}\right. \ , \ 
\left\{\begin{aligned}
&\Delta\phi_{x,wj}-\pi Q_x=(\phi_{x,j}-\phi_{x,w})-\pi Q_x
                    \ ,\ \hbox{if\ }\phi_{x,j}>\phi_{x,w}\\
&\Delta\phi_{x,wj}-\pi Q_x=(\phi_{x,j}-\phi_{x,w})+\pi Q_x
                    \ ,\ \hbox{if\ }\phi_{x,j}<\phi_{x,w}
\end{aligned}\right. \ .\qquad\quad
\end{eqnarray}
Even though Eq.~\eqref{eq:orm_classic} is 
found in the literature with $\cos{(|\Delta\phi_{x,wj}|-\pi Q_x)}$, 
which smartly accounts for both cases, since $\cos{(x)}=\cos{(-x)}$, 
it is no longer 
convenient for the more general formula to be derived. Hence, the 
definition of $\Delta\phi_{x,wj}$ given in Eq.~\eqref{eq:deltaphisign}
is kept throughout the paper.

The orbit response being linear in $\Theta_w$, if several sources 
of dipole perturbations are present, a sum over $w$ shall be 
included in the above equations. In the vertical plane identical 
relations apply after substituting $x$ with $y$.\\

In the presence of focusing errors and linear coupling the above procedure 
does not apply, since the two planes are no longer decoupled, neither in 
the Cartesian nor in the C-S coordinates. Even including focusing errors 
in the model ($\delta K_1=0$) betatron coupling induced by skew quadrupole 
fields $\delta J_1\ne0$ requires a more careful approach. The generalization 
of the C-S coordinates in the presence of betatron coupling (and 
nonlinearities) is represented by the normal form coordinates. As the 
C-S transformation {\sl absorbs} the envelope 
modulation induced by the ideal focusing lattice and reshape the 
$s$-dependent elliptical phase space portraits in an invariant circle, 
the (non-resonant) normal form transformation {\sl absorbs} 
focusing errors, betatron 
coupling and, with some precautions, lattice nonlinearities, retrieving 
circular orbits in phase space from the distorted curves in the original 
Cartesian phase space. In normal forms, the two planes are also decoupled.
Such transformation is a polynomial function $F$
\begin{eqnarray}\label{eq:Fdef}
F=\sum_{n}\sum_{pqrt}^{n=p+q+r+t}{f_{pqrt}
             \zeta_{x,+}^p\zeta_{x,-}^q\zeta_{y,+}^r\zeta_{y,-}^t}\ ,
\end{eqnarray}
where $n$ denotes the multipole order, $f_{pqrt}$ are 
RDTs and $\zeta_{z,\pm}=\sqrt{2I_z}e^{\mp i(\psi_z+\psi_{z,0})}$ 
are the new complex normal form coordinates ($z$ stands for 
either $x$ or $y$), which are the decoupled and nonlinear 
generalization of the complex C-S complex 
variable $h_{z,\pm}=\tilde{z}\pm i\tilde{p}_z =\sqrt{2J_z}
e^{\mp i(\phi_z+\phi_{z,0})}$. The equation establishing the change 
of coordinates in normal form at a generic point $j$ may be written 
in terms of Lie operators and Poisson brackets $[\ ,\ ]$ 
\begin{eqnarray}
\vec{\zeta}_j&=&\label{eq:NormFormTrans}
e^{:-F_j:}\vec{h}_j=\vec{h}_j +[-F_j,\vec{h}_j] +O(F^2) 
\ ,\qquad
\vec{h}_j=
e^{:F_j:}\vec{\zeta}_j=\vec{\zeta}_j +[F_j,\vec{\zeta}_j] +O(F^2)\ ,
\end{eqnarray}
where $\vec{h}=(h_{x,-},h_{x,+},h_{y,-},h_{y,+})^T$ and $\vec{\zeta}$ its 
equivalent in normal form, whereas $e^{: :}$ denotes the Lie operator. 
The remainder $O(F^2)$ contains nested Poisson Brackets and scales 
with the RDTs squared. The above transformations imply that to 
the first order in the RDTs, $F(\vec{\zeta})=F(\vec{h})+O(f^2)$ since 
the two variables $\vec{\zeta}$ and $\vec{h}$ are tangent, i.e. 
$\vec{\zeta}=\vec{h}+O(f)$.
In the presence of focusing errors and sources of betatron coupling 
only terms such that $p+q+r+t=2$ (i.e. normal and skew quadrupolar 
$\propto x^2,\ y^2$ and $xy$) are to be selected in Eq.~\eqref{eq:Fdef} 
in order to remove the dependence upon them in the normal forms 
coordinates:
\begin{eqnarray}\nonumber
F=f_{2000}\zeta_{x,+}^2+f_{2000}^*\zeta_{x,-}^2 +
  f_{0020}\zeta_{y,+}^2+f_{0020}^*\zeta_{y,-}^2 +
  f_{1001}\zeta_{x,+}\zeta_{y,-} +f_{1001}^*\zeta_{x,-}\zeta_{y,+}+
  f_{1010}\zeta_{x,+}\zeta_{y,+} +f_{1010}^*\zeta_{x,-}\zeta_{y,-} \ ,
\end{eqnarray}
where the relation (valid to first order only~\cite{Andrea-arxiv}) 
$f_{pqrt}=f_{qptr}^*$ has been used. To first order, the RDTs 
at a location $j$ read~\cite{Andrea-arxiv}
\begin{equation}\label{eq:RDT-1st}
f_{pqrt}(j) =\frac{\sum\limits_m^M
h_{m,pqrt}e^{i[(p-q)\Delta\phi_{x,mj}+(r-t)\Delta\phi_{y,mj}]}}
                     {1-e^{2\pi i[(p-q)Q_x+(r-t)Q_y]}}\ \ .
\end{equation}
The coefficients $h_{m,pqrt}$ derive from the 
Hamiltonian term in the complex C-S coordinates 
generated by a generic magnet $m$
\begin{eqnarray}\label{eq:Hdef}
\tilde{H}_m=\sum_{pqrt}^{n=p+q+r+t}{h_{m,pqrt}
             h_{m,x,+}^p h_{m,x,-}^q h_{m,y,+}^r h_{m,y,-}^t}\ ,
\end{eqnarray}
and read 
\begin{eqnarray}
h_{m,pqrt}&=&-\displaystyle
\frac{\bigl[K_{m,n-1}\Omega(r+t)+
           iJ_{m,n-1}\Omega(r+t+1)\bigr]}
     {p!\quad q!\quad r!\quad t!\quad 2^{\hspace{.2mm}p+q+r+t}} 
     \ i^{r+t} \bigl(\beta_{m,x}\bigr)^{\frac{p+q}{2}}
     \bigl(\beta_{m,y}\bigr)^{\frac{r+t}{2}},\label{eq:h_Vs_KJ} \nonumber\\
     \nonumber \\
     \Omega&(i)&=1 \hbox{ if } i \hbox{ is even},\quad
                      \Omega(i)=0 \hbox{ if } i \hbox{ is odd}
     \ .
\end{eqnarray}
$\Omega(i)$ is introduced to select either the 
normal or the skew multipoles. $K_{m,n-1}$ and 
$J_{m,n-1}$ are the integrated magnet strengths 
of the multipole expansion (MADX definition)
\begin{equation}
-\Re\left[\sum_{n}{(K_{m,n-1}+iJ_{m,n-1})
            \frac{(x_m+iy_m)^n}{n!}}\right]\ ,
\end{equation}
from which Eqs.~\eqref{eq:Hdef} and ~\eqref{eq:h_Vs_KJ}  
are derived when moving from the Cartesian coordinates 
to the complex Courant-Snyder's: 
$x_m=\sqrt{\beta_{m,x}}(h_{m,x,-}+h_{m,x,+})/2$ and 
$y_m=\sqrt{\beta_{m,y}}(h_{m,y,-}+h_{m,y,+})/2$. 
By recalling that 
\begin{eqnarray}\label{eq:Poisson-bracket}
[h_{z,+}^p,h_{z,-}^q]=-2i(pq)h_{z,+}^{p-1}h_{z,-}^{q-1}
			     =-[h_{z,-}^q,h_{z,+}^p]\ ,\qquad
\end{eqnarray}
and that all other combinations yield zero Poisson brackets, 
Eq.~\eqref{eq:NormFormTrans} truncated to first order reads
\begin{eqnarray}\label{eq:RDT-1st-front}
\vec{\zeta}_j&=&\mathbf{B}_j\vec{h}_j +O(f^2)\quad ,\quad
\mathbf{B}_j=\left(
\begin{array}{c c c c}
    1             & 4if_{2000,j}    &\ \ 2if_{1001,j}&\ \ 2if_{1010,j} \\
 -4if_{2000,j}^*  &        1        &-2if_{1010,j}^* &-2if_{1001,j}^*  \\
\ \ 2if_{1001,j}^*&\ \  2if_{1010,j}&       1        &\ \  4if_{0020,j}\\
-2if_{1010,j}^*   &-2if_{1001,j}    &-4if_{0020,j}^* &       1
\end{array}
\right)\ +O(f^2)\ .
\end{eqnarray}
The inverse transformation reads
\begin{eqnarray}\label{eq:RDT-1st-back}
\vec{h}_j&=&\mathbf{B}_j^{-1}\vec{\zeta}_j +O(f^2)\quad ,\quad
\mathbf{B}_j^{-1}=\left(
\begin{array}{c c c c}
    1             & -4if_{2000,j}  &-2if_{1001,j}     &-2if_{1010,j}     \\
\ \ 4if_{2000,j}^*&        1       &\ \ 2if_{1010,j}^*&\ \ 2if_{1001,j}^*\\
-2if_{1001,j}^*   &-2if_{1010,j}   &       1          &-4if_{0020,j}     \\
\ \ 2if_{1010,j}^*&\ \ 2if_{1001,j}&\ \ 4if_{0020,j}^*&       1 
\end{array}
\right) + O(f^2)\ .
\end{eqnarray}
Since in normal forms the motion is decoupled and the phase space 
trajectories are circles rotating with the betatron phase, the 
closed-orbit condition of Eq.~\eqref{eq:CO_standard} at a generic 
orbit corrector $w$ becomes
\begin{eqnarray}\label{eq:coNF}
e^{i\mathbf{Q}}\vec{\zeta}_w=\vec{\zeta}_w-\delta\vec{\zeta}_w\quad,\quad
\vec{\zeta}_w=\frac{\delta\vec{\zeta}_w}{\mathbf{1}-e^{i\mathbf{Q}}}\ ,
\end{eqnarray}
where $e^{i\mathbf{Q}}=
\hbox{diag}(e^{2\pi iQ_x},e^{-2\pi iQ_x},e^{2\pi iQ_y},e^{-2\pi iQ_y})$, 
$\mathbf{1}$ is a $4\times4$ identity matrix,  and 
$\delta\vec{\zeta}_w$ denotes the orbit perturbation in normal forms, of 
which more later. The closed orbit at a generic position $j$ is computed 
by rotating the normal form coordinates by the phase advance between the 
source of distortion $w$ and $j$, as done for the ideal case in the C-S 
coordinates. 
\begin{eqnarray}\label{eq:coNF2}
\vec{\zeta}_j=e^{i\mathbf{\Delta\phi}_{wj}}\frac{\delta\vec{\zeta}_w}
               {\mathbf{1}-e^{i\mathbf{Q}}}\ ,
\end{eqnarray}
where $e^{i\mathbf{\Delta\phi}_{wj}}=\hbox{diag}(e^{i\Delta\phi_{x,wj}},
e^{-i\Delta\phi_{x,wj}},e^{i\Delta\phi_{y,wj}},e^{-i\Delta\phi_{y,wj}})$ 
is the diagonal matrix describing the phase advance rotation in the two 
normal form planes, which are uncoupled and described by circular 
trajectories in phase space. In practice it is of interest to transform 
Eq.~\eqref{eq:coNF2} in the C-S coordinates, first, and Cartesian, 
then, in order to derive measurable quantities. The transformations of 
Eqs.~\eqref{eq:RDT-1st-front} and~\eqref{eq:RDT-1st-back} may be applied 
to Eq.\eqref{eq:coNF2}, yielding
\begin{eqnarray}\label{eq:co1}
\vec{h}_j&=&\mathbf{B}_j^{-1}\vec{\zeta}_j=
\mathbf{B}_j^{-1}e^{i\mathbf{\Delta\phi}_{wj}}\vec{\zeta}_w=
\mathbf{B}_j^{-1}e^{i\mathbf{\Delta\phi}_{wj}}\frac{\delta\vec{\zeta}_w}{\mathbf{1}-e^{i\mathbf{Q}}}=
\mathbf{B}_j^{-1}\frac{e^{i\mathbf{\Delta\phi}_{wj}}}{\mathbf{1}-e^{i\mathbf{Q}}}\mathbf{B}_w\delta\vec{h}_w\ .
\end{eqnarray}
When composing the three matrices in the the above relation, only 
terms linear in the RDTs are to be kept, their product going in the 
remainder $O(f^2)$. The generalization to several sources of distortion may 
be carried out by introducing a sum over $w$ in the r.h.s.  
\begin{eqnarray}\label{eq:co1B}
\vec{h}_j&=&\mathbf{B}_j^{-1}\sum_{w=1}^W\left\{
\frac{e^{i\mathbf{\Delta\phi}_{wj}}}{\mathbf{1}-e^{i\mathbf{Q}}}
\mathbf{B}_w\ \delta\vec{h}_w\right\}\ .
\end{eqnarray}
The perturbation $\delta\vec{h}_w$ is generated by orbit 
correctors via the dipole terms $\delta K_{w,0}$ and $\delta J_{w,0}$ 
($n=1$) and the Hamiltonian
\begin{eqnarray}\nonumber
\tilde{H}_w=h_{w,1000}h_{w,x,+}+h_{w,0100}h_{w,x,-}+ 
              h_{w,0010}h_{w,y,+}+h_{w,0001}h_{w,y,-}
\quad , \ 
\left\{\begin{aligned}
h_{1000}&=h_{0100}=-\frac{\sqrt{\beta_x}}{2}\delta K_0 =
            -\frac{\sqrt{\beta_x}}{2}\Theta_x\\
h_{0010}&=h_{0001}=\ \  \frac{\sqrt{\beta_y}}{2}\delta J_0 =
            -\frac{\sqrt{\beta_y}}{2}\Theta_y
\end{aligned}\right. \ ,
\\ \label{eq:HamDisp}
\end{eqnarray}
where the definitions of the Hamiltonian terms derive from 
Eq.~\eqref{eq:h_Vs_KJ}. Note that if a positive horizontal 
field $\delta K_0>0$ induces a positive deflection $\Theta_x>0$ 
a negative vertical field $\delta J_0<0$ is needed for a 
positive deflection $\Theta_y>0$.
The Hamilton's equations in the Lie algebra read
\begin{eqnarray}\label{eq:deltah}
\vec{h}_{w+\epsilon}=\vec{h}_w-[\tilde{H}_w,\vec{h}_w]\quad\Rightarrow\quad
\delta \vec{h}_{w}=-[\tilde{H}_w,\vec{h}_w]\quad ,
\end{eqnarray}
where $\epsilon$ the 
infinitesimal step downstream the position $w$. By making use of 
Eq.~\eqref{eq:Poisson-bracket}, Eq.~\eqref{eq:deltah} reads
\begin{eqnarray}\label{eq:deltah2}
\delta\vec{h}_{w}=\left(\begin{array}{c}\delta h_{w,x,-}\\\delta h_{w,x,+}\\
                    \delta h_{w,y,-} \\ \delta h_{w,y,+}\end{array}\right)=
2i\left(\begin{array}{r}h_{w,1000}\\-h_{w,0100}\\h_{w,0010}\\-h_{w,0001}
   \end{array}\right)= 
i\left(\begin{array}{r} 
-\sqrt{\beta_{w,x}}\Theta_{w,x}\\ \sqrt{\beta_{w,x}}\Theta_{w,x} \\
-\sqrt{\beta_{w,y}}\Theta_{w,y}\\ \sqrt{\beta_{w,y}}\Theta_{w,y}
   \end{array}\right)\ .
\end{eqnarray}
The orbit response matrix of Eq.~\eqref{eq:ORM_01} can be 
then derived from Eqs.~\eqref{eq:co1B} and~\eqref{eq:deltah2}, 
recalling that orbit at a BPM $j$ is just 
$O_j=\sqrt{\beta_j}\Re\{h_j\}$: 
\begin{eqnarray}\label{eq:co2}
\begin{aligned}
&O^{(xx)}_{wj}=\sqrt{\beta_{j,x}\beta_{w,x}}\Re\left\{i
                  \mathbf{B}_j^{-1}
\frac{e^{i\mathbf{\Delta\phi}_{wj}}}{\mathbf{1}-e^{i\mathbf{Q}}}
\mathbf{B}_w\right\}^{(1,1\rightarrow2)}\ ,\quad
O^{(xy)}_{wj}=\sqrt{\beta_{j,x}\beta_{w,y}}\Re\left\{i
                  \mathbf{B}_j^{-1}
\frac{e^{i\mathbf{\Delta\phi}_{wj}}}{\mathbf{1}-e^{i\mathbf{Q}}}
\mathbf{B}_w\right\}^{(1,3\rightarrow4)}\ , \\
&O^{(yx)}_{wj}=\sqrt{\beta_{j,y}\beta_{w,x}}\Re\left\{i
                  \mathbf{B}_j^{-1}
\frac{e^{i\mathbf{\Delta\phi}_{wj}}}{\mathbf{1}-e^{i\mathbf{Q}}}
\mathbf{B}_w\right\}^{(3,1\rightarrow2)}\ ,\quad
O^{(yy)}_{wj}=\sqrt{\beta_{j,y}\beta_{w,y}}\Re\left\{i
                  \mathbf{B}_j^{-1}
\frac{e^{i\mathbf{\Delta\phi}_{wj}}}{\mathbf{1}-e^{i\mathbf{Q}}}
\mathbf{B}_w\right\}^{(3,3\rightarrow4)}\ .
\end{aligned}\qquad
\end{eqnarray}
In the above notation, given a $4\times4$ matrix $\mathbf{P}$, 
$\mathbf{P}^{(a,b\rightarrow c)}=P_{ac}-P_{ab}$, where the 
minus sign stems from the opposite sign between neighbor 
elements in the $\delta\vec{h}_{w}$ of Eq.~\eqref{eq:deltah2}. Indeed, 
the second and fourth row of the $4\times4$ matrix within the curly 
brackets in Eq.~\eqref{eq:co2} are just the complex 
conjugate of the first and third rows, respectively, 
and do not contribute to the ORM. 
For an explicit evaluation of the complete $4\times4$ complex ORM  
it is convenient to use in Eq.~\eqref{eq:co2} 
the actual C-S parameters (i.e. including the focusing 
error). By doing so $f_{2000,w}=f_{2000,j}=0$, the 
upper diagonal blocks of $\mathbf{B}^{-1}_j$ and 
$\mathbf{B}_w$ are a $2\times2$ identify matrix. The dependence 
on the focusing errors will be then restored via the 
C-S parameters. The complex ORM then reads
\begin{eqnarray}\label{eq:ORM1}
\mathbf{P}_{wj}=&&\mathbf{B}_j^{-1}
\frac{e^{i\mathbf{\Delta\phi}_{wj}}}{\mathbf{1}-e^{i\mathbf{Q}}}
\mathbf{B}_w \\ \nonumber
&&\hspace{-8mm}=\mathbf{B}_j^{-1}
\left(
\begin{array}{c c c c}
\hspace{-1mm}\frac{e^{i\Delta\phi_{x,wj}^{(mod)}}}{1-e^{i2\pi Q_x^{(mod)}}} & 0 & 0 & 0 \\
0 &\hspace{-3mm}\frac{e^{-i\Delta\phi_{x,wj}^{(mod)}}}{1-e^{-i2\pi Q_x^{(mod)}}} & 0 & 0 \\
0 & 0 &\hspace{-3mm}\frac{e^{i\Delta\phi_{y,wj}^{(mod)}}}{1-e^{i2\pi Q_y^{(mod)}}} & 0 \\
0 & 0 & 0 &\hspace{-3mm}\frac{e^{-i\Delta\phi_{y,wj}^{(mod)}}}{1-e^{-i2\pi Q_y^{(mod)}}} 
\end{array}\hspace{-1mm}\right)\hspace{-2mm}
\left(
\begin{array}{c c c c}
    1             &        0        &\hspace{-1mm}\ \ 2if_{1001,w}&\hspace{-1mm}\ \ 2if_{1010,w}  \\
    0             &        1        &\hspace{-1mm} -2if_{1010,w}^*&\hspace{-1mm}  -2if_{1001,w}^* \\
\hspace{-2mm}\ \ 2if_{1001,w}^*&\hspace{-1mm}\ \ 2if_{1010,w} &       1        &       0          \\
\hspace{-2mm}-2if_{1010,w}^*   &\hspace{-1mm}-2if_{1001,w}    &       0        &       1          
\end{array}\hspace{-1mm}
\right)\hspace{-.5mm}+\hspace{-.5mm}O(f^2) \\ \nonumber 
&&\hspace{-8mm}=\hspace{-1mm}\left(
\begin{array}{c c c c}
    1             &       0          &-2if_{1001,j}     &  -2if_{1010,j}   \\
    0             &       1          &\ \ 2if_{1010,j}^*&\ \ 2if_{1001,j}^*\\
-2if_{1001,j}^*   &-2if_{1010,j}     &         1        &       0          \\
\ \ 2if_{1010,j}^*&\ \ 2if_{1001,j}  &         0        &       1
\end{array}
\right) \times \\ \nonumber 
&&\hspace{-5mm}\left(
\begin{array}{c c c c}
\frac{e^{i\Delta\phi_{x,wj}^{(mod)}}}{1-e^{i2\pi Q_x^{(mod)}}} & 0 &
\ 2if_{1001,w}\frac{e^{i\Delta\phi_{x,wj}^{(mod)}}}{1-e^{i2\pi Q_x^{(mod)}}}&
\ 2if_{1010,w}\frac{e^{i\Delta\phi_{x,wj}^{(mod)}}}{1-e^{i2\pi Q_x^{(mod)}}}\\
0 & \frac{e^{-i\Delta\phi_{x,wj}^{(mod)}}}{1-e^{-i2\pi Q_x^{(mod)}}} & 
-2if_{1010,w}^*\frac{e^{-i\Delta\phi_{x,wj}^{(mod)}}}{1-e^{-i2\pi Q_x^{(mod)}}}&
-2if_{1001,w}^*\frac{e^{-i\Delta\phi_{x,wj}^{(mod)}}}{1-e^{-i2\pi Q_x^{(mod)}}}\\
\ 2if_{1001,w}^*\frac{e^{i\Delta\phi_{y,wj}^{(mod)}}}{1-e^{i2\pi Q_y^{(mod)}}}&
\ 2if_{1010,w}\frac{e^{i\Delta\phi_{y,wj}^{(mod)}}}{1-e^{i2\pi Q_y^{(mod)}}}&
\frac{e^{i\Delta\phi_{y,wj}^{(mod)}}}{1-e^{i2\pi Q_y^{(mod)}}} & 0 \\
-2if_{1010,w}^*\frac{e^{-i\Delta\phi_{y,wj}^{(mod)}}}{1-e^{-i2\pi Q_y^{(mod)}}}&
-2if_{1001,w}\frac{e^{-i\Delta\phi_{y,wj}^{(mod)}}}{1-e^{-i2\pi Q_y^{(mod)}}}&
0 &\frac{e^{-i\Delta\phi_{y,wj}^{(mod)}}}{1-e^{-i2\pi Q_y^{(mod)}}} 
\end{array}
\right)\hspace{-.5mm}+\hspace{-.5mm}O(f^2)\ ,
\end{eqnarray}
resulting in 
\begin{eqnarray}\label{eq:ORM2}
\begin{aligned}
P_{11,wj}&=&\frac{e^{i\Delta\phi_{x,wj}^{(mod)}}}{1-e^{i2\pi Q_x^{(mod)}}}
            \hspace{55mm}+O(f^2) \quad , \quad P_{21}=P_{12}^*\quad ,\\
P_{12,wj}&=&0\hspace{75mm}+O(f^2)\quad , \quad P_{22}=P_{11}^* \quad ,\\
P_{13,wj}&=&2if_{1001,w}\frac{e^{i\Delta\phi_{x,wj}^{(mod)}}}{1-e^{i2\pi Q_x^{(mod)}}}
           -2if_{1001,j}\frac{e^{i\Delta\phi_{y,wj}^{(mod)}}}{1-e^{i2\pi Q_y^{(mod)}}}
           \hspace{0mm}+O(f^2)\quad , \quad P_{23}=P_{14}^*\quad ,\\
P_{14,wj}&=&2if_{1010,w}\frac{e^{i\Delta\phi_{x,wj}^{(mod)}}}{1-e^{i2\pi Q_x^{(mod)}}}
           -2if_{1010,j}\frac{e^{-i\Delta\phi_{y,wj}^{(mod)}}}{1-e^{-i2\pi Q_y^{(mod)}}} 
           \hspace{0mm}+O(f^2)\quad , \quad P_{24}=P_{13}^*\quad , \\
P_{31,wj}&=&-2if_{1001,j}^*\frac{e^{i\Delta\phi_{x,wj}^{(mod)}}}{1-e^{i2\pi Q_x^{(mod)}}}
            +2if_{1001,w}^*\frac{e^{i\Delta\phi_{y,wj}^{(mod)}}}{1-e^{i2\pi Q_y^{(mod)}}}
            \hspace{0mm}+O(f^2)\quad , \quad P_{41}=P_{32}^*\quad ,\\
P_{32,wj}&=&-2if_{1010,j}\frac{e^{-i\Delta\phi_{x,wj}^{(mod)}}}{1-e^{-i2\pi Q_x^{(mod)}}}
            +2if_{1010,w}\frac{e^{i\Delta\phi_{y,wj}^{(mod)}}}{1-e^{i2\pi Q_y^{(mod)}}}
            \hspace{0mm}+O(f^2)\quad , \quad P_{42}=P_{31}^*\quad ,\\
P_{33,wj}&=&\frac{e^{i\Delta\phi_{y,wj}^{(mod)}}}{1-e^{i2\pi Q_y^{(mod)}}}
            \hspace{55mm}+O(f^2) \quad , \quad P_{43}=P_{34}^*\quad ,\\
P_{34,wj}&=&0\hspace{75mm}+O(f^2)\quad , \quad P_{44}=P_{33}^* \quad .
\end{aligned}
\end{eqnarray}
The ORM of Eq.~\eqref{eq:co2} then reads
\begin{eqnarray}\label{eq:co3}
\begin{aligned}
&O^{(xx)}_{wj}=\sqrt{\beta_{j,x}\beta_{w,x}}\Re\left\{-i
                  P_{11,wj}\right\}\ ,\quad
O^{(xy)}_{wj}=\sqrt{\beta_{j,x}\beta_{w,y}}\Re\left\{i
                  (P_{14,wj}-P_{13,wj})\right\}\ , \\
&O^{(yx)}_{wj}=\sqrt{\beta_{j,y}\beta_{w,x}}\Re\left\{i
                  (P_{32,wj}-P_{31,wj})\right\}\ ,\quad
O^{(yy)}_{wj}=\sqrt{\beta_{j,y}\beta_{w,y}}\Re\left\{-i
                  P_{33,wj}\right\}\ .
\end{aligned}\qquad
\end{eqnarray}

\subsection{ORM response due to quadrupole errors}
In this section explicit formulas for the evaluation of 
the impact of a focusing error $\delta K_1$ on the diagonal blocks 
of the ORM $O^{(xx)}_{wj}$ and $O^{(yy)}_{wj}$
are derived. These allow the direct computation of the 
matrix $\mathbf{N}$ of Eq.~\eqref{eq:ORM_04} from 
the ideal C-S parameters with no need of computing 
numerically the derivative of the ORM with respect to 
$\delta K_1$. The detailed mathematical derivation is 
carried out for the horizontal block, the calculations 
for the vertical one being identical. Since the actual 
C-S parameters (i.e. including the focusing error) are 
used, $O^{(xx)}_{wj}$ simplifies to
\begin{eqnarray}\label{eq:co3B}
O^{(xx)}_{wj}=\sqrt{\beta_{j,x}\beta_{w,x}}\Re\left\{i                 
\frac{e^{i\mathbf{\Delta\phi}_{wj}}}{\mathbf{1}-e^{i\mathbf{Q}}}
\right\}^{(1,1\rightarrow2)} + O(f^2_{1001},f^2_{1010})\ .
\end{eqnarray}
This complex notation reduces to the standard formulas 
in the ideal case (with model C-S parameters and no 
betatron coupling, $f_{1001}=f_{1010}=0$) since
\begin{eqnarray}\nonumber
O^{(xx,mod)}_{wj}&=&\sqrt{\beta_{j,x}^{(mod)}\beta_{w,x}^{(mod)}}
       \Re\left\{i\frac{e^{i\mathbf{\Delta\phi}_{wj}^{(mod)}}}
                           {\mathbf{1}-e^{i\mathbf{Q}^{(mod)}}}
\right\}^{(1,1\rightarrow2)}\\ \nonumber &=&
\sqrt{\beta_{j,x}^{(mod)}\beta_{w,x}^{(mod)}}\Re\left\{-i                 
\frac{e^{i\Delta\phi_{x,wj}^{(mod)}}}{1-e^{i2\pi Q_x^{(mod)}}}\right\}= 
\frac{\sqrt{\beta_{j,x}^{(mod)}\beta_{w,x}^{(mod)}}}{2\sin{(\pi Q_x^{(mod)})}}
\Re\left\{e^{i(\Delta\phi_{x,wj}^{(mod)}-\pi Q_x^{(mod)})}\right\}\\  &=&
\frac{\sqrt{\beta_{j,x}^{(mod)}\beta_{w,x}^{(mod)}}}{2\sin{(\pi Q_x^{(mod)})}}
\cos{(\Delta\phi_{x,wj}^{(mod)}-\pi Q_x^{(mod)})}\ ,\label{eq:ORMIde}
\end{eqnarray}
where the following identity has been used $-i/(1-e^{i2\pi Q_x})=
e^{-i\pi Q_x}/(2\sin{\pi Q_x})$. In Ref.~\cite{Andrea-Linear-arxiv} 
analytic formulas relating the actual C-S parameters to the 
ideal ones (from the model) and the RDTs were derived:
\begin{eqnarray}\label{insertion-xy} 
&\hspace{-5mm}\left\{\begin{aligned}
\beta_{x_j}&=\beta_{x,j}^{(mod)}
             \left(1 +8\Im\left\{f_{2000,j}\right\}\right)+O(f_{2000}^2)\\
\alpha_{x,j}&=\alpha_{x,j}^{(mod)}\left(1+8\Im\{f_{2000,j}\}\right)
             -8\Re\{f_{2000,j}\} +O(f_{2000}^2)\\
\Delta\phi_{x,wj}&=\Delta\phi_{x,wj}^{(mod)}
                 \hspace{-.7mm}-\hspace{-.7mm}2h_{1100,wj} 
                 +4\Re\left\{f_{2000,j}-f_{2000,w}\right\}+O(f_{2000}^2)\\
h_{1100,wj}&=-\frac{1}{4}\sum_{w<m<j}{\beta_{m,x}^{(mod)}\delta K_{m,1}}
             +O(\delta K_1^2) \\
\end{aligned}\right. \ .
\end{eqnarray}
In the vertical plane identical relations apply, with the 
only difference that the detuning term $h_{1100,wj}$ is replaced 
by $h_{0011,wj}=+\frac{1}{4}\sum{\beta_{m,y}^{(mod)}\delta K_{m,1}}
+O(\delta K_1^2)$, the sum in both coefficient being over all 
quadrupole errors between the positions $j$ and $w$. The 
above definitions of $h_{wj}$ require that the two positions 
are such that $s_j>s_m>s_w$. If $s_j<s_w$ they 
are no longer valid and need to be tweaked, as shown later 
in Eq.~\eqref{eq:detun2}. By replacing the C-S parameters of 
Eq.~\eqref{insertion-xy} in the elements of Eq.~\eqref{eq:co3B}, 
we obtain: 
\begin{itemize}
\item $\sqrt{\beta_{j,x}\beta_{w,x}}=
        \sqrt{\beta_{j,x}^{(mod)}\beta_{w,x}^{(mod)}}\left(1+
        4\Im\left\{f_{2000,j}+f_{2000,w}\right\}\right)+O(f_{2000}^2)\ .$
\item $e^{i\Delta\phi_{x,wj}}= e^{i\Delta\phi_{x,wj}^{(mod)}}
        \left[1-2ih_{1100,wj} +4i\Re\left\{f_{2000,j}-f_{2000,w}\right\}
        \right]+O(h_{1100}^2,f_{2000}^2)\ .$
\item $\displaystyle \frac{1}{1-e^{i2\pi Q_x}}=
      \frac{1}{1-e^{i2\pi Q_x^{(mod)}-2ih_{1100}}}=
      \frac{1}{1-e^{i2\pi Q_x^{(mod)}}(1-2ih_{1100})}+O(h_{1100}^2)$ \\
      $$\hspace{-3.6cm}\displaystyle=
        \frac{1}{1-e^{i2\pi Q_x^{(mod)}}}\left[1-
        \frac{i2h_{1100\ }e^{i2\pi Q_x^{(mod)}}}{1-e^{i2\pi Q_x^{(mod)}}}\right]
        +O(h_{1100}^2)$$ 
      $$\hspace{-4.0cm}\displaystyle=
        \frac{1}{1-e^{i2\pi Q_x^{(mod)}}}\left[1+
        \frac{h_{1100\ }e^{i\pi Q_x^{(mod)}}}{\sin{(\pi Q_x^{(mod)})}}\right]
        +O(h_{1100}^2)\ .$$
\end{itemize}
The detuning coefficient $h_{1100}$ is the same of 
Eq.~\eqref{insertion-xy}, with the only difference that the sums 
extends over all quadrupole errors along the ring. 
Since $\left\{\frac{e^{i\mathbf{\Delta\phi}_{wj}}}{\mathbf{1}-e^{i\mathbf{Q}}}\right\}$ 
is a diagonal matrix, 
\begin{equation}
\left\{i\frac{e^{i\mathbf{\Delta\phi}_{wj}}}{\mathbf{1}-e^{i\mathbf{Q}}}\right\}^{(1,1\rightarrow2)}=
-\left\{i\frac{e^{i\mathbf{\Delta\phi}_{wj}}}{\mathbf{1}-e^{i\mathbf{Q}}}\right\}^{(1,1)}=
\frac{-ie^{i\Delta\phi_{x,wj}}}{1-e^{i2\pi Q_x}}\ ,
\end{equation}
and Eq.~\eqref{eq:co3B} reads 
\begin{eqnarray}\nonumber 
O^{(xx)}_{wj}&=&\sqrt{\beta_{j,x}^{(mod)}\beta_{w,x}^{(mod)}}
             \left(1+4\Im\left\{f_{2000,j}+f_{2000,w}\right\}\right)
             \Re\left\{\frac{-ie^{i\Delta\phi_{x,wj}^{(mod)}}}
                       {1-e^{i2\pi Q_x^{(mod)}}}\right.\times \\
&&\quad\left.\left[1-2ih_{1100,wj} +4i\Re\left\{f_{2000,j}-f_{2000,w}\right\}\right]
  \left[1+\frac{h_{1100\ }e^{i\pi Q_x^{(mod)}}}{\sin{(\pi Q_x^{(mod)})}}\right]
   \right\}+O(h_{1100}^2,f_{2000}^2)\nonumber \\
&=&\sqrt{\beta_{j,x}^{(mod)}\beta_{w,x}^{(mod)}}
             \left(1+4\Im\left\{f_{2000,j}+f_{2000,w}\right\}\right)
             \Re\left\{\frac{-ie^{i\Delta\phi_{x,wj}^{(mod)}}}
                       {1-e^{i2\pi Q_x^{(mod)}}}\right.\times \nonumber  \\
&&\quad\left.\left[1-2ih_{1100,wj} +4i\Re\left\{f_{2000,j}-f_{2000,w}\right\}
  +\frac{h_{1100\ }e^{i\pi Q_x^{(mod)}}}{\sin{(\pi Q_x^{(mod)})}}\right]
   \right\}+O(h_{1100}^2,f_{2000}^2)\nonumber  \\
&=&\sqrt{\beta_{j,x}^{(mod)}\beta_{w,x}^{(mod)}}
             \Re\left\{\frac{-ie^{i\Delta\phi_{x,wj}^{(mod)}}}
                       {1-e^{i2\pi Q_x^{(mod)}}}\right.\times \nonumber  \\
&&\quad\left.\left[1-2ih_{1100,wj} +4i\Re\left\{f_{2000,j}-f_{2000,w}\right\}
  +\frac{h_{1100\ }e^{i\pi Q_x^{(mod)}}}{\sin{(\pi Q_x^{(mod)})}}
  +4\Im\left\{f_{2000,j}+f_{2000,w}\right\}\right]
   \right\}+O(h_{1100}^2,f_{2000}^2)\nonumber \\
&=&\frac{\sqrt{\beta_{j,x}^{(mod)}\beta_{w,x}^{(mod)}}}
              {2\sin{(\pi Q_x^{(mod)})}}
             \Re\left\{e^{i(\Delta\phi_{x,wj}^{(mod)}-\pi Q_x^{(mod)})}
               \right.\times \nonumber  \\
&&\quad\left.\left[1-2ih_{1100,wj}+4i\Re\left\{f_{2000,j}-f_{2000,w}\right\}
  +\frac{h_{1100\ }e^{i\pi Q_x^{(mod)}}}{\sin{(\pi Q_x^{(mod)})}}
  +4\Im\left\{f_{2000,j}+f_{2000,w}\right\}\right]
   \right\}+O(h_{1100}^2,f_{2000}^2)\nonumber  \\
&=&\frac{\sqrt{\beta_{j,x}^{(mod)}\beta_{w,x}^{(mod)}}}
              {2\sin{(\pi Q_x^{(mod)})}}
             \Re\left\{e^{i(\Delta\phi_{x,wj}^{(mod)}-\pi Q_x^{(mod)})}
               \Big[1+4\Im\left\{f_{2000,j}+f_{2000,w}\right\}
    -i\big(2h_{1100,wj}-4\Re\left\{f_{2000,j}-f_{2000,w}\right\}\big)
   \Big] \right. \nonumber \\
&&\hspace{3.2cm}\left.
  +\frac{h_{1100\ }e^{i\Delta\phi_{x,wj}^{(mod)}}}{\sin{(\pi Q_x^{(mod)})}}
   \right\}+O(h_{1100}^2,f_{2000}^2)\ ,\label{eq:co4}
\end{eqnarray}
where the remainder is always proportional to $h_{1100}^2$ and 
$f_{2000}^2$, and hence to the square of quadrupole error field 
$\delta K_1^2$. Making explicit in the above expression the real 
part of the above curly brackets results in
\begin{eqnarray}\nonumber 
O^{(xx)}_{wj}&=&\frac{\sqrt{\beta_{j,x}^{(mod)}\beta_{w,x}^{(mod)}}}
              {2\sin{(\pi Q_x^{(mod)})}}\Bigg\{
        \ \cos{(\Delta\phi_{x,wj}^{(mod)}-\pi Q_x^{(mod)})}
              \Big[1+4\Im\left\{f_{2000,j}+f_{2000,w}\right\}\Big]
              \\ \nonumber &&\hspace{2.55cm}
        +\sin{(\Delta\phi_{x,wj}^{(mod)}-\pi Q_x^{(mod)})}
             \Big[2h_{1100,wj}-4\Re\left\{f_{2000,j}-f_{2000,w}\right\}\Big]
              \\ &&\hspace{2.55cm}
        +\cos{(\Delta\phi_{x,wj}^{(mod)})}
              \frac{h_{1100}}{\sin{(\pi Q_x^{(mod)})}}\Bigg\}+O(\delta K_1^2)
\ .\label{eq:co4B}
\end{eqnarray}
The first term within the first square brackets is the ideal 
ORM block $O^{(xx,mod)}_{wj}$ of Eq.~\eqref{eq:ORMIde}. Hence 
the difference $\delta O^{(xx)}_{wj}$ of Eq.~\eqref{eq:ORM_04} reads 
\begin{eqnarray}\nonumber 
\delta O^{(xx)}_{wj}&=&\frac{\sqrt{\beta_{j,x}^{(mod)}\beta_{w,x}^{(mod)}}}
              {2\sin{(\pi Q_x^{(mod)})}}\Bigg\{
        \ \cos{(\Delta\phi_{x,wj}^{(mod)}-\pi Q_x^{(mod)})}
              \Big[4\Im\left\{f_{2000,j}+f_{2000,w}\right\}\Big]
              \\ \nonumber &&\hspace{2.55cm}
        +\sin{(\Delta\phi_{x,wj}^{(mod)}-\pi Q_x^{(mod)})}
             \Big[2h_{1100,wj}-4\Re\left\{f_{2000,j}-f_{2000,w}\right\}\Big]
              \\ &&\hspace{2.55cm}
        +\cos{(\Delta\phi_{x,wj}^{(mod)})}
              \frac{h_{1100}}{\sin{(\pi Q_x^{(mod)})}}\Bigg\}+O(\delta K_1^2)
\ .\label{eq:co5}
\end{eqnarray}
The same algebra applied to the vertical diagonal block yields 
\begin{eqnarray}\nonumber 
\delta O^{(yy)}_{wj}&=&\frac{\sqrt{\beta_{j,y}^{(mod)}\beta_{w,y}^{(mod)}}}
              {2\sin{(\pi Q_y^{(mod)})}}\Bigg\{
        \ \cos{(\Delta\phi_{y,wj}^{(mod)}-\pi Q_y^{(mod)})}
              \Big[4\Im\left\{f_{0020,j}+f_{0020,w}\right\}\Big]
              \\ \nonumber &&\hspace{2.55cm}
        +\sin{(\Delta\phi_{y,wj}^{(mod)}-\pi Q_y^{(mod)})}
             \Big[2h_{0011,wj}-4\Re\left\{f_{0020,j}-f_{0020,w}\right\}\Big]
               \\ &&\hspace{2.55cm}
        +\cos{(\Delta\phi_{y,wj}^{(mod)})}
              \frac{h_{0011}}{\sin{(\pi Q_y^{(mod)})}}\Bigg\}+O(\delta K_1^2)
\ .\label{eq:co6}
\end{eqnarray}
The next step is to make explicit the focusing error RDTs and 
the detuning terms so to factorize the dependence on the 
quadrupole errors $\delta K_1$. To first order, the RDTs 
at a location $j$ read~\cite{Andrea-arxiv}
\begin{eqnarray} 
\left\{\begin{aligned}
f_{2000,w}&=-\frac{\sum\limits_{m=1}^M \delta K_{m,1}\beta_{m,x}^{(mod)}
        	e^{2i\Delta\phi_{x,mw}^{(mod)}}}
            {8(1-e^{4\pi iQ_x})}+O(\delta K_1^2)
           =-\frac{i}{16\sin{(2\pi Q_x)}}
           \sum\limits_{m=1}^M \delta K_{m,1}\beta_{m,x}^{(mod)}
       	e^{i(2\Delta\phi_{x,mw}^{(mod)}-2\pi Q_x)}+O(\delta K_1^2)\\
f_{0020,w}&=+\frac{\sum\limits_{m=1}^M \delta K_{m,1}\beta_{m,y}^{(mod)}
	        e^{2i\Delta\phi_{y,mw}^{(mod)}}}
        {8(1-e^{4\pi iQ_y})}+O(\delta K_1^2) 
           =+\frac{i}{16\sin{(2\pi Q_y)}}
           \sum\limits_{m=1}^M \delta K_{m,1}\beta_{m,y}^{(mod)}
       	e^{i(2\Delta\phi_{y,mw}^{(mod)}-2\pi Q_y)}+O(\delta K_1^2)
\end{aligned}\right. \ , \nonumber \\\label{eq:def_f2000}
\end{eqnarray}
where $M$ is the number of all sources of quadrupolar errors along 
the ring. The corresponding real and imaginary parts then are
\begin{eqnarray} 
\begin{aligned}
\Re\left\{f_{2000,w}\right\}&=\quad \sum\limits_{m=1}^M{
   \frac{\delta K_{m,1}\beta_{m,x}^{(mod)}}{16\sin{(2\pi Q_x)}}
   \sin{(2\Delta\phi_{x,mw}^{(mod)}-2\pi Q_x)}}+O(\delta K_1^2)\quad ,\\
\Im\left\{f_{2000,w}\right\}&=-\sum\limits_{m=1}^M{
   \frac{\delta K_{m,1}\beta_{m,x}^{(mod)}}{16\sin{(2\pi Q_x)}}
   \cos{(2\Delta\phi_{x,mw}^{(mod)}-2\pi Q_x)}}+O(\delta K_1^2)\quad ,\\
\Re\left\{f_{0020,w}\right\}&=-\sum\limits_{m=1}^M{
   \frac{\delta K_{m,1}\beta_{m,y}^{(mod)}}{16\sin{(2\pi Q_y)}}
   \sin{(2\Delta\phi_{y,mw}^{(mod)}-2\pi Q_y)}}+O(\delta K_1^2)\quad ,\\
\Im\left\{f_{0020,w}\right\}&=\quad\sum\limits_{m=1}^M{
   \frac{\delta K_{m,1}\beta_{m,y}^{(mod)}}{16\sin{(2\pi Q_y)}}
   \cos{(2\Delta\phi_{y,mw}^{(mod)}-2\pi Q_y)}}+O(\delta K_1^2)\quad .
\end{aligned} \label{eq:ReIm_f2000}
\end{eqnarray}
The following quantities can be hence evaluated
\begin{eqnarray} 
\begin{aligned}
\Im\left\{f_{2000,j}+f_{2000,w}\right\}&=-\sum\limits_{m=1}^M{
       \frac{\delta K_{m,1}\beta_{m,x}^{(mod)}}{16\sin{(2\pi Q_x)}}
       \left[\cos{(2\Delta\phi_{x,mj}^{(mod)}-2\pi Q_x)}+
             \cos{(2\Delta\phi_{x,mw}^{(mod)}-2\pi Q_x)}\right]}
       +O(\delta K_1^2)\quad ,\qquad \\
\Re\left\{f_{2000,j}-f_{2000,w}\right\}&=+\sum\limits_{m=1}^M{
       \frac{\delta K_{m,1}\beta_{m,x}^{(mod)}}{16\sin{(2\pi Q_x)}}
       \left[\sin{(2\Delta\phi_{x,mj}^{(mod)}-2\pi Q_x)}-
             \sin{(2\Delta\phi_{x,mw}^{(mod)}-2\pi Q_x)}\right]}
      +O(\delta K_1^2)\quad ,\qquad \\
\Im\left\{f_{0020,j}+f_{0020,w}\right\}&=+\sum\limits_{m=1}^M{
       \frac{\delta K_{m,1}\beta_{m,y}^{(mod)}}{16\sin{(2\pi Q_y)}}
       \left[\cos{(2\Delta\phi_{y,mj}^{(mod)}-2\pi Q_y)}+
             \cos{(2\Delta\phi_{y,mw}^{(mod)}-2\pi Q_y)}\right]}
       +O(\delta K_1^2)\quad ,\qquad  \\
\Re\left\{f_{0020,j}-f_{0020,w}\right\}&=-\sum\limits_{m=1}^M{
       \frac{\delta K_{m,1}\beta_{m,y}^{(mod)}}{16\sin{(2\pi Q_y)}}
       \left[\sin{(2\Delta\phi_{y,mj}^{(mod)}-2\pi Q_y)}-
             \sin{(2\Delta\phi_{y,mw}^{(mod)}-2\pi Q_y)}\right]}
       +O(\delta K_1^2)\quad .\qquad  
\end{aligned}\label{eq:ReIm_SumDif}
\end{eqnarray}
The detuning terms in Eqs.~\eqref{eq:co4}-\eqref{eq:co5} descend 
from Eq.~\eqref{insertion-xy}
\begin{eqnarray}\label{eq:detun2} 
\begin{aligned}
h_{1100,wj}&=-\frac{1}{4}\sum_{m=1}^M{\beta_{m,x}^{(mod)}\delta K_{m,1}
              \left[\Pi(m,j)-\Pi(m,w)+\Pi(j,w)\right]}+O(\delta K_1^2)
\ ,\  
h_{1100}=-\frac{1}{4}\sum_{m=1}^M{\beta_{m,x}^{(mod)}\delta K_{m,1}}
             +O(\delta K_1^2)\ ,\\
h_{0011,wj}&=+\frac{1}{4}\sum_{m=1}^M{\beta_{m,y}^{(mod)}\delta K_{m,1}
              \left[\Pi(m,j)-\Pi(m,w)+\Pi(j,w)\right]}+O(\delta K_1^2)
\ ,\ 
h_{0011}=+\frac{1}{4}\sum_{m=1}^M{\beta_{m,y}^{(mod)}\delta K_{m,1}}
             +O(\delta K_1^2)\ ,
\end{aligned}
\end{eqnarray}
where the function $\Pi$ is introduced so to have the same sum index 
in $h_{wj}$ and $h$, while accounting for the limited range of $h_{wj}$, 
and is defined as 
\begin{eqnarray}\label{eq:Theta} 
\Pi(a,b)=1\quad\hbox{if\ }s_a<s_b\quad,\qquad
\Pi(a,b)=0\quad\hbox{if\ }s_a\ge s_b\quad ,
\end{eqnarray}
$s_a$ and $s_b$ being the longitudinal position of the elements 
$a$ and $b$, respectively. The function $\Pi(j,w)$ is included 
in the definition of $h_{wj}$ of Eq.~\eqref{eq:detun2} to account 
for the case in which $s_w>s_j$. This issue was already encountered 
in the computation of the phase advance $\Delta\phi_{x,wj}$ of
Eq.~\eqref{eq:deltaphisign} from the betatron phases and was 
fixed by adding $2\pi Q$ each time $s_w>s_j$. As for the 
phase advance, the subscript $wj$ delimits a region with the 
second element $j$ downstream the first element $w$: If $s_w>s_j$ 
care needs to be taken in definition of the correct region. 
The sketches of Fig.~\ref{fig_detuning1} should clarify 
the concept. If $s_j>s_w$ (left drawing), $\Pi(j,w)=0$ and $h_{wj}$ 
is correctly defined by the quadrupole errors $m=3,4,5,6$ between the 
element $w$ and $j$. Without $\Pi(j,w)$, if $s_j<s_w$ (center drawing) 
$h_{wj}$ would be wrongly defined by the elements $m=7,8,9$ and with the 
wrong sign. To compute the correct $h_{wj}$ with the element $m=10,11$ 
and $m=1,2,3,4,5,6$ the whole detuning term $h$ is to be added, which 
is equivalent to include $\Pi(j,w)$ in Eq.~\eqref{eq:detun2} (right 
drawing).

\begin{figure}[!b]
\rule{0mm}{3mm}
\centerline{\includegraphics[width=5.3cm]{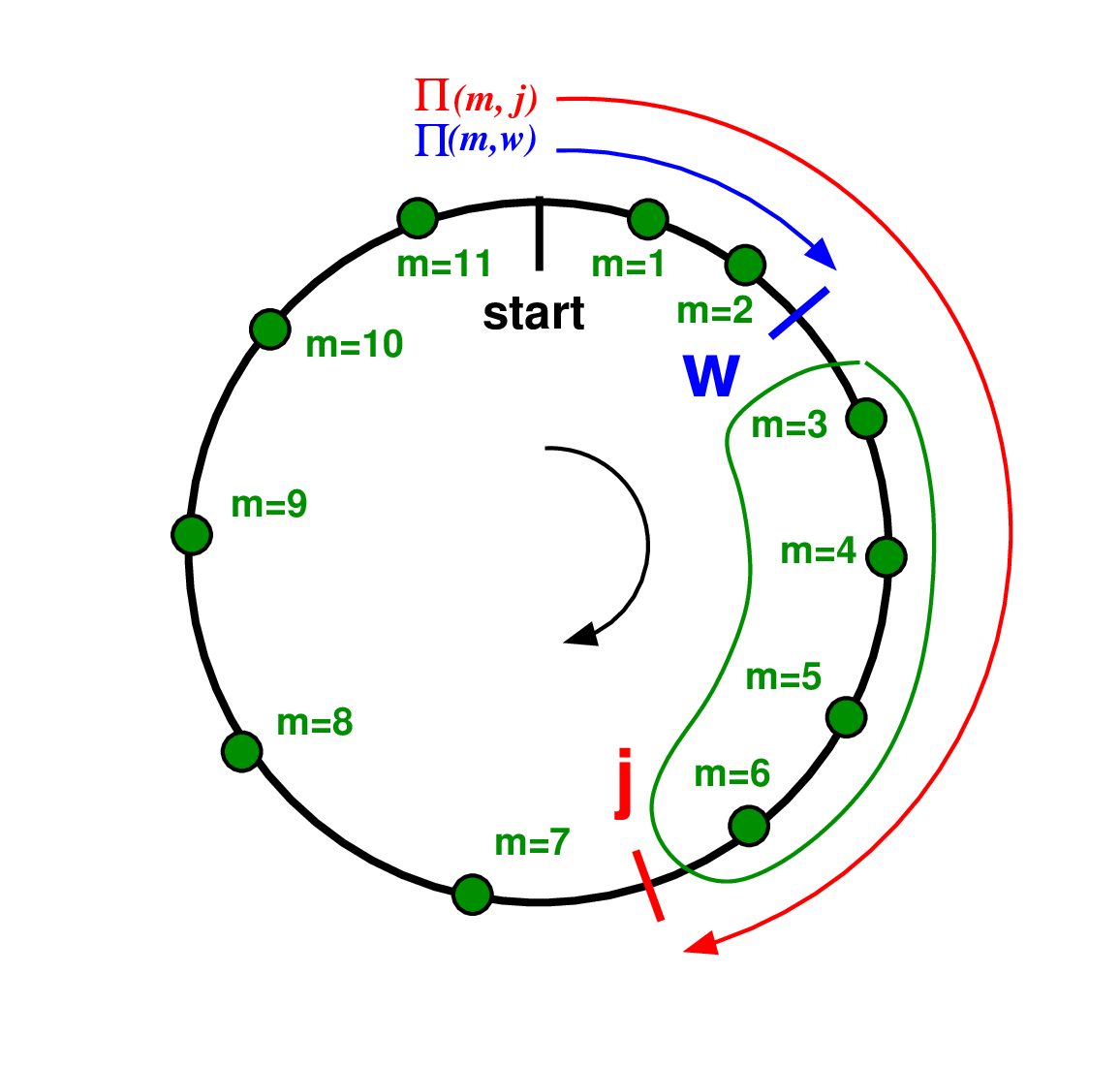}
\ \ \ \     \includegraphics[width=5.3cm]{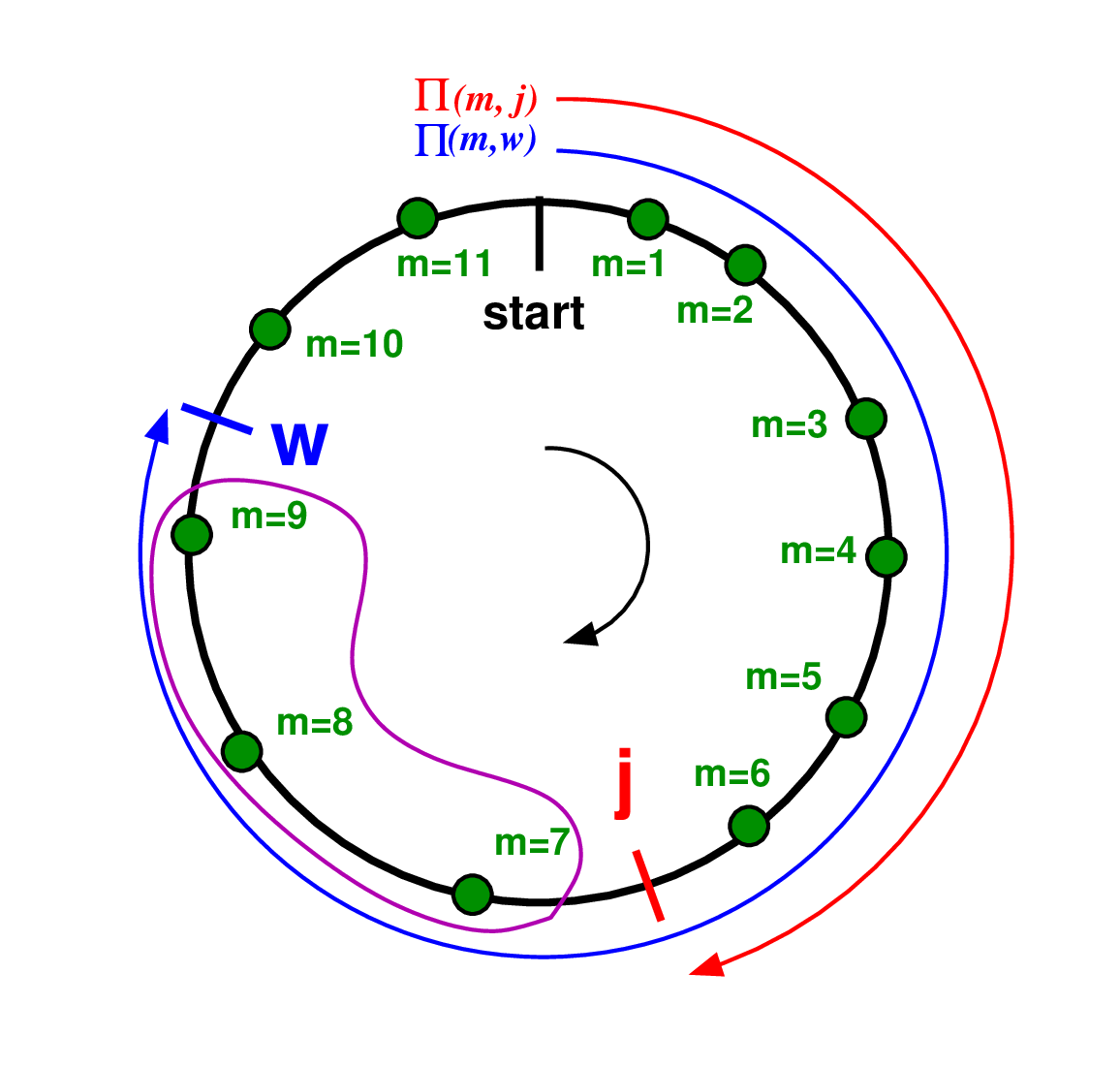}
\ \ \ \     \includegraphics[width=5.3cm]{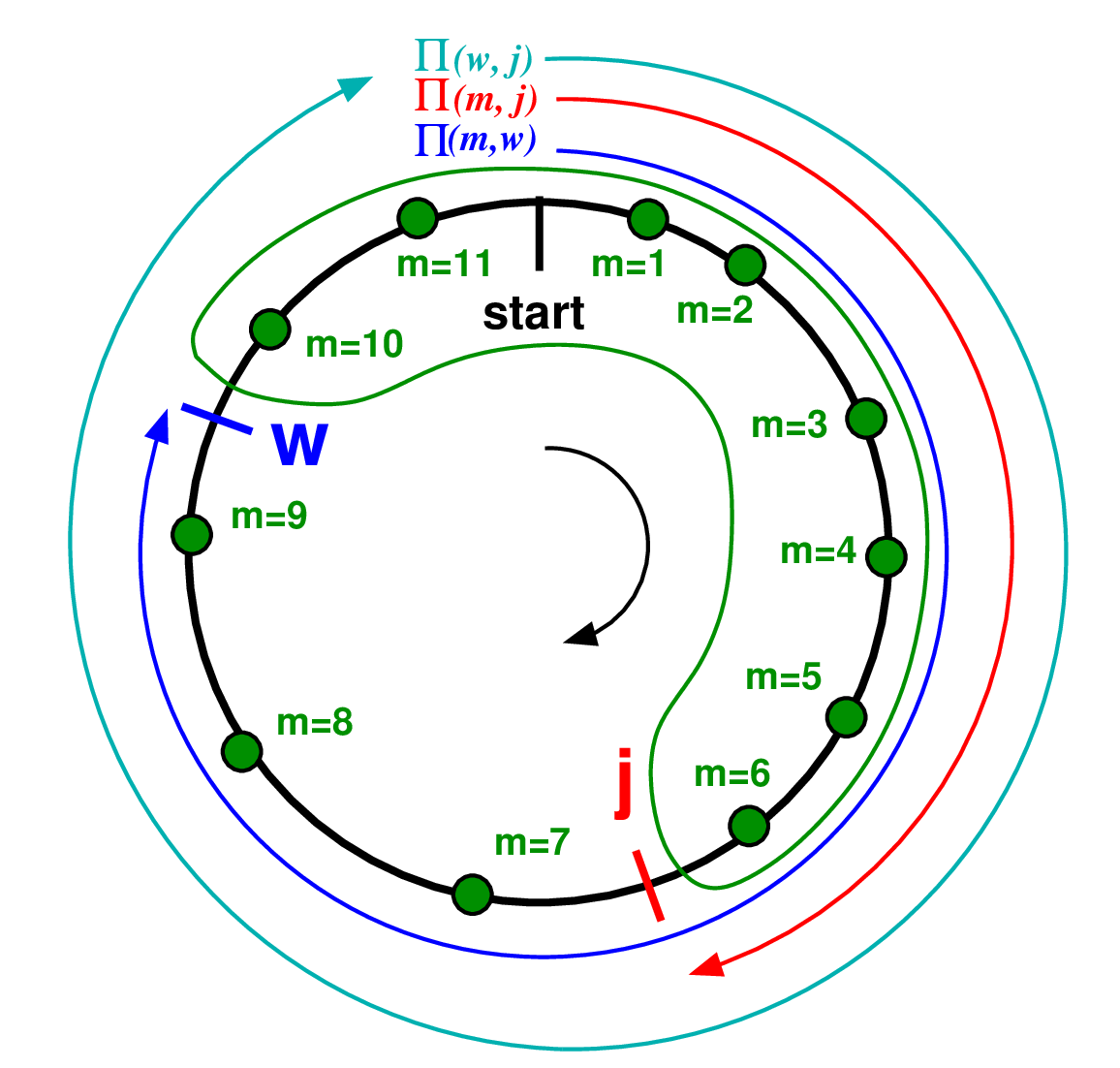}}
  \caption{\label{fig_detuning1} (Color) Two possible configurations 
           with $s_j>s_w$ (left) and $s_w>s_j$ (center, wrong, and 
           right, correct). See text for a detailed explanation.}
\rule{0mm}{3mm}
\end{figure}
By inserting Eqs.~\eqref{eq:detun2}-\eqref{eq:detun2} into 
Eqs.~\eqref{eq:co5}-\eqref{eq:co6} the explicit dependence of the 
ORM diagonal blocks upon the quadrupole error is derived, namely
\begin{eqnarray}\nonumber 
\delta O^{(xx)}_{wj}&\simeq&-\sum\limits_{m=1}^M{
  \frac{\sqrt{\beta_{j,x}^{(mod)}\beta_{w,x}^{(mod)}}\beta_{m,x}^{(mod)}}
       {2\sin{(\pi Q_x^{(mod)})}}
 \Bigg\{\frac{\cos{(\Delta\phi_{x,wj}^{(mod)}-\pi Q_x^{(mod)})}}
              {4\sin{(2\pi Q_x^{(mod)})}}
              \Big[\cos{(2\Delta\phi_{x,mj}^{(mod)}-2\pi Q_x)}
                  +\cos{(2\Delta\phi_{x,mw}^{(mod)}-2\pi Q_x)}\Big]}
              \\ \nonumber &&\hspace{4.3cm}
        +\frac{\ \sin{(\Delta\phi_{x,wj}^{(mod)}-\pi Q_x^{(mod)})}}
              {4\sin{(2\pi Q_x^{(mod)})}}
             \Big[\sin{(2\Delta\phi_{x,mj}^{(mod)}-2\pi Q_x)}
               -\ \sin{(2\Delta\phi_{x,mw}^{(mod)}-2\pi Q_x)}\Big]
              \\ \nonumber &&\hspace{3.2cm}
       +\frac{1}{2}\sin{(\Delta\phi_{x,wj}^{(mod)}-\pi Q_x^{(mod)})}
        \left[\Pi(m,j)-\Pi(m,w)+\Pi(j,w)\right]
       +\frac{\cos{(\Delta\phi_{x,wj}^{(mod)})}}
            {4\sin{(\pi Q_x^{(mod)})}}\Bigg\}\delta K_{m,1}\ ,
\\ &&\hspace{0.4cm} \label{eq:co7} \\ \nonumber
\delta O^{(yy)}_{wj}&\simeq&+\sum\limits_{m=1}^M{
  \frac{\sqrt{\beta_{j,y}^{(mod)}\beta_{w,y}^{(mod)}}\beta_{m,y}^{(mod)}}
       {2\sin{(\pi Q_y^{(mod)})}}
 \Bigg\{\frac{\cos{(\Delta\phi_{y,wj}^{(mod)}-\pi Q_y^{(mod)})}}
              {4\sin{(2\pi Q_y^{(mod)})}}
              \Big[\cos{(2\Delta\phi_{y,mj}^{(mod)}-2\pi Q_y)}
                  +\cos{(2\Delta\phi_{y,mw}^{(mod)}-2\pi Q_y)}\Big]}
              \\ \nonumber &&\hspace{4.3cm}
        +\frac{\ \sin{(\Delta\phi_{y,wj}^{(mod)}-\pi Q_y^{(mod)})}}
              {4\sin{(2\pi Q_y^{(mod)})}}
             \Big[\sin{(2\Delta\phi_{y,mj}^{(mod)}-2\pi Q_y)}
               -\ \sin{(2\Delta\phi_{y,mw}^{(mod)}-2\pi Q_y)}\Big]
              \\ \nonumber &&\hspace{3.2cm}
       +\frac{1}{2}\sin{(\Delta\phi_{y,wj}^{(mod)}-\pi Q_y^{(mod)})}
        \left[\Pi(m,j)-\Pi(m,w)+\Pi(j,w)\right]
       +\frac{\cos{(\Delta\phi_{y,wj}^{(mod)})}}
            {4\sin{(\pi Q_y^{(mod)})}}\Bigg\}\delta K_{m,1}\ ,
\end{eqnarray}
where the remainder $O(\delta K_1^2)$ has been omitted. From the 
above equations, analytic expressions for the betatronic part of 
the response matrix $\mathbf{N}$ of Eq.~\eqref{eq:ORM_04}, i.e. 
of the derivative of $\delta O^{(xx)}_{wj}$ and $\delta O^{(yy)}_{wj}$ 
with respect to $\delta K_{m,1}$, are derived
\begin{eqnarray}\nonumber 
N^{(xx)}_{wj,m}&\simeq&-
  \frac{\sqrt{\beta_{j,x}^{(mod)}\beta_{w,x}^{(mod)}}\beta_{m,x}^{(mod)}}
       {2\sin{(\pi Q_x^{(mod)})}}
 \Bigg\{\frac{\cos{(\Delta\phi_{x,wj}^{(mod)}-\pi Q_x^{(mod)})}}
              {4\sin{(2\pi Q_x^{(mod)})}}
              \Big[\cos{(2\Delta\phi_{x,mj}^{(mod)}-2\pi Q_x)}
                  +\cos{(2\Delta\phi_{x,mw}^{(mod)}-2\pi Q_x)}\Big]
              \\ \nonumber &&\hspace{3.6cm}
        +\frac{\ \sin{(\Delta\phi_{x,wj}^{(mod)}-\pi Q_x^{(mod)})}}
              {4\sin{(2\pi Q_x^{(mod)})}}
             \Big[\sin{(2\Delta\phi_{x,mj}^{(mod)}-2\pi Q_x)}
               -\ \sin{(2\Delta\phi_{x,mw}^{(mod)}-2\pi Q_x)}\Big]
              \\ \nonumber &&\hspace{3.6cm}
       +\frac{1}{2}\sin{(\Delta\phi_{x,wj}^{(mod)}-\pi Q_x^{(mod)})}
        \left[\Pi(m,j)-\Pi(m,w)+\Pi(j,w)\right]
       +\frac{\cos{(\Delta\phi_{x,wj}^{(mod)})}}
            {4\sin{(\pi Q_x^{(mod)})}}\Bigg\}\ ,
\\ &&\hspace{0.4cm} \label{eq:co8} \\ \nonumber
N^{(yy)}_{wj,m}&\simeq&+
  \frac{\sqrt{\beta_{j,y}^{(mod)}\beta_{w,y}^{(mod)}}\beta_{m,y}^{(mod)}}
       {2\sin{(\pi Q_y^{(mod)})}}
 \Bigg\{\frac{\cos{(\Delta\phi_{y,wj}^{(mod)}-\pi Q_y^{(mod)})}}
              {4\sin{(2\pi Q_y^{(mod)})}}
              \Big[\cos{(2\Delta\phi_{y,mj}^{(mod)}-2\pi Q_y)}
                  +\cos{(2\Delta\phi_{y,mw}^{(mod)}-2\pi Q_y)}\Big]
              \\ \nonumber &&\hspace{3.6cm}
        +\frac{\ \sin{(\Delta\phi_{y,wj}^{(mod)}-\pi Q_y^{(mod)})}}
              {4\sin{(2\pi Q_y^{(mod)})}}
             \Big[\sin{(2\Delta\phi_{y,mj}^{(mod)}-2\pi Q_y)}
               -\ \sin{(2\Delta\phi_{y,mw}^{(mod)}-2\pi Q_y)}\Big]
              \\ \nonumber &&\hspace{3.6cm}
       +\frac{1}{2}\sin{(\Delta\phi_{y,wj}^{(mod)}-\pi Q_y^{(mod)})}
        \left[\Pi(m,j)-\Pi(m,w)+\Pi(j,w)\right]
       +\frac{\cos{(\Delta\phi_{y,wj}^{(mod)})}}
            {4\sin{(\pi Q_y^{(mod)})}}\Bigg\}\ ,
\end{eqnarray}
where again the remainder, this time linear in $\delta K_1$, has been 
omitted. The function $\Pi(a,b)$ is defined in Eq.~\eqref{eq:Theta}, 
whereas the (always positive) phase advance $\Delta\phi_{ab}$ is 
to be computed according to Eq.~\eqref{eq:deltaphisign}.

\subsection{ORM response due to skew quadrupole fields}
In this section explicit formulas for the evaluation of 
the impact of a skew quadrupole integrated strength 
$J_1$ on the off-diagonal blocks of the ORM, $O^{(xy)}_{wj}$ 
and $O^{(yx)}_{wj}$ of Eq.~\eqref{eq:co3}, are derived. 
These equations allow the direct computation of the 
matrix $\mathbf{S}$ of Eq.~\eqref{eq:ORM_05} from 
the C-S parameters with no need of computing 
numerically the derivative of the ORM with respect to 
$J_1$. It is assumed that the analysis of the ORM diagonal 
blocks is already carried out and a model comprising focusing 
errors is available, so to be able to compute the 
actual C-S parameters. These, and not the ideal ones, are 
to be used in the final formulas to ensure an RMS error 
within a few percents (numerical simulations showed that if 
the ideal C-S are used the discrepancy may increase up to 
$20\%$ for the old ESRF storage ring with the same beta-beating 
of Fig.~\ref{fig_Oxxyy1}). If large betatron coupling is 
present in the machine, the C-S parameters are affected by 
coupling RDTs, as shown in Ref.~\cite{prstab_esr_coupling}: 
This will corrupt the overall analysis and an iterative 
process of measurement and correction of linear lattice 
errors (focusing and coupling) is required. 

From Eqs.~\eqref{eq:ORM2},~\eqref{eq:co3}, the off-diagonal 
block corresponding to the horizontal orbit response 
to a vertical deflection reads
\begin{eqnarray}\label{eq:co9}
O^{(xy)}_{wj}&=&\sqrt{\beta_{j,x}\beta_{w,y}}\Re\left\{i
                  (P_{14,wj}-P_{13,wj})\right\}\ , \\ \nonumber
             &\simeq&\sqrt{\beta_{j,x}\beta_{w,y}}\Re\left\{i\left[
               2if_{1010,w}\frac{e^{ i\Delta\phi_{x,wj}}}{1-e^{ i2\pi Q_x}}
              -2if_{1010,j}\frac{e^{-i\Delta\phi_{y,wj}}}{1-e^{-i2\pi Q_y}}
              -2if_{1001,w}\frac{e^{ i\Delta\phi_{x,wj}}}{1-e^{ i2\pi Q_x}}
              +2if_{1001,j}\frac{e^{ i\Delta\phi_{y,wj}}}{1-e^{ i2\pi Q_y}}
              \right]\right\}\ ,
\end{eqnarray}
where higher-order terms $\propto O(f^2)$ have been neglected. 
By making use of the following identities and definitions
\begin{eqnarray}\label{eq:def_acc1}
\frac{2i}{1-e^{\pm 2iz}}=\mp\frac{e^{\mp iz}}{\sin{z}}\qquad , \hskip 1cm
\tau_{z,ab}=\Delta\phi_{z,ab}-\pi Q_z\ ,\quad z=x,y\quad .
\end{eqnarray}
Eq.~\eqref{eq:co9} simplifies to 
\begin{eqnarray}\label{eq:co10}
O^{(xy)}_{wj}&\simeq&\sqrt{\beta_{j,x}\beta_{w,y}}\Re\left\{i\left[
               -f_{1010,w}\frac{e^{ i\tau_{x,wj}}}{\sin{\pi Q_x}}
               -f_{1010,j}\frac{e^{-i\tau_{y,wj}}}{\sin{\pi Q_y}}
               +f_{1001,w}\frac{e^{ i\tau_{x,wj}}}{\sin{\pi Q_x}}
               -f_{1001,j}\frac{e^{ i\tau_{y,wj}}}{\sin{\pi Q_y}}
              \right]\right\}\ .
\end{eqnarray}
The coupling RDTs of Eq.~\eqref{eq:f1001} can be also written as 
\begin{eqnarray}
f_{{\tiny\begin{array}{c} 1001 \\ 1010\end{array}},\ j}\simeq
        \frac{\sum\limits_{m=1}^M J_{m,1}\sqrt{\beta_{m,x}\beta_{m,y}} 
        e^{i(\Delta\phi_{x,mj}\mp \Delta\phi_{y,mj})}}
        {4(1-e^{2\pi i(Q_u\mp Q_v)})}
    =\frac{i}{8\sin{[\pi(Q_x\mp Q_y)]}}
     \sum\limits_{m=1}^M J_{m,1}{\sqrt{\beta_{m,x}\beta_{m,y}}
          e^{i(\tau_{x,mj}\mp \tau_{y,mj})}} \ ,\quad
\end{eqnarray}
where again higher order terms $\propto O(J_1^2)$ are ignored. 
After replacing the RDTs in Eq.~\eqref{eq:co10} with the above 
expression, the off-diagonal block can be eventually written 
as a function of the skew quadrupole strength $J_{m,1}$:
\begin{eqnarray}
O^{(xy)}_{wj}\hskip -.05cm&\simeq&\hskip -.15cm\sum\limits_{m=1}^M\frac{J_{m,1}}{8}
      \sqrt{\beta_{j,x}\beta_{w,y}\beta_{m,x}\beta_{m,y}}
      \left\{\frac{1}{\sin{[\pi(Q_x-Q_y)]}}\left[
             \frac{\cos{(\tau_{x,mj}-\tau_{y,mj}+\tau_{y,wj})}}{\sin{\pi Q_y}}
            -\frac{\cos{(\tau_{x,mw}-\tau_{y,mw}+\tau_{x,wj})}}{\sin{\pi Q_x}}
      \right]\right.\nonumber \\ 
     &&\hskip 4.2cm\left.
            +\frac{1}{\sin{[\pi(Q_x+Q_y)]}}\left[
             \frac{\cos{(\tau_{x,mj}+\tau_{y,mj}-\tau_{y,wj})}}{\sin{\pi Q_y}}
            +\frac{\cos{(\tau_{x,mw}+\tau_{y,mw}+\tau_{x,wj})}}{\sin{\pi Q_x}}
      \right]\right\}\nonumber . \\ \label{eq:co11}
\end{eqnarray}
The same procedure applied to the vertical orbit response 
to a horizontal steerer results in 
\begin{eqnarray}
O^{(yx)}_{wj}\hskip -.05cm&\simeq&\hskip -.15cm\sum\limits_{m=1}^M\frac{J_{m,1}}{8}
      \sqrt{\beta_{j,y}\beta_{w,x}\beta_{m,x}\beta_{m,y}}
      \left\{\frac{1}{\sin{[\pi(Q_x-Q_y)]}}\left[
            -\frac{\cos{(\tau_{x,mj}-\tau_{y,mj}-\tau_{x,wj})}}{\sin{\pi Q_x}}
            +\frac{\cos{(\tau_{x,mw}-\tau_{y,mw}-\tau_{y,wj})}}{\sin{\pi Q_y}}
      \right]\right.\nonumber \\ 
     &&\hskip 4.2cm\left.
            +\frac{1}{\sin{[\pi(Q_x+Q_y)]}}\left[
             \frac{\cos{(\tau_{x,mj}+\tau_{y,mj}-\tau_{x,wj})}}{\sin{\pi Q_x}}
            +\frac{\cos{(\tau_{x,mw}+\tau_{y,mw}+\tau_{y,wj})}}{\sin{\pi Q_y}}
      \right]\right\}\nonumber . \\ \label{eq:co12}
\end{eqnarray}
From the above equations, analytic expressions for the betatronic 
part of the response matrix $\mathbf{S}$ of Eq.~\eqref{eq:ORM_05}, i.e. 
of the derivative of $\delta O^{(xy)}_{wj}$ and $\delta O^{(yx)}_{wj}$ 
with respect to $J_{m,1}$, are derived
\begin{eqnarray}\nonumber 
S^{(xy)}_{wj,m}&\simeq&
      \frac{1}{8}\sqrt{\beta_{j,x}\beta_{w,y}\beta_{m,x}\beta_{m,y}}
      \left\{\frac{1}{\sin{[\pi(Q_x-Q_y)]}}\left[
             \frac{\cos{(\tau_{x,mj}-\tau_{y,mj}+\tau_{y,wj})}}{\sin{\pi Q_y}}
            -\frac{\cos{(\tau_{x,mw}-\tau_{y,mw}+\tau_{x,wj})}}{\sin{\pi Q_x}}
      \right]\right.\nonumber \\ 
     &&\hskip 3.25cm\left.
            +\frac{1}{\sin{[\pi(Q_x+Q_y)]}}\left[
             \frac{\cos{(\tau_{x,mj}+\tau_{y,mj}-\tau_{y,wj})}}{\sin{\pi Q_y}}
            +\frac{\cos{(\tau_{x,mw}+\tau_{y,mw}+\tau_{x,wj})}}{\sin{\pi Q_x}}
      \right]\right\}\ ,
\nonumber\\ &&\hspace{0.4cm} \label{eq:co13} \\ \nonumber
S^{(yx)}_{wj,m}&\simeq&
      \frac{1}{8}\sqrt{\beta_{j,y}\beta_{w,x}\beta_{m,x}\beta_{m,y}}
      \left\{\frac{1}{\sin{[\pi(Q_x-Q_y)]}}\left[
            -\frac{\cos{(\tau_{x,mj}-\tau_{y,mj}-\tau_{x,wj})}}{\sin{\pi Q_x}}
            +\frac{\cos{(\tau_{x,mw}-\tau_{y,mw}-\tau_{y,wj})}}{\sin{\pi Q_y}}
      \right]\right.\nonumber \\ 
     &&\hskip 3.25cm\left.
            +\frac{1}{\sin{[\pi(Q_x+Q_y)]}}\left[
             \frac{\cos{(\tau_{x,mj}+\tau_{y,mj}-\tau_{x,wj})}}{\sin{\pi Q_x}}
            +\frac{\cos{(\tau_{x,mw}+\tau_{y,mw}+\tau_{y,wj})}}{\sin{\pi Q_y}}
      \right]\right\}\ ,\nonumber
\end{eqnarray}
where again the remainder, this time linear in $J_1$, 
has been omitted. $\tau_{ab}$ is defined in 
Eq.~\eqref{eq:def_acc1} from the phase advance 
$\Delta\phi_{ab}$ which is to be computed according 
to Eq.~\eqref{eq:deltaphisign}. The C-S parameters 
refer to the linear lattice including focusing errors.

\subsection{Impact of sextupoles in the ORM measurement}
\label{sec:app_sext}
Equations~\eqref{eq:co7} and~\eqref{eq:co12} have been 
derived ignoring the presence of sextupoles in the 
lattice. In reality, the orbit distortion at sextupoles 
induced by steerer magnets generates normal and skew 
quadrupole feed-down fields, $\delta K_{m,1}=-K_{m,2}x_{m,\rm c.o.}$ 
and $J_{m,1}=K_{m,2}y_{m,\rm c.o.}$, where $K_{m,2}$ 
denotes the integrated strength of the sextupole $m$
and $(x_{m,\rm c.o.},\ y_{m,\rm c.o.})$ is the 
corresponding closed orbit. Horizontal and vertical 
dipolar feed-down fields proportional to 
$(x_{m,\rm c.o.}^2,\ y_{m,\rm c.o.}^2)$ are also generated. 

From Eqs.~\eqref{eq:ORM_01} and~\eqref{eq:ORM-formula1} 
the closed orbit at a generic BPM $j$ induced by a steerer 
kick $\theta_w$ can be written as
\begin{equation}
x_j=O_{wj}\theta_w\quad\Rightarrow\quad
x_j=\left[O_{wj}^{(mod)}+\sum_m{N_{wj,m}\delta K_{m,1}}\right]\theta_w\ ,
\end{equation}
where $O_{wj}^{(mod)}$ is the ORM element for the 
ideal lattice. 
The focusing errors would then stem from quadrupole 
imperfections and from the feed-down (quadrupolar 
and dipolar) generated by sextupoles, namely
\begin{equation}
x_j=\left[O_{wj}^{(mod)}
        +\sum_m{N_{wj,m}^{(Q)}\delta K_{m,1}}
        -\sum_m{N_{wj,m}^{(S)}(K_{m,2}x_{m,\rm c.o.})}
        +O(x_{m,\rm c.o.}^2)\right]\theta_w\ .
\end{equation}
The ORM is usually measured by recording the orbit 
distortion $x_{j,\pm}$ generated by two opposite 
steerer strengths $\pm\theta_w$, with $Q_{wj}=(x_{j,+}-x_{j,-})/(2\theta_w)$. 
In the presence of sextupoles, the two orbits 
read
\begin{eqnarray}
x_{j,+}&=&\left[O_{wj}^{(mod)}
        +\sum_m{N_{wj,m}^{(Q)}\delta K_{m,1}}
        -\sum_m{N_{wj,m}^{(S)}(K_{m,2}x_{m,\rm c.o.})}
        +O(x_{m,\rm c.o.}^2)\right]\theta_w\ , \\
x_{j,-}&=&-\left[O_{wj}^{(mod)}
        +\sum_m{N_{wj,m}^{(Q)}\delta K_{m,1}}
        +\sum_m{N_{wj,m}^{(S)}(K_{m,2}x_{m,\rm c.o.})}
        +O(x_{m,\rm c.o.}^2)\right]\theta_w\ ,
\end{eqnarray}
where it is assumed that lattice errors are sufficiently 
small to have $x_{m,\rm c.o.,+}\simeq-x_{m,\rm c.o.,-}$. 
The measured ORM then is
\begin{equation}
Q_{wj}=O_{wj}^{(mod)}
        +\sum_m{N_{wj,m}^{(Q)}\delta K_{m,1}}
        +O(x_{m,\rm c.o.}^2)\ .
\end{equation}
Since $x_{m,\rm c.o.}\propto\theta_w$, the error is 
proportional to $\theta_w^2$. Equivalent considerations 
apply to the vertical orbit.

It is worthwhile noticing that the cancellation of 
the quadrupolar terms generated by sextupoles does 
not disappear if the orbit distortion is measured 
with an asymmetric perturbation, i.e. 
$\theta_{w,+}\ne-\theta_{w,-}$.



 
\section{Derivation of chromatic functions}
\label{app:2}
In this appendix analytical formulas for the 
chromatic functions (linear and nonlinear 
dispersion, chromaticity, chromatic beating and 
chromatic coupling) are derived. In order to 
greatly simplify the mathematics, it is assumed 
that focusing errors $\delta K_1$ are included 
in the model and in the computation of the C-S 
parameters, as done for the evaluation of betatron 
coupling. This requires that the analysis of the 
diagonal blocks of the ORM be carried out before 
evaluating the chromatic functions. 

Another assumption made here is that the edge 
focusing provided by nonzero dipole pole-face 
angles is negligible, the magnetic modelling 
being based on the multipolar expansion of 
Eq.~\eqref{eq:MADX}. This may introduce a 
systematic error in the evaluation of the 
chromatic functions from the following 
analytic formulas.  However, for the calibration 
of sextupole magnets and the correction of the 
chromatic functions, these formulas can still be 
effectively used, since any systematic error is 
canceled out. 

The Hamiltonian in complex C-S coordinates of 
Eq.~\eqref{eq:Hdef} is the starting point for 
the study of the 4D betatron motion. Chromatic 
effects may be inferred from the same Hamiltonian 
after replacing the betatron coordinates with the 
ones including dispersive terms:
\begin{eqnarray}
\left\{
\begin{aligned}
x   \rightarrow x+D_x\delta\\
p_x\rightarrow p_x+D_x'\delta
\end{aligned}
\right.   
\hskip 2mm \Rightarrow\hskip 2mm 
\left\{
\begin{aligned}
\tilde{x}   \rightarrow \tilde{x} +\tilde{D}_x\delta\\
\tilde{p}_x\rightarrow \tilde{p}_x+\tilde{D}_x'\delta
\end{aligned}
\right.  
\hskip 5mm ,\hskip 5mm 
\left\{
\begin{aligned}
y   \rightarrow y+D_y\delta\\
p_y\rightarrow p_y+D_y'\delta
\end{aligned}
\right. 
\hskip 2mm \Rightarrow\hskip 2mm
\left\{
\begin{aligned}
\tilde{y}   \rightarrow \tilde{y} +\tilde{D}_y\delta\\
\tilde{p}_y\rightarrow \tilde{p}_y+\tilde{D}_y'\delta
\end{aligned}
\right. \quad , \qquad
\end{eqnarray}
where $\delta=(p-p_0)/p_0$ represents the relative deviation 
from the reference momentum, $D$ and $D'$ denote the dispersion 
and its derivative in Cartesian coordinates, whereas $\tilde{D}$ 
and $\tilde{D}'$ are the equivalent in the C-S coordinates. The 
above relations result in 
\begin{eqnarray}
\left\{
\begin{aligned}
h_{x,\pm}   \rightarrow h_{x,\pm}+d_{x,\pm}\delta\\
h_{y,\pm}   \rightarrow h_{y,\pm}+d_{y,\pm}\delta
\end{aligned}
\right.  \label{eq:hdisp1} 
\hskip 5mm &,&\hskip 5mm 
\left\{
\begin{aligned}
d_{x,\pm}=\tilde{D}_x\pm i\tilde{D}_x'\\
d_{y,\pm}=\tilde{D}_y\pm i\tilde{D}_y'
\end{aligned}
\right. \ . \quad\qquad 
\end{eqnarray}
$d_{\pm}$ represents hence the dispersion in the complex C-S 
coordinates. The dependence on the particle energy is contained 
also in the Hamiltonian coefficients of Eq.~\eqref{eq:Hdef} 
$h_{m,pqrt}$ through the magnetic rigidity and reads
\begin{eqnarray}\label{eq:hdisp2} 
\left\{
\begin{aligned}
K_{m,n-1}\rightarrow \frac{K_{m,n-1}}{1+\delta}\\
J_{m,n-1}\rightarrow \frac{J_{m,n-1}}{1+\delta}
\end{aligned}
\right.  
\hskip 5mm \Rightarrow\hskip 5mm
h_{m,pqrt} \rightarrow\displaystyle{\frac{h_{m,pqrt}}{1+\delta}}=
           h_{m,pqrt}(1-\delta+\delta^2 + ...\ ) \ .\qquad
\end{eqnarray}
By substituting Eqs.~\eqref{eq:hdisp1}-\eqref{eq:hdisp2} 
in Eq.~\eqref{eq:Hdef} the energy-dependent Hamiltonian term 
(up to second order in $\delta$) reads
\begin{eqnarray}
\tilde{H}_{m,pqrtd} &\rightarrow& 
         h_{m,pqrtd}h_{m,x,+}^ph_{m,x,-}^qh_{m,y,+}^rh_{m,y,-}^t
         \delta^d \\&\rightarrow&
         h_{m,lkno}(1-\delta+\delta^2)
	(h_{m,x,+}+d_{m,x,+}\delta)^l(h_{m,x,-}+d_{m,x,-}\delta)^k
        (h_{m,y,+}+d_{m,y,+}\delta)^n(h_{m,y,-}+d_{m,y,-}\delta)^o
        \nonumber \ \ .
\end{eqnarray}
Note that in the last row generic indices $lkno$ replace 
the initial ones $pqrt$ because several combinations of 
the former may contribute to generate the Hamiltonian 
term $\tilde{H}_{pqrtd}$ when going off momentum $d>0$. 
The binomials may indeed be expanded as 
\begin{eqnarray}
\tilde{H}_{m,pqrtd}&\rightarrow& 
         h_{m,lkno}(1-\delta+\delta^2)
         \sum_{a=0}^l{\left(\begin{array}{c}l \\ a\end{array}\right)}
               h_{m,x,+}^{l-a}(d_{m,x,+}\delta)^a
         \sum_{b=0}^k{\left(\begin{array}{c}k \\ b\end{array}\right)}
               h_{m,x,-}^{k-b}(d_{m,x,-}\delta)^b \times \nonumber\\
 &&\hskip 29mm
         \sum_{c=0}^n{\left(\begin{array}{c}n \\ c\end{array}\right)}
               h_{m,y,+}^{n-c}(d_{m,y,+}\delta)^c
         \sum_{e=0}^o{\left(\begin{array}{c}o \\ e\end{array}\right)}
               h_{m,y,-}^{o-e}(d_{m,y,-}\delta)^e \label{eq:Hpqrtd1}
               \qquad ,
\end{eqnarray}
where $\left(\begin{array}{c}l \\ a\end{array}\right)=\displaystyle\frac{l!}{a!(l-a)!}$ 
is the binomial coefficient. Before deriving the chromatic 
observables (chromaticity, chromatic beta-beating, 
chromatic coupling and dispersion) from the Hamiltonian 
terms $\tilde{H}_{m,pqrtd}$, it is worthwhile to distinguish 
the different nature of the Hamiltonian terms of 
Eq.~\eqref{eq:Hpqrtd1}. 
\begin{itemize}
\item {$\mathbf{p+q+r+t=1,\ d=0}$,\ \bf orbit-like terms}: The magnetic 
      elements corresponding to these terms define the orbit. Since 
      the reference orbit is assumed to be known, only dipole errors 
      $\delta K_0$ and $J_0$ (for planar rings $\delta J_0=J_0$) 
      inducing orbit distortion shall be used in the definition of 
      $\tilde{H}_{pqrtd}$.
      \begin{eqnarray}\label{eq:Sum1A}
        \left\{
        \begin{aligned}\displaystyle
        p+q+r+t=1 \\
        d=0
        \end{aligned}
        \right. \quad\Rightarrow\quad \delta K_0,\ J_0\ 
        \hbox{ in Eq.~\eqref{eq:h_Vs_KJ}}\ .
      \end{eqnarray}
\item {$\mathbf{p+q+r+t=1,\ d=1}$,\ \bf dispersion-like terms}: The 
      magnetic elements corresponding to these terms define the 
      linear dependence of the orbit on $\delta$. This is generated 
      by the linear dependence of the bending angles $K_0$ (including 
      possible field errors $\delta K_0$) on the beam 
      energy and on the linear optics. The latter depends on 
      $\delta$ too, though to first order ($d=1$) this dependence 
      is to be ignored (it may not be neglected when $d=2$). It is 
      assumed that the linear optics is known through the C-S 
      parameters and that focusing errors are included in the 
      model, which is equivalent to say that with respect to 
      the the used C-S parameters they are zero, $\delta K_1=0$. 
      Betatron coupling may be instead non-zero, as well as 
      vertical deflections $J_0$, if any.
      \begin{eqnarray}\label{eq:Sum1B}
        \left\{
        \begin{aligned}\displaystyle
        p+q+r+t=1 \\
        d=1
        \end{aligned}
        \right. \quad\Rightarrow\quad  K_0,\ J_0,\ \delta K_1(=0),\ J_1
        \hbox{ in Eq.~\eqref{eq:h_Vs_KJ}} \ .
      \end{eqnarray}
\item {$\mathbf{p+q+r+t=2,\ d=1}$,\ \bf betatron-like terms}: The magnetic 
      elements corresponding to these terms define the linear dependence 
      of the betatron motion on $\delta$. This is generated by the dependence 
      of the normalized quadrupole strengths on the beam energy and on the 
      additional focusing provided by the quadrupolar feed-down field 
      experienced by the beam when entering the sextupoles off axis. 
      There is no dependence on the dipolar fields, the betatron-like 
      terms describing only the motion around the closed orbit. 
      \begin{eqnarray}\label{eq:Sum2}
        \left\{
        \begin{aligned}\displaystyle
        p+q+r+t=2 \\
        d=1
        \end{aligned}\right.
        \quad\Rightarrow\quad K_1,\ J_1, \ K_2,\ J_2\ 
        \hbox{in Eq.~\eqref{eq:h_Vs_KJ}}\ .
      \end{eqnarray}
\item {$\mathbf{p+q+r+t=1,\ d=2}$,\ \bf second-order dispersion-like 
      terms}: This higher-order dependence of the beam orbit on the 
      energy imposes the inclusion of the dependence of the 
      focusing lattice on $\delta$, i.e. quadrupole and sextupole 
      strengths
      \begin{eqnarray}\label{eq:Sum3}
        \left\{
        \begin{aligned}\displaystyle
        p+q+r+t=1 \\
        d=2
        \end{aligned}\right.
        \quad\Rightarrow\quad K_0,\ J_0,\ K_1,\ J_1\ ,\ K_2,\ J_2\ 
        \hbox{in Eq.~\eqref{eq:h_Vs_KJ}}\ .
      \end{eqnarray}
\end{itemize}

\subsection{First-order chromatic terms (d=1)} 
\label{sec:d1}
Among all elements in the r.h.s. of Eq.~\eqref{eq:Hpqrtd1} 
only those proportional to $\delta$ are kept, along with 
those proportional to $h_{m,x,+}^ph_{m,x,-}^qh_{m,y,+}^rh_{m,y,-}^t$. 
The Hamiltonian terms linear in $\delta$ read
\begin{eqnarray}
\tilde{H}_{m,pqrt1}&\rightarrow& 
         h_{m,lkno}(1-\delta)
         \sum_{a=0}^l\sum_{b=0}^k\sum_{c=0}^n\sum_{e=0}^o
         {\left(\hskip -1mm\begin{array}{c} l\\ a\end{array}\hskip -1mm\right)}
         {\left(\hskip -1mm\begin{array}{c} k\\ b\end{array}\hskip -1mm\right)}
         {\left(\hskip -1mm\begin{array}{c} n\\ c\end{array}\hskip -1mm\right)}
         {\left(\hskip -1mm\begin{array}{c} o\\ e\end{array}\hskip -1mm\right)}
               h_{m,x,+}^{l-a}h_{m,x,-}^{k-b}
               h_{m,y,+}^{n-c}h_{m,y,-}^{o-e}
               d_{m,x,+}^{\ a}d_{m,x,-}^{\ b}d_{m,y,+}^{\ c}d_{m,y,-}^{\ e}
               \times\nonumber \\ 
        &&\hspace{7.3cm}\delta^{a+b+c+e},\label{eq:Hpqrtd2} 
\end{eqnarray}
where all sets of indices $abce$ and $lkno$ satisfying 
the following systems are kept: 
\begin{eqnarray}\label{eq:Hpqrtd3}
\left\{
\begin{array}{l}
l-a=p \\
k-b=q \\
n-c=r \\
o-e=t \\
a+b+c+e=1  
\end{array}
\right.
\qquad\hbox{and}\qquad
\left\{
\begin{array}{l}
l-a=p \\
k-b=q \\
n-c=r \\
o-e=t \\
a+b+c+e=0 
\end{array}
\right. \quad . 
\end{eqnarray}
The two systems stem from the magnetic rigidity term 
$(1-\delta)$. After some algebra, the Hamiltonian terms 
linear in $\delta$ read 
\begin{eqnarray}\label{eq:Hpqrtd4A}
\tilde{H}_{m,pqrt1}&\rightarrow& 
         h_{m,pqrt1}h_{m,x,+}^ph_{m,x,-}^qh_{m,y,+}^rh_{m,y,-}^t\delta \\
h_{m,pqrt1}&=&(p+1)h_{m,(p+1)qrt}d_{m,x,+}+
              (q+1)h_{m,p(q+1)rt}d_{m,x,-}+ \nonumber \\\label{eq:Hpqrtd4}
            &&(r+1)h_{m,pq(r+1)t}d_{m,y,+}+
              (t+1)h_{m,pqr(t+1)}d_{m,y,-}- h_{m,pqrt}\quad . 
\end{eqnarray}

\subsection{Linear dispersion}
\label{sec:lindisp}
The on-momentum Hamiltonian of Eq.~\eqref{eq:HamDisp} 
needs to be extended to include a dependence on the 
energy deviation $\delta$. A second-order expansion 
reads
\begin{eqnarray}\nonumber
\tilde{H}_m&=&h_{m,1000 }h_{m,x,+}+h_{m,0100 }h_{m,x,-}+ 
              h_{m,0010 }h_{m,y,+}+h_{m,0001 }h_{m,y,-}+ \\\nonumber
&&           (h_{m,10001}h_{m,x,+}+h_{m,01001}h_{m,x,-}+
              h_{m,00101}h_{m,y,+}+h_{m,00011}h_{m,y,-})\delta + \\
&&           (h_{m,10002}h_{m,x,+}+h_{m,01002}h_{m,x,-}+
              h_{m,00102}h_{m,y,+}+h_{m,00012}h_{m,y,-})\delta^2 
              +O(\delta^3)\ .
\end{eqnarray}
Hereafter, the subscript $m$ corresponding to a generic magnet 
replaces here the label $w$ of a generic orbit corrector
in Eq.~\eqref{eq:HamDisp}, since we are no longer interested in 
the evaluation of an ORM, but rather of chromatic functions 
dependent on the strengths of magnets of different order (dipole, 
quadrupole and sextupole). 
The off-momentum generalization of Eq.~\eqref{eq:deltah2} 
then reads
\begin{eqnarray}\label{eq:deltah2B}
\delta \vec{h}_{m}=\left(\begin{array}{c}\delta h_{m,x,-}\\\delta h_{m,x,+}\\
                    \delta h_{m,y,-} \\ \delta h_{m,y,+}\end{array}\right)=
2i\left(\begin{array}{r} h_{m,1000}\\-h_{m,0100}\\h_{m,0010}\\-h_{m,0001}
   \end{array}\right)\ 
+2i\left(\begin{array}{r}h_{m,10001}\\-h_{m,01001}\\h_{m,00101}\\-h_{m,00011}
   \end{array}\right)\delta\ 
+2i\left(\begin{array}{r}h_{m,10002}\\-h_{m,01002}\\h_{m,00102}\\-h_{m,00012}
   \end{array}\right)\delta^2\ \ +O(\delta^3).
\end{eqnarray}
The first terms in the r.h.s. are responsible for the 
orbit distortion, the second terms proportional to 
$\delta$ modify the linear dispersion $D$, whereas the last 
elements account for the derivative of the dispersion 
with respect to $\delta$, $D'$.
For the evaluation of linear dispersion, the Hamiltonian 
terms in the second vector of the r.h.s. of 
Eq.~\eqref{eq:deltah2B} are to be computed. These are 
evaluated from Eq.~\eqref{eq:Hpqrtd4} ($d=1$), yielding 
\begin{eqnarray}
\left\{
\begin{aligned}\displaystyle\label{eq:Hamdisp1}
h_{m,10001}&=2h_{m,2000\ }d_{m,x,+}+ h_{m,1100\ }d_{m,x,-}+
              h_{m,1010\ }d_{m,y,+}+ h_{m,1001\ }d_{m,y,-}-h_{m,1000}\\
h_{m,01001}&= h_{m,1100\ }d_{m,x,+}+2h_{m,2000\ }d_{m,x,-}+
              h_{m,0110\ }d_{m,y,+}+ h_{m,0101\ }d_{m,y,-}-h_{m,0100}\\
h_{m,00101}&= h_{m,1010\ }d_{m,x,+}+ h_{m,0110\ }d_{m,x,-}+
             2h_{m,0020\ }d_{m,y,+}+ h_{m,0011\ }d_{m,y,-}-h_{m,0010}\\
h_{m,00011}&= h_{m,1001\ }d_{m,x,+}+ h_{m,0101\ }d_{m,x,-}+
              h_{m,0011\ }d_{m,y,+}+2h_{m,0002\ }d_{m,y,-}-h_{m,0001}
\end{aligned}
\right. \ .
\end{eqnarray}
The Hamiltonian coefficients $h_{m,pqrt}$ are computed from 
Eq.~\eqref{eq:h_Vs_KJ}:
\begin{eqnarray}\label{eq:Hamdisp1b}
\left\{
\begin{aligned}\displaystyle
2h_{m,2000}&=h_{m,1100}=-\frac{1}{4}\delta K_{m,1}\beta_{m,x} = 0 \\
2h_{m,0020}&=h_{m,0011}=+\frac{1}{4}\delta K_{m,1}\beta_{m,y} = 0 \\
 h_{m,1010}&=h_{m,1001}=h_{m,0110}=+\frac{1}{4} J_{m,1}
                         \sqrt{\beta_{m,x}\beta_{m,y}}\\
 h_{m,1000}&=h_{m,0100}=-\frac{1}{2}K_{m,0}\sqrt{\beta_{m,x}}\\
 h_{m,0010}&=h_{m,0001}=+\frac{1}{2}J_{m,0}\sqrt{\beta_{m,y}}
\end{aligned}
\right. \ .
\end{eqnarray}
Since Hamiltonian terms with $p+q+r+t=1$ and $d=1$ 
are evaluated in Eq.~\eqref{eq:Hamdisp1}, 
Eq.~\eqref{eq:Sum1B} applies and the main bending 
magnet horizontal angles $K_0$ (and vertical $J_0$, 
if any) are used, whereas the focusing errors are 
assumed to be included in the computation of the 
beta functions, hence $\delta K_1=0$. 
Equation~\eqref{eq:Hamdisp1} then reads
\begin{eqnarray}
\left\{
\begin{aligned}\displaystyle
h_{m,10001}&=h_{m,1010\ }2\Re\left\{d_{m,y,\pm}\right\}-h_{m,1000}
            =+\frac{1}{2}\left(K_{m,0}+J_{m,1}D_{m,y}\right)\sqrt{\beta_{m,x}}\\
h_{m,01001}&=h_{m,10001}\\
h_{m,00101}&=h_{m,1010\ }2\Re\left\{d_{m,x,\pm}\right\}-h_{m,0010}
            =-\frac{1}{2}\left(J_{m,0}-J_{m,1}D_{m,x}\right)\sqrt{\beta_{m,y}}\\
h_{m,00011}&=h_{m,00101}
\end{aligned}
\right. \ ,
\end{eqnarray}
since $\Re\left\{d_{m,q,\pm}\right\}=D_{m,q}/\sqrt{\beta_{m,q}}$. 
Thus, the off-momentum closed orbit of 
Eq.~\eqref{eq:co1B} becomes
\begin{eqnarray}\label{eq:co2C}
\vec{h}(j)&=&\mathbf{B}_j^{-1}\sum_{m=1}^M\left\{
\frac{e^{i\mathbf{\Delta\phi}_{mj}}}{1-e^{i\mathbf{Q}}}
\mathbf{B}_m\ i
\left(\begin{array}{r} \left(K_{m,0}+J_{m,1}D_{m,y}\right)\sqrt{\beta_{m,x}}\\
                      -\left(K_{m,0}+J_{m,1}D_{m,y}\right)\sqrt{\beta_{m,x}}\\
                      -\left(J_{m,0}-J_{m,1}D_{m,x}\right)\sqrt{\beta_{m,y}}\\
                       \left(J_{m,0}-J_{m,1}D_{m,x}\right)\sqrt{\beta_{m,y}}
   \end{array}\right)\right\}\delta \quad .
\end{eqnarray}
The above expression simplifies greatly, since for 
the linear dispersion the effects of the normal 
form transformations $\mathbf{B}_j^{-1}$ and 
$\mathbf{B}_m$ are of higher order and shall be 
ignored here, hence leaving 
\begin{eqnarray}\label{eq:co2B}
\vec{h}(j)&\simeq&\sum_{m=1}^M\left\{
\frac{e^{i\mathbf{\Delta\phi}_{mj}}}{1-e^{i\mathbf{Q}}}
\ i
\left(\begin{array}{r} \left(K_{m,0}+J_{m,1}D_{m,y}\right)\sqrt{\beta_{m,x}}\\
                      -\left(K_{m,0}+J_{m,1}D_{m,y}\right)\sqrt{\beta_{m,x}}\\
                      -\left(J_{m,0}-J_{m,1}D_{m,x}\right)\sqrt{\beta_{m,y}}\\
                       \left(J_{m,0}-J_{m,1}D_{m,x}\right)\sqrt{\beta_{m,y}}
   \end{array}\right)\right\}\delta \quad ,
\end{eqnarray}
where now all C-S parameters and dispersion refer 
to the ideal lattice with no focusing errors and 
betatron coupling, though with possible vertical 
dispersion induced by vertical dipole terms.
The complex dispersion vector 
$\vec{d}=(\tilde{d}_{x,-},\tilde{d}_{x,+},
\tilde{d}_{y,-},\tilde{d}_{y,+})^T$ hence reads
\begin{eqnarray}\label{eq:co3C}
\vec{d}(j)&=&\frac{\partial\vec{h}(j)}{\partial \delta}\simeq\sum_{m=1}^M\left\{
\frac{e^{i\mathbf{\Delta\phi}_{mj}}}{1-e^{i\mathbf{Q}}}\ i
\left(\begin{array}{r} \left(K_{m,0}+J_{m,1}D_{m,y}\right)\sqrt{\beta_{m,x}}\\
                      -\left(K_{m,0}+J_{m,1}D_{m,y}\right)\sqrt{\beta_{m,x}}\\
                      -\left(J_{m,0}-J_{m,1}D_{m,x}\right)\sqrt{\beta_{m,y}}\\
                       \left(J_{m,0}-J_{m,1}D_{m,x}\right)\sqrt{\beta_{m,y}}
   \end{array}\right)\right\}\ ,
\end{eqnarray}
from which both the horizontal and the vertical 
dispersion at a location $j$ can be inferred, 
since $D_{x}=\Re\left\{\tilde{d}_{x,-}\right\}\sqrt{\beta_{x}}$ 
and $D_{y}=\Re\left\{\tilde{d}_{y,-}\right\}\sqrt{\beta_{y}}$: 
\begin{eqnarray}\label{eq:co4D}
\vec{d}(j)&\simeq&i\sum_{m=1}^M\left\{
\left(\begin{array}{c c c c}
           \frac{e^{ i\Delta\phi_{x,mj}}}{1-e^{ i2\pi Q_x}} & 0 & 0 & 0\\
    0&     \frac{e^{-i\Delta\phi_{x,mj}}}{1-e^{-i2\pi Q_x}} & 0 & 0 \\
0 & 0 &    \frac{e^{ i\Delta\phi_{y,mj}}}{1-e^{ i2\pi Q_y}} & 0 \\
0 & 0 & 0 &\frac{e^{-i\Delta\phi_{y,mj}}}{1-e^{-i2\pi Q_y}} 
\end{array}\right)
\left(\begin{array}{r} \left(K_{m,0}+J_{m,1}D_{m,y}\right)\sqrt{\beta_{m,x}}\\
                      -\left(K_{m,0}+J_{m,1}D_{m,y}\right)\sqrt{\beta_{m,x}}\\
                      -\left(J_{m,0}-J_{m,1}D_{m,x}\right)\sqrt{\beta_{m,y}}\\
                       \left(J_{m,0}-J_{m,1}D_{m,x}\right)\sqrt{\beta_{m,y}}
   \end{array}\right)
\right\} \ ,
\end{eqnarray}
resulting in 
\begin{eqnarray}
\left\{
\begin{aligned}
D_x(j)&\simeq+\frac{\sqrt{\beta_{j,x}}}{2\sin{(\pi Q_x)}}\sum_{m=1}^M
\left(K_{m,0}+J_{m,1}D_{m,y}\right)\sqrt{\beta_{m,x}}\cos(\Delta\phi_{x,mj}-\pi Q_x)
\\
D_y(j)&\simeq-\frac{\sqrt{\beta_{j,y}}}{2\sin{(\pi Q_y)}}\sum_{m=1}^M
\left(J_{m,0}-J_{m,1}D_{m,x}\right)\sqrt{\beta_{m,y}}\cos(\Delta\phi_{y,mj}-\pi Q_y)
\end{aligned}\right. \ .\label{eq:co6B}
\end{eqnarray} 
For consistency with the nomenclature used 
throughout this paper, the phase advance 
$\Delta\phi_{mj}$ is to be computed as in 
Eq.~\eqref{eq:deltaphisign}. If the mere 
difference between the two betatron phases 
at the positions $m$ and $j$ is used, the 
absolute value $|2\Delta\phi_{x,mj}|$ 
shall then be used, as done in textbooks, 
whose formulas for the ideal case are 
retrieved from Eq.~\eqref{eq:co6B} after
removing betatron coupling ($J_1=0$) and vertical 
dispersion ($J_0=0$, $D_{m,y}=0$). Note that the 
above equations are still valid in the presence 
of focusing errors and betatron coupling, provided 
that the corresponding C-S parameters $\beta$ 
and $\phi$ and dispersion are used. 
$D_{m,y}$ in the r.h.s. of Eq.~\eqref{eq:co6B} shall 
then be the one generated by the vertical dipole 
fields $J_0$ (if any) but not by betatron coupling. 
Indeed, it is Eq.~\eqref{eq:co6B} that describes the 
entanglement between the horizontal and vertical 
dispersion functions due to skew quadrupole fields.

\subsection{linear chromaticity}
\label{sec:ap:LinChrom}
As second example of application of Eq.~\eqref{eq:Hpqrtd4} 
the linear detuning Hamiltonian terms proportional 
to $\delta$ are evaluated. They are linked to the 
linear chromaticity. As shown in Ref.~\cite{Andrea-arxiv}, 
the Hamiltonian term at a generic position $j$ generated 
by all $M$ magnets reads
\begin{eqnarray}\label{eq:Hdef2}
\tilde{H}_{pqrt}(j)=\sum_{m=1}^M\tilde{H}_m(j)=\sum_{m=1}^M{h_{m,pqrt}
             e^{i[(p-q)\Delta\phi_{x,mj}+(r-t)\Delta\phi_{y,mj}]}
             h_{m,x,+}^p h_{m,x,-}^q h_{m,y,+}^r h_{m,y,-}^t}\quad .
\end{eqnarray}
Without loss of generality we can expand the 
Hamiltonian terms at a location $j$ in a 
power series of $\delta$
\begin{eqnarray}\label{eq:Hdef3}
\tilde{H}_{pqrt}(j,\delta)=\sum_{d\ge0}\tilde{H}_{pqrtd}(j)\delta^d=
                    \tilde{H}_{pqrt}(j)+\tilde{H}_{pqrt1}\delta 
                    +\tilde{H}_{pqrt2}(j)\delta^2 +O(\delta^3)\ ,
\end{eqnarray}
where $\tilde{H}_{pqrt}(j)$ is the geometric Hamiltonian 
of Eq.~\eqref{eq:Hdef2}, whereas $\tilde{H}_{pqrt1}(j)$ 
is the corresponding first chromatic Hamiltonian, 
\begin{eqnarray}\label{eq:Hdef4}
\tilde{H}_{pqrt1}(j)=\sum_{m=1}^M{h_{m,pqrt1}
             e^{i[(p-q)\Delta\phi_{x,mj}+(r-t)\Delta\phi_{y,mj}]}
             h_{m,x,+}^p h_{m,x,-}^q h_{m,y,+}^r h_{m,y,-}^t}\delta\ ,
\end{eqnarray}
whose coefficients $h_{m,pqrt1}$ are those of 
Eq.~\eqref{eq:Hpqrtd4}.

Detuning terms are those with $p=q$ and $r=t$, hence 
independent of the betatron phases, since in the 
above expression the phases are all equal to 1 and 
the product of all coordinates is invariant, 
$h_{m,x,+}^p h_{m,x,-}^p h_{m,y,+}^r h_{m,y,-}^r=(2I_x)^p(2I_y)^r$. 
The first chromatic non-zero detuning coefficients are 
$h_{m,11001}$ and $h_{m,00111}$ in the horizontal and 
vertical planes, respectively. The substitution of those 
indices in Eq.~\eqref{eq:Hpqrtd4} yields 
\begin{eqnarray}
\left\{
\begin{array}{l}
h_{m,11001}= 2h_{m,2100}\ d_{m,x,+}+ 2h_{m,1200}\ d_{m,x,-}+
            \ h_{m,1110}\ d_{m,y,+}+\ h_{m,1101}\ d_{m,y,-}-h_{m,1100} \\
h_{m,00111}=\ h_{m,1011}\ d_{m,x,+}+\ h_{m,0111}\ d_{m,x,-}+
             2h_{m,0021}\ d_{m,y,+}+ 2h_{m,0012}\ d_{m,y,-}-h_{m,0011}
\end{array}
\right. \quad .\label{eq:chrom1}
\end{eqnarray}
The Hamiltonian coefficients in the above r.h.s. 
may be made explicit via Eq.~\eqref{eq:h_Vs_KJ}: 
\begin{eqnarray}\label{eq:h1200}
h_{m,2100}&=&h_{m,1200}=-\frac{1}{16}K_{m,2}\beta_{m,x}^{3/2}\quad ,\quad
h_{m,1110}=h_{m,1101}=+\frac{1}{8}J_{m,2}\beta_{m,x}\sqrt{\beta_{m,y}}
\quad ,\quad h_{m,1100}=-\frac{1}{4}K_{m,1}\beta_{m,x}  \quad ,\quad \qquad \\
h_{m,0021}&=&h_{m,0012}=-\frac{1}{16}J_{m,2}\beta_{m,y}^{3/2}\quad ,\quad
h_{m,1011}=h_{m,0111}=+\frac{1}{8}K_{m,2}\sqrt{\beta_{m,x}}\beta_{m,y}
\quad ,\quad h_{m,0011}=+\frac{1}{4}K_{m,1}\beta_{m,y} \quad .\quad\qquad  
\end{eqnarray}
Eq.~\eqref{eq:chrom1} then reads
\begin{eqnarray}
\left\{
\begin{array}{l}
h_{m,11001}= 4h_{m,2100}\Re\left\{d_{m,x,+}\right\}+
             2h_{m,1110}\Re\left\{d_{m,y,+}\right\}-h_{m,1100} \\
h_{m,00111}= 2h_{m,1011}\Re\left\{d_{m,x,+}\right\}+
             4h_{m,0021}\Re\left\{d_{m,y,+}\right\}-h_{m,0011}
\end{array}
\right. \quad .\label{eq:chrom2}
\end{eqnarray}
Since $\Re\left\{d_{m,q,+}\right\}=D_{m,q}/\sqrt{\beta_{m,q}}$, 
the two Hamiltonian coefficients become 
\begin{eqnarray}
\left\{
\begin{array}{l}
h_{m,11001}=+\displaystyle\frac{1}{4}\left(K_{m,1}-K_{m,2}D_{m,x}+ J_{m,2}D_{m,y}
                        \right)\beta_{m,x}  \vspace{ 1.5mm}\\
h_{m,00111}=-\displaystyle\frac{1}{4}\left(K_{m,1}-K_{m,2}D_{m,x}+ J_{m,2}D_{m,y}
                        \right)\beta_{m,y}
\end{array}
\right. \quad .\label{eq:chrom3}
\end{eqnarray}
As expected, both quantities are real and differ 
only by the sign and the beta functions, the 
argument within the parenthesis being the same 
in both planes. These indeed represent the effective 
quadrupole forces experienced by off-energy 
particles. The Hamiltonian accounting for all 
magnets is derived from Eq.~\eqref{eq:Hdef4},
\begin{eqnarray}
\left\{
\begin{array}{l}
\tilde{H}_{11001}=+\displaystyle\frac{1}{4}\sum\limits_{m=1}^M\left(K_{m,1}-
                        K_{m,2}D_{m,x}+ J_{m,2}D_{m,y}\right)\beta_{m,x}
                        \ h_{m,x,+}h_{m,x,-}\delta  \vspace{ 1.5mm}\\
\tilde{H}_{00111}=-\displaystyle\frac{1}{4}\sum\limits_{m=1}^M\left(K_{m,1}-
                        K_{m,2}D_{m,x}+ J_{m,2}D_{m,y}\right)\beta_{m,y}
                        \ h_{m,y,+}h_{m,y,-}\delta
\end{array}
\right. \quad ,\label{eq:chrom4}
\end{eqnarray}
or equivalently
\begin{eqnarray}
\left\{
\begin{array}{l}\displaystyle
\tilde{H}_{11001}=h_{11001}(2J_x)\delta=\sum\limits_{m=1}^Mh_{m,11001}(2J_x)\delta\\
\displaystyle
\tilde{H}_{00111}=h_{00111}(2J_y)\delta=\sum\limits_{m=1}^Mh_{m,00111}(2J_y)\delta
\end{array}
\right. \quad .\label{eq:chrom5}
\end{eqnarray}
As expected, neither term depends on the betatron 
phases and hence on the longitudinal position $j$. 
The linear chromaticity is defined as 
\begin{eqnarray}
\left\{
\begin{array}{l}\displaystyle\hspace{-1mm}
Q_x'=\frac{\partial Q_x}{\partial \delta}\bigg|_{\delta=0}\hskip-2mm=
     \frac{\partial}{\partial \delta}\left(\hspace{-1mm}-{\frac{1}{2\pi}}
       \frac{\partial \left(\tilde{H}_{11000}+\tilde{H}_{11001}\delta\right)}
            {\partial J_x}\right)_{\hspace{-1mm}\delta=0}\hskip-2mm=-\frac{h_{11001}}{\pi}
    \hspace{-1mm}=-\displaystyle\frac{1}{4\pi}\sum\limits_{m=1}^M
              \hspace{-1mm}\left(K_{m,1}-K_{m,2}D_{m,x}+ J_{m,2}D_{m,y}
                        \right)\beta_{m,x}  \vspace{ 1.5mm}\nonumber\\
\displaystyle\hspace{-1mm}
Q_y'=\frac{\partial Q_y}{\partial \delta}\bigg|_{\delta=0}\hskip-2mm=
     \frac{\partial}{\partial \delta}\left(\hspace{-1mm}-{\frac{1}{2\pi}}
       \frac{\partial \left(\tilde{H}_{00110}+\tilde{H}_{00111}\delta\right)}
            {\partial J_y}\right)_{\hspace{-1mm}\delta=0}\hskip-2mm=-\frac{h_{00111}}{\pi}
    \hspace{-1mm}=+\displaystyle\frac{1}{4\pi}\sum\limits_{m=1}^M
              \hspace{-1mm}\left(K_{m,1}-K_{m,2}D_{m,x}+ J_{m,2}D_{m,y}
                        \right)\beta_{m,y}  \nonumber
\end{array}
\right. \ .\\ \label{eq:chrom6}
\end{eqnarray}
$\tilde{H}_{11000}$ and $\tilde{H}_{00110}$ are 
both zero because focusing errors $\delta K_1$ 
are either zero or included in the model to 
compute the C-S parameters. The above relations 
require some comments. First, textbook 
formulas are retrieved when removing either 
vertical dispersion or the skew sextupole 
strengths $J_{2}$. Second, skew quadrupole 
fields $J_{1}$ do not influence explicitly 
linear chromaticity, at least to first 
order in the Hamiltonian truncation, of which 
more in Sec.~\ref{sec:app-SecOrdHam}. Betatron
coupling enters indirectly in Eq.~\eqref{eq:chrom6} 
through vertical dispersion $D_y$. 

\subsection{Chromatic beating}
\label{sec:ap:ChromBeat}
Off-energy particles experience a non-zero 
closed orbit described by the dispersion function. 
When crossing normal sextupoles off axis, those 
particles are subjected to quadrupolar feed-down 
fields. This additional focusing results 
in modulated beta functions. Even an ideal lattice 
with no focusing error (i.e. no on-momentum 
{\sl geometric} beta-beating) is unavoidably 
subjected to this chromatic modulation of the 
beta functions and hence to the corresponding 
half-integer resonance. The geometric beta-beating 
is described by the two RDTs $f_{2000}$ and 
$f_{0020}$. These in turn are generated by the 
geometric Hamiltonian coefficients $h_{2000}$ 
and $h_{0020}$, see Eq.~\eqref{eq:RDT-1st}. It is 
then natural to seek the source of chromatic 
beating in the two chromatic Hamiltonian 
coefficients $h_{20001}$ and $h_{00201}$ and 
to derive expressions for the linear dependence 
of the beta functions on $\delta$, i.e. 
$\partial\beta/\partial\delta$. From 
Eq.~\eqref{eq:Hpqrtd4} we obtain
\begin{eqnarray}
\left\{
\begin{array}{l}
h_{m,20001}= 3h_{m,3000}\ d_{m,x,+}+\ h_{m,2100}\ d_{m,x,-}+
            \ h_{m,2010}\ d_{m,y,+}+\ h_{m,2001}\ d_{m,y,-}-h_{m,2000} \\
h_{m,00201}=\ h_{m,1020}\ d_{m,x,+}+\ h_{m,0120}\ d_{m,x,-}+
             3h_{m,0030}\ d_{m,y,+}+\ h_{m,0021}\ d_{m,y,-}-h_{m,0020}
\end{array}
\right. \quad .\label{eq:beat1}
\end{eqnarray}
Since in both cases $p+q+r+t=2$ and $d=1$, 
Eq.~\eqref{eq:Sum2} applies and all Hamiltonian 
coefficients are to be computed from 
Eq.~\eqref{eq:h_Vs_KJ} using the total 
focusing strengths $K_1$ (nominal plus 
errors) as well as the sextupole strengths 
(normal and skew):
\begin{eqnarray}
h_{m,2100}&=&3h_{m,3000}=-\frac{1}{16}K_{m,2}\beta_{m,x}^{3/2}\ ,\quad
h_{m,2010}=h_{m,2001}=+\frac{1}{16}J_{m,2}\beta_{m,x}\sqrt{\beta_{m,y}}
\ ,\quad h_{m,2000}=-\frac{1}{8}K_{m,1}\beta_{m,x}  \ ,\ \qquad \\
h_{m,0021}&=&3h_{m,0030}=-\frac{1}{16}J_{m,2}\beta_{m,y}^{3/2}\ ,\quad
h_{m,1020}=h_{m,0120}=+\frac{1}{16}K_{m,2}\sqrt{\beta_{m,x}}\beta_{m,y}
\ ,\quad h_{m,0020}=+\frac{1}{8}K_{m,1}\beta_{m,y} \ .\ \qquad  
\end{eqnarray}
Eq.~\eqref{eq:beat1} then reads
\begin{eqnarray}
\left\{
\begin{array}{l}
h_{m,20001}= 6h_{m,3000}\Re\left\{d_{m,x,+}\right\}+
             2h_{m,2010}\Re\left\{d_{m,y,+}\right\}-h_{m,2000} \\
h_{m,00201}= 2h_{m,1020}\Re\left\{d_{m,x,+}\right\}+
             6h_{m,0030}\Re\left\{d_{m,y,+}\right\}-h_{m,0020}
\end{array}
\right. \quad .\label{eq:beat2}
\end{eqnarray}
Since $\Re\left\{d_{m,q,+}\right\}=D_{m,q}/\sqrt{\beta_{m,q}}$, 
the two Hamiltonian coefficients become 
\begin{eqnarray}
\left\{
\begin{array}{l}
h_{m,20001}=+\displaystyle\frac{1}{8}\left(K_{m,1}-K_{m,2}D_{m,x}+ J_{m,2}D_{m,y}
                        \right)\beta_{m,x}  \vspace{ 1.5mm}\\
h_{m,00201}=-\displaystyle\frac{1}{8}\left(K_{m,1}-K_{m,2}D_{m,x}+ J_{m,2}D_{m,y}
                        \right)\beta_{m,y}
\end{array}
\right. \quad .\label{eq:beat3}
\end{eqnarray}
Note how the arguments in the above parenthesis 
are the same and equal to those of 
Eq.~\eqref{eq:chrom3}, this being the effective 
quadrupole strength experienced by off-energy 
particles. The Hamiltonian at a generic location 
$j$ accounting for all magnets is derived from 
Eq.~\eqref{eq:Hdef4} and reads
\begin{eqnarray}
\left\{
\begin{array}{l}
\tilde{H}_{20001}(j)=\sum\limits_{m=1}^Mh_{m,20001}\ h_{m,x,+}^2\delta=
       +\displaystyle\frac{1}{8}\sum\limits_{m=1}^M\left(K_{m,1}-
                         K_{m,2}D_{m,x}+ J_{m,2}D_{m,y}\right)\beta_{m,x}
                         e^{2i\Delta\phi_{x,mj}}
                         \ h_{m,x,+}^2\delta\vspace{ 1.5mm}\\
\tilde{H}_{00201}(j)=\sum\limits_{m=1}^Mh_{m,00201}\ h_{m,y,+}^2\delta=
       -\displaystyle\frac{1}{8}\sum\limits_{m=1}^M\left(K_{m,1}-
                         K_{m,2}D_{m,x}+ J_{m,2}D_{m,y}\right)\beta_{m,y}
                        e^{2i\Delta\phi_{y,mj}}\ h_{m,y,+}^2\delta
\end{array}
\right. .\label{eq:beat4}
\end{eqnarray}
Conversely to the invariant detuning terms 
of Eq.~\eqref{eq:chrom4}, the chromatic beating 
terms are modulated at twice the betatron phase, 
as the geometric beta-beating. According to 
Eq.~\eqref{eq:RDT-1st}, the on-momentum beta-beating 
RDTs ($d=0$) read 
\begin{equation}\label{eq:RDT-beat0}
f_{2000}(j) =\frac{\sum\limits_{m=1}^M 
            h_{m,2000}e^{2i\Delta\phi_{x,mj}}}{1-e^{4\pi iQ_x}}\qquad ,\qquad
f_{0020}(j) =\frac{\sum\limits_{m=1}^M 
            h_{m,0020}e^{2i\Delta\phi_{y,mj}}}{1-e^{4\pi iQ_y}}\quad ,
\end{equation}
which are both zero, since focusing errors 
$\delta K_1$ are assumed to be included in 
the model and hence $h_{m,2000}=h_{m,0020}=0$. 
The extension of the above relations to the 
off-momentum dynamics reads
\begin{equation}\label{eq:RDT-beat1}
\left\{
\begin{array}{l}
f_{2000}(j,\delta) =f_{2000}(j)+f_{20001}(j)\delta= \displaystyle
\frac{\sum\limits_{m=1}^M  h_{m,20001}e^{2i\Delta\phi_{x,mj}}}
                     {1-e^{4\pi iQ_x}}\delta \quad ,\qquad 
f_{20001}(j) =\displaystyle\frac{\partial f_{2000}(j,\delta)}{\partial\delta}
              \bigg|_{\delta=0}\\ \ \\ \displaystyle
f_{0020}(j,\delta) =f_{0020}(j)+f_{00201}(j)\delta= 
\frac{\sum\limits_{m=1}^M  h_{m,00201}e^{2i\Delta\phi_{y,mj}}}
                     {1-e^{4\pi iQ_y}}\delta\quad ,\qquad 
f_{00201}(j)=\displaystyle\frac{\partial f_{0020}(j,\delta)}{\partial\delta}
              \bigg|_{\delta=0}
\end{array}
\right. \quad .
\end{equation}
The derivative of the two RDTs with respect 
to $\delta$ can then be written as 
\begin{equation}\label{eq:RDT-beat2}
\left\{
\begin{array}{l}
f_{20001}(j) =\displaystyle\hskip -3mm
=\frac{\sum\limits_{m=1}^M  h_{m,20001}e^{2i\Delta\phi_{x,mj}}}
                     {1-e^{4\pi iQ_x}}
=\frac{1}{8\left(1-e^{4\pi iQ_x}\right)}
 \sum\limits_{m=1}^M\left(K_{m,1}-
                         K_{m,2}D_{m,x}+ J_{m,2}D_{m,y}\right)\beta_{m,x}
                         e^{2i\Delta\phi_{x,mj}} \\ \ \\ \displaystyle
f_{00201}(j)=\displaystyle\hskip -3mm
=\frac{\sum\limits_{m=1}^M  h_{m,00201}e^{2i\Delta\phi_{y,mj}}}
                     {1-e^{4\pi iQ_y}}
=\frac{-1}{8\left(1-e^{4\pi iQ_y}\right)}
 \sum\limits_{m=1}^M\left(K_{m,1}-
                         K_{m,2}D_{m,x}+ J_{m,2}D_{m,y}\right)\beta_{m,y}
                         e^{2i\Delta\phi_{y,mj}}
\end{array}
\right. \ .
\end{equation}
The dependence of the betatron tune on 
$\delta$, i.e. the linear chromaticity, is 
neglected since both $f_{20001}$ and 
$f_{00201}$ are multiplied by $\delta$ 
in Eq.~\eqref{eq:RDT-beat1}. The on-momentum 
beta-beating reads~\cite{Andrea-arxiv,Andrea-Linear-arxiv}
\begin{eqnarray}\label{eq:RDT-beat3}
\left\{
\begin{array}{l}\displaystyle
\frac{\Delta\beta_x}{\beta_{x}}=2\sinh{(4|f_{2000}|)}\Big[
                \sinh{(4|f_{2000}|)}
                +\cosh{(4|f_{2000}|)}\sin{q_{2000}}\Big] 
                \simeq 8|f_{2000}|\sin{q_{2000}} +O(|f_{2000}|^2) 
\\ \ \\ \displaystyle
\frac{\Delta\beta_y}{\beta_{y}}=2\sinh{(4|f_{0020}|)}\Big[
                \sinh{(4|f_{0020}|)}
                +\cosh{(4|f_{0020}|)}\sin{q_{0020}}\Big]
                \simeq 8|f_{0020}|\sin{q_{0020}} +O(|f_{0020}|^2)
\end{array}
\right. \quad ,
\end{eqnarray}
where $q_{2000}$ and $q_{0020}$ represent the 
phase of the two RDTs, $f=|f|e^{iq}$, and a 
first-order truncation has been performed, 
valid as long as $|f_{2000}|,\ |f_{0020}|\ll 1$, 
i.e. for weak beating. By noting that 
$|f|\sin{q}=\Im\left\{f\right\}$, the above 
formulas may be rewritten at a generic location 
$j$ as
\begin{eqnarray}\label{eq:RDT-beat3B}
\left\{
\begin{array}{l}\displaystyle
\beta_x(j)\simeq \beta_{x}^{(mod)}(j) 
      +8\beta_{x}^{(mod)}(j)\Im\left\{f_{2000}(j)\right\}
\\ \ \\ \displaystyle
\beta_y(j)\simeq \beta_{y}^{(mod)}(j)
      +8\beta_{y}^{(mod)}(j)\Im\left\{f_{0020}(j)\right\}
\end{array}
\right. \quad .
\end{eqnarray}
Off-momentum particles will then experience the 
following beta functions 
\begin{eqnarray}\label{eq:RDT-beat3C}
\left\{
\begin{array}{l}\displaystyle
\beta_x(j,\delta)\simeq \beta_{x}^{(mod)}(j) 
      +8\beta_{x}^{(mod)}(j)\Im\left\{f_{2000}(j,\delta)\right\}
\\ \ \\ \displaystyle
\beta_y(j,\delta)\simeq \beta_{y}^{(mod)}(j)
      +8\beta_{y}^{(mod)}(j)\Im\left\{f_{0020}(j,\delta)\right\}
\end{array}
\right. \quad .
\end{eqnarray}
By making use of Eq.~\eqref{eq:RDT-beat1} with 
$f_{2000}=f_{0020}=0$, the above expressions truncated 
to first order in $\delta$ read
\begin{eqnarray}\label{eq:RDT-beat3D}
\left\{
\begin{array}{l}\displaystyle
\beta_x(j,\delta)\simeq \beta_{x}^{(mod)}(j)
      +8\beta_{x}^{(mod)}(j)\Im\left\{f_{20001}(j)\right\}\delta+O(\delta^2)
\\ \ \\ \displaystyle
\beta_y(j,\delta)\simeq \beta_{y}^{(mod)}(j)
      +8\beta_{y}^{(mod)}(j)\Im\left\{f_{00201}(j)\right\}\delta+O(\delta^2)
\end{array}
\right. \quad ,
\end{eqnarray}
When going off momentum, the beta function is not the only 
quantity to change the linear phase space orbit and geometry, 
whose maximum normalized amplitude in the horizontal plane is 
given by $|\tilde{x}(j)_{\rm max}|=\sqrt{2J_x}$.
Indeed, the invariant too is deformed by the change of 
energy, since $\sqrt{2J_x}(\delta)\propto\delta K_0(\delta)$, where 
$\delta K_0(\delta)=\delta K_0/(1+\delta)$ represents the normalized 
kick received by the particle which depends on the 
its magnetic rigidity and hence on $\delta$. For example, 
particles with $\delta>0$ will be deviated (by steerer 
magnets) or excited (by dipole kickers) less than 
the one at nominal energy, generating  phase space portraits 
of lower amplitudes. Therefore, the dependence of 
the invariant on $\delta$ can be written as 
$\sqrt{2J_x}(\delta)=\sqrt{2J_x}/(1+\delta)=\sqrt{2J_x}(1-\delta) +O(\delta^2)$. 
The entire chromatic phase space deformation can be 
described by an effective chromatic beating via the 
following definition
\begin{eqnarray}
\frac{\partial\beta_x}{\partial\delta}&=&\lim_{\sqrt{2J_x}\rightarrow0}
\frac{1}{\sqrt{2J_x}}\frac{\partial\sqrt{2J_x}\beta_x}{\partial\delta}=
\lim_{\sqrt{2J_x}\rightarrow0}\frac{1}{\sqrt{2J_x}}
\left(\frac{\partial\sqrt{2J_x}}{\partial\delta}\beta_x+
\sqrt{2J_x}\frac{\partial\beta_x}{\partial\delta}\right) 
\nonumber \\ &\simeq&
\lim_{\sqrt{2J_x}\rightarrow0}\frac{1}{\sqrt{2J_x}}
\left(-\sqrt{2J_x}\beta_x+
      \sqrt{2J_x}8\beta_x\Im\left\{f_{20001}\right\}\right)+O(\delta)
\label{eq:RDT-beat4A}
\end{eqnarray}
Identical considerations apply to the vertical plane. 
The effective chromatic beating at a generic location $j$, 
then reads 
\begin{eqnarray}\label{eq:RDT-beat4}
\left\{
\begin{array}{l}\displaystyle
\frac{\partial\beta_x(j)}{\partial\delta}\bigg|_{\delta=0}\simeq 
  -\beta_x(j)+8\beta_x(j)\Im\left\{f_{20001}(j)\right\}
\\ \ \\ \displaystyle
\frac{\partial\beta_y(j)}{\partial\delta}\bigg|_{\delta=0}\simeq  
  -\beta_y(j)+8\beta_y(j)\Im\left\{f_{00201}(j)\right\}
\end{array}
\right. \quad .
\end{eqnarray}
The imaginary parts are easily computed 
once noticing that
\begin{eqnarray}
\Im\left\{\frac{e^{2i\Delta\phi_{mj}}}{1-e^{4\pi iQ}} \right\}=
\Im\left\{\frac{2i}{e^{-2\pi iQ}-e^{2\pi iQ}}
          \frac{e^{2i\Delta\phi_{mj}-2\pi iQ}}{2i}\right\}=
\frac{\Re\left\{e^{2i\Delta\phi_{mj}-2\pi iQ}
                                \right\}}{2\sin{(2\pi Q)}}=
\frac{\cos{(2\Delta\phi_{mj}-2\pi Q)}}{2\sin{(2\pi Q)}}
\ ,\nonumber 
\end{eqnarray}
Eq.~\eqref{eq:RDT-beat4} then reads
\begin{eqnarray}\label{eq:RDT-beat5}
\left\{
\begin{array}{l}\displaystyle
\frac{\partial\beta_x(j)}{\partial\delta}\bigg|_{\delta=0}\simeq 
-\beta_x(j)+\frac{\beta_x(j)}{2\sin{(2\pi Q_x)}}
\sum\limits_{m=1}^M\left(K_{m,1}-K_{m,2}D_{m,x}+ J_{m,2}D_{m,y}\right)
                  \beta_{m,x}\cos{(2\Delta\phi_{x,mj}-2\pi Q_x)}
\\ \ \\ \displaystyle
\frac{\partial\beta_y(j)}{\partial\delta}\bigg|_{\delta=0}\simeq 
-\beta_y(j)-\frac{\beta_y(j)}{2\sin{(2\pi Q_y)}}
\sum\limits_{m=1}^M\left(K_{m,1}-K_{m,2}D_{m,x}+ J_{m,2}D_{m,y}\right)
                  \beta_{m,y}\cos{(2\Delta\phi_{y,mj}-2\pi Q_y)}
\end{array}
\right. \ .\qquad
\end{eqnarray}
Note that the above expressions differ from the ones 
found in the literature~\cite{Luo-chrom,Tevatron}
for the presence of the $-\beta$ in the r.h.s., 
which stems from the invariant. This term does not 
affect the construction of a response matrix to 
correct the chromatic beating with sextupoles, 
since it cancels out. Similarly, it does not affect 
the evaluation of the difference between the model 
and the measured chromatic beating, provided that 
the former is computed by an optics code, such as 
MADX or PTC, which includes automatically this term.

For consistency with the nomenclature used 
throughout this paper, the phase advance 
$\Delta\phi_{mj}$ is to be computed as in 
Eq.~\eqref{eq:deltaphisign}. If the mere 
difference between the two betatron phases 
at the positions $m$ and $j$ is used, the 
absolute value $|2\Delta\phi_{x,mj}|$ 
shall then be used, as done in textbooks.
The normalized (dimensionless) chromatic beating 
eventually reads
\begin{eqnarray}\label{eq:RDT-beat6}
\left\{
\begin{array}{l}\displaystyle 
\left(\frac{1}{\beta_x}\frac{\partial\beta_x}{\partial\delta}\right)(j)\simeq 
\ \ \frac{1}{2\sin{(2\pi Q_x)}}
\sum\limits_{m=1}^M\left(K_{m,1}-K_{m,2}D_{m,x}+ J_{m,2}D_{m,y}\right)
                  \beta_{m,x}\cos{(2\Delta\phi_{x,mj}-2\pi Q_x)}-1
\\ \ \\ \displaystyle
\left(\frac{1}{\beta_y}\frac{\partial\beta_y}{\partial\delta}\right)(j)\simeq 
-\frac{1}{2\sin{(2\pi Q_y)}}
\sum\limits_{m=1}^M\left(K_{m,1}-K_{m,2}D_{m,x}+ J_{m,2}D_{m,y}\right)
                  \beta_{m,y}\cos{(2\Delta\phi_{y,mj}-2\pi Q_y)}-1
\end{array}
\right. \ .\qquad
\end{eqnarray}

\subsection{Chromatic phase advance shift}
\label{sec:ap:ChromPhAdvShift}
Quadrupole errors induce a betatron phase shift to 
particles with the nominal energy. When going off 
momentum the additional focusing provided by 
the off-axis closed orbit across sextupoles generate 
a similar {\sl chromatic} phase shift. 
In Ref.~\cite{Andrea-Linear-arxiv} 
analytic formulas relating the actual on-momentum 
betatron phase advance to the ideal one (from the 
model), detuning terms and the RDTs were derived:
\begin{eqnarray}
\left\{\begin{aligned}
\Delta\phi_{x,wj}&\simeq\Delta\phi_{x,wj}^{(mod)}
                 -2h_{1100,wj} 
                 +4\Re\left\{f_{2000,j}-f_{2000,w}\right\}\\
\Delta\phi_{y,wj}&\simeq\Delta\phi_{y,wj}^{(mod)}
                 -2h_{0011,wj} 
                 +4\Re\left\{f_{0020,j}-f_{0020,w}\right\}
\end{aligned}\right. , \label{eq:ChromPhAd_1} 
\end{eqnarray}
where the detuning terms $h_{wj}$ are the same of 
Eq.~\eqref{eq:detun2},
\begin{eqnarray}\label{eq:detun2B} 
\begin{aligned}
h_{1100,wj}&\simeq\sum_{m=1}^M{h_{m,1100}
              \Big[\Pi(m,j)-\Pi(m,w)+\Pi(j,w)\Big]}\quad , \quad 
h_{m,1100}=-\frac{1}{4}\beta_{m,x}^{(mod)}\delta K_{m,1}\quad ,\\
h_{0011,wj}&=\sum_{m=1}^M{h_{m,0011}
              \Big[\Pi(m,j)-\Pi(m,w)+\Pi(j,w)\Big]}\quad , \quad 
h_{m,0011}=+\frac{1}{4}\beta_{m,y}^{(mod)}\delta K_{m,1}\quad .
\end{aligned}
\end{eqnarray}
The binary function $\Pi$ is defined in Eq.~\eqref{eq:Theta}. 
The dependence on $\delta$ in $\Delta\phi_{x,wj}$ of 
Eq.~\eqref{eq:ChromPhAd_1} (identical relations apply 
in the vertical plane) can be made explicit according to
\begin{eqnarray}\label{eq:ChromPhAd_2}
\Delta\phi_{x,wj}(\delta)&\simeq&\Delta\phi_{x,wj}+
   \frac{\partial\Delta\phi_{x,wj}}{\partial\delta}\Bigg|_{\delta=0}\delta
+O(\delta^2)\\
  &=&\Delta\phi_{x,wj}^{(mod)} 
   -2\left[h_{1100,wj}+h_{11001,wj}\delta\right]
   +4\Re\left\{\left(f_{2000,j}-f_{2000,w}\right)
               +\left(f_{20001,j}-f_{20001,w}\right)\delta\right\}
+O(\delta^2)\ . \nonumber
\end{eqnarray}
From Eqs.~\eqref{eq:chrom3}, \eqref{eq:beat3},
\eqref{eq:RDT-beat1} and \eqref{eq:ReIm_SumDif} the 
following expressions for the chromatic phase advance 
shift, $\Delta\phi_{wj}'=\partial\Delta\phi_{wj}/\partial\delta|_{\delta=0}$ 
are derived
\begin{eqnarray}
\Delta\phi_{x,wj}'&\simeq&
   -\hspace{-2mm}\sum_{m=1}^M\left(K_{m,1}-\hspace{-1mm}K_{m,2}D_{m,x}+\hspace{-1mm}J_{m,2}D_{m,y}
   \right)\frac{\beta_{m,x}}{4}\left\{2\Big[\Pi(m,j)-\hspace{-.5mm}\Pi(m,w)+\hspace{-.5mm}\Pi(j,w)\Big]
   +\frac{\sin{(2\tau_{x,mj})}-\hspace{-.5mm}\sin{(2\tau_{x,mw})}}
         {\sin{(2\pi Q_x)}}\right\} , \nonumber\\
\Delta\phi_{y,wj}'&\simeq&
   +\hspace{-2mm}\sum_{m=1}^M\left(K_{m,1}-\hspace{-1mm}K_{m,2}D_{m,x}+\hspace{-1mm}J_{m,2}D_{m,y}
   \right)\frac{\beta_{m,y}}{4}\left\{2\Big[\Pi(m,j)-\hspace{-.5mm}\Pi(m,w)+\hspace{-.5mm}\Pi(j,w)\Big]
   +\frac{\sin{(2\tau_{y,mj})}-\hspace{-.5mm}\sin{(2\tau_{y,mw})}}
         {\sin{(2\pi Q_y)}}\right\} ,
\nonumber \\ \label{eq:ChromPhAd_3} 
\end{eqnarray}
where $\tau_{ab}$ is a mere shifted (on-momentum) phase 
advance between two locations $a$ and $b$,  
\begin{eqnarray}\label{eq:tau}
\tau_{z,ab}=\Delta\phi_{z,ab}-\pi Q_z\ ,\quad z=x,y\quad ,
\end{eqnarray}
and the (on-momentum) phase advance $\Delta\phi_{wj}$ 
is evaluated as usual according to 
Eq.~\eqref{eq:deltaphisign}. Interestingly, Eq.~\eqref{eq:ChromPhAd_1} 
can be made more explicit via Eqs.~\eqref{eq:ReIm_SumDif}, 
\eqref{eq:detun2B} to compute the on-momentum phase 
advance shift generated by quadrupole errors 
$\delta K_{1}$
\begin{eqnarray}
\begin{aligned}
\Delta\phi_{x,wj}&\simeq&\Delta\phi_{x,wj}^{(mod)}
   +\sum_{m=1}^M \delta K_{m,1}\frac{\beta_{m,x}^{(mod)}}{4}
   \left\{2\Big[\Pi(m,j)-\hspace{-.5mm}\Pi(m,w)+\hspace{-.5mm}\Pi(j,w)\Big]
   +\frac{\sin{(2\tau_{x,mj}^{(mod)})}-\hspace{-.5mm}\sin{(2\tau_{x,mw}^{(mod)})}}
         {\sin{(2\pi Q_x^{(mod)})}}\right\}\quad ,\hspace{8mm}\\
\Delta\phi_{y,wj}'&\simeq&\Delta\phi_{y,wj}^{(mod)}
   -\sum_{m=1}^M \delta K_{m,1}\frac{\beta_{m,y}^{(mod)}}{4}
    \left\{2\Big[\Pi(m,j)-\hspace{-.5mm}\Pi(m,w)+\hspace{-.5mm}\Pi(j,w)\Big]
   +\frac{\sin{(2\tau_{y,mj}^{(mod)})}-\hspace{-.5mm}\sin{(2\tau_{y,mw}^{(mod)})}}
         {\sin{(2\pi Q_y^{(mod)})}}\right\}\quad ,\hspace{8mm}
\end{aligned} \label{eq:PhAd_1} 
\end{eqnarray}
where $(mod)$ refers to the lattice model not including 
the quadrupole errors $\delta K_{1}$. If $w$ is the ring 
origin and $j$ its end, $\Pi(m,j)=1$, $\Pi(m,w)=\Pi(j,w)=0$, 
$\Delta\phi_{mw}=-\phi_{m}+2\pi Q=\Delta\phi_{mj}$ (see 
definition of $\Delta\phi$ in Eq.~\eqref{eq:deltaphisign}) 
and hence $\tau_{mj}=\tau_{mw}$, resulting in the standard
formulas for the tune shifts 
\begin{eqnarray}
\begin{aligned}
\delta Q_x=\frac{1}{2\pi}\left(\Delta\phi_{x,wj}-\Delta\phi_{x,wj}^{(mod)}\right)
    \simeq+\frac{1}{4\pi}\sum_{m=1}^M \delta K_{m,1}\beta_{m,x}^{(mod)}\quad ,\\
\delta Q_y=\frac{1}{2\pi}\left(\Delta\phi_{y,wj}-\Delta\phi_{y,wj}^{(mod)}\right)
    \simeq-\frac{1}{4\pi}\sum_{m=1}^M \delta K_{m,1}\beta_{m,y}^{(mod)}\quad .
\end{aligned} \label{eq:PhAd_1B} 
\end{eqnarray}

\subsection{Chromatic RDTs}
\label{sec:ap:ChromRDTs}
In Eqs.~\eqref{eq:RDT-beat2} and~\eqref{eq:RDT-beat3} 
the derivative of the beta-beating RDTs 
with respect to $\delta$ could be easily 
evaluated thanks to the fact the the geometric 
RDTs, i.e. those with $d=0$ are zero. 
Geometric coupling and higher-order RDTs are 
intrinsically non-zero quantities and some 
care is required in evaluating the derivative, 
since even the geometric RDTs depend on 
$\delta$ through the tunes in the denominator 
(i.e. chromaticity), the C-S parameters of the 
Hamiltonian coefficients $h_{pqrt}$ and the 
betatron phases in the numerator (i.e. chromatic 
beating and phase modulation). A first-order 
expansion in $\delta$ of Eq.~\eqref{eq:RDT-1st} 
reads
\begin{equation}\label{eq:RDTd-1}
f_{pqrt}(j,\delta) =f_{pqrt}(j)+f_{pqrt1}(j)\delta=
         \frac{\sum\limits_{m=1}^M \left(h_{m,pqrt}+h_{m,pqrt1}\delta\right)
               e^{i[(p-q)\Delta\phi_{x,mj}+(r-t)\Delta\phi_{y,mj}]}}
                 {1-e^{2\pi i[(p-q)Q_x+(r-t)Q_y]}}\ +O(\delta^2)\ .
\end{equation}
The dependence on $\delta$ is implicit in 
$h_{m,pqrt}$ through the beta functions (see 
Eq.~\eqref{eq:h_Vs_KJ}), whereas is ignored 
in $h_{m,pqrt1}$ since this term is already 
multiplied by $\delta$ and any additional 
dependence goes into the remainder proportional 
to $\delta^2$. The betatron phase advance 
$\Delta\phi_{mj}$ depends on the beam energy, 
but this dependence is kept only when the 
corresponding exponential term multiplies 
$h_{m,pqrt}$, while it is ignored when coupled 
with $h_{m,pqrt1}$. For sake of clarity the 
above definition may be rewritten ad
\begin{equation}\label{eq:RDTd-2a}
f(j,\delta) =\frac{\mathcal{A}(j,\beta_{\delta},\phi_{\delta})
              +\mathcal{B}(j)\delta}
              {1-e^{i(\mathcal{C}+\mathcal{D}\delta)}}\ +O(\delta^2)\ ,
\end{equation}
where
\begin{eqnarray}\label{eq:RDTd-2b}
\left\{
\begin{array}{l}
\mathcal{A}(j,\beta_{\delta},\phi_{\delta})=\displaystyle\hspace{-1.5mm}
\sum\limits_{m=1}^M \hspace{-1mm}
h_{m,pqrt}e^{i[(p-q)\Delta\phi_{x,mj}+(r-t)\Delta\phi_{y,mj}]}=\hspace{-1.5mm}
\sum\limits_{m=1}^M \hspace{-1mm}
\mathcal{G}_{m,pqrt}\bigl(\beta_{x,m,\delta}\bigr)^{\frac{p+q}{2}}
               \bigl(\beta_{y,m,\delta}\bigr)^{\frac{r+t}{2}}\hspace{-1mm}
e^{i[(p-q)\Delta\phi_{x,mj,\delta}+(r-t)\Delta\phi_{y,mj,\delta}]} 
\hskip-6mm\  \\
\mathcal{B}(j)=\displaystyle\sum\limits_{m=1}^M 
h_{m,pqrt1}e^{i[(p-q)\Delta\phi_{x,mj}+(r-t)\Delta\phi_{y,mj}]} \\
\mathcal{C}\hskip0.75mm=\displaystyle2\pi[(p-q)Q_x+(r-t)Q_y] \\
\mathcal{D}=\displaystyle2\pi[(p-q)Q_x'+(r-t)Q_y'] \\
\mathcal{G}_{m,pqrt}=\displaystyle\frac{h_{m,pqrt}}
                      {(\beta_{x,m,\delta})^{\frac{p+q}{2}}
                      (\beta_{y,m,\delta})^{\frac{r+t}{2}}}=
                    \frac{\bigl[K_{m,n-1}\Omega(r+t)+
                      iJ_{m,n-1}\Omega(r+t+1)\bigr]}
                      {p!\quad q!\quad r!\quad t!\quad 2^{p+q+r+t}}
\end{array}
\right. \hskip -1.0cm .\hskip 1.2cm  
\end{eqnarray}
$\mathcal{G}_{m,pqrt}$ is then nothing else 
than the Hamiltonian coefficient $h_{m,pqrt}$ 
normalized by the beta functions so to make 
this dependence explicit, see Eq.~\eqref{eq:h_Vs_KJ}. 
The dependence of $\mathcal{A}$ on $\delta$ 
is implicit in the C-S parameters, whereas 
$\mathcal{B}$, $\mathcal{C}$ and $\mathcal{D}$ 
are all $\delta$-independent to first 
order. The derivative evaluated at $\delta=0$ 
then reads
\begin{eqnarray}\label{eq:RDTd-3}
f_{pqrt1}(j)=
\frac{\partial f_{pqrt}(j,\delta)}{\partial \delta}\Bigg|_{\delta=0} 
   &=&\frac{\mathcal{A}'(j,\beta_{\delta},\phi_{\delta})+\mathcal{B}(j)}
               {1-e^{i(\mathcal{C}+\mathcal{D}\delta)}}\Bigg|_{\delta=0}
-\frac{\mathcal{A}(j,\beta_{\delta},\phi_{\delta})+\mathcal{B}(j)\delta}
                 {\left(1-e^{i(\mathcal{C}+\mathcal{D}\delta)}\right)^2}
                  \left(-e^{i(\mathcal{C}+\mathcal{D}\delta)}\right)
                  \left(i\mathcal{D}\right)
                  \Bigg|_{\delta=0} \nonumber \\ \nonumber \\
&=&\frac{\mathcal{A}'(j,\beta_{\delta},\phi_{\delta})|_{\delta=0}}
                     {1-e^{i\mathcal{C}}}
  +\frac{\mathcal{B}(j)}{1-e^{i\mathcal{C}}}
  +\frac{\mathcal{A}(j)}{1-e^{i\mathcal{C}}}
         \left(\frac{i\mathcal{D}}{e^{-i\mathcal{C}}-1}\right) \ .
\end{eqnarray}
The last term in the r.h.s. of the above equation 
is the geometrical RDT ($\mathcal{A}/(1-e^{i\mathcal{C}})$) 
multiplied by a factor proportional to the 
linear chromaticity $\mathcal{D}$. The second term 
is the one generated by the first-order chromatic 
Hamiltonian coefficients $h_{m,pqrt1}$ contained in 
$\mathcal{B}$, whereas the first contains the 
derivative of the geometric Hamiltonian coefficients 
$h_{m,pqrt}$ with respect to $\delta$. From the 
definition of $\mathcal{A}$ in Eq.~\eqref{eq:RDTd-3}, 
its derivative is
\begin{eqnarray}\label{eq:RDTd-4}
\mathcal{A}'(j,\beta_{\delta},\phi_{\delta})&=&
            \displaystyle\sum\limits_{m=1}^M\mathcal{G}_{m,pqrt}
            \bigl(\beta_{m,x}\bigr)^{\frac{p+q}{2}}
            \bigl(\beta_{m,y}\bigr)^{\frac{r+t}{2}}
e^{i[(p-q)\Delta\phi_{x,mj}+(r-t)\Delta\phi_{y,mj}]} 
\times \\ &&\quad \left\{
 \left(\frac{p+q}{2}\right)\left(\frac{1}{\beta_{m,x}}\frac{\partial \beta_{m,x}}{\partial\delta}\right)
+\left(\frac{r+t}{2}\right)\left(\frac{1}{\beta_{m,y}}\frac{\partial \beta_{m,y}}{\partial\delta}\right)
+i\left[(p-q)\Delta\phi_{x,mj}'+(r-t)\Delta\phi_{y,mj}'\right]
   \right\}\ .\nonumber
\end{eqnarray}
We notice that $\mathcal{G}_{m,pqrt}(\beta_{m,x})^{\frac{p+q}{2}}
(\beta_{m,y})^{\frac{r+t}{2}}$ is the original 
Hamiltonian coefficient $h_{m,pqrt}$, whereas 
$(1/\beta)\partial\beta/\partial\delta$ is the 
normalized chromatic beating at the magnet 
$m$ of  Eq.~\eqref{eq:RDT-beat5} 
and the chromatic phase advance 
shift $\Delta\phi'=\partial\Delta\phi/\partial\delta$ 
is the same of Eq.~\eqref{eq:ChromPhAd_3}. 
A first rude approximation for $\Delta\phi'$ can be 
also made assuming that such a shift be linear 
with chromaticity, i.e. 
\begin{eqnarray}\label{eq:RDTd-5}
\Delta\phi_{x,mj}'\simeq\Delta\phi_{x,mj}\frac{Q_x'}{Q_x}\qquad , \qquad
\Delta\phi_{y,mj}'\simeq\Delta\phi_{y,mj}\frac{Q_y'}{Q_y} \quad .
\end{eqnarray}
Equation~\eqref{eq:RDTd-3} eventually reads
\begin{eqnarray}\label{eq:RDTd-6}
f_{pqrt1}(j)&=&
 \frac{\sum\limits_{m=1}^M\mathcal{H}_{m,pqrt}(j)
       e^{i[(p-q)\Delta\phi_{x,mj}+(r-t)\Delta\phi_{y,mj}]}}
      {1-e^{2\pi i[(p-q)Q_x+(r-t)Q_y]}}
+ \frac{\sum\limits_{m=1}^Mh_{m,pqrt1}
       e^{i[(p-q)\Delta\phi_{x,mj}+(r-t)\Delta\phi_{y,mj}]}}
      {1-e^{2\pi i[(p-q)Q_x+(r-t)Q_y]}} + \nonumber \\ &&
  f_{pqrt}\left(\frac{2\pi i[(p-q)Q_x'+(r-t)Q_y']}{
                e^{-2\pi i[(p-q)Q_x+(r-t)Q_y]}-1}\right)\ ,
\end{eqnarray} 
where $f_{pqrt}$ is the geometrical RDT of 
Eq.~\eqref{eq:RDT-1st} and $\mathcal{H}_{m,pqrt}$ 
is defined as 
\begin{eqnarray}
\mathcal{H}_{m,pqrt}(j)\simeq h_{m,pqrt}\left\{
 \frac{p+q}{2}\left(\frac{1}{\beta_{m,x}}\frac{\partial \beta_{m,x}}{\partial\delta}\right)
+\frac{r+t}{2}\left(\frac{1}{\beta_{m,y}}\frac{\partial \beta_{m,y}}{\partial\delta}\right)
+i\left[(p-q)\Delta\phi_{x,mj}'+(r-t)\Delta\phi_{y,mj}'\right]
   \right\}\ . \nonumber \\\label{eq:RDTd-7}
\end{eqnarray}
In the case of the chromatic beating the 
derivative reduces to the term generated by 
$h_{pqrt1}$ only, as indicated by 
Eq.~\eqref{eq:RDT-beat2}, since $h_{m,pqrt}=0$ 
and hence $ f_{pqrt}=0$, .

\subsection{Chromatic coupling}
\label{sec:ap:ChromCoup}
Betatron coupling between the two transverse 
planes is generated by tilted quadrupoles, 
$J_1=-K_1\sin{(2\theta)}$, where $\theta$ is the 
rotation angle, and non-zero vertical closed 
orbit inside sextupole magnets, whose feed-down 
field is of the skew-quadrupole type. Betatron 
coupling induces some vertical dispersion, on 
top of the one generated by any source of vertical 
deflection along the ring. When going off momentum, 
vertical dispersion adds an additional vertical 
beam displacement across the sextupoles, hence 
generating a new {\sl chromatic} coupling. If 
skew sextupole fields are also present, the 
horizontal displacements induced by the natural 
horizontal dispersion contribute also to 
coupling. The two RDTs describing betatron coupling 
are $f_{1001}$ and $f_{1010}$ for the difference 
and sum resonances, respectively. The first 
step in evaluating their derivative with respect 
to $\delta$, i.e. chromatic coupling, is to 
compute the Hamiltonian coefficients $h_{m,10011}$ 
and $h_{m,10101}$ from Eq.~\eqref{eq:Hpqrtd4}.
\begin{eqnarray}
\left\{
\begin{array}{l}
h_{m,10011}= 2h_{m,2001}\ d_{m,x,+}+\ h_{m,1101}\ d_{m,x,-}+
            \ h_{m,1011}\ d_{m,y,+}+ 2h_{m,1002}\ d_{m,y,-}-h_{m,1001} \\
h_{m,10101}= 2h_{m,2010}\ d_{m,x,+}+\ h_{m,1110}\ d_{m,x,-}+
             2h_{m,1020}\ d_{m,y,+}+\ h_{m,1011}\ d_{m,y,-}-h_{m,1010}
\end{array}
\right. \quad .\label{eq:chrCoup1}
\end{eqnarray}
Once again, the Hamiltonian coefficients in 
the above r.h.s. may be made explicit via 
Eq.~\eqref{eq:h_Vs_KJ}: 
\begin{eqnarray}\label{eq:chrCoup1b}
h_{m,1101}&=&h_{m,1110}=2h_{m,2010}=2h_{m,2001}=+\frac{1}{8}J_{m,2}\beta_{m,x}\sqrt{\beta_{m,y}}\quad ,\quad \\
h_{m,1001}&=&h_{m,1010}=+\frac{1}{4}J_{m,1}\sqrt{\beta_{m,x}\beta_{m,y}}\quad ,\qquad 
h_{m,1011} =2h_{m,1002}= 2h_{m,1020}=+\frac{1}{8}K_{m,2}\sqrt{\beta_{m,x}}\beta_{m,y}\quad .\ \qquad
\end{eqnarray}
Equation~\eqref{eq:chrCoup1} then becomes
\begin{eqnarray}
\left\{
\begin{array}{l}
h_{m,10011}=2h_{m,1101\ }\Re\left\{d_{m,x,\pm}\right\}
           +2h_{m,1011\ }\Re\left\{d_{m,y,\pm}\right\}-h_{m,1001}\\
h_{m,10101}=2h_{m,1110\ }\Re\left\{d_{m,x,\pm}\right\}
           +2h_{m,1011\ }\Re\left\{d_{m,y,\pm}\right\}-h_{m,1010}
\end{array}
\right. \quad .\label{eq:chrCoup2}
\end{eqnarray}
Since $\Re\left\{d_{m,q,+}\right\}=D_{m,q}/\sqrt{\beta_{m,q}}$, the above 
equations reduce to
\begin{eqnarray}
\left\{
\begin{array}{l}
h_{m,10011}=-\displaystyle\frac{1}{4}
             \left(J_{m,1}-K_{m,2}D_{m,y}-J_{m,2}D_{m,x}\right)
             \sqrt{\beta_{m,x}\beta_{m,y}}  \vspace{ 1.5mm}\\
h_{m,10101}=-\displaystyle\frac{1}{4}
             \left(J_{m,1}-K_{m,2}D_{m,y}-J_{m,2}D_{m,x}\right)
             \sqrt{\beta_{m,x}\beta_{m,y}} 
\end{array}
\right. \quad .\label{eq:chrCoup3}
\end{eqnarray}
As expected, the two terms are identical, 
$h_{m,10011}=h_{m,10101}$, as equal are the 
geometrical coefficients $h_{m,1001}=h_{m,1010}$. 
The Hamiltonian term at a generic position $j$ 
accounting for all magnets is derived from 
Eq.~\eqref{eq:Hdef4},
\begin{eqnarray}
\left\{
\begin{array}{l}
\tilde{H}_{10011}(j)=-\displaystyle\frac{1}{4}\sum\limits_{m=1}^M
                   \left(J_{m,1}-K_{m,2}D_{m,y}- J_{m,2}D_{m,x}\right)
                   \sqrt{\beta_{m,x}\beta_{m,y}}
                   e^{i(\Delta\phi_{x,mj}-\Delta\phi_{y,mj})}
                        \ h_{m,x,+}h_{m,y,-}\delta  \vspace{ 1.5mm}\\
\tilde{H}_{10101}(j)=-\displaystyle\frac{1}{4}\sum\limits_{m=1}^M
                   \left(J_{m,1}-K_{m,2}D_{m,y}- J_{m,2}D_{m,x}\right)
                   \sqrt{\beta_{m,x}\beta_{m,y}}
                   e^{i(\Delta\phi_{x,mj}+\Delta\phi_{y,mj})}
                        \ h_{m,x,+}h_{m,y,+}\delta
\end{array}
\right. \quad ,\label{eq:chrCoup4}
\end{eqnarray}
From Eq.~\eqref{eq:RDTd-6} the following 
expressions for the chromatic coupling are 
eventually derived:
\begin{eqnarray}\label{eq:chrCoup5}
\left\{
\begin{array}{l}\displaystyle
f_{10011}(j)= \frac{\partial f_{1001}(j)}{\partial\delta}\Bigg|_{\delta=0} \simeq
 \frac{\sum\limits_{m=1}^M\left(\mathcal{H}_{m,1001}(j)+h_{m,10011}\right)
       e^{i(\Delta\phi_{x,mj}-\Delta\phi_{y,mj})}}
      {1-e^{2\pi i(Q_x-Q_y)}}
 +f_{1001}(j)\left(\frac{2\pi i(Q_x'-Q_y')}{e^{-2\pi i(Q_x-Q_y)}-1}\right) 
\\\displaystyle
f_{10101}(j)= \frac{\partial f_{1010}(j)}{\partial\delta}\Bigg|_{\delta=0} \simeq
 \frac{\sum\limits_{m=1}^M\left(\mathcal{H}_{m,1010}(j)+h_{m,10101}\right)
       e^{i(\Delta\phi_{x,mj}+\Delta\phi_{y,mj})}}
      {1-e^{2\pi i(Q_x+Q_y)}}
 +f_{1010}(j)\left(\frac{2\pi i(Q_x'+Q_y')}{e^{-2\pi i(Q_x+Q_y)}-1}\right) 
\end{array}
\right. \hskip-0.3cm ,\hskip 1.2cm
\end{eqnarray}
where the Hamiltonian terms $h_{m,10011}$ and 
$h_{m,10101}$ are those of Eq.~\eqref{eq:chrCoup3}, 
whereas the geometrical RDTs are derived from 
Eq.~\eqref{eq:RDT-1st} and read
\begin{eqnarray}
f_{1001}(j)=\displaystyle\frac{\sum\limits_{m=1}^M  J_{m,1}\sqrt{\beta_{m,x}\beta_{m,y}} 
                    e^{i(\Delta\phi_{x,mj} - \Delta\phi_{y,mj})}}
               {4\left[1-e^{2\pi i(Q_x-Q_y)}\right]} \quad , \quad
f_{1010}(j)=\displaystyle\frac{\sum\limits_{m=1}^M  J_{m,1}\sqrt{\beta_{mx,}\beta_{m,y}} 
                    e^{i(\Delta\phi_{x,mj} + \Delta\phi_{y,mj})}}
               {4\left[1-e^{2\pi i(Q_x+Q_y)}\right]} \quad . \qquad
\end{eqnarray}
According to Eq.~\eqref{eq:RDTd-7}, the 
modified chromatic Hamiltonian terms $\mathcal{H}$ 
are
\begin{eqnarray}
\left\{
\begin{array}{l}\displaystyle
\mathcal{H}_{m,1001}(j)\simeq\frac{1}{4}J_{m,1}\sqrt{\beta_{m,x}\beta_{m,y}} \left\{
 \frac{1}{2}\left(\frac{1}{\beta_{m,x}}\frac{\partial \beta_{m,x}}{\partial\delta}\right)
+\frac{1}{2}\left(\frac{1}{\beta_{m,y}}\frac{\partial \beta_{m,y}}{\partial\delta}\right)
+i\left[\Delta\phi_{x,mj}'-\Delta\phi_{y,mj}'\right]
   \right\} \\ \ \\\displaystyle
\mathcal{H}_{m,1010}(j)\simeq\frac{1}{4}J_{m,1}\sqrt{\beta_{m,x}\beta_{m,y}} \left\{
 \frac{1}{2}\left(\frac{1}{\beta_{m,x}}\frac{\partial \beta_{m,x}}{\partial\delta}\right)
+\frac{1}{2}\left(\frac{1}{\beta_{m,y}}\frac{\partial \beta_{m,y}}{\partial\delta}\right)
+i\left[\Delta\phi_{x,mj}'+\Delta\phi_{y,mj}'\right]
   \right\} 
\end{array}
\right. \quad .
\end{eqnarray}
The chromatic beating at the location of the 
magnet $(1/\beta_m\partial\beta_m/\partial\delta)$ 
can be computed from Eq.~\eqref{eq:RDT-beat5} 
after replacing $j$ with $m$. Note that, even if 
chromatic beating terms and linear chromaticity $Q'$  
scale with the strengths of normal quadrupoles 
and sextupoles (normal and skew), see 
Eqs.~\eqref{eq:RDT-beat5} and~\eqref{eq:chrom6}, 
they do not depend on the skew quadrupole strengths 
$J_{1}$. This implies that all terms in the r.h.s. of 
Eq.~\eqref{eq:chrCoup5} are linear in $J_{1}$. 

For practical purpose, since the sextupole correctors, 
$\delta K_2$ and $\delta J_2$, will be used for the 
simultaneous correction of all chromatic terms, 
Eq.~\eqref{eq:chrCoup5} may be rewritten as 
\begin{eqnarray}\label{eq:chrCoup6}
\left\{
\begin{array}{l}\displaystyle
f_{10011}(j)= \frac{\partial f_{1001}(j)}{\partial\delta}\Bigg|_{\delta=0} =
 \frac{\sum\limits_{m=1}^M h_{m,10011}
       e^{i(\Delta\phi_{x,mj}-\Delta\phi_{y,mj})}}
      {1-e^{2\pi i(Q_x-Q_y)}}
 +F_{10011}(j,J_1)
\\\displaystyle
f_{10101}(j)= \frac{\partial f_{1010}(j)}{\partial\delta}\Bigg|_{\delta=0} =
 \frac{\sum\limits_{m=1}^M h_{m,10101}
       e^{i(\Delta\phi_{x,mj}+\Delta\phi_{y,mj})}}
      {1-e^{2\pi i(Q_x+Q_y)}}
 +F_{10101}(j,J_1)
\end{array}
\right. \quad ,
\end{eqnarray}
with both $h_{m,10011}$ and $h_{m,10101}$ scaling 
linearly with the sextupole fields, see 
Eq.~\eqref{eq:chrCoup3}. On the other hand 
the two functions $F_{10011}$ and 
$F_{10101}$ do not depend on the sextupole 
fields to first order (provided that 
sextupole correctors do not alter the 
linear chromaticity $Q'$) and are linear in 
the skew quadrupole strengths (which are 
assumed to be fixed to correct betatron coupling 
and vertical dispersion)
\begin{eqnarray}
\left\{
\begin{array}{l}
\begin{array}{l}\displaystyle
F_{10011}(j)\simeq\sum\limits_{m=1}^M 
            \frac{J_{m,1}\sqrt{\beta_{m,x}\beta_{m,y}}}
              {4\left[1-e^{2\pi i(Q_x-Q_y)}\right]}\left\{
 \frac{1}{2}\left(\frac{1}{\beta_{m,x}}\frac{\partial \beta_{m,x}}{\partial\delta}\right)
+\frac{1}{2}\left(\frac{1}{\beta_{m,y}}\frac{\partial \beta_{m,y}}{\partial\delta}\right)
+i\left[\Delta\phi_{x,mj}'-\Delta\phi_{y,mj}'\right]+\right.  \\
       \left.\hskip 5.75cm\displaystyle
       \left(\frac{2\pi i(Q_x'-Q_y')}{e^{-2\pi i(Q_x-Q_y)}-1}\right)\right\}
         e^{i(\Delta\phi_{x,mj}-\Delta\phi_{y,mj})} 
\end{array} \\ \ \\ 
\begin{array}{l}\displaystyle
F_{10101}(j)\simeq\sum\limits_{m=1}^M 
              \frac{J_{m,1}\sqrt{\beta_{m,x}\beta_{m,y}}}
              {4\left[1-e^{2\pi i(Q_x+Q_y)}\right]}\left\{
 \frac{1}{2}\left(\frac{1}{\beta_{m,x}}\frac{\partial \beta_{m,x}}{\partial\delta}\right)
+\frac{1}{2}\left(\frac{1}{\beta_{m,y}}\frac{\partial \beta_{m,y}}{\partial\delta}\right)
+i\left[\Delta\phi_{x,mj}'+\Delta\phi_{y,my}'\right]+\right.  \\
       \left.\hskip 5.75cm\displaystyle
       \left(\frac{2\pi i(Q_x'+Q_y')}{e^{-2\pi i(Q_x+Q_y)}-1}\right)\right\}
         e^{i(\Delta\phi_{x,mj}+\Delta\phi_{y,mj})} 
\end{array}
\end{array}
\right. . \nonumber \\ \label{eq:chrCoup7}
\end{eqnarray}
The sextupole correctors may actually change
$F_{10011}$ and $F_{10101}$, through the 
chromatic beating $(1/\beta_m\partial\beta_m/\partial\delta)$. 
This variation is however of second order, 
being proportional to product of small 
quantities $J_1\times \delta K_2$, compared to 
the baseline values which scales with the 
nominal chromatic beating, and hence with 
 $J_1\times K_2$, where $K_2$ refer to the 
chromatic and harmonic sextupoles, whose strengths 
are usually order of magnitudes greater than 
the corrector strengths $\delta K_2$. The 
first terms in the r.h.s. of Eq.~\eqref{eq:chrCoup6} 
can be then effectively used to compute the 
chromatic coupling response to sextupole 
correctors.

\subsection{Second-order Hamiltonian contribution to the first-order 
                      chromatic terms ($d=1$)}
\label{sec:app-SecOrdHam}
In dealing with chromatic functions such as dispersion, 
chromaticity, chromatic beating and chromatic 
coupling, all expansions in $\delta$ are truncated 
to first order, i.e. only contributions 
linear in $\delta$ are kept ($d=1$ in the set 
of indices $pqrtd$), as reported in 
Sec.~\ref{sec:d1}. Nevertheless an implicit 
and hidden truncation has been performed at the 
very beginning when deriving 
Eqs.~\eqref{eq:Hdef2}-\eqref{eq:Hdef4}, which needs to be 
further investigated in order to account for all 
sources of chromatic terms proportional to $\delta$. 
The Hamiltonian of Eqs.~\eqref{eq:Hdef2}-\eqref{eq:Hdef4} 
results indeed from the truncation to first 
order of the one-turn map describing the betatron 
motion. Back in the '90s~\cite{Bengtsson} and more 
recently~\cite{Andrea-arxiv} it has been shown 
how the second-order contribution to the Hamiltonian 
terms and to the corresponding RDTs are
\begin{eqnarray}\label{eq:HamSecOrd}
H^{(2)}=\tilde{H}^{(2)}
+\frac{1}{2}\left[<\tilde{H}^{(1)}>_{\phi}\ ,\ 
                  \frac{I+R}{I-R}\tilde{H}^{(1)\ddagger}\right]
+\frac{1}{2}\left[\tilde{H}^{(1)\ddagger}\ ,\ 
                  \frac{\tilde{H}^{(1)\ddagger}}{I-R}\right] \quad ,\qquad
F^{(2)}=\frac{H^{(2)\ddagger}}{(I-R)}\ ,
\end{eqnarray}
where $[A,B]$ denotes the Poisson brackets 
between two operators $A$ and $B$, and 
\begin{itemize}
\item $\tilde{H}^{(2)}=\displaystyle
        \frac{1}{2}\sum\limits_{m=1}^M{\sum\limits_{u=1}^{m-1}
        {\left[\tilde{H}_u,\tilde{H}_m\right]}}$ is the 
      second-order contribution stemming from the 
      Poisson brackets between the first-order 
      Hamiltonians of Eq.~\eqref{eq:Hdef2}; 
\item $<\tilde{H}^{(1)}>_{\phi}$ is the first-order 
      Hamiltonian of Eq.~\eqref{eq:Hdef2} containing 
      terms independent of the betatron phase, i.e. 
      with $p=q$ and $r=t$.
\item $\tilde{H}^{(1)\ddagger}$ is the same Hamiltonian 
      of Eq.~\eqref{eq:Hdef2}, but containing terms 
      dependent of the betatron phase, i.e. 
      $p\ne q$ or $r\ne t$.
\item $R\rightarrow e^{2\pi i\left[(p-q)Q_x+(r-t)Q_y)\right]}$ 
      is the rotational term for each set of index $pqrt$.
\item $H^{(2)\ddagger}$ is the phase-dependent part 
      of the above second-order Hamiltonian 
      $H^{(2)}$, i.e. with $p\ne q$ and $r\ne t$.
\end{itemize}
First we analyze in details $\tilde{H}^{(2)}$. 
Eq.~\eqref{eq:Hdef3} reads
\begin{eqnarray}
\tilde{H}_{u}=\sum_d\tilde{H}_{u,d}\delta^d=
                    \tilde{H}_{u,0}+\tilde{H}_{u,1}\delta 
                    +O(\delta^2)\ ,
\end{eqnarray}
where $\tilde{H}_{u,0}$ is the geometric Hamiltonian 
of Eq.~\eqref{eq:Hdef2} and $\tilde{H}_{u,1}$ is the 
corresponding first chromatic Hamiltonian, whose 
terms are those of Eq.~\eqref{eq:Hpqrtd4}. 
$\tilde{H}^{(2)}$ then reads
\begin{eqnarray}
\tilde{H}^{(2)}&=&\displaystyle
        \frac{1}{2}\sum\limits_{m=1}^M{\sum\limits_{u=1}^{m-1}
        {\left[\tilde{H}_u\ ,\ \tilde{H}_m\right]}}=
        \frac{1}{2}\sum\limits_{m=1}^M{\sum\limits_{u=1}^{m-1}
        {\left[\tilde{H}_{u,0}+\tilde{H}_{u,1}\delta\ ,\ 
               \tilde{H}_{m,0}+\tilde{H}_{m,1}\delta\right]}} + O(\delta^2) 
\nonumber \\ &=&\displaystyle\nonumber 
        \frac{1}{2}\sum\limits_{m=1}^M{\sum\limits_{u=1}^{m-1}
        {\left[\tilde{H}_{u,0}\ , \tilde{H}_{m,0}\right] +
         \left\{\left[\tilde{H}_{u,0}\ ,\ \tilde{H}_{m,1}\right]
               +\left[\tilde{H}_{u,1}\ ,\ \tilde{H}_{m,0}\right]
               \right\}\delta}} + O(\delta^2) \quad .\\
&&\hskip 25mm\big\Downarrow\hskip 38mm \big\Downarrow \nonumber \\
&&\hskip 25mm\tilde{H}_{0}^{(2)}\hskip 33mm \tilde{H}_{1}^{(2)}
\label{eq:H2tildeA}
\end{eqnarray}
The first Poisson bracket contains purely 
geometric Hamiltonian terms, whose explicit 
expressions can be found in Ref.~\cite{Andrea-arxiv}, 
and are of no interest in the context of 
this paper. The last two Poisson brackets 
are instead to be evaluated, as their contribution 
is linear in $\delta$. In Ref.~\cite{Andrea-arxiv} 
a procedure for their evaluation has been 
derived. For each chromatic second-order 
Hamiltonian term $\tilde{H}_{pqrt1}^{(2)}$ 
($d=1$, i.e. linear in $\delta$), the 
double summation and the Poisson brackets 
results in the following systems to be solved
\begin{eqnarray}\label{eq:select0}
\tilde{H}^{(2)}_{pqrt1}(j)=\left\{\tilde{h}^{(2)}_{pqrt1}(j)\right\}
                   h_{x,+}^ph_{x,-}^qh_{y,+}^rh_{y,-}^t\delta
\quad \Rightarrow\hspace{0.5cm}
S1:\ 
\left\{
\begin{array}{l}
a+l-1=p \\
b+k-1=q \\
c+n=r   \\
e+o=t   
\end{array}
\right.
\quad\hbox{and}\quad
S2:\ 
\left\{
\begin{array}{l}
a+l=p   \\
b+k=q   \\
c+n-1=r \\
e+0-1=t 
\end{array}
\right. \quad , \hskip 1cm
\end{eqnarray}
\begin{eqnarray}
\tilde{h}^{(2)}_{pqrt1}(j)&=&i
       \sum\limits_{m=1}^M{\sum\limits_{u=1}^{m-1}\Bigg\{
	{\sum_{abce}\sum_{lkno}
         \left(h_{u,abce}h_{m,lkno1}+h_{u,abce1}h_{m,lkno}\right)
	 e^{i[(a-b)\Delta\phi_{x,uj}+(c-e)\Delta\phi_{y,uj}
	     +(l-k)\Delta\phi_{x,mj}+(n-o)\Delta\phi_{y,mj}]}
}}     \nonumber\\ &&\hskip 3 cm	\times
\bigg[(lb-ka)\Big|_{S1}+ (ne-oc)\Big|_{S2}\bigg]\Bigg\}\ .\label{eq:H2tilde}
\end{eqnarray}
The subscripts $\big|_{S1}$ and $\big|_{S2}$ 
mean that when selecting the two sets of 
indices $abce$ and $lkno$ from either of 
the two systems of Eq.~\eqref{eq:select0}, 
only the corresponding parenthesis is to be 
computed and the other ignored. In 
Eq.~\eqref{eq:H2tilde} then, $h_{u,abce}$ 
and $h_{m,lkno}$ are the geometric Hamiltonian 
coefficients of Eq.~\eqref{eq:h_Vs_KJ}, 
whereas $h_{u,abce1}$ and $h_{m,lkno1}$ are 
the chromatic Hamiltonian terms of 
Eq.~\eqref{eq:Hpqrtd4}. 
To complete the evaluation of second-order 
contribution to the linear chromatic functions, 
the other two elements in the r.h.s. of 
Eq.~\eqref{eq:HamSecOrd} need to be computed. 
Both are Poisson brackets similar to those of 
$\tilde{H}^{(2)}$ and the same selection rules 
of Eq.~\eqref{eq:select0} apply. The difference 
will be only the explicit form. 
The dependence on $\delta$ of the first Poisson 
bracket in Eq.~\eqref{eq:HamSecOrd} reads
\begin{eqnarray}
\bar{H}^{(2)}\displaystyle
&=&\frac{1}{2}\left[<\tilde{H}^{(1)}>_{\phi}\ ,\ 
                  \frac{I+R}{I-R}\tilde{H}^{(1)\ddagger}\right]
=\frac{1}{2}\sum\limits_{m=1}^M{\sum\limits_{u=1}^{M}
{\left[<\tilde{H}_u^{(1)}>\ ,\ \frac{I+R}{I-R}\tilde{H}_m^{\ddagger}\right]}}
\nonumber \\
&=&\frac{1}{2}\sum\limits_{m=1}^M{\sum\limits_{u=1}^{M}
{\left[<\tilde{H}_{u,0}^{(1)}>+<\tilde{H}_{u,1}^{(1)}>\delta\ ,\ 
 \frac{I+R_0+R'\delta}{I-R_0-R'\delta}\ 
  (\tilde{H}_{m,0}^{\ddagger}+\tilde{H}_{m,1}^{\ddagger}\delta)\right]}}
  +O(\delta^2)\ ,
\end{eqnarray}
where the linear dependence on $\delta$ of the 
rotation $R$ includes the chromatic rotation $R'$ 
(which is the linear chromaticity of Eq.~\eqref{eq:chrom6}), 
since $R=R_0+R'\delta +O(\delta^2)$.
The average over the phases maintains only those 
(detuning) Hamiltonian terms independent on 
them, i.e. with $p=q$ and $r=t$, such as $h_{1100}$ 
and $h_{0011}$ (from quadrupolar errors 
$\delta k_1$) and $h_{2200}$, $h_{1111}$ and 
$h_{0022}$ (from octupole magnets). As we assume 
to include all focusing errors in the linear model 
to evaluate the C-S parameters, $h_{1100}=h_{0011}=0$. 
Since the chromatic terms studied in this 
paper are not affected by octupolar terms, 
their corresponding first-order Hamiltonian 
coefficients are zero too (note that octupolar-like 
Hamiltonian terms are generated by sextupoles, 
though to second order only, i.e. through a 
non-zero $<\tilde{H}_{u,0}^{(2)}>_{\phi}$). Hence 
$<\tilde{H}^{(1)}_{u,0}>_{\phi}=0$, whereas 
$<\tilde{H}^{(1)}_{u,1}>_{\phi}$ contains only two 
phase-independent Hamiltonian terms, $h_{11001}$ 
and $h_{00111}$, i.e. the two non-zero chromaticities.
$\bar{H}^{(2)}$ then reads
\begin{eqnarray}
\bar{H}^{(2)}\displaystyle
&=&\frac{1}{2}\sum\limits_{m=1}^M{\sum\limits_{u=1}^{M}
{\left[<\tilde{H}_{u,1}^{(1)}>\ ,\ 
 \frac{I+R_0}{I-R_0}\ \tilde{H}_{m,0}^{\ddagger}\right]}}
  \delta\ +\ O(\delta^2)\ ,
\end{eqnarray}
The same procedure applied for 
Eq.~\eqref{eq:H2tildeA} is repeated here, 
with the addition of the factor $(1+R_0)/(I-R_0)$, 
\begin{eqnarray}\nonumber 
\bar{h}^{(2)}_{pqrt1}(j)&=&\frac{1}{2}\left[<\tilde{H}^{(1)}>_{\phi}\ ,\ 
                  \frac{I+R_0}{I-R_0}\tilde{H}^{(1)\ddagger}\right]_{abce1} 
\\\nonumber &=&i\sum\limits_{m=1}^M\sum\limits_{u=1}^{M}\Bigg\{
		\sum_{lkno}
              \left(h_{u,11001}+h_{u,00111}\right)h_{m,lkno}
              \frac{1+e^{-2\pi i[(l-k)Q_x+(n-o)Q_y]}}
                   {1-e^{-2\pi i[(l-k)Q_x+(n-o)Q_y]}}\times\\
&&\hskip 26mm e^{i[(l-k)\Delta\phi_{x,uj}+(n-o)\Delta\phi_{y,uj}]}
     \bigg[(l-k)\Big|_{S1}+ (n-o)\Big|_{S2}\bigg]\Bigg\} \ ,
     \nonumber\\
&&\hskip 0mm
\hbox{with }(abce=1100\ ,\ abce=0011)
\hbox{ and }(l\ne k\hbox{ or }n\ne o)
.\label{eq:H2bar}
\end{eqnarray}
The same algebra can be applied to the last Poisson 
bracket in Eq.~\eqref{eq:HamSecOrd}, 
\begin{eqnarray}
\hat{H}^{(2)}\displaystyle
&=&\frac{1}{2}\left[\tilde{H}^{(1)\ddagger}\ ,\ 
   \frac{\tilde{H}^{(1)\ddagger}}{I-R}\right]
=\frac{1}{2}\sum\limits_{m=1}^M{\sum\limits_{u=1}^{M}
{\left[\tilde{H}_u^{\ddagger}\ ,\ \frac{\tilde{H}_m^{\ddagger}}{I-R}\right]}}
\nonumber \\
&=&\frac{1}{2}\sum\limits_{m=1}^M{\sum\limits_{u=1}^{M}
        {\left[\tilde{H}_{u,0}^{\ddagger}+\tilde{H}_{u,1}^{\ddagger}\delta\ ,\ 
               \frac{\tilde{H}_{m,0}^{\ddagger}}{I-R_0}+\left(\frac{\tilde{H}_{m,1}^{\ddagger}}{I-R_0}
                +\frac{\tilde{H}_{m,0}^{\ddagger}}{(I-R_0)^2}R'\right)\delta\right]}} 
      + O(\delta^2) \nonumber  \\
&=&\displaystyle\nonumber 
  \frac{1}{2}\sum\limits_{m=1}^M{\sum\limits_{u=1}^{M}
  {\left[\tilde{H}_{u,0}^{\ddagger}\ , 
          \frac{\tilde{H}_{m,0}^{\ddagger}}{I-R_0}\right] +
  \left\{\left[\tilde{H}_{u,0}^{\ddagger}\ ,\ 
          \frac{\tilde{H}_{m,1}^{\ddagger}}{I-R_0}\right]
  +\left[\tilde{H}_{u,1}^{\ddagger}\ ,\ 
          \frac{\tilde{H}_{m,0}^{\ddagger}}{I-R_0}\right]
  +\left[\tilde{H}_{u,0}^{\ddagger}\ ,\ 
          \frac{\tilde{H}_{m,0}^{\ddagger}}{(I-R_0)^2}R'\right]
  \right\}\delta}} + O(\delta^2) \quad . \label{eq:H2hat0} \\
&&\hskip 25mm\big\Downarrow\hskip 65mm \big\Downarrow \nonumber \\
&&\hskip 25mm\hat{H}_{0}^{(2)}\hskip 60mm \hat{H}_{1}^{(2)}
\end{eqnarray}
The first sum, $\hat{H}_{0}^{(2)}$, in the above expression 
can be ignored, as it contains purely 
geometric terms not impacting the chromatic 
functions. The second block, $\hat{H}_{1}^{(2)}$, 
is instead proportional to $\delta$ and 
shall be made explicit: 
\begin{eqnarray}\nonumber 
\hat{h}^{(2)}_{pqrt1}(j)&=&\frac{1}{2}\left[\tilde{H}^{(1)\ddagger}\ ,\ 
                  \frac{\tilde{H}^{(1)\ddagger}}{I-R}\right]_{abce1}
\\\nonumber &=&i\sum\limits_{m=1}^M\sum\limits_{u=1}^{M}\Bigg\{
		\sum_{abce}\sum_{lkno}
              \left(h_{u,abce}h_{m,lkno1}+h_{u,abce1}h_{m,lkno}+
              \frac{h_{u,abce}h_{m,lkno}2\pi i[(l-k)Q_x'+(n-o)Q_y']}
                   {e^{-2\pi i[(l-k)Q_x+(n-o)Q_y]}-1}\right)\times\\
&&\hskip 30mm\frac{e^{i[(l-k)\Delta\phi_{x,uj}+(n-o)\Delta\phi_{y,uj}
     +(a-b)\Delta\phi_{x,mj}+(c-e)\Delta\phi_{y,mj}]}}
     {1-e^{2\pi i[(l-k)Q_x+(n-o)Q_y]}}
     \bigg[(lb-ka)\Big|_{S1}+ (ne-oc)\Big|_{S2}\bigg]\Bigg\} \ ,
     \nonumber\\
&&\hskip 20mm
\hbox{with }(a\ne b \hbox{ or } c\ne e)
\hbox{ and }(l\ne k\hbox{ or }n\ne o)
.\label{eq:H2hat}
\end{eqnarray}
As in Eq.~\eqref{eq:H2tilde} 
the subscripts $\big|_{S1}$ and $\big|_{S2}$ 
refer to the two systems of Eq.~\eqref{eq:select0}, 
whose sets of solution $abce$ and $lkno$ are 
to be used in the above expression. Since 
only phase-dependent terms are to be considered 
in the Poisson bracket, only phase-dependent 
Hamiltonian coefficients ($a\ne b \hbox{ or } 
c\ne e)\hbox{ and }(l\ne k\hbox{ or }n\ne o$) 
are to be used. Equations~\eqref{eq:H2hat} 
and~\eqref{eq:H2bar} are 
similar to Eq.~\eqref{eq:H2tilde} with 
two notable differences. The summations 
here extend over the total number of magnets 
$M$, whereas the two are nested in 
Eq.~\eqref{eq:H2tilde}. Second, there is 
and additional dependence on the rotational 
term $R_0$ and $R'$.

In summary, the second-order Hamiltonian 
contribution to to the first-order chromatic 
terms ($d=1$) may be written as 
\begin{eqnarray}\label{eq:HamSecOrd2}
H^{(2)}_{pqrt1}(j)&=&\tilde{H}^{(2)}_{pqrt1}(j)+
                   \bar{H}^{(2)}_{pqrt1}(j)\hat{H}^{(2)}_{pqrt1}(j) 
\\ \nonumber 
&=&\left\{\tilde{h}^{(2)}_{pqrt1}(j)+\bar{h}^{(2)}_{pqrt1}(j)
   +\hat{h}^{(2)}_{pqrt1}(j)\right\}
    h_{x,+}^ph_{x,-}^qh_{y,+}^rh_{y,-}^t\delta\ ,
\end{eqnarray}
where $\tilde{h}^{(2)}_{pqrt1}$ is computed 
from Eq.~\eqref{eq:H2tilde}, $\bar{h}^{(2)}_{pqrt1}$ 
from Eq.~\eqref{eq:H2bar} and 
$\hat{h}^{(2)}_{pqrt1}$ from Eq.~\eqref{eq:H2hat}, 
along with the systems of Eq.~\eqref{eq:select0}. 
Explicit formulas risk of being too long and 
of little help: Numerical solutions of 
Eq.~\eqref{eq:HamSecOrd2} from a lattice 
model may be computed if such second-order 
contributions are of importance. A zero linear 
chromaticity ensures that $\bar{H}^{(2)}_{pqrt1}=0$ 
along the ring and removes the terms proportional 
to $Q'_{x,y}$ in $\hat{H}^{(2)}_{pqrt1}$.

It is worthwhile noticing that if focusing 
errors are included in the model, they do 
not contribute to the second-order Hamiltonian 
$H^{(2)}$. On the other hand it can be shown 
that skew quadrupole sources $J_1$ are the 
main ingredients of $H^{(2)}$ along with linear 
chromaticity. With $Q_{x,y}'=0$ and no coupling 
in the machine $H^{(2)}$ is zero. Hence, it 
is of interest to minimize the geometric 
betatron coupling, along with the geometric 
beta beating, before undertaking any correction 
of chromatic functions in order to enlarge the 
range of validity of the analytic formulas 
presented here.

\subsection{Second-order chromatic terms (d=2)} 
\label{sec:d2}
Among all elements in the r.h.s. of 
Eq.~\eqref{eq:Hpqrtd1} only those proportional 
to $\delta^2$ are kept, along with those 
proportional to $h_{m,x,+}^ph_{m,x,-}^qh_{m,y,+}^rh_{m,y,-}^t$. 
The Hamiltonian terms quadratic in $\delta$ read
\begin{eqnarray}
\tilde{H}_{m,pqrt2}&\rightarrow&  \nonumber
         h_{m,lkno}(1-\delta+\delta^2)\times\\&&
         \sum_{a=0}^l\sum_{b=0}^k\sum_{c=0}^n\sum_{e=0}^o
         {\left(\hskip -1mm\begin{array}{c} l\\ a\end{array}\hskip -1mm\right)}
         {\left(\hskip -1mm\begin{array}{c} k\\ b\end{array}\hskip -1mm\right)}
         {\left(\hskip -1mm\begin{array}{c} n\\ c\end{array}\hskip -1mm\right)}
         {\left(\hskip -1mm\begin{array}{c} o\\ e\end{array}\hskip -1mm\right)}
               h_{m,x,+}^{l-a}h_{m,x,-}^{k-b}
               h_{m,y,+}^{n-c}h_{m,y,-}^{o-e}
               d_{m,x,+}^{\ a}d_{m,x,-}^{\ b}
               d_{m,y,+}^{\ c}d_{m,y,-}^{\ e}
               \delta^{a+b+c+e},\nonumber \\ \label{eq:Hd2-1} 
\end{eqnarray}
where all sets of indices $abce$ and $lkno$ 
satisfying the following systems are kept: 
\begin{eqnarray}\label{eq:Hd2-2} 
\left\{
\begin{array}{l}
l-a=p \\
k-b=q \\
n-c=r \\
o-e=t \\
a+b+c+e=2
\end{array}
\right.
\qquad,\qquad
\left\{
\begin{array}{l}
l-a=p \\
k-b=q \\
n-c=r \\
o-e=t \\
a+b+c+e=1 
\end{array}
\right.
\qquad\hbox{and}\qquad
\left\{
\begin{array}{l}
l-a=p \\
k-b=q \\
n-c=r \\
o-e=t \\
a+b+c+e=0 
\end{array}
\right. \quad , 
\end{eqnarray}
where the three systems stem from the magnetic 
rigidity term $(1-\delta+\delta^2)$. After some 
algebra, the Hamiltonian terms quadratic in 
$\delta$ read 
\begin{eqnarray}
\tilde{H}_{m,pqrt2}&\rightarrow& 
       h_{m,pqrt2}h_{m,x,+}^ph_{m,x,-}^qh_{m,y,+}^rh_{m,y,-}^l\delta^2\\
h_{m,pqrt2}&=&
  {\left(\begin{array}{c} p+2\\ 2\end{array}\right)}h_{m,(p+2)qrt}d_{m,x,+}^2+
  {\left(\begin{array}{c} q+2\\ 2\end{array}\right)}h_{m,p(q+2)rt}d_{m,x,-}^2+
  \nonumber \\
&&{\left(\begin{array}{c} r+2\\ 2\end{array}\right)}h_{m,pq(r+2)t}d_{m,y,+}^2+
  {\left(\begin{array}{c} t+2\\ 2\end{array}\right)}h_{m,pqr(t+2)}d_{m,y,-}^2+
  \nonumber \\
&&(p+1)(q+1)h_{m,(p+1)(q+1)rt}d_{m,x,+}d_{m,x,-}+
  (p+1)(r+1)h_{m,(p+1)q(r+1)t}d_{m,x,+}d_{m,y,+}+\nonumber \\
&&(p+1)(t+1)h_{m,(p+1)qr(t+1)}d_{m,x,+}d_{m,y,-}+
  (q+1)(r+1)h_{m,p(q+1)(r+1)t}d_{m,x,-}d_{m,y,+}+\nonumber \\
&&(q+1)(t+1)h_{m,p(q+1)r(t+1)}d_{m,x,-}d_{m,y,-}+
  (r+1)(t+1)h_{m,pq(r+1)(t+1)}d_{m,y,+}d_{m,y,-}\nonumber \\
&& -\left\{(p+1)h_{m,(p+1)qrt}d_{m,x,+}+(q+1)h_{m,p(q+1)rt}d_{m,x,-}\right. 
  \nonumber \\
&&\hspace{5mm}
   \left. (r+1)h_{m,pq(r+1)t}d_{m,y,+}+(t+1)h_{m,jqr(t+1)}d_{m,y,-}\right\}
+ h_{m,pqrt}\quad . \label{eq:Hd2-3} 
\end{eqnarray}

\subsection{Second-order dispersion}
\label{sec:quaddisp}
For the evaluation of the second-order dispersion $D'=\partial D/
\partial\delta$, the Hamiltonian terms in the 
third vector of the r.h.s. of Eq.~\eqref{eq:deltah2B} are to be 
computed. These are evaluated from Eq.~\eqref{eq:Hd2-3} ($d=2$), 
yielding 
\begin{eqnarray}
\left\{
\begin{aligned}\displaystyle
h_{m,10002}&=3h_{m,3000\ }d_{m,x,+}^2+2h_{m,2100\ }d_{m,x,+}d_{m,x,-}+
             2h_{m,2010\ }d_{m,x,+}d_{m,y,+}+
             2h_{m,2001\ }d_{m,x,+}d_{m,y,-}
\\ \displaystyle&\ \ 
         +h_{m,1200\ }d_{m,x,-}^2+
          h_{m,1110\ }d_{m,x,-}d_{m,y,+}+h_{m,1101\ }d_{m,x,-}d_{m,y,-}+
          h_{m,1020\ }d_{m,y,+}^2 
\\ \displaystyle&\ \ 
         +h_{m,1011\ }d_{m,y,+}d_{m,y,-}+h_{m,1002\ }d_{m,y,-}^2
        -\left\{2h_{m,2000\ }d_{m,x,+}+h_{m,1100\ }d_{m,x,-}\right.
\\ \displaystyle&\ \ \left.
        +h_{m,1010\ }d_{m,y,+}+h_{m,1001\ }d_{m,y,-}\right\}
        +h_{m,1000\ }
\\ \displaystyle
h_{m,00102}&=h_{m,2010\ }d_{m,x,+}^2+h_{m,1110\ }d_{m,x,+}d_{m,x,-}+
            2h_{m,1020\ }d_{m,x,+}d_{m,y,+}+
             h_{m,1011\ }d_{m,x,+}d_{m,y,-}
\\ \displaystyle&\ \ 
            +h_{m,0210\ }d_{m,x,-}^2
            2h_{m,0120\ }d_{m,x,-}d_{m,y,+}+h_{m,0111\ }d_{m,x,-}d_{m,y,-}+
            3h_{m,0030\ }d_{m,y,+}^2
\\ \displaystyle&\ \ 
           +2h_{m,0021\ }d_{m,y,+}d_{m,y,-}+
             h_{m,0012\ }d_{m,y,-}^2
        -\left\{ h_{m,1010\ }d_{m,x,+}+h_{m,0110\ }d_{m,x,-}\right.
\\ \displaystyle&\ \ \left.
               +2h_{m,0020\ }d_{m,y,+}+h_{m,0011\ }d_{m,y,-}\right\}
        +h_{m,0010\ }
\end{aligned}
\right. \ . \qquad \label{eq:Hamdisp2}
\end{eqnarray}
The coefficients in the above r.h.s. are computed from 
Eq.~\eqref{eq:h_Vs_KJ}:
\begin{eqnarray}\label{eq:Hamdisp2b}
\left\{
\begin{aligned}\displaystyle
h_{m,1100}&=2h_{m,2000}=-\frac{1}{4}K_{m,1}\beta_{m,x} \\
h_{m,0011}&=2h_{m,0020}=+\frac{1}{4}K_{m,1}\beta_{m,y} \\
h_{m,1010}&=h_{m,1001}=h_{m,0110}=+\frac{1}{4} J_{m,1}
                         \sqrt{\beta_{m,x}\beta_{m,y}}\\
 h_{m,1000}&=h_{m,0100}=-\frac{1}{2}K_{m,0}\sqrt{\beta_{m,x}}\\
 h_{m,0010}&=h_{m,0001}=+\frac{1}{2}J_{m,0}\sqrt{\beta_{m,y}} \\
\end{aligned}
\right. \ ,\quad
\left\{
\begin{aligned}\displaystyle
h_{m,2100}&=3h_{m,3000}=-\frac{1}{16}K_{m,2}\beta_{m,x}^{3/2}  \\
h_{m,1110}&=h_{m,1101}=+\frac{1}{8}J_{m,2}\beta_{m,x}\sqrt{\beta_{m,y}} \\
h_{m,0012}&=3h_{m,0030}=-\frac{1}{16}J_{m,2}\beta_{m,y}^{3/2} \\
h_{m,1011}&=h_{m,0111}=+\frac{1}{8}K_{m,2}\sqrt{\beta_{m,x}}\beta_{m,y} 
\end{aligned}
\right. \ .\qquad
\end{eqnarray}
From the above expressions we can rewrite 
Eq.~\eqref{eq:Hamdisp2} in more convenient forms
\begin{eqnarray}
\left\{
\begin{aligned}\displaystyle
h_{m,10002}&=
     3h_{m,3000}\left[d_{m,x,-}^2+2d_{m,x,-}d_{m,x,+}+d_{m,x,+}^2\right]+
     2h_{m,2010}(d_{m,x,-}+d_{m,x,+})(d_{m,y,-}+d_{m,y,+})+h_{m,1000\ }+ 
\\ &\qquad
      h_{m,1002}\left[d_{m,y,-}^2+2d_{m,y,-}d_{m,y,+}+d_{m,y,+}^2\right]
    -2h_{m,2000}(d_{m,x,-}+d_{m,x,+})
     -h_{m,1010}(d_{m,y,-}+d_{m,y,+}) \\
h_{m,00102}&=
      h_{m,2010}\left[d_{m,x,-}^2+2d_{m,x,-}d_{m,x,+}+d_{m,x,+}^2\right]+
      h_{m,1011}(d_{m,x,-}+d_{m,x,+})(d_{m,y,-}+d_{m,y,+})+h_{m,0010\ }+ 
\\ &\ \ \ 
     3h_{m,0030}\left[d_{m,y,-}^2+2d_{m,y,-}d_{m,y,+}+d_{m,y,+}^2\right]
     -h_{m,1010}(d_{m,x,-}+d_{m,x,+})
    -2h_{m,0020}(d_{m,y,-}+d_{m,y,+})
\end{aligned}
\right. \nonumber \ ,
\end{eqnarray}
and hence
\begin{eqnarray}
\left\{
\begin{aligned}\displaystyle
h_{m,10002}&=
     3h_{m,3000}\left(2\Re\{d_{m,x,\pm}\}\right)^2+
     2h_{m,2010}(2\Re\{d_{m,x,\pm}\}2\Re\{d_{m,x,\pm}\})+h_{m,1000\ }+ 
\\ &\qquad
      h_{m,1002}\left(2\Re\{d_{m,y,\pm}\}\right)^2
    -2h_{m,2000}(2\Re\{d_{m,x,\pm}\})
     -h_{m,1010}(2\Re\{d_{m,y,\pm}\})\\
h_{m,00102}&=
      h_{m,2010}\left(2\Re\{d_{m,x,\pm}\}\right)^2+
      h_{m,1011}(2\Re\{d_{m,x,\pm}\}2\Re\{d_{m,x,\pm}\})+h_{m,0010\ }+ 
\\ &\ \ \ 
     3h_{m,0030}\left(2\Re\{d_{m,y,\pm}\}\right)^2
     -h_{m,1010}(2\Re\{d_{m,x,\pm}\})
    -2h_{m,0020}(2\Re\{d_{m,y,\pm}\})
\end{aligned}
\right. \label{eq:Hamdisp2c}\ .
\end{eqnarray}
The Hamiltonian coefficients of Eq.~\eqref{eq:Hamdisp2b} 
may be substituted in the above expressions. 
After recalling that 
$\Re\left\{d_{m,q,\pm}\right\}=D_{m,q}/\sqrt{\beta_{m,q}}$ 
and applying some algebra, Eq.~\eqref{eq:Hamdisp2c} 
eventually reads
\begin{eqnarray}
\left\{
\begin{aligned}\displaystyle
h_{m,10002}&=\frac{1}{2}\left[-K_{m,0}-J_{m,1}D_{m,y}+ K_{m,1}D_{m,x}
              -\frac{1}{2}K_{m,2}\left(D_{m,x}^2-D_{m,y}^2\right) 
              +J_{m,2}D_{m,x}D_{m,y}\right]\sqrt{\beta_{m,x}} \\
h_{m,00102}&=\frac{1}{2}\left[\ \ J_{m,0}-J_{m,1}D_{m,x}- K_{m,1}D_{m,y}
              +\frac{1}{2}J_{m,2}\left(D_{m,x}^2-D_{m,y}^2\right) 
              +K_{m,2}D_{m,x}D_{m,y}\right]\sqrt{\beta_{m,y}}
\end{aligned}
\right. \label{eq:Hamdisp2d}\ .
\end{eqnarray}
Since these two terms are real quantities, 
$h_{m,01002}=h_{m,10002}$ and $h_{m,00102}=h_{m,00012}$. 
The most right vector of Eq.~\eqref{eq:deltah2B}
and hence the vector $\delta\vec{h}_m$ are 
then defined. The complex second-order 
dispersion vector at a generic location $j$ 
is defined as 
$\vec{d}'(j)=\frac{\partial\vec{d}(j)}{\partial\delta}
                =\frac{\partial^2\vec{h}(j)}{\partial\delta^2}$,
where the complex closed-orbit vector $\vec{h}$ 
is evaluated from Eq.~\eqref{eq:co1B}:
\begin{eqnarray}
\frac{\partial^2\vec{h}(j)}{\partial\delta^2}&\simeq&
\mathbf{B}_j^{-1}\sum_{m=1}^M\left\{
\frac{e^{i\mathbf{\Delta\phi}_{mj}}}{1-e^{i\mathbf{Q}}}
\mathbf{B}_m\ \frac{\partial^2\delta\vec{h}_m}{\partial\delta^2}\right\}\ ,
\end{eqnarray}
where the dependence of the matrices $\mathbf{B}$, 
$e^{i\mathbf{\Delta\phi}_{mj}}$ and $e^{i\mathbf{Q}}$ 
on $\delta$ has been ignored. The complex 
second-order dispersion vector and their real 
Cartesian counterparts then read 
\begin{eqnarray}\label{eq:Hamdisp2e}
\vec{d}'(j)=
\left(\begin{array}{r}\tilde{d}_{x,-}'\\\tilde{d}_{x,+}'\\
                      \tilde{d}_{y,-}'\\\tilde{d}_{y,+}'\end{array}\right)
\simeq\mathbf{B}_j^{-1}\sum_{m=1}^M\left\{
\frac{e^{i\mathbf{\Delta\phi}_{mj}}}{1-e^{i\mathbf{Q}}}
\mathbf{B}_m\ 4i\  
\left(\begin{array}{r} h_{m,10002}\\-h_{m,10002}\\h_{m,00102}\\-h_{m,00102}
   \end{array}\right)\right\}\ ,\qquad
\left\{
\begin{aligned}\displaystyle
D_{x}'&=\Re\left\{\tilde{d}_{x,-}'\right\}\sqrt{\beta_{x}}\\
D_{y}'&=\Re\left\{\tilde{d}_{y,-}'\right\}\sqrt{\beta_{y}}
\end{aligned}
\right.\ .\qquad
\end{eqnarray}
Note that conversely to the linear dispersion 
of Eq.~\eqref{eq:co3C} the matrices $\mathbf{B}_m$, 
and $\mathbf{B}_j^{-1}$ with the coupling RDTs are 
to be kept, because the correct dependence 
of the linear optics and its errors on $\delta$ 
may not be neglected anymore. Textbook formulas 
are retrieved for the ideal uncoupled lattice, 
with $\mathbf{B}_m=\mathbf{B}_j^{-1}
=\mathbf{I}$, $D_{m,y}=0$, $J_{m,0}=J_{m,1}=J_{m,2}=0$ 
and hence $h_{m,00102}=h_{m,00012}=0$: 
\begin{eqnarray}
\left\{
\begin{aligned}\displaystyle
D_{x}'(j)&=4\sqrt{\beta_{j,x}}
             \Re\left\{\frac{e^{-i\Delta\phi_{x,mj}}}{1-e^{i2\pi Q_x}}
             i\ h_{m,01002}\right\}\\
D_{y}'&=0
\end{aligned}
\right.\ .\qquad
\end{eqnarray}
The ideal second-order horizontal dispersion then reads
\begin{eqnarray}
D_{x}'(j)&=&\frac{\sqrt{\beta_{j,x}}}{\sin{(\pi Q_x)}}
\sum_{m=1}^M \left[-K_{m,0}+ K_{m,1}D_{m,x}-\frac{1}{2}K_{m,2}D_{m,x}^2\right]
       \sqrt{\beta_{m,x}}\cos{(\Delta\phi_{x,mj}-\pi Q_x)}\\
&=&-2D_x(j)+\frac{\sqrt{\beta_{j,x}}}{\sin{(\pi Q_x)}}
\sum_{m=1}^M \left[K_{m,1}-\frac{1}{2}K_{m,2}D_{m,x}\right]
       D_{m,x}\sqrt{\beta_{m,x}}\cos{(\Delta\phi_{x,mj}-\pi Q_x)}\quad ,
\end{eqnarray}
corresponding to Eq.(112) of Ref.~\cite{Bengtsson} 
multiplied by a factor two. In the above equation, 
the linear dispersion $D_x{(j)}$ of Eq.~\eqref{eq:co3C} 
has been extracted from the summation. For a lattice 
with errors the more general Eqs.~\eqref{eq:Hamdisp2d} 
and~\eqref{eq:Hamdisp2e} shall be used and 
numerically evaluated. 
For consistency with the nomenclature used 
throughout this paper, the phase advance 
$\Delta\phi_{mj}$ is to be computed as in 
Eq.~\eqref{eq:deltaphisign}. If the mere 
difference between the two betatron phases 
at the positions $m$ and $j$ is used, the 
absolute value $|\Delta\phi_{x,mj}|$ 
shall then be used, as done in textbooks.

\section{Corrections for the variation of the lattice parameters
                 across magnets}
\label{app:3}
All equations derived in the previous appendices have been 
derived assuming constant values for the beta and 
dispersion functions ($\beta_m$ and $D_m$) across a 
generic magnet $m$, usually computed at its center. 
The phase advance $\Delta\phi_{mj}$ between 
the magnet and a generic location $j$ refers to 
its center too. This approximation may not be sufficiently 
accurate in general and in particular for lattices 
comprising combined-function 
magnets, along which the beta function varies 
considerably. It is then of interest to evaluate 
corrections to the final equations of 
Secs.~\ref{ORM-Formulas}-\ref{CHROM-Formulas}
accounting for that variation. In doing so, another 
approximation is implicitly performed: Magnets  
are assumed to be hard-edged, i.e. their effective 
strengths are constant along the magnets and fall 
immediately to zero at their (magnetic) ends. The 
edge focusing at the dipole ends is also neglected. 
In Eq.~\eqref{eq:Nmatrix1} steerers $w$ are also assumed 
to be thin elements. In ring-based light sources orbit 
correctors are usually obtained from trim coils installed 
on sextupole magnets, and this approximation is usually 
rather robust. On the other hand, in rings making use 
of trim coils on bending magnets the variation of the 
lattice parameters across them may be no longer neglected. 
At the end of this section an approximate generalization 
accounting for thick steerers and sextupoles is 
eventually presented. 

In order to account for variation of the lattice 
parameters across the magnets, all terms dependent 
on $m$ in the formulas of 
Secs.~\ref{ORM-Formulas}-\ref{CHROM-Formulas} need 
to be replaced by their integral forms
\begin{eqnarray}
\beta_{m}\quad &\longrightarrow&\quad I_{\beta,m}=\frac{1}{L_m}
 \int_0^{L_m}{\beta(s)_{\ }ds}\ ,\qquad\label{eq:Ib1}\\
\beta_{m}\sin{(2\tau_{mj})}\quad &\longrightarrow&\quad\label{eq:IS1}
I_{S,mj}=\frac{1}{L_m}
 \int_0^{L_m}{\beta(s)\sin{(2\tau_{sj})}_{\ }ds}=\frac{1}{L_m}
 \int_0^{L_m}{\beta(s)\sin{(2\tau_{s_{m}j}- 2\Delta\phi_{s})}_{\ }ds}\ ,\qquad\\
\beta_{m}\cos{(2\tau_{mj})}\quad &\longrightarrow&\quad\label{eq:IC1}
I_{C,mj}=\frac{1}{L_m}
 \int_0^{L_m}{\beta(s)\cos{(2\tau_{sj})}_{\ }ds}=\frac{1}{L_m}
 \int_0^{L_m}{\beta(s)\cos{(2\tau_{s_{m}j}- 2\Delta\phi_{s})}_{\ }ds}\ ,\qquad\\
\sqrt{\beta_{m}}\sin{(\tau_{mj})}\quad &\longrightarrow&\quad\label{eq:JS1}
J_{C,mj}=\frac{1}{L_m}\int_0^{L_m}
{\hskip-2mm\sqrt{\beta(s)}\sin{(\tau_{sj})}_{\ }ds}=\frac{1}{L_m}
\int_0^{L_m}{\hskip-2mm\sqrt{\beta(s)}\sin{(\tau_{s_{m}j}-\Delta\phi_s)}_{\ }ds}\ ,\qquad\\
\sqrt{\beta_{m}}\cos{(\tau_{mj})}\quad &\longrightarrow&\quad\label{eq:JC1}
J_{C,mj}=\frac{1}{L_m}\int_0^{L_m}
{\hskip-2mm\sqrt{\beta(s)}\cos{(\tau_{sj})}_{\ }ds}=\frac{1}{L_m}
\int_0^{L_m}{\hskip-2mm\sqrt{\beta(s)}\cos{(\tau_{s_{m}j}-\Delta\phi_s)}_{\ }ds}\ ,\qquad\\
\sqrt{\beta_{m,x}}D_{m,y}\cos{(\tau_{x,mj})}\quad &\longrightarrow&\quad\label{eq:JCDy}
J_{C,mj}^{(D_y)}=\frac{1}{L_m}\int_0^{L_m}
{\sqrt{\beta_x(s)}D_y(s)\cos{(\tau_{x,s_{m}j}-\Delta\phi_{x,s})}_{\ }ds}\ ,\qquad\\
\sqrt{\beta_{m,y}}D_{m,x}\cos{(\tau_{y,mj})}\quad &\longrightarrow&\quad\label{eq:JCDx}
J_{C,mj}^{(D_x)}=\frac{1}{L_m}\int_0^{L_m}
{\sqrt{\beta_y(s)}D_x(s)\cos{(\tau_{y,s_{m}j}-\Delta\phi_{y,s})}_{\ }ds}\ ,\qquad
\end{eqnarray}
where $L_m$ denotes the length of the magnet $m$, 
$s_{m}$ represents the position along the ring of its entrance 
whereas $\Delta\phi_{s}$ is the phase advance along the magnet. 

\subsection{$I_{\beta,m},\ I_{S,mj}$ and $I_{C,mj}$ for quadrupoles}
The above integrals enter in Eq.~\eqref{eq:Nmatrix1} to 
account for the variation of the C-S parameters across 
quadrupoles in the evaluation of the diagonal ORM blocks. 
They can be solved by making use of the two 
representations of the transfer matrix along the magnet, 
i.e. from it entrance $s_{m}=0$ and the generic position $s$
\begin{eqnarray}
\mathbf{A_m}(s)=\left\{
\begin{aligned}
\left(\begin{array}{c c}
\cos{(\sqrt{k_m}s)}            & \frac{1}{\sqrt{k_m}}\sin{(\sqrt{k_m}s)} \\
-\sqrt{k_m}\sin{(\sqrt{k_m}s)} & \cos{(\sqrt{k_m}s)}
\end{array} \right) \qquad&\hbox{focusing plane}\\
\left(\begin{array}{c c}
\cosh{(\sqrt{|k_m|}s)}           & \frac{1}{\sqrt{|k_m|}}\sinh{(\sqrt{|k_m|}s)} \\
\sqrt{|k_m|}\sinh{(\sqrt{|k_m|}s)} & \cosh{(\sqrt{|k_m|}s)}
\end{array} \right) \qquad&\hbox{defocusing plane}\
\end{aligned}\right. \quad ,\label{eq:Adef1}\\
\mathbf{C_m}(s)=
\left(\begin{array}{c c}
\sqrt{\frac{\beta(s)}{\beta_{m0}}}\left(\cos{\Delta\phi_s}+
\alpha_{m0}\sin{\Delta\phi_s}\right) & 
\sqrt{\beta(s)\beta_{m0}}\sin{\Delta\phi_s} \\
\frac{\alpha_{m0}-\alpha(s)}{\sqrt{\beta(s)\beta_{m0}}}\cos{\Delta\phi_s}
-\frac{1-\alpha_{m0}\alpha(s)}{\sqrt{\beta(s)\beta_{m0}}}\sin{\Delta\phi_s} &
\sqrt{\frac{\beta_{m0}}{\beta(s)}}\left(\cos{\Delta\phi_s}+
\alpha(s)\sin{\Delta\phi_s}\right)
\end{array} \right)\quad ,\label{eq:Cdef1}
\end{eqnarray}
where $k_m$ represents the model non-integrated quadrupole 
strength, whereas $\beta_{m0}$ and $\alpha_{m0}$ are 
the C-S parameters at the quadrupole entrance, both computed 
from the ideal model. By imposing 
that $A_{m,11}=C_{m,11}$ and $A_{m,12}=C_{m,12}$ we obtain
\begin{eqnarray}\left\{
\begin{aligned}
\sqrt{\beta(s)}\sin{\Delta\phi_s}&=\frac{1}{\sqrt{|k_m|\beta_{m0}}}
\left(\begin{array}{c}\sin{(\sqrt{k_m}s)} \\ \sinh{(\sqrt{|k_m|}s)}
\end{array}\right) \\
\sqrt{\beta(s)}\cos{\Delta\phi_s}&=\sqrt{\beta_{m0}}
\left(\begin{array}{c}\cos{(\sqrt{k_m}s)} \\ \cosh{(\sqrt{|k_m|}s)}
\end{array}\right)
-\frac{\alpha_{m0}}{\sqrt{|k_m|\beta_{m0}}}
\left(\begin{array}{c}\sin{(\sqrt{k_m}s)} \\ \sinh{(\sqrt{|k_m|}s)}
\end{array}\right)
\end{aligned}\right. \quad ,\label{eq:ACmatrix}
\end{eqnarray}
where the upper and lower terms refer to the focusing and 
defocusing planes, respectively. By summing the square of 
the above equations the following expression for the beta 
function along the quadrupole is obtained
\begin{eqnarray}
\beta(s)=\frac{\gamma_{m0}}{|k_m|}\left(\begin{array}{c}
         \sin^2{(\sqrt{k_m}s)} \\ \sinh^2{(\sqrt{|k_m|}s)}\end{array}\right)
        +\beta_{m0}\left(\begin{array}{c}
         \cos^2{(\sqrt{k_m}s)} \\ \cosh^2{(\sqrt{|k_m|}s)}\end{array}\right)
        -\frac{\alpha_{m0}}{\sqrt{|k_m|}}\left(\begin{array}{c}
         \sin{(2\sqrt{k_m}s)} \\ \sinh{(2\sqrt{|k_m|}s)}\end{array}\right)
\quad ,\label{eq:beta_quad}
\end{eqnarray}
where $\gamma_{m0}=(1+\alpha_{m0}^2)/\beta_{m0}$ is the 
third C-S parameter. After inserting the above expression 
in the integral of Eq.~\eqref{eq:Ib1} and applying the 
following relations, 
\begin{eqnarray} \label{eq:integr1}
\begin{aligned}
\int_0^{L_m}{\sin^2{(\sqrt{k_m}s)} ds}=\frac{L_m}{2}-
       \frac{\sin{(2\sqrt{k_m}L_m)}}{4\sqrt{k_m}}\quad&,\quad
\int_0^{L_m}{\sinh^2{(\sqrt{|k_m|}s)} ds}=-\frac{L_m}{2}+
       \frac{\sinh{(2\sqrt{|k_m|}L_m)}}{4\sqrt{|k_m|}}\quad ,  \\
\int_0^{L_m}{\cos^2{(\sqrt{k_m}s)} ds}=\frac{L_m}{2}+
       \frac{\sin{(2\sqrt{k_m}L_m)}}{4\sqrt{k_m}}\quad&,\quad
\int_0^{L_m}{\cosh^2{(\sqrt{|k_m|}s)} ds}=\ \frac{L_m}{2}+
       \frac{\sinh{(2\sqrt{|k_m|}L_m)}}{4\sqrt{|k_m|}}\quad ,\\
\int_0^{L_m}{\sin{(2\sqrt{k_m}s)} ds}=\frac{1}{2\sqrt{k_m}}-
       \frac{\cos{(2\sqrt{k_m}L_m)}}{2\sqrt{k_m}}\quad&,\quad
\int_0^{L_m}{\sinh{(2\sqrt{|k_m|}s)} ds}=-\frac{1}{2\sqrt{|k_m|}}+
       \frac{\cosh{(2\sqrt{|k_m|}L_m)}}{2\sqrt{|k_m|}}\quad ,
\end{aligned}
\end{eqnarray}
the following expression for the correction term 
$I_{\beta,m}$ is obtained
\begin{eqnarray}
I_{\beta,m}=\left\{
\begin{aligned}
&\frac{1}{2}\left[\beta_{m0}+\frac{\gamma_{m0}}{k_m}\right]
+\hskip 1mm \frac{\sin{(2\sqrt{k_m}L_m)}}{4\sqrt{k_m}L_m}\hskip 1mm
      \left[\beta_{m0}-\frac{\gamma_{m0}}{k_m}\right]
+\hskip 1mm \frac{\alpha_{m0}}{2k_mL_m}\hskip 1mm
       \left[\cos{(2\sqrt{k_m}L_m)}-1\right] 
\hskip8mm \hbox{focusing plane}\\
&\frac{1}{2}\left[\beta_{m0}-\frac{\gamma_{m0}}{|k_m|}\right]
+\frac{\sinh{(2\sqrt{|k_m|}L_m)}}{4\sqrt{|k_m|}L_m}
      \left[\beta_{m0}+\frac{\gamma_{m0}}{|k_m|}\right]
-\frac{\alpha_{m0}}{2|k_m|L_m}
       \left[\cosh{(2\sqrt{|k_m|}L_m)}-1\right]
\ \hbox{defocusing plane}
\end{aligned}\right. .\nonumber
\end{eqnarray}
The two cases, focusing ($k_m>0$) and defocusing ($k_m<0$) quadrupoles,
can be actually described by a single formula
\begin{eqnarray}\displaystyle
I_{\beta,m}=\frac{1}{2}\left[\beta_{m0}+\frac{\gamma_{m0}}{k_m}\right]
+\hskip 1mm \frac{\sin{(2\sqrt{k_m}L_m)}}{4\sqrt{k_m}L_m}\hskip 1mm
      \left[\beta_{m0}-\frac{\gamma_{m0}}{k_m}\right]
+\hskip 1mm \frac{\alpha_{m0}}{2k_mL_m}\hskip 1mm
       \left[\cos{(2\sqrt{k_m}L_m)}-1\right] \ , \label{eq:Ib2}
\end{eqnarray}
since for $k_m<0$ $\sin{(2\sqrt{k_m}L_m)}/\sqrt{k_m}=
\sinh{(2\sqrt{|k_m|}L_m)}/\sqrt{|k_m|}$ and $\cos{(2\sqrt{k_m}L_m)}=
\cosh{(2\sqrt{|k_m|}L_m)}$. 
The trigonometric integrals of Eqs.~\eqref{eq:IS1}-\eqref{eq:IC1}
can be rewritten as
\begin{eqnarray}\label{eq:IS2}
\begin{aligned}
I_{S,mj}&=\sin{(2\tau_{s_{m}j})}I_{C,m}-\cos{(2\tau_{s_{m}j})}I_{S,m}
\quad ,\quad I_{C,m}=\frac{1}{L_m}
 \int_0^{L_m}{\beta(s)\cos{(2\Delta\phi_{s})}_{\ }ds}\ ,\\
I_{C,mj}&=\cos{(2\tau_{s_{m}j})}I_{C,m}+\sin{(2\tau_{s_{m}j})}I_{S,m}
\quad ,\quad I_{S,m}=\frac{1}{L_m}
 \int_0^{L_m}{\beta(s)\sin{(2\Delta\phi_{s})}_{\ }ds}\ .
\end{aligned} 
\end{eqnarray}
In the above expressions $\tau_{s_{m}j}=\Delta\phi_{s_{m}j}-\pi Q$ 
represents the shifted phase advance between the 
generic location $j$ and the entrance of the 
magnet $m$ ($s_{m}$), whereas the integrals $I_{C,m}$ 
and $I_{S,m}$ depend on the magnet $m$ only. By making use 
of the relation $\cos{(2\Delta\phi_{s})}=1-2\sin^2{\Delta\phi_{s}}$, 
replacing $\beta(s)\sin^2{\Delta\phi_{s}}$ with the first 
expression in Eq.~\eqref{eq:ACmatrix}, and applying the 
integrals of Eq.~\eqref{eq:integr1}, the first integral reads
\begin{eqnarray}\label{eq:IC2}
  I_{C,m}=I_{\beta,m}-
  \frac{1}{k_m\beta_{m0}}\left[\hskip1mm 1-
        \frac{\sin{(2\sqrt{k_m}L_m)}}{2\sqrt{k_m}L_m}\right] 
         \quad ,
\end{eqnarray}
which is valid for both focusing and defocusing quadrupoles,
as for Eq.~\eqref{eq:Ib2} ($\sin{(2\sqrt{k_m}L_m)}/\sqrt{k_m}=
\sinh{(2\sqrt{|k_m|}L_m)}/\sqrt{|k_m|}$ for $k_m<0$).
The second integral can be computed by replacing 
$\beta(s)\sin{(2\Delta\phi_{s})}=2(\sqrt{\beta(s)}\sin{\Delta\phi_{s}})
(\sqrt{\beta(s)}\cos{\Delta\phi_{s}})$, by substituting the 
terms in the parenthesis with the expressions of 
Eq.~\eqref{eq:ACmatrix}, and by making use of the 
integrals of Eq.~\eqref{eq:integr1}. The result is 
\begin{eqnarray}\label{eq:IS3}
\begin{aligned}
I_{S,m}&=\frac{1}{k_mL_m}\left[\hskip2mm\frac{1}{2}\left(1-\cos{(2\sqrt{k_m}L_m)}
       \right)+\frac{\alpha_{m0}}{\beta_{m0}}\left(
       \frac{\sin{(2\sqrt{k_m}L_m)}}{2\sqrt{k_m}}-L_m\right)\right]\ ,\\
&=\frac{1}{2k_mL_m}\left[1-\cos{(2\sqrt{k_m}L_m)}\right]+
  \alpha_{m0}\left(I_{C,m}-I_{\beta,m}\right)\ .
\end{aligned}
\end{eqnarray}

\subsection{$I_{\beta,m},\ I_{S,mj}$ and $I_{C,mj}$ for   thick orbit deflectors and sextupoles}
Even though the index {\sl m} refers to a quadrupole whose integrated field or strength error $\delta K_{m,1}$ induces an optic distortion, the latter can be generated or modelled at an arbitrary location or magnet, such as a steerer or a sextupole. The three integrals {$I_{\beta,m},\ I_{S,mj}$ and $I_{C,mj}$} shall then be evaluated for a drift, under the realistic assumption that the latter well describes steerers and nonlinear magnets as far as the linear optics is concerned. 

The integrals can be evaluated in two ways. The first it to repeat the same calculations of the previous sections after replacing the $\mathbf{C_m}$ matrix of \eqref{eq:Cdef1} with the one of the drift:
\begin{eqnarray}
\mathbf{C_m}(s)=
\left(\begin{array}{c c}
1 & s \\  0 & 1 
\end{array} \right)\quad .\label{eq:Cdef2}
\end{eqnarray}
The system of Eq.~\eqref{eq:ACmatrix} then reads
\begin{eqnarray}\left\{
\begin{aligned}
\sqrt{\beta(s)}\sin{\Delta\phi_s}&=\frac{s}{\sqrt{\beta_{m0}}} \\
\sqrt{\beta(s)}\cos{\Delta\phi_s}&=\sqrt{\beta_{m0}}-
     \frac{\alpha_{m0}}{\sqrt{\beta_{m0}}}s
\end{aligned}\right. \quad .\label{eq:ACmatrix3}
\end{eqnarray}
As in the previous section, by summing the square of 
the above equations the following expression for the beta 
function along the quadrupole is obtained
\begin{eqnarray}
\beta(s)=\gamma_{m0}s^2
        +\beta_{m0}
        -2\alpha_{m0}\ s
\quad ,\label{eq:beta_drift}
\end{eqnarray}
and the integral $I_{\beta,m}$ eventually reads
\begin{eqnarray}
I_{\beta,m}=\frac{1}{L_m}\int_0^{L_m}{\beta(s)_{\ }ds} 
           \ =\  \frac{1}{L_m}\int_0^{L_m}{\left(\gamma_{m0}s^2
               +\beta_{m0}-2\alpha_{m0}\ s\right) ds}
           \ =\ \frac{\gamma_{m0}}{3}L_m^2+\beta_{m0}
               -\alpha_{m0}L_m \quad .\label{eq:Ibeta_drift}
\end{eqnarray}
The same result is obtained from Eq.~\eqref{eq:Ib2} in the limit for $k_m\rightarrow 0 $, with thus  $\sin{(2\sqrt{k_m}L_m)}/(4\sqrt{k_m}L_m)\rightarrow 1$ and $\cos{(2\sqrt{k_m}L_m)}\rightarrow 1$ (the same limits apply for the hyperbolic functions).
By multiplying the two equations of Eq.~\eqref{eq:ACmatrix3} we obtain
\begin{eqnarray}\label{eq:Id_drift}
\beta(s)\sin{(2\Delta\phi_s)}=2\left(s-\frac{\alpha_{m0}}{\beta_{m0}}s^2\right)
\qquad\Rightarrow\qquad 
I_{S,m}=\frac{1}{L_m}\int_0^{L_m}{\beta(s)\sin{(2\Delta\phi_{s})}_{\ }ds}
       =L_m-\frac{2}{3}\frac{\alpha_{m0}}{\beta_{m0}}L_m^2\quad . 
\end{eqnarray}
Next, by taking the difference of the square of both equations of Eq.~\eqref{eq:ACmatrix3} the cosine term reads
\begin{eqnarray}
\beta(s)\cos{(2\Delta\phi_s)}=\beta_{m0}-2\alpha_{m0}s+\gamma_{m0}s^2-
                    \frac{2}{\beta_{m0}}s^2=\beta(s)-\frac{2}{\beta_{m0}}s^2
\quad ,
\end{eqnarray}
where Eq.~\eqref{eq:beta_drift} is used to introduce explicitly $\beta(s)$, thus further simplifying the last integral:
\begin{eqnarray}
\begin{aligned}
I_{C,m}&=\frac{1}{L_m}\int_0^{L_m}{\beta(s)\cos{(2\Delta\phi_{s})}_{\ }ds}
        =I_{\beta,m}-\frac{2}{3\beta_{m0}}L_m^2\ , \\
I_{S,m}&=L_m+\alpha_{m0}\left(I_{C,m}-I_{\beta,m}\right)\ ,
\end{aligned} \label{eq:Ic_drift}
\end{eqnarray}
Where the last expression for $I_{S,m}$ is obtained from Eq.~\eqref{eq:Id_drift} and the definition of $I_{C,m}$. Both $I_{S,m}$ and $I_{C,m}$ can be inserted in Eq.~\eqref{eq:IS2} to eventually compute the two integrals $I_{S,mj}$ and $I_{C,mj}$.

\subsection{$J_{S,mj}$ and $J_{C,mj}$ for sector dipoles}
$J_{C,mj}$  enters in Eq.~\eqref{eq:Text_Disp1} to account for 
the evolution of the C-S parameters across a dipole magnet in the 
evaluation of the dispersion function. $J_{S,mj}$ is evaluated 
for completeness.

The transfer matrix $\mathbf{A_m}$ for a sector dipole reads
\begin{eqnarray}
\mathbf{A_m}(s)=\left(\begin{array}{c c}
\cos{(K_{m,0})}                             & \rho\sin{(K_{m,0})} \\
\displaystyle-\frac{1}{\rho}\sin{(K_{m,0})} & \cos{(K_{m,0})}
\end{array} \right) \quad = \quad
\left(\begin{array}{c c}\displaystyle
\cos{(K_{m,0})}  & \displaystyle\frac{s}{K_{m,0}}\sin{(K_{m,0})} \\
\displaystyle-\frac{K_{m,0}}{s}\sin{(K_{m,0})} & \cos{(K_{m,0})}
\end{array} \right) \qquad .\label{eq:Adef2}
\end{eqnarray}
As in the previous case, by imposing that 
$A_{m,11}=C_{m,11}$ and $A_{m,12}=C_{m,12}$, 
the elements of the $\mathbf{C_m}$ matrix 
being the same of Eq.~\eqref{eq:Cdef1}, we obtain
\begin{eqnarray}\left\{
\begin{aligned}
\sqrt{\beta(s)}\sin{\Delta\phi_s}&=
             \frac{\sin{K_{m,0}}}{K_{m,0}\sqrt{\beta_{m0}}}\ s\\
\sqrt{\beta(s)}\cos{\Delta\phi_s}&=\sqrt{\beta_{m0}}\cos{K_{m,0}}-
             \frac{\alpha_{m0}\sin{K_{m,0}}}{K_{m,0}\sqrt{\beta_{m0}}}\ s
\end{aligned}\right. \quad .\label{eq:ACmatrix2}
\end{eqnarray}
By integrating the above expressions across the magnet yields
\begin{eqnarray}
\begin{aligned}\label{eq:JCmj0}
T_{S,m}&=\frac{1}{L_m}\int_0^{L_m}{\sqrt{\beta(s)}\sin{\Delta\phi_s}\ ds}=
             \frac{L_m\sin{K_{m,0}}}{2K_{m,0}\sqrt{\beta_{m0}}}\quad ,\\
T_{C,m}&=\frac{1}{L_m}\int_0^{L_m}{\sqrt{\beta(s)}\cos{\Delta\phi_s}\ ds}=
\sqrt{\beta_{m0}}\cos{K_{m,0}}-
             \frac{\alpha_{m0}L_m\sin{K_{m,0}}}{2K_{m,0}\sqrt{\beta_{m0}}}
=\sqrt{\beta_{m0}}\cos{K_{m,0}}-\alpha_{m0}T_{S,m}\quad .
\end{aligned} 
\end{eqnarray}
$J_{S,mj}$ and $J_{C,mj}$ for sector dipoles can be eventually 
written and computed as 
\begin{eqnarray}\label{eq:JCmj1}
\begin{aligned}
J_{S,mj}&=\sin{(\tau_{s_{m}j})}T_{C,m}-\cos{(\tau_{s_{m}j})}T_{S,m}\quad , \\
J_{C,mj}&=\cos{(\tau_{s_{m}j})}T_{C,m}+\sin{(\tau_{s_{m}j})}T_{S,m}\quad ,
\end{aligned}
\end{eqnarray}
where, as usual, $\tau_{s_{m}j}$ is the shifted phase 
advance between the entrance of the magnet $m$ and 
the observation point $j$.

\subsection{$J_{S,mj}$ and $J_{C,mj}$ for quadrupoles and combined-function dipoles}\label{sec:app-J}
$J_{C,mj}$  enters in Eq.~\eqref{eq:Text_Disp1} to account for 
the evolution of the C-S parameters across a combined-function 
dipole magnet in the evaluation of the dispersion function. 
Both $J_{S,mj}$ and $J_{C,mj}$ are also needed in the evaluation 
of the integrals of Eq.~\eqref{eq:Smatrix1-thick} for the 
computation of the off-diagonal ORM block. 

As far as the betatron motion, the transfer matrix of a 
combined-function dipole is the one of simple quadrupole 
(the same is not true for dispersive terms). 
The integration of both sides and equations of 
Eq.~\eqref{eq:ACmatrix} reads
\begin{eqnarray}
\begin{aligned}
G_{S,m}&=\frac{1}{L_m}\int_0^{L_m}{\sqrt{\beta(s)}\sin{\Delta\phi_s}\ ds}=
         \frac{1}{L_m\sqrt{k_m\beta_{m0}}}\int_0^{L_m}{\sin{(\sqrt{k_m}s)}
         \ ds}   \quad ,\\
G_{C,m}&=\frac{1}{L_m}\int_0^{L_m}{\sqrt{\beta(s)}\cos{\Delta\phi_s}\ ds}=
         \frac{\sqrt{\beta_{m0}}}{L_m}\int_0^{L_m}{\cos{(\sqrt{k_m}s)}\ ds}
         -\frac{\alpha_{m0}}{L_m\sqrt{k_m\beta_{m0}}}
          \int_0^{L_m}{\sin{(\sqrt{k_m}s)}\ ds}\quad .
\end{aligned} 
\end{eqnarray}
Once again the above expressions holds both for focusing and 
defocusing quadrupoles, since for $k_m<0$ $\sin{(2\sqrt{k_m}L_m)}/\sqrt{k_m}=
\sinh{(2\sqrt{|k_m|}L_m)}/\sqrt{|k_m|}$ and $\cos{(2\sqrt{k_m}L_m)}=
\cosh{(2\sqrt{|k_m|}L_m)}$. The result is 
\begin{eqnarray}
\begin{aligned}\label{eq:JCmj2}
G_{S,m}&=\frac{1}{L_m k_m\sqrt{\beta_{m0}}}\left[1-\cos{(\sqrt{k_m}L_m)}
         \right]\quad ,\\
G_{C,m}&=\frac{\sqrt{\beta_{m0}}}{L_m\sqrt{k_m}}\sin{(\sqrt{k_m}L_m)}
         -\frac{\alpha_{m0}}{L_m k_m\sqrt{\beta_{m0}}}
          \left[1-\cos{(\sqrt{k_m}L_m)}\right]
         =\frac{\sqrt{\beta_{m0}}}{L_m\sqrt{k_m}}\sin{(\sqrt{k_m}L_m)}
         -\alpha_{m0}G_{S,m} \quad .
\end{aligned} 
\end{eqnarray}
$J_{S,mj}$ and $J_{C,mj}$ for quadrupoles and combined-function 
magnets can be eventually written and computed as 
\begin{eqnarray}\label{eq:JCmj3}
\begin{aligned}
J_{S,mj}&=\sin{(\tau_{s_{m}j})}G_{C,m}-\cos{(\tau_{s_{m}j})}G_{S,m}\quad , \\
J_{C,mj}&=\cos{(\tau_{s_{m}j})}G_{C,m}+\sin{(\tau_{s_{m}j})}G_{S,m}\quad ,
\end{aligned}
\end{eqnarray}
where $\tau_{s_{m}j}$ is again the shifted phase 
advance between the entrance of the magnet $m$ and 
the observation point $j$.

\subsection{$J_{C,mj}^{(D_y)}$  and $J_{C,mj}^{(D_x)}$  for quadrupoles}
These integrals can be used  in the evaluation of 
the dispersion function from Eq.~\eqref{eq:Text_Disp1} 
to account for the evolution of the C-S parameters and 
dispersion across a skew quadrupole magnet. 

Differently from the integrals $I$ and $J$ discussed in 
the previous sections, the $J^{(D)}$ integrals 
contain the dispersion function (in the orthogonal 
plane), whose propagation along the quadrupole 
needs to be evaluated separately. The following 
expression applies for a generic combined-function 
magnet
\begin{eqnarray}\label{eq:Transport1}
\left(\begin{array}{c} D(s) \\ D'(s) \\ 1\end{array}\right)=
\left(\begin{array}{c c}\mathbf{A_m}(s)&\vec{d}\\ 0 & 1\end{array}\right)
\left(\begin{array}{c} D_{m,0} \\ D'_{m,0} \\ 1\end{array}\right)
\quad ,
\end{eqnarray}
where $D_{m,0}$ and $D'_{m,0}$ are the dispersion 
and its derivative at the entrance of the magnet, 
$\mathbf{A_m}$ is the betatron matrix of 
Eq.~\eqref{eq:Adef2} for a pure sector dipole, 
of Eq.~\eqref{eq:Adef1} otherwise. For 
quadrupoles $\vec{d}=(0 \ 0)^T$ and dispersion is 
propagated according to 
\begin{eqnarray}
D(s)&=D_{m0}\left(\begin{array}{c}
             \cos{(\sqrt{k_m}s)} \\ \cosh{(\sqrt{|k_m|}s)}
             \end{array}\right) 
     +\frac{D'_{m0}}{\sqrt{|k_m|}}\left(\begin{array}{c}
             \sin{(\sqrt{k_m}s)} \\ \sinh{(\sqrt{|k_m|}s)}
             \end{array}\right)\ ,
\end{eqnarray}
where, once again, the upper and lower terms 
refer to the focusing and defocusing planes, 
respectively. By merging the above expression 
with the ones in Eq.~\eqref{eq:ACmatrix}, 
the following relations are obtained
\begin{eqnarray}\label{eq:T1}
\begin{aligned}
T_{S,x,m}&=\frac{1}{L_m}\int_0^{L_m}{\hskip -2mm\sqrt{\beta_x(s)}D_y(s)
            \sin{\Delta\phi_{x,s}}\ ds}= 
            \frac{D_{y,m0}}{L_m\sqrt{|k_m|\beta_{x,m0}}}
            \int_0^{L_m}\hskip -1mm\left(\begin{array}{l}
             \cosh{(\sqrt{|k_m|}s)}\sin{(\sqrt{|k_m|}s)} \\ 
             \sinh{(\sqrt{|k_m|}s)}\cos{(\sqrt{|k_m|}s)}
             \end{array}\right) ds\\ &\hskip3mm
          +\frac{D'_{y,m0}}{|k_m|L_m\sqrt{\beta_{x,m0}}}
           \int_0^{L_m}\hskip -3mm
           \sin{(\sqrt{|k_m|}s)}\sinh{(\sqrt{|k_m|}s)} ds\quad ,\\
T_{C,x,m}&=\frac{1}{L_m}\int_0^{L_m}{\hskip -2mm\sqrt{\beta_x(s)}D_y(s)
           \cos{\Delta\phi_{x,s}}\ ds}= 
           \frac{\sqrt{\beta_{x,m0}}D_{y,m0}}{L_m}\int_0^{L_m}\hskip -3mm
           \cosh{(\sqrt{|k_m|}s)}\cos{(\sqrt{|k_m|}s)}\ ds\\&\hskip3mm
          -\frac{\alpha_{x,m0}D'_{y,m0}}{L_m|k_m|\sqrt{\beta_{x,m0}}}
           \int_0^{L_m}\hskip -3mm
           \sinh{(\sqrt{|k_m|}s)}\sin{(\sqrt{|k_m|}s)}\ ds
          -\frac{\alpha_{x,m0}D_{y,m0}}{L_m\sqrt{|k_m|\beta_{x,m0}}}
           \int_0^{L_m}\hskip -1mm\left(\begin{array}{c}
             \cosh{(\sqrt{|k_m|}s)}\sin{(\sqrt{|k_m|}s)} \\ 
             \sinh{(\sqrt{|k_m|}s)}\cos{(\sqrt{|k_m|}s)}
             \end{array}\right) ds\\&\hskip3mm
          +\frac{\sqrt{\beta_{x,m0}}D'_{y,m0}}{L_m\sqrt{|k_m|}}
           \int_0^{L_m}\hskip -1mm\left(\begin{array}{c}
             \sinh{(\sqrt{|k_m|}s)}\cos{(\sqrt{|k_m|}s)} \\
             \cosh{(\sqrt{|k_m|}s)}\sin{(\sqrt{|k_m|}s)}
             \end{array}\right) ds \quad ,\\
T_{S,y,m}&=\frac{1}{L_m}\int_0^{L_m}{\hskip -2mm\sqrt{\beta_y(s)}D_x(s)
            \sin{\Delta\phi_{y,s}}\ ds}= 
            \frac{D_{x,m0}}{L_m\sqrt{|k_m|\beta_{y,m0}}}
            \int_0^{L_m}\hskip -1mm\left(\begin{array}{l}
             \sinh{(\sqrt{|k_m|}s)}\cos{(\sqrt{|k_m|}s)} \\
             \cosh{(\sqrt{|k_m|}s)}\sin{(\sqrt{|k_m|}s)} 
             \end{array}\right) ds\\ &\hskip3mm
          +\frac{D'_{x,m0}}{|k_m|L_m\sqrt{\beta_{y,m0}}}
           \int_0^{L_m}\hskip -3mm
           \sin{(\sqrt{|k_m|}s)}\sinh{(\sqrt{|k_m|}s)} ds\quad ,\\
T_{C,y,m}&=\frac{1}{L_m}\int_0^{L_m}{\hskip -2mm\sqrt{\beta_y(s)}D_x(s)
            \cos{\Delta\phi_{y,s}}\ ds}=  
           \frac{\sqrt{\beta_{y,m0}}D_{x,m0}}{L_m}\int_0^{L_m}\hskip -3mm
           \cosh{(\sqrt{|k_m|}s)}\cos{(\sqrt{|k_m|}s)}\ ds\\&\hskip3mm
          -\frac{\alpha_{y,m0}D'_{x,m0}}{L_m|k_m|\sqrt{\beta_{y,m0}}}
           \int_0^{L_m}\hskip -3mm
           \sinh{(\sqrt{|k_m|}s)}\sin{(\sqrt{|k_m|}s)}\ ds
          -\frac{\alpha_{y,m0}D_{x,m0}}{L_m\sqrt{|k_m|\beta_{y,m0}}}
           \int_0^{L_m}\hskip -1mm\left(\begin{array}{c}
             \sinh{(\sqrt{|k_m|}s)}\cos{(\sqrt{|k_m|}s)} \\
             \cosh{(\sqrt{|k_m|}s)}\sin{(\sqrt{|k_m|}s)}  
             \end{array}\right) ds\\&\hskip3mm
          +\frac{\sqrt{\beta_{y,m0}}D'_{x,m0}}{L_m\sqrt{|k_m|}}
           \int_0^{L_m}\hskip -1mm\left(\begin{array}{c}
             \cosh{(\sqrt{|k_m|}s)}\sin{(\sqrt{|k_m|}s)} \\
             \sinh{(\sqrt{|k_m|}s)}\cos{(\sqrt{|k_m|}s)} 
             \end{array}\right) ds \quad .
\end{aligned}
\end{eqnarray}
Note that because of the presence of the dispersion 
of the orthogonal plane, the above integrals in the 
two planes have a different functional dependence on 
the quadrupole gradients $|k_m|$ (the integrals in the 
r.h.s. are swapped). The absolute value is introduced in 
the trigonometric functions too, in order account for the fact 
that when one plane is focusing, the other is defocusing though 
both the trigonometric and the hyperbolic functions will 
have always a positive argument: The use of the absolute 
value then avoids any conflict with the sign of $k_m$.

The four terms of Eq.~\eqref{eq:T1}
can be made explicit after computing and replacing 
four integrals
\begin{eqnarray}
\begin{aligned}
\int_0^{L_m}\hskip -3mm \cos{(\sqrt{|k_m|}s)}\cosh{(\sqrt{|k_m|}s)} ds &= 
  \frac{1}{2\sqrt{|k_m|}}\left[\sin{(\sqrt{|k_m|}L_m)}\cosh{(\sqrt{|k_m|}L_m)}
           +\cos{(\sqrt{|k_m|}L_m)}\sinh{(\sqrt{|k_m|}L_m)}\right]\ ,\\
\int_0^{L_m}\hskip -3mm \sin{(\sqrt{|k_m|}s)}\sinh{(\sqrt{|k_m|}s)} ds &= 
  \frac{1}{2\sqrt{|k_m|}}\left[\sin{(\sqrt{|k_m|}L_m)}\cosh{(\sqrt{|k_m|}L_m)}
           -\cos{(\sqrt{|k_m|}L_m)}\sinh{(\sqrt{|k_m|}L_m)}\right]\ ,\\
\int_0^{L_m}\hskip -3mm \cosh{(\sqrt{|k_m|}s)}\sin{(\sqrt{|k_m|}s)} ds &= 
  \frac{1}{2\sqrt{|k_m|}}\left[\sin{(\sqrt{|k_m|}L_m)}\sinh{(\sqrt{|k_m|}L_m)}
           -\cos{(\sqrt{|k_m|}L_m)}\cosh{(\sqrt{|k_m|}L_m)}+1\right]\ ,\\
\int_0^{L_m}\hskip -3mm \sinh{(\sqrt{|k_m|}s)}\cos{(\sqrt{|k_m|}s)} ds &=
  \frac{1}{2\sqrt{|k_m|}}\left[\sin{(\sqrt{|k_m|}L_m)}\sinh{(\sqrt{|k_m|}L_m)}
           +\cos{(\sqrt{|k_m|}L_m)}\cosh{(\sqrt{|k_m|}L_m)}-1\right]\ ,\\
\end{aligned}\nonumber 
\end{eqnarray}
resulting in 
\begin{eqnarray}\label{eq:T2}
\begin{aligned}
T_{S,x,m}&= \frac{D_{y,m0}}{2|k_m|L_m\sqrt{\beta_{x,m0}}}
            \left(\begin{array}{l}
             \sin{(\sqrt{|k_m|}L_m)}\sinh{(\sqrt{|k_m|}L_m)}
           -\cos{(\sqrt{|k_m|}L_m)}\cosh{(\sqrt{|k_m|}L_m)}+1 \\ 
             \sin{(\sqrt{|k_m|}L_m)}\sinh{(\sqrt{|k_m|}L_m)}
           +\cos{(\sqrt{|k_m|}L_m)}\cosh{(\sqrt{|k_m|}L_m)}-1
             \end{array}\right)\\ &\hskip3mm
          +\frac{D'_{y,m0}}{2|k_m|^{3/2}L_m\sqrt{\beta_{x,m0}}}
           \left[\sin{(\sqrt{|k_m|}L_m)}\cosh{(\sqrt{|k_m|}L_m)}
           -\cos{(\sqrt{|k_m|}L_m)}\sinh{(\sqrt{|k_m|}L_m)}\right]\quad ,\\
T_{C,x,m}&= 
           \frac{\sqrt{\beta_{x,m0}}D_{y,m0}}{2\sqrt{|k_m|}L_m}
           \left[\sin{(\sqrt{|k_m|}L_m)}\cosh{(\sqrt{|k_m|}L_m)}
           +\cos{(\sqrt{|k_m|}L_m)}\sinh{(\sqrt{|k_m|}L_m)}\right] \\&\hskip3mm
          -\frac{\alpha_{x,m0}D'_{y,m0}}{2|k_m|^{3/2}L_m\sqrt{\beta_{x,m0}}}
           \left[\sin{(\sqrt{|k_m|}L_m)}\cosh{(\sqrt{|k_m|}L_m)}
                -\cos{(\sqrt{|k_m|}L_m)}\sinh{(\sqrt{|k_m|}L_m)}\right]
           \\&\hskip3mm
-\frac{\alpha_{x,m0}D_{y,m0}}{2|k_m|L_m\sqrt{\beta_{x,m0}}}
           \left(\begin{array}{c}
             \sin{(\sqrt{|k_m|}L_m)}\sinh{(\sqrt{|k_m|}L_m)}
            -\cos{(\sqrt{|k_m|}L_m)}\cosh{(\sqrt{|k_m|}L_m)}+1 \\
             \sin{(\sqrt{|k_m|}L_m)}\sinh{(\sqrt{|k_m|}L_m)}
            +\cos{(\sqrt{|k_m|}L_m)}\cosh{(\sqrt{|k_m|}L_m)}-1
           \end{array}\right)
           \\&\hskip3mm
          +\frac{\sqrt{\beta_{x,m0}}D'_{y,m0}}{2|k_m|L_m}
           \left(\begin{array}{c}
             \sin{(\sqrt{|k_m|}L_m)}\sinh{(\sqrt{|k_m|}L_m)}
            +\cos{(\sqrt{|k_m|}L_m)}\cosh{(\sqrt{|k_m|}L_m)}-1 \\
             \sin{(\sqrt{|k_m|}L_m)}\sinh{(\sqrt{|k_m|}L_m)}
            -\cos{(\sqrt{|k_m|}L_m)}\cosh{(\sqrt{|k_m|}L_m)}+1
           \end{array}\right)
 \quad ,\\
T_{S,y,m}&=\frac{D_{x,m0}}{2|k_m|L_m\sqrt{\beta_{y,m0}}}
            \left(\begin{array}{l}
            \sin{(\sqrt{|k_m|}L_m)}\sinh{(\sqrt{|k_m|}L_m)}
           +\cos{(\sqrt{|k_m|}L_m)}\cosh{(\sqrt{|k_m|}L_m)}-1 \\ 
            \sin{(\sqrt{|k_m|}L_m)}\sinh{(\sqrt{|k_m|}L_m)}
           -\cos{(\sqrt{|k_m|}L_m)}\cosh{(\sqrt{|k_m|}L_m)}+1
             \end{array}\right)\\ &\hskip3mm
          +\frac{D'_{x,m0}}{2|k_m|^{3/2}L_m\sqrt{\beta_{y,m0}}}
           \left[\sin{(\sqrt{|k_m|}L_m)}\cosh{(\sqrt{|k_m|}L_m)}
           -\cos{(\sqrt{|k_m|}L_m)}\sinh{(\sqrt{|k_m|}L_m)}\right]\quad ,\\
T_{C,y,m}&= 
           \frac{\sqrt{\beta_{y,m0}}D_{x,m0}}{2\sqrt{|k_m|}L_m}
           \left[\sin{(\sqrt{|k_m|}L_m)}\cosh{(\sqrt{|k_m|}L_m)}
           +\cos{(\sqrt{|k_m|}L_m)}\sinh{(\sqrt{|k_m|}L_m)}\right] \\&\hskip3mm
          -\frac{\alpha_{y,m0}D'_{x,m0}}{2|k_m|^{3/2}L_m\sqrt{\beta_{y,m0}}}
           \left[\sin{(\sqrt{|k_m|}L_m)}\cosh{(\sqrt{|k_m|}L_m)}
                -\cos{(\sqrt{|k_m|}L_m)}\sinh{(\sqrt{|k_m|}L_m)}\right]
           \\&\hskip3mm
-\frac{\alpha_{y,m0}D_{x,m0}}{2|k_m|L_m\sqrt{\beta_{y,m0}}}
           \left(\begin{array}{c}
             \sin{(\sqrt{|k_m|}L_m)}\sinh{(\sqrt{|k_m|}L_m)}
            +\cos{(\sqrt{|k_m|}L_m)}\cosh{(\sqrt{|k_m|}L_m)}-1 \\
             \sin{(\sqrt{|k_m|}L_m)}\sinh{(\sqrt{|k_m|}L_m)}
            -\cos{(\sqrt{|k_m|}L_m)}\cosh{(\sqrt{|k_m|}L_m)}+1
           \end{array}\right)
           \\&\hskip3mm
          +\frac{\sqrt{\beta_{y,m0}}D'_{x,m0}}{2|k_m|L_m}
           \left(\begin{array}{c}
             \sin{(\sqrt{|k_m|}L_m)}\sinh{(\sqrt{|k_m|}L_m)}
            -\cos{(\sqrt{|k_m|}L_m)}\cosh{(\sqrt{|k_m|}L_m)}+1 \\
             \sin{(\sqrt{|k_m|}L_m)}\sinh{(\sqrt{|k_m|}L_m)}
            +\cos{(\sqrt{|k_m|}L_m)}\cosh{(\sqrt{|k_m|}L_m)}-1
           \end{array}\right)
 \quad .
\end{aligned}
\end{eqnarray}
Once again, the upper and lower terms in the brackets 
refer to the focusing and defocusing planes, respectively.
By decomposing the cosine term within the integrals 
of Eqs.~\eqref{eq:JCDy}-\eqref{eq:JCDx}, $J_{C,mj}^{(D_y)}$ 
and $J_{C,mj}^{(D_x)}$ at the quadrupoles eventually 
read
\begin{eqnarray}\label{eq:JCDxy}
\begin{aligned}
J_{C,mj}^{(D_y)}&=\cos{(\tau_{x,s_mj})}T_{C,x,m}+
                  \sin{(\tau_{x,s_mj})}T_{S,x,m}\quad , \\
J_{C,mj}^{(D_x)}&=\cos{(\tau_{y,s_mj})}T_{C,y,m}+
                  \sin{(\tau_{y,s_mj})}T_{S,y,m}\quad ,
\end{aligned}
\end{eqnarray}
where the $T$ functions are those of Eq.~\eqref{eq:T2}, and 
$\tau_{s_mj}=\Delta\phi_{s_mj}-\pi Q$ 
represents the usual shifted phase advance between the 
generic location $j$ and the entrance of the 
magnet $m$ ($s_{m}$).

\subsection{$J_{C,mj}^{(D_y)}$  and $J_{C,mj}^{(D_x)}$  for 
            combined-function magnets}
For the evaluation of the two integrals across 
combined-function magnets (i.e. with both dipole 
and quadrupole fields), the same calculations 
carried out in the previous section need to be 
repeated from Eq.~\eqref{eq:Transport1}, with a 
nonzero ($1\times2$) vector 
\begin{eqnarray}\label{eq:vectd-comb}
\vec{d}=
\left\{\begin{aligned}
&\left(\frac{1}{\rho k_m}(1-\cos{\sqrt{k_m}s})\quad 
      \frac{1}{\rho \sqrt{k_m}}\sin{\sqrt{k_m}s}\right)^T \\
&\left(\frac{1}{\rho|k_m|}(-1+\cosh{\sqrt{|k_m|}s})\quad 
      \frac{1}{\rho\sqrt{|k_m|}}\sinh{\sqrt{|k_m|}s}\right)^T
\end{aligned}\right. \qquad .
\end{eqnarray}
For the old ESRF storage ring, this calculation 
is not needed, though it shall be performed for 
machines such as the present ALBA and the new ESRF 
storage rings (EBS), both comprising defocusing 
combined-function dipoles, whose tilts may affect 
the dispersion function. 

\subsection{Corrections for thick orbit deflectors and sextupoles}
\label{app:thicksteerers}
So far, integrals replacing constant terms have been 
derived for quadrupoles and combined-function magnets $m$, 
only, i.e. assuming thin steerers $w$ (along with thin 
BPMs $j$) for the ORM of Eq.~\eqref{eq:Nmatrix1} 
and thin sextupoles $m$ for the chromatic functions of 
Sec.~\ref{CHROM-Formulas}. In order to account for 
the variation of the C-S parameters across all magnets $m$ 
and deflectors $w$ in Eqs.~\eqref{eq:Nmatrix1},~\eqref{eq:Text_Chrom1}, 
\eqref{eq:Text_ChromBeat3},~\eqref{eq:Text-ChromPhAdx} 
and~\eqref{eq:Text-ChromPhAdy} the following terms 
need to be replaced
\begin{eqnarray}
\sqrt{\beta_{w}}\cos{(\tau_{wj})}\quad &\longrightarrow&\quad\label{eq:ICwj}
J_{C,wj}=\frac{1}{L_w}\int_0^{L_w}{\sqrt{\beta(s)}\cos{(\tau_{sj})}_{\ }ds}\ ,\\
\sqrt{\beta_{w}}\sin{(\tau_{wj})}\quad &\longrightarrow&\quad\label{eq:ISwj}
J_{S,wj}=\frac{1}{L_w}\int_0^{L_w}{\sqrt{\beta(s)}\sin{(\tau_{sj})}_{\ }ds}\ ,\\
\sqrt{\beta_{w}}\cos{(\Delta\phi_{wj})}\quad &\longrightarrow&\quad\label{eq:ICDwj}
J_{C_\Delta,wj}=\frac{1}{L_w}\int_0^{L_w}{\sqrt{\beta(s)}
                                          \cos{(\Delta\phi_{sj})}_{\ }ds}\ , \\
\sqrt{\beta_{w}}\beta_{m}\cos{(\tau_{wj})}\cos{(2\tau_{mw})}\ &\longrightarrow&\ 
   P_{C,mwj}=\frac{1}{L_mL_w}\label{eq:PC1}
   \int_0^{L_w}{\sqrt{\beta(s')}\cos{(\tau_{s'j})}
   \int_0^{L_m}{\beta(s'')\cos{(2\tau_{s''s'})}_{\ }ds''}\ ds'}\ ,\hskip1cm\\
\sqrt{\beta_{w}}\beta_{m}\sin{(\tau_{wj})}\sin{(2\tau_{mw})}\ &\longrightarrow&\ 
   P_{S,mwj}=\frac{1}{L_mL_w}\label{eq:PS1}
   \int_0^{L_w}{\sqrt{\beta(s')}\sin{(\tau_{s'j})}
   \int_0^{L_m}{\beta(s'')\sin{(2\tau_{s''s'})}_{\ }ds''}\ ds'}\ ,\\
\beta_{m}D_{m}\quad &\longrightarrow&\quad\label{eq:LBD}
L_{\beta D,m}=\frac{1}{L_m}\int_0^{L_m}{\beta(s)D(s)_{\ }ds}\ ,\\
\beta_{m,p}D_{m,q}\cos{(2\tau_{p,mj})}\quad &\longrightarrow&\quad\label{eq:LBDC}
L_{C_p,D_q,mj}=\frac{1}{L_m}\int_0^{L_m}{\beta_p(s)D_q(s)\cos{(2\tau_{p,sj})}_{\ }ds}\ ,
\qquad p,q=x,y\\
\beta_{m,p}D_{m,q}\sin{(2\tau_{p,mj})}\quad &\longrightarrow&\quad\label{eq:LBDS}
L_{S_p,D_q,mj}=\frac{1}{L_m}\int_0^{L_m}{\beta_p(s)D_q(s)\sin{(2\tau_{p,sj})}_{\ }ds}\ ,
\qquad p,q=x,y
\end{eqnarray}
where $L_w$ and $L_m$ denote the magnetic lengths,  
$s'$ and $s''$ represent the position inside the 
two magnets, and $\Delta\phi_{s}$ is the phase advance 
along either the steerer $w$ or the magnet $m$ ($\tau$ 
is the usual shifted phase advance).  

The same algebra carried out in the previous sections can 
be repeated by noting that an approximated transfer matrix 
for an orbit corrector or a sextupole reads
\begin{eqnarray}
\mathbf{A}(s)=\left(\begin{array}{c c}
1  & s \\ 0 & 1\end{array} \right) \quad  .\label{eq:AdefW}
\end{eqnarray}
This approximation is valid for sextupoles, as they do not 
alter the linear optics, whereas for steerers it holds 
for small deflection angles only. The few tens of 
$\mu$rad usually imparted during ORM measurements 
definitively meet this condition. From the same $C$ matrix 
of Eq.~\eqref{eq:Cdef1} the following relations hold
\begin{eqnarray}
\left\{\begin{aligned} A_{w,11}&=C_{w,11} \\ A_{w,12}&=C_{w,12}
       \end{aligned}\right. 
\Longrightarrow\left\{
\begin{aligned}
\sqrt{\beta(s)}\sin{\Delta\phi_s}&=\frac{s}{\sqrt{\beta_{w0}}} \\
\sqrt{\beta(s)}\cos{\Delta\phi_s}&=\sqrt{\beta_{w0}}
-\frac{\alpha_{w0}}{\sqrt{\beta_{w0}}}s
\end{aligned}\right. \quad ,\label{eq:CWmatrix}
\end{eqnarray}
where $\beta_{w0}$ and $\alpha_{w0}$ are the C-S parameters 
at the entrance of the steerer (as well as of the sextupole after 
replacing $w$ with $m$). The integrals across the 
magnet then reads
\begin{eqnarray}
\begin{aligned}
T_{S,w}&=\frac{1}{L_w}\int_0^{L_w}\frac{s}{\sqrt{\beta_{w0}}}\ ds
       =\frac{L_w}{2\sqrt{\beta_{w0}}}\quad ,  \\
T_{C,w}&=\frac{1}{L_w}\int_0^{L_w}\left(\sqrt{\beta_{w0}}
              -\frac{\alpha_{w0}}{\sqrt{\beta_{w0}}}s\right)\ ds
        =\sqrt{\beta_{w0}}-\frac{\alpha_{w0}}{2\sqrt{\beta_{w0}}}L_w
        =\sqrt{\beta_{w0}}-\alpha_{w0}T_{S,w}\quad .
\end{aligned}\label{eq:IWCS}
\end{eqnarray}
The integrals of Eqs.~\eqref{eq:ICwj}-\eqref{eq:ICDwj} can 
be then computed after noting that 
\begin{eqnarray}
\begin{aligned}
J_{S,wj}&=\sin{(\tau_{s_wj})}T_{C,w}-\cos{(\tau_{s_wj})}T_{S,w}\quad , \\
J_{C,wj}&=\cos{(\tau_{s_wj})}T_{C,w}+\sin{(\tau_{s_wj})}T_{S,w}\quad , \\
J_{C_\Delta,wj}&=\cos{(\Delta\phi_{s_wj})}T_{C,w}+\sin{(\Delta\phi_{s_wj})}
                 T_{S,w}\quad , \\
\end{aligned}\label{eq:IWCS2}
\end{eqnarray}
where $\Delta\phi_{s_wj}$ and $\tau_{s_{m}j}$ are the phase 
advance and the shifted phase advance, respectively, between 
the entrance of the steerer $w$ and the BPM $j$.

After some algebra the integrals $P_{C,mwj}$ and $P_{S,mwj}$ 
of Eqs.~\eqref{eq:PC1}-\eqref{eq:PS1} can be written as
\begin{eqnarray}
   P_{C,mwj}&=&\frac{1}{L_w}\nonumber 
   \int_0^{L_w}{\hskip-4mm\sqrt{\beta(s')}\Bigl[
      I_{C,mw_0}\cos{(2\Delta\phi_{s'})}
     -I_{S,mw_0}\sin{(2\Delta\phi_{s'})}\Bigr]\Bigl[
      \cos{(\tau_{w_0,j})}\cos{(\Delta\phi_{s'})}
     +\sin{(\tau_{w_0,j})}\sin{(\Delta\phi_{s'})}\Bigr]ds'}\ ,\\
   P_{S,mwj}&=&\frac{1}{L_w}\nonumber 
   \int_0^{L_w}{\hskip-4mm\sqrt{\beta(s')}\Bigl[
      I_{S,mw_0}\cos{(2\Delta\phi_{s'})}
     +I_{C,mw_0}\sin{(2\Delta\phi_{s'})}\Bigr]\Bigl[
      \sin{(\tau_{w_0,j})}\cos{(\Delta\phi_{s'})}
     -\cos{(\tau_{w_0,j})}\sin{(\Delta\phi_{s'})}\Bigr]ds'}\ ,
\end{eqnarray}
where $I_{C,mw_0}$ and  $I_{C,mw_0}$ are the same integrals 
across the magnet $m$ of Eq.~\eqref{eq:IS2}, with the location 
of the BPM $j$ replaced by $w_0$ which is the entrance of 
the steerer magnet $w$. It remains hence to integrate the 
trigonometric terms of $\Delta\phi_{s'}$, i.e. of the phase 
advance along the steerer $w$. To this end some approximations 
are necessary, by assuming that total phase advance the orbit 
corrector is sufficiently small ($\Delta\phi_w\ll 1$) so 
to have
\begin{eqnarray}
\cos{(2\Delta\phi_{s'})}\cos{(\Delta\phi_{s'})}\simeq&\nonumber
           1\hskip7mm +O(\Delta\phi_{s'}^2)\ &\simeq\cos{(\Delta\phi_{s'})}\ , \\
\sin{(2\Delta\phi_{s'})}\cos{(\Delta\phi_{s'})}\simeq&\ \nonumber
           2\Delta\phi_s+O(\Delta\phi_{s'}^3)\ &\simeq2\sin{(\Delta\phi_{s'})}\ , \\
\cos{(2\Delta\phi_{s'})}\sin{(\Delta\phi_{s'})}\simeq&\ \nonumber
           \ \Delta\phi_s+O(\Delta\phi_{s'}^2)\ &\simeq\sin{(\Delta\phi_{s'})}\ , \\
\sin{(2\Delta\phi_{s'})}\sin{(\Delta\phi_{s'})}\simeq&\nonumber
           0\hskip7mm +O(\Delta\phi_{s'}^2)\ &\simeq 0\ .
\end{eqnarray}
With these approximations the above integrals simplify to
\begin{eqnarray}
\begin{aligned}
   P_{C,mwj}&\simeq&\frac{1}{L_w}
   \int_0^{L_w}I_{C,mw_0}\Bigl[
      \cos{(\tau_{w_0,j})}\left(\sqrt{\beta(s')}\cos{(\Delta\phi_{s'})}\right)
     +\sin{(\tau_{w_0,j})}\left(\sqrt{\beta(s')}\sin{(\Delta\phi_{s'})}\right)
     \Bigr]- \\ && \hskip 12mm 2I_{S,mw_0}
      \cos{(\tau_{w_0,j})}\left(\sqrt{\beta(s')}\sin{(\Delta\phi_{s'})}\right)
     ds'  \\
   P_{S,mwj}&\simeq&\frac{1}{L_w}
   \int_0^{L_w}I_{S,mw_0}\Bigl[
      \sin{(\tau_{w_0,j})}\left(\sqrt{\beta(s')}\cos{(\Delta\phi_{s'})}\right)
     -\cos{(\tau_{w_0,j})}\left(\sqrt{\beta(s')}\sin{(\Delta\phi_{s'})}\right)
     \Bigr]+ \\ && \hskip 12mm 2I_{C,mw_0}
      \sin{(\tau_{w_0,j})}\left(\sqrt{\beta(s')}\sin{(\Delta\phi_{s'})}\right)
     ds'\ .
\end{aligned}\ . \label{eq:PC2}
\end{eqnarray}
The integrands in $s'$ within the large parenthesis are 
the same of Eq.~\eqref{eq:CWmatrix} and the above 
quantities read
\begin{eqnarray}
\begin{aligned}
   P_{C,mwj}&\simeq&
   I_{C,mw_0}\Bigl[\cos{(\tau_{w_0,j})}T_{C,w}
                  +\sin{(\tau_{w_0,j})}T_{S,w}\Bigr]
  -2I_{S,mw_0}T_{S,w}\cos{(\tau_{w_0,j})} \\
   P_{S,mwj}&\simeq&
   I_{S,mw_0}\Bigl[\sin{(\tau_{w_0,j})}T_{C,w}
                  -\cos{(\tau_{w_0,j})}T_{S,w}\Bigr]
  +2I_{C,mw_0}T_{S,w}\sin{(\tau_{w_0,j})}
\end{aligned}\ , \label{eq:PC3}
\end{eqnarray}
where $\tau_{w_0,j}$ is the same shifted phase advance between 
the entrance of the steerer $w$ and the BPM $j$ of 
Eq.~\eqref{eq:def_tau}, $T_{C,w}$ and $T_{S,w}$ are 
computed in Eq.~\eqref{eq:IWCS}, while $I_{C,mw_0}$ 
and $I_{S,mw_0}$ are to be evaluated via Eq.~\eqref{eq:IS2}.

Eq.~\eqref{eq:Nmatrix1} can be now rewritten in a more compact 
notation accounting for thick quadrupole magnets $m$ and orbit 
correctors $w$.
\begin{eqnarray}\nonumber 
N_{wj,m}&\simeq&\mp
  \frac{\sqrt{\beta_{j}^{(mod)}}}{2\sin{(\pi Q^{(mod)})}}
 \Bigg\{\frac{1}{4\sin{(2\pi Q^{(mod)})}}
              \Big[J_{C,wj}I_{C,mj}+P_{C,mwj}+J_{S,wj}I_{S,mj}-P_{S,mwj}\Big]
              \\ \nonumber &&\hspace{3.2cm}
        +\frac{1}{2}J_{S,wj}I_{\beta,m}\left[\Pi(m,j)-\Pi(m,w)+\Pi(j,w)\right]
       +\frac{J_{C_\Delta,wj}I_{\beta,m}}
            {4\sin{(\pi Q^{(mod)})}}\Bigg\}\ ,
\end{eqnarray}
where the sign is negative in the horizontal plane, positive in 
the vertical plane. $I_{\beta,m}$, $I_{C,mj}$ and $I_{S,mj}$ are 
the integrals across the quadrupoles of Eqs.~\eqref{eq:Ib2}-\eqref{eq:IS2}. 
$J_{C,wj}$, $J_{s,wj}$ and $J_{C_\Delta,wj}$ are the integrals across 
the steerer magnets of Eq.~\eqref{eq:IWCS2}, whereas the double 
integrals $P_{C,mwj}$ and $P_{S,mwj}$ are those of Eq.~\eqref{eq:PC3}.

The integral $L_{\beta D,m}$ can be used in the evaluation of linear 
chromaticity of Eq.~\eqref{eq:Text_Chrom1} when the 
variation of beta function and dispersion 
across a sextupole cannot be neglected. The magnet is here 
approximated as a drift. By summing up the square of the two 
equations in Eq.~\eqref{eq:CWmatrix} we obtain
\begin{eqnarray}\label{eq:beta}
\beta(s)=\beta_{m0}-2\alpha_{m0}s+\gamma_{m0}s^2 \quad,
\end{eqnarray}
whereas from Eqs.~\eqref{eq:Transport1} and~\eqref{eq:AdefW}
(for a drift $\vec{d}=(0 \ 0)^T$) the dispersion evolves linearly as 
\begin{eqnarray}\label{eq:Disp}
D(s)=D_{m0}+D'_{m0}s\quad .
\end{eqnarray}
The integral is then easily computed, resulting in 
\begin{eqnarray}
L_{\beta D,m}=\beta_{m0}D_{m0}
           +\left(\beta_{m0}D_{m0}'-2\alpha_{m0}D_{m0}\right)\frac{L_m}{2}
           +\left(\gamma_{m0}D_{m0}-2\alpha_{m0}D_{m0}'\right)\frac{L_m^2}{3}
           +\gamma_{m0}D_{m0}'\frac{L_m^3}{4}\quad . \label{eq:LBD2}
\end{eqnarray}

The integrals $L_{C_p,D_q,mj}$ and $L_{S_p,D_q,mj}$ (with $p$ 
and $q$ either $x$ or $y$) can be used in the evaluation of the 
chromatic beating of Eq.~\eqref{eq:Text_ChromBeat3} and 
of the chromatic phase advance shift of 
Eqs.~\eqref{eq:Text-ChromPhAdx}-\eqref{eq:Text-ChromPhAdy}, 
respectively. From Eqs.~\eqref{eq:CWmatrix},~\eqref{eq:beta} 
and~\eqref{eq:Disp} we can write
\begin{eqnarray}
\begin{aligned}
D_{q,m}(s)\beta_{p,m}(s)\cos{(2\Delta\phi_{p,s})}
     &=&D_{q,m}(s)\left[\beta_{p,m}(s)-2\beta_{p,m}(s)\sin{(\Delta\phi_{p,s})}^2\right]
        \hskip 3.8cm\\
     &=&\left(D_{m0,q}+D_{m0,q}'s\right)\left(\beta_{p,m0}-2\alpha_{p,m0}s
        +\gamma_{p,m0}s^2-\frac{2}{\beta_{p,m0}}s^2\right) 
     \hskip 0.6cm\quad ,\\
D_{q,m}(s)\beta_{p,m}(s)\sin{(2\Delta\phi_{p,s})}
     &=&2D_{q,m}(s)\left(\sqrt{\beta_{p,m}(s)}\sin{(\Delta\phi_{p,s})}\right)
        \left(\sqrt{\beta_{p,m}(s)}\cos{(\Delta\phi_{p,s})}\right)
     \hskip 1.2cm\\
     &=&2\left(D_{m0,q}+D_{m0,q}'s\right)\left(\frac{s}{\sqrt{\beta_{p,m0}}}\right)
        \left(\sqrt{\beta_{p,m0}}-\frac{\alpha_{p,m0}}{\sqrt{\beta_{p,m0}}}s\right)
     \hskip 1.2cm\quad .
\end{aligned}
\end{eqnarray}
The integrals along the magnetic length 
\begin{eqnarray}
\begin{aligned}
\mathcal{D}_{C_p,q,m}=\frac{1}{L_m}\int_0^{L_m}
        {D_{q,m}(s)\beta_{p,m}(s)\cos{(2\Delta\phi_{p,s})}_{\ }ds}\quad ,\\
\mathcal{D}_{S_p,q,m}=\frac{1}{L_m}\int_0^{L_m}
        {D_{q,m}(s)\beta_{p,m}(s)\sin{(2\Delta\phi_{p,s})}_{\ }ds}\quad ,
\end{aligned}
\end{eqnarray}
are then easily computed
\begin{eqnarray}
\begin{aligned}
\mathcal{D}_{C_p,q,m}&=D_{m0,q}\beta_{p,m0}
                     +\frac{L_m}{2}\left(D_{m0,q}'\beta_{p,m0}-2\alpha_{p,m0}D_{m0,q}\right)
                     +\frac{L_m^2}{3}\left(D_{m0,q}\frac{\alpha_{p,m0}^2-1}{\beta_{p,m0}}-2\alpha_{p,m0}D_{m0,q}'\right)
     \\ &\hskip 0.3cm+\frac{L_m^3}{4}\left(D_{m0,q}'\frac{\alpha_{p,m0}^2-1}{\beta_{p,m0}}\right)
             \quad ,\\
\mathcal{D}_{S_p,q,m}&=D_{m0,q}L_m
                     +\frac{2L_m^2}{3}\left(D_{m0,q}'-\frac{\alpha_{p,m0}}{\beta_{p,m0}}D_{m0,q}\right)
                     +\frac{L_m^3}{2}\left(D_{m0,q}'\frac{\alpha_{p,m0}}{\beta_{p,m0}}\right)
             \quad ,
\end{aligned}
\end{eqnarray}
The integrals of Eqs.~\eqref{eq:LBDC}-\eqref{eq:LBDS} can 
be then computed after noting that 
\begin{eqnarray}
\begin{aligned}
L_{S_p,D_q,mj}&=\sin{(2\tau_{s_mj})}\mathcal{D}_{C_p,q,m}
              -\cos{(2\tau_{s_mj})}\mathcal{D}_{S_p,q,m}\quad , \\
L_{C_p,D_q,mj}&=\cos{(2\tau_{s_mj})}\mathcal{D}_{C_p,q,m}
              +\sin{(2\tau_{s_mj})}\mathcal{D}_{S_p,q,m}\quad , \\
\end{aligned}\label{eq:LBDCS}
\end{eqnarray}
where, once again, $p$ and $q$ can be either 
$x$ or $y$.

\subsection{Impact of corrections for the ORM of the FCC-ee}
\label{app:EBS-ORM}
In the analytic evaluation of the ORM for the old ESRF storage ring lattice the corrections accounting for the variation of the optical functions along the magnets had a very limited impact, as shown in Figs.~\ref{fig_Oxxyy1} and~\ref{fig_Oxyyx1}. This is not the case for the new EBS storage ring, with its stronger focusing and to greater extend to the proposed Future Circular Collider FCC-ee. To test the applicability of the analytic formulas for the ORM Jacobian $N$ of Eq.~\eqref{eq:Nmatrix1}, the latter was compared with the one obtained from the numerical simulation of the orbit distortion generated by 8 steerers and recorded at 1600 BPMs, after introducing a tiny error in one quadrupole. The left plot of Fig.~\ref{fig_EBS-ORM1} shows one column of $N^{(xx)}$ and $N^{(yy)}$ over a selected region of the FCC-ee ring, as computed numerically (red curve) and analytically (red curve) by using the thin-lens approximations for quadrupoles and steerers. The difference (green curve is sizeable) and is greatly reduced when including the {\sl thick} corrections accounting for the variation of the optical parameters across the magnets of the previous sections, as shown in the right plot of Fig.~\ref{fig_EBS-ORM1}. The difference (green curve) is barely visible: indeed is it about $10^-4$ smaller than the one resulting from the thin approximation (right plot of Fig.~\ref{fig_EBS-ORM2}). 

\begin{figure}[]
\rule{0mm}{0mm}
\centerline{\includegraphics[width=9.3cm]{ORM_Nxxyy_NumThin.eps}
\ \         \includegraphics[width=9.3cm]{ORM_Nxxyy_NumThick.eps}}
  \caption{\label{fig_EBS-ORM1} (Color) Example of column of the FCC-ee ORM Jacobian $N$ computed numerically (red) and analytically (blue). A zoom over only 150 BPMs (among the 1600) is displayed for a better visualization. Left: the analytic $N$ is computed in the thin-lens approximation with constant optical parameters, yielding a sizeable discrepancy (green curve). Right: the analytic solution is computed by replacing the optical functions by the corresponding integrals of the previous sections, showing an almost perfect agreement.}
\rule{0mm}{0mm}
\end{figure}

\begin{figure}[]
\rule{0mm}{0mm}
\centerline{\includegraphics[width=9.3cm, height=7cm]{QuadQC1L1_1_analytic.eps}
\ \         \includegraphics[width=9.3cm]{ORM_Nxxyy_NumThickx10000.eps}}
  \caption{\label{fig_EBS-ORM2} (Color) Right: Same plot as the right chart of Fig.~\ref{fig_EBS-ORM1} with the difference between numerical and analytical solution multiplied by $10^4$ in order to visualize the residual difference. Left: rms and maximum relative difference from the (true) numerical response and the analytical ones computed along both the matrix columns and rows.}
\rule{0mm}{3mm}
\end{figure}

As global figures of merit to asses the increase in accuracy for the entire Jacobian $N$, the rms and maximum relative difference from the (true) numerical response and the analytical ones have been computed along both the matrix columns and rows. The left plot of Fig.~\ref{fig_EBS-ORM2} shows the results, confirming the substantial gain in accuracy provided by these corrections.

\subsection{Corrections for the tune shift induced by a quadrupolar error}
\label{app:tuneshift-thick}
The standard first-order formula used to evaluate the betatron tune shift $\Delta Q$ induced by quadrupole excitation $\Delta K_m$ reads~\cite{Zimmermann-book}
\begin{eqnarray} 
\frac{\Delta Q_{q}}{\Delta K_m}=\pm\frac{\beta_{q,m}}{4\pi}\ ,
\hspace{1.0cm}\hbox{with }\ q=x,y\ ,\label{eq:tuneshift1}
\end{eqnarray} 
where the $\pm$ sign refers to the horizontal and vertical planes, respectively. The beta function is here considered as its mean value across the quadrupole $m$. We can thus replace it with the integral $I_{\beta,m}$ defined in Eq.~\eqref{eq:Ib1} and evaluated via Eq.~\eqref{eq:Ib2} in both planes. 
\begin{eqnarray} 
\frac{\Delta Q_{q}}{\Delta K_m}=\pm\frac{I_{\beta_{q},m}}{4\pi}\ .
\label{eq:tuneshift2}
\end{eqnarray}
Even though the above equation remains a first-order (improved) approximation, its accuracy increases dramatically. A test was carried out with the lattice of the ESRF EBS by varying one focusing quadrupole by $\Delta K_m=1^{-4}$ 1/m. The relative error $\delta_q=\left(\Delta Q_q^{\hbox{(true)}}-\Delta Q_q^{\hbox{(formula)}}\right)/\Delta Q_q^{\hbox{(true)}}$ in evaluating the tune shifts with the two formulas were
\begin{eqnarray}\left\{ 
\begin{aligned}
\delta_x&=-5.9\% \\
\delta_y&=12.8\% 
\end{aligned}\right.
\hspace{1.0cm}\hbox{with Eq.~\eqref{eq:tuneshift1}}\ , \qquad
\left\{ 
\begin{aligned}
\delta_x&=<10^{-3\ }\% \\
\delta_y&=<10^{-3\ }\% 
\end{aligned}\right.
\hspace{1.0cm}\hbox{with Eq.~\eqref{eq:tuneshift2}}\ , 
\end{eqnarray}
which is dramatic, despite the fact that Eq.~\eqref{eq:tuneshift2} remains a first-order approximation.

\subsection{Correction for the coupling and focusing RDTs}
\label{app:RDT-thick}
All the above integrals accounting for the variation of the optical functions across magnets improve dramatically the accuracy of the analytic formulas for the ORM, chromatic functions and tune shift. It is therefore natural to explore their impact in the evaluation of the coupling and focusing RDTs from the analytic formulas to first order 
\begin{eqnarray} 
f_{2000,j}&=&-\frac{\sum\limits_{m=1}^M \delta K_{m,1}\beta_{m,x}
        	e^{2i\Delta\phi_{x,mj}}}
            {8(1-e^{4\pi iQ_x})}\ ,\hspace{2.1cm}
f_{0020,j}=+\frac{\sum\limits_{m=1}^M \delta K_{m,1}\beta_{m,y}
	        e^{2i\Delta\phi_{y,mj}}}
        {8(1-e^{4\pi iQ_y})}\ , \nonumber \\
f_{1001,\j}&=&\frac{\sum\limits_{m=1}^M J_{m,1}\sqrt{\beta_{m,x}\beta_{m,y}} 
        e^{i(\Delta\phi_{x,mj}-\Delta\phi_{y,mj})}}
        {4(1-e^{2\pi i(Q_u-Q_v)})}  \ ,\quad
f_{1010,\j}=\frac{\sum\limits_{m=1}^M J_{m,1}\sqrt{\beta_{m,x}\beta_{m,y}} 
        e^{i(\Delta\phi_{x,mj}+\Delta\phi_{y,mj})}}
        {4(1-e^{2\pi i(Q_u+Q_v)})}  \ ,
\nonumber \\\quad\label{eq-app:def_rdt1}
\end{eqnarray}
where $j$ denotes the usual generic location. The above definitions assume thin quadrupoles (normal and skew) or constant optical parameters $\beta$ and $\phi$. In order to account for their variation across the magnets, the following terms need to be replaced by their integrals
\begin{eqnarray}
\beta_{m,q}e^{2i\Delta\phi_{q,mj}}\quad&\rightarrow&\quad R_{m,j}^{(2,q)}=
\frac{1}{L_m}\int_0^{L_m}{\beta_q(s)e^{2i\Delta\phi(s)_{q,j}}ds},
\hspace{1.0cm}\hbox{with }\ q=x,y\ ; \label{eq-app:rdt1A}\\
\sqrt{\beta_{m,x}\beta_{m,y}}e^{i(\Delta\phi_{x,mj}\pm\Delta\phi_{y,mj})}
\quad&\rightarrow&\quad R_{m,j}^{(11,\pm)}=
\frac{1}{L_m}\int_0^{L_m}{\sqrt{\beta_{x}(s)\beta_{y}(s)}
                      e^{i[\Delta\phi(s)_{x,m}\pm\Delta\phi(s)_{y,m}]}\ ds}
\label{eq-app:rdt1B}\ .
\end{eqnarray}
The positive sign in the latter refers to the RDT $f_{1010}$, while the negative to $f_{1001}$. All ingredients to evaluate the above integrals have been already derived in the previous sections. Indeed, the left term of Eq.~\eqref{eq-app:rdt1A} reads
\begin{eqnarray}
\beta_{m,q}e^{2i\Delta\phi_{q,mj}}=\beta_{m,q}[\cos{(2\Delta\phi_{x,mj})}+
                                              i\sin{(2\Delta\phi_{x,mj})}]
\ , \label{eq-app:rdt2A}
\end{eqnarray}
which comprises the same integrals of Eqs.~\eqref{eq:IS1},~\eqref{eq:IC1} and~\eqref{eq:IS2}, after replacing the shifted phase advance between the location $j$ and the entrance of the magnet $m$, $\tau_{s_{m}j}$, with the standard (unshifted) phase advance $\Delta\phi_{s_{m}j}$.
\begin{eqnarray}
\begin{aligned}
\beta_{m,q}\sin{(2\Delta\phi_{x,mj})}&\rightarrow \sin{(2\Delta\phi_{s_{m}j})}I_{C,m}-\cos{(2\Delta\phi_{s_{m}j})}I_{S,m}
\quad ,\quad I_{C,m}=\frac{1}{L_m}
 \int_0^{L_m}{\beta(s)\cos{(2\Delta\phi_{s})}_{\ }ds}\ ,\\
\beta_{m,q}\cos{(2\Delta\phi_{x,mj})}&\rightarrow \cos{(2\Delta\phi_{s_{m}j})}I_{C,m}+\sin{(2\Delta\phi_{s_{m}j})}I_{S,m}
\quad ,\quad I_{S,m}=\frac{1}{L_m}
 \int_0^{L_m}{\beta(s)\sin{(2\Delta\phi_{s})}_{\ }ds}\ .\label{eq-app:rdt3A}
\end{aligned} 
\end{eqnarray}
By insetting Eq.~\eqref{eq-app:rdt3A} into Eq.~\eqref{eq-app:rdt2A} and the latter into Eq.~\eqref{eq-app:rdt1A}, we obtain
\begin{eqnarray}
R_{m,j}^{(2)}=\cos{(2\Delta\phi_{s_{m}j})}\left[I_{C,m}-iI_{S,m}\right]
               +\sin{(2\Delta\phi_{s_{m}j})}\left[I_{S,m}+iI_{C,m}\right]\ ,
\end{eqnarray}
where $I_{C,m}$ and $I_{S,m}$ for quadrupoles are defined in Eqs.~\eqref{eq:IC2} and~\eqref{eq:IS3}, respectively. The plane label $q=x,y$ has been omitted here for sake of notation, though the above expression shall be evaluated in the horizontal plane for $f_{2000}$ and in the vertical for $f_{0020}$, since 

\begin{eqnarray} 
f_{2000,j}=-\frac{\sum\limits_{m=1}^M \delta K_{m,1}R_{m,j}^{(2,x)}}
            {8(1-e^{4\pi iQ_x})}\quad ,\hspace{1.1cm}
f_{0020,j}=+\frac{\sum\limits_{m=1}^M \delta K_{m,1}R_{m,j}^{(2,y)}}
            {8(1-e^{4\pi iQ_y})}\ .
\quad\label{eq-app:def_rdt2}
\end{eqnarray}
The coupling terms $R_{m,j}^{(11,\pm)}$ require a bit of trigonometric gymnastics in order to isolate and evaluate the integrals across the quadrupole length. The first step is 
\begin{eqnarray}
e^{i(\Delta\phi_{x,mj}\pm\Delta\phi_{y,mj})}=&&\ 
     \cos{(\Delta\phi_{x,mj}\pm\Delta\phi_{y,mj})}
   +i\sin{(\Delta\phi_{x,mj}\pm\Delta\phi_{y,mj})}\nonumber\\
=&&\hspace{0.5cm}
   \left[\cos{(\Delta\phi_{x,mj})}\cos{(\Delta\phi_{y,mj})}\mp
          \sin{(\Delta\phi_{x,mj})}\sin{(\Delta\phi_{y,mj})}\right] \nonumber\\
 &&+i\left[\sin{(\Delta\phi_{x,mj})}\cos{(\Delta\phi_{y,mj})}\pm
          \cos{(\Delta\phi_{x,mj})}\sin{(\Delta\phi_{y,mj})}\right]
\label{eq-app:rdt2B}\ .
\end{eqnarray}
By replacing the constant terms with their integrals
\begin{eqnarray}
\begin{aligned}
\sqrt{\beta_{m,q}}\sin{(\Delta\phi_{q,mj})}\ &\rightarrow\ 
M_{S,mj}^{(q)}=\frac{1}{L_m}
 \int_0^{L_m}{\sqrt{\beta_q(s)}\sin{[\Delta\phi_{q,j}(s)]}_{\ }ds}\\
\sqrt{\beta_{m,q}}\cos{(\Delta\phi_{q,mj})}\ &\rightarrow\ 
M_{C,mj}^{(q)}=\frac{1}{L_m}
 \int_0^{L_m}{\sqrt{\beta_q(s)}\cos{[\Delta\phi_{q,j}(s)]}_{\ }ds}
\end{aligned}\label{eq-app:rdt3B}\ ,
\end{eqnarray}
it turns out that the above integrals are similar to the ones of Eqs.~\eqref{eq:JS1} and~\eqref{eq:JC1}, after replacing the shifted phase advance between the location $j$ and the entrance of the magnet $m$, $\tau_{s_{m}j}$, with the standard (unshifted) phase advance $\Delta\phi_{s_{m}j}$. They are then computed by applying the same derivation of Sec.~\ref{sec:app-J}, yielding
\begin{eqnarray}\label{eq-app:rdt4B}
\begin{aligned}
M_{S,mj}&=\sin{(\Delta\phi_{s_{m}j})}G_{C,m}-\cos{(\Delta\phi_{s_{m}j})}G_{S,m}\quad , \\
M_{C,mj}&=\cos{(\Delta\phi_{s_{m}j})}G_{C,m}+\sin{(\Delta\phi_{s_{m}j})}G_{S,m}\quad ,
\end{aligned}
\end{eqnarray}
where $G_{S,m}$ and $G_{C,m}$ are the same of Eq.~\eqref{eq:JCmj2} and shall be evaluated for both planes, whose label $q=x,y$ has been dropped here for sake of notation. By replacing Eq.~\eqref{eq-app:rdt3B} into Eq.~\eqref{eq-app:rdt2B} the coupling terms $R_{m,j}^{(11,\pm)}$ read
\begin{eqnarray}\label{eq-app:rdt5B}
R_{m,j}^{(11,\pm)}= M_{C,mj}^{(x)}M_{C,mj}^{(y)}\mp
                    M_{S,mj}^{(x)}M_{S,mj}^{(y)}
            +i\left[M_{S,mj}^{(x)}M_{C,mj}^{(y)}\pm
                    M_{C,mj}^{(x)}M_{S,mj}^{(y)}\right]\ ,
\end{eqnarray}
and the RDTs
\begin{eqnarray} 
f_{1001,\j}=\frac{\sum\limits_{m=1}^M J_{m,1}R_{m,j}^{(11,-)}}
        {4(1-e^{2\pi i(Q_u-Q_v)})}\quad ,\hspace{1.1cm}
f_{1010,\j}=\frac{\sum\limits_{m=1}^M J_{m,1}R_{m,j}^{(11,+)}}
        {4(1-e^{2\pi i(Q_u+Q_v)})}  \ .
\quad\label{eq-app:def_rdt6B}
\end{eqnarray}

\subsection{Inferring the focusing and coupling RDTs from the one-turn matrix}
\label{app:RDT-OTM}

RDTs are usually either computed from the analytic formulas presented in the previous section, or extracted from the harmonic analysis of turn-by-turn data (see Ref.~\cite{Andrea-Linear-arxiv} and references therein). Additionally, coupling RDTs can be inferred from the coupling matrix $\mathbf{C}$, which in turn is derived from the one-turn matrix, as discussed in Ref.~\cite{merging}. This multitude of sources allows a handy crosscheck of any code implementation to evaluate them. In this section we derive yet another way to compute the RDTs directly from the one-turn matrix (OTM). The focusing RDTs $f_{2000}$ and $f_{0020}$ will be inferred from the difference between the ideal on-diagonal blocks of the OTM and the one obtained after including focusing errors. The off-diagonal blocks will be instead used to extract the coupling RDTs $f_{1001}$ and $f_{1010}$. 

The starting point the OTM in Cartesian coordinates $\mathbf{M}$ at an arbitrary location along the ring, which is the one readily available in output of optics codes such as MADX or Accelerator Toolbox:
\begin{eqnarray} 
\vec{X}^{(N+1)}=\mathbf{M}\vec{X}^{(N)}\ , \qquad 
\vec{X}=\left(\begin{array}{c} x\\ p_{x} \\ y \\p_y \end{array}\right)\quad ,
\quad\label{OTM-1}
\end{eqnarray}
where $N$ denotes a generic turn. The OTM $\mathbf{M}$ is a 4x4 matrix with non-zero off-diagonal blocks, which differs from the ideal block-diagonal 4x4 OTM $\mathbf{M_0}$. 

In the normal form coordinates $\vec{\zeta}$ of Eq.~\eqref{eq:NormFormTrans}, the OTM is a diagonal matrix describing a pure rotation defined by the eigen-tunes (which correspond to the betatron tunes as long as the they are sufficiently far away from the resonance stop-band~\cite{prstab_coup}) $e^{i\mathbf{Q}}=
\hbox{diag}(e^{2\pi iQ_x},e^{-2\pi iQ_x},e^{2\pi iQ_y},e^{-2\pi iQ_y})$ of Eq.~\eqref{eq:coNF}: 
\begin{eqnarray}\label{OTM-2}
\vec{\zeta}^{(N+1)}=e^{i\mathbf{Q}}\vec{\zeta}^{(N)}\ , \qquad 
\vec{\zeta}^{(N)}=\left(\begin{array}{c} \zeta_{x,-}\\ \zeta_{x,+} \\ \zeta_{y,-}\\ \zeta_{y,+} \end{array}\right)\quad .
\end{eqnarray}
By walking through the three changes of coordinates that transform the Cartesian vector in the one in normal form, the relation between the RDTs and the OTM $\mathbf{M}$ will be derived. The first step is the standard Courant-Snyder (C-S) transformation matrix $\mathbf{C}$  
\begin{eqnarray} 
\vec{\tilde{X}}=\mathbf{C}\vec{X}\quad , \quad 
\mathbf{C}=
 \left(\begin{array}{c c c c } 
   \displaystyle\frac{1}{\sqrt{\beta_x}} & 0 & 0 & 0 \\
   \displaystyle\frac{\alpha_x}{\sqrt{\beta_x}} & \displaystyle\sqrt{\beta_x}  & 0 & 0 \\
   0 & 0 & \displaystyle\frac{1}{\sqrt{\beta_y}}& 0 \\
   0 & 0 & \displaystyle \frac{\alpha_y}{\sqrt{\beta_y}} & \displaystyle\sqrt{\beta_y} 
 \end{array}\right) \quad\Rightarrow\quad\qquad\quad \\ 
\ \nonumber \\\ \nonumber \\
\quad\Rightarrow\quad
\left(\mathbf{C^{-1}}\vec{\tilde{X}}^{(N+1)}\right)=
\mathbf{M}\left(\mathbf{C^{-1}}\vec{\tilde{X}}^{(N)}\right) 
\quad\Rightarrow\quad
\vec{\tilde{X}}^{(N+1)}=\left(\mathbf{C}\mathbf{M}\mathbf{C^{-1}}\right)\vec{\tilde{X}}^{(N)} \quad . 
\label{OTM-3}
\end{eqnarray}
The second step is to move from the real C-S to the complex C-S coordinates $h_{z,\pm}=\tilde{z}\pm i\tilde{p}_z$ of Eq.~\eqref{eq:NormFormTrans}
\begin{eqnarray} 
\vec{h}=\mathbf{T}\vec{\tilde{X}}\quad , \quad 
\mathbf{C}=
\vec{h}=\left(\begin{array}{c} h_{x,-}\\h_{x,+}\\h_{y,-}\\h_{y,+}\end{array}\right) \quad , \quad 
\mathbf{T}=\left(\begin{array}{c c c c } 
   1 & -i & 0 &  0 \\
   1 & +i & 0 &  0 \\
   0 &  0 & 1 & -i \\
   0 &  0 & 1 & +i 
\end{array}\right) \quad\Rightarrow\quad\qquad\qquad \\ 
\ \nonumber \\\ \nonumber \\
\quad\Rightarrow\quad
\left(\mathbf{T^{-1}}\vec{h}^{(N+1)}\right)=
\left(\mathbf{C}\mathbf{M}\mathbf{C^{-1}}\right)\left(\mathbf{T^{-1}}\vec{h}^{(N)}\right) 
\quad\Rightarrow\quad
\vec{h}^{(N+1)}=\left(\mathbf{T}\mathbf{C}\mathbf{M}\mathbf{C^{-1}}\mathbf{T^{-1}}\right)\vec{h}^{(N)} \quad . 
\label{OTM-4}
\end{eqnarray}
Note that $\mathbf{T}$ is not symplectic, each diagonal block having a determinant equal to 1/2. The last step is to eventually move into normal form via the transformation of Eq.~\eqref{eq:RDT-1st-front}
\begin{eqnarray}\label{OTM-5A}
\vec{\zeta}=\mathbf{B}\vec{h} +O(f^2)\quad ,\quad
\mathbf{B}=\left(
\begin{array}{c c c c}
    1             & 4if_{2000}    &\ \ 2if_{1001}&\ \ 2if_{1010} \\
   -4if_{2000}^*  &      1        &-2if_{1010}^* &-2if_{1001}^*  \\
  \ \ 2if_{1001}^*&\ \  2if_{1010}&     1        &\ \  4if_{0020}\\
  -2if_{1010}^*   &-2if_{1001}    &-4if_{0020}^* &       1
\end{array}
\right)\ +O(f^2) \quad\Rightarrow\quad\qquad\qquad \\ 
\ \nonumber \\\ \nonumber \\
\quad\Rightarrow\quad
\left(\mathbf{B^{-1}}\vec{\zeta}^{(N+1)}\right)=
\left(\mathbf{T}\mathbf{C}\mathbf{M}\mathbf{C^{-1}}\mathbf{T^{-1}}\right)\left(\mathbf{B^{-1}}\vec{\zeta}^{(N)}\right) 
\quad\Rightarrow\quad
\vec{\zeta}^{(N+1)}=\left(\mathbf{B}\mathbf{T}\mathbf{C}\mathbf{M}\mathbf{C^{-1}}\mathbf{T^{-1}}\mathbf{B^{-1}}\right)\vec{\zeta}^{(N)} \quad . 
\label{OTM-5}
\end{eqnarray}
Since the last relation in the above equation must be equal to the first of Eq.~\eqref{OTM-2}, we can write
\begin{eqnarray}
e^{i\mathbf{Q}}=\mathbf{B}\mathbf{T}\mathbf{C}\mathbf{M}\mathbf{C^{-1}}\mathbf{T^{-1}}\mathbf{B^{-1}}
\quad\Rightarrow\quad
\mathbf{M}=\mathbf{C^{-1}}\mathbf{T^{-1}}\mathbf{B^{-1}}e^{i\mathbf{Q}}\mathbf{B}\mathbf{T}\mathbf{C}\ .
\label{OTM-6}
\end{eqnarray}
Interestingly, even though four complex matrices appear in the r.h.s. of the last equation, namely $\mathbf{T^{-1}},\ \mathbf{B^{-1}},\ e^{i\mathbf{Q}},\ \mathbf{B}$ and $\mathbf{T}$, the entire r.h.d. must be a real 4x4 matrix which in the limit of no focusing errors (i.e. $f_{2000}=f_{0020}=0$) and of no coupling (i.e. $f_{1001}=f_{1010}=0$) shall be equal to the ideal OTM
\begin{eqnarray} 
\mathbf{M_0}=
\left(
\begin{array}{c c c c}
\cos{\mu_x}+\alpha_x\sin{\mu_x} & \beta_x\sin{\mu_x} & 0 & 0 \\
-\gamma_x\sin{\mu_x}&\cos{\mu_x}-\alpha_x\sin{\mu_x} & 0 & 0 \\
0 & 0 & \cos{\mu_y}+\alpha_y\sin{\mu_y} & \beta_y\sin{\mu_y} \\
0 & 0 &-\gamma_y\sin{\mu_y}&\cos{\mu_y}-\alpha_y\sin{\mu_y}
\end{array}
\right)\quad ,
\quad\label{OTM-0}
\end{eqnarray}
where $\mu=2\pi Q$ and $\gamma=(1+\alpha^2)/\beta$.

Two intermediary steps will help deriving more accurate and readable formulas for the focusing RDTs. The first step is to actually develop the last relation of Eq.~\eqref{OTM-6} in the C-S rather than Cartesian coordinates, namely
\begin{eqnarray}
\mathbf{\tilde{M}}=\mathbf{T^{-1}}\mathbf{B^{-1}}e^{i\mathbf{Q}}\mathbf{B}\mathbf{T}\ ,\quad\hbox{where }\quad 
\mathbf{\tilde{M}}=\mathbf{C}\mathbf{M}\mathbf{C^{-1}}\ .
\label{OTM-6B}
\end{eqnarray}
The second step is to replace the first-order transformation $\mathbf{B}$ of Eq.~\eqref{OTM-5A} with one including all orders of the focusing RDTS $f_{2000}$ and $f_{0020}$, while keeping the coupling part truncated to the first order. Even though analytic formulas including all orders of the coupling RDTs do exist (see Appendix A of Ref.~\cite{prstab_esr_coupling} and Appendix C of Ref.~\cite{Sext-RDT}), they have been derived assuming no focusing error. The cross-talk between higher order focusing and coupling RDTs renders more general analytic expressions cumbersome to derive and to write. For this reason, the priority is given here to an all-order analysis of focusing RDTs, while keeping the coupling part truncated to first order, i.e. to weak coupling ($|f_{1001}|,|f_{1010}|\ll|f_{2000}|,|f_{0020}|$) with tunes outside the coupling stop-bands ($|Q_x\pm Q_y|\gg |C_{\pm}|$). By making use of Eqs.(C8) and (C9) of Ref.~\cite{Sext-RDT}), the transformation matrix $\mathbf{B}$ can be written as 
\begin{eqnarray}
\mathbf{B}=
\left(
\begin{array}{c c}
\mathbf{B_{xx}} & \mathbf{B_{xy}} \\ \mathbf{B_{yx}} & \mathbf{B_{yy}} 
\end{array}
\right)\quad ,\qquad
\mathbf{B_{xx}}=
\left(
\begin{array}{c c}
\cosh{(4|f_{2000}|)}                & i\sinh{(4|f_{2000}|)}e^{iq_{2000}} \\
-i\sinh{(4|f_{2000}|)}e^{iq_{2000}} & \cosh{(4|f_{2000}|)} 
\end{array}
\right)\ ,
\label{OTM-6C}
\end{eqnarray}
where $q_{2000}=\hbox{arg}\{f_{2000}\}$ is the RDT phase. The vertical diagonal block $\mathbf{B_{yy}}$ comprises the same elements, after replacing $f_{2000}$ with $f_{0020}$. The off-diagonal blocks $\mathbf{B_{xy}}$ and $\mathbf{B_{xy}}$ are the same first-order truncated expression of Eq.~\eqref{OTM-5A}.

By making explicit the r.h.s of the first relation in Eq.~\eqref{OTM-6B}, the horizontal diagonal block of the OTM in C-S coordinates reads
\begin{eqnarray}
\mathbf{\tilde{M}}=\mathbf{\tilde{M_0}} +\mathbf{\delta \tilde{M}}
\quad ,\quad 
\mathbf{\tilde{M_0}}=
\left(
\begin{array}{c c}
\Re\{e^{i\mu_x}\}  &  \Im\{e^{i\mu_x}\} \\
-\Im\{e^{i\mu_x}\} & \Re\{e^{i\mu_x}\}
\end{array}
\right)\quad , \hspace{4cm}\label{OTM-6D1}\\
\mathbf{\tilde{\delta M}}=
\sin{\mu_x}
\left(
\begin{array}{c c}
-\sinh{(8|f_{2000}|)}\cos{q_{2000}} & 
2\sinh^2{(4|f_{2000}|)}+\sinh{(8|f_{2000}|)}\sinh{q_{2000}}\\
-2\sinh^2{(4|f_{2000}|)}+\sinh{(8|f_{2000}|)}\sinh{q_{2000}}&
\sinh{(8|f_{2000}|)}\cos{q_{2000}}
\end{array}
\right)\ .\quad
\label{OTM-6D2}
\end{eqnarray}
$\mathbf{\tilde{M_0}}$ is the unperturbed ideal OTM in C-S coordinates, whereas 
$\mathbf{\delta\tilde{M}}$ is the difference between the actual and the ideal OTM in C-S coordinates. The latter represents the observable, as it can be computed by any optics code. Other higher-order terms stemming from the normal form transformations (not discussed here, but derivable from Appendix A of Ref.~\cite{Sext-RDT})) generate phase-independent terms in the matrix $\mathbf{\delta\tilde{M}}$ which are not accounted for here. A simple, though not perfect, way to remove their contribution is to evaluate the average of $\mathbf{\delta\tilde{M}}$ along the ring circumference $C$ and to subtract it from $\mathbf{\delta\tilde{M}}$, so to have only phase-dependent terms and to be more consistent with the present scheme, namely
\begin{eqnarray}
\mathbf{\delta \tilde{M}}\rightarrow\mathbf{\delta \tilde{M}}-<\mathbf{\delta \tilde{M}}>
\quad ,\quad\hbox{where}\quad 
<\mathbf{\delta \tilde{M}}>=\frac{1}{C}\oint{\mathbf{\delta \tilde{M}}(s)\ ds}\ .
\label{OTM-6D2B}
\end{eqnarray}
This manipulation is sufficient in the presence of a single quadrupole error, i.e. with $|f_{2000}|$ constant along the ring, and thus in the evaluation of any RDT response matrix, though for a more generic lattice with distributed quadrupole errors (and possibly coupling), the resulting variation of $|f_{2000}|$ along the ring would still introduce an error, which can be estimated by the later Eq.~\eqref{OTM-6D}.

The system of Eq.~\eqref{OTM-6D2} provides the following relations for the horizontal focusing RDT
\begin{eqnarray}
\begin{aligned}
|f_{2000}|&=\frac{1}{8}\sinh^{-1}{\Biggl(\frac{\sqrt{(\delta\tilde{M}_{12}+\delta\tilde{M}_{21})^2+4\delta\tilde{M}_{11}^2}}{2\sin{\mu_x}}\ \Biggr)}\ , \\
\cos{q_{2000}}&=\frac{-2\delta\tilde{M}_{11}}{\sqrt{(\delta\tilde{M}_{12}+\delta\tilde{M}_{21})^2+4\delta\tilde{M}_{11}^2}}\ , \\
\sin{q_{2000}}&=\frac{\delta\tilde{M}_{12}+\delta\tilde{M}_{21}}{\sqrt{(\delta\tilde{M}_{12}+\delta\tilde{M}_{21})^2+4\delta\tilde{M}_{11}^2}}\ ,
\end{aligned}
\label{OTM-6C}
\end{eqnarray}
where $\sinh^{-1}{(x)}=\log{(\sqrt{x^2+1})}+x$ is the inverse hyperbolic function. The above expressions suffice to compute both the imaginary and real parts of $f_{2000}$ a any location. Equation~\eqref{OTM-6D2} provides also a relation that shall be satisfied by $\mathbf{\tilde{\delta M}}$ which in turn can be used to estimate the accuracy in the above calculations, namely
\begin{eqnarray}
0\simeq \hbox{Err}=\left|\frac{\delta\tilde{M}_{11}+\delta\tilde{M}_{22}}{16\sin{\mu_x}}\right|\ll
|f_{2000}|\ .  
\label{OTM-6D}
\end{eqnarray}
By inserting Eq.~\eqref{OTM-6D1} into the right expression of Eq.~\eqref{OTM-6B} the matrix elements $\delta\tilde{M}_{ij}$ are readily computed from the C-S unperturbed parameters and the difference between the actual and unperturbed (or ideal) OTMs in Cartesian coordinates, $\mathbf{\delta M}$, both to be computed by any optics code:
\begin{eqnarray}
\left\{
\begin{aligned}
\delta\tilde{M}_{11}&=\delta M_{11} - \frac{\alpha_x}{\beta_x}\delta M_{12} \\
\delta\tilde{M}_{12}&=\alpha_x\delta M_{11}-\frac{\alpha_x^2}{\beta_x}\delta M_{12}+\beta_x\delta M_{21} - \alpha_x\delta M_{22}\\
\delta\tilde{M}_{21}&=\frac{1}{\beta_x}\delta M_{21} \\
\delta\tilde{M}_{22}&=\frac{\alpha_x}{\beta_x}\delta M_{12}+\delta M_{22}
\end{aligned}
\right. \quad .\label{OTM-E}
\end{eqnarray}
It is reminded that the subtraction of Eq.~\eqref{OTM-6D2B} shall be performed prior to the insertion of the above terms into Eq.~\eqref{OTM-6C}. \\

The same considerations applied to the vertical diagonal block yields to the following relations for the focusing RDT $f_{0020}$:
\begin{eqnarray}
\begin{aligned}
|f_{0020}|&=\frac{1}{8}\sinh^{-1}{\Biggl(\frac{\sqrt{(\delta\tilde{M}_{34}+\delta\tilde{M}_{43})^2+4\delta\tilde{M}_{33}^2}}{2\sin{\mu_y}}\ \Biggr)}\ , \\
\cos{q_{0020}}&=\frac{-2\delta\tilde{M}_{33}}{\sqrt{(\delta\tilde{M}_{34}+\delta\tilde{M}_{43})^2+4\delta\tilde{M}_{33}^2}}\ , \\
\sin{q_{0020}}&=\frac{\delta\tilde{M}_{34}+\delta\tilde{M}_{43}}{\sqrt{(\delta\tilde{M}_{34}+\delta\tilde{M}_{43})^2+4\delta\tilde{M}_{33}^2}}\ ,
\end{aligned}
\label{OTM-7C}
\end{eqnarray}
with the same error estimation
\begin{eqnarray}
0\simeq \hbox{Err}=\left|\frac{\delta\tilde{M}_{33}+\delta\tilde{M}_{44}}{\sin{\mu_y}}\right|\ll
|f_{0020}||\ ,  
\label{OTM-7D}
\end{eqnarray}
and computation of $\mathbf{\delta \tilde{M}_{yy}}$
\begin{eqnarray}
\left\{
\begin{aligned}
\delta\tilde{M}_{33}&=\delta M_{33} - \frac{\alpha_y}{\beta_y}\delta M_{34} \\
\delta\tilde{M}_{34}&=\alpha_y\delta M_{33}-\frac{\alpha_y^2}{\beta_y}\delta M_{34}+\beta_y\delta M_{43} - \alpha_y\delta M_{44}\\
\delta\tilde{M}_{43}&=\frac{1}{\beta_y}\delta M_{43} \\
\delta\tilde{M}_{44}&=\frac{\alpha_y}{\beta_y}\delta M_{34}+\delta M_{44}
\end{aligned}
\right. \quad .\label{OTM-7E}
\end{eqnarray}

In Fig.~\ref{fig_OTM1} an example of different computations of $f_{2000}$ at the 224 BPMs of the old ESRF storage ring, with a single error at a focusing quadrupole is shown. The RDT is calculated from the lattice formula of Eq.~\eqref{eq-app:def_rdt1} (black), from the FFT of single particle tracking data and Eq.(C8) of Ref.~\cite{Sext-RDT} (green) and from the OTM via Eq.~\eqref{OTM-6C}. The error functions $\hbox{Err}_{\Re}$ (left plot) and $\hbox{Err}_{\Im}$ (right plot) are displayed in blue, indicating an excellent applicability of the OTM formulas. The agreement between the lattice and OTM formulas is indeed remarkable. Analogous results are obtained for $f_{0020}$ (not shown here).

In order to explore the range of validity of the OTM formulas and the {\sl pollution} introduced by betatron coupling, the same computation has been carried out after including a typical linear lattice error model comprising distributed quadrupole errors and sources of coupling. Results for $f_{2000}$  are shown in Fig.~\ref{fig_OTM2} (similar results apply to $f_{0020}$, not shown here). The agreement between the different formulas is poorer, and the sizeable amplitude of the error function $\hbox{Err}_{\Im}$ (blue curve, right plot), casts doubts on the accuracy of the derivation of the focusing RDTs from the OTM.

\begin{figure}[]
\rule{0mm}{0mm}
\centerline{\includegraphics[width=8.3cm]{f2000_Single_Re.eps}
\ \ \ \  \  \includegraphics[width=8.3cm]{f2000_Single_Im.eps}}
  \caption{\label{fig_OTM1} (Color) Example of different computations of $f_{2000}$ at the 224 BPMs of the old ESRF storage ring, with a single error at a focusing quadrupole: from the lattice formula of Eq.~\eqref{eq-app:def_rdt1} (black), from the FFT of single particle tracking data and Eq.(C8) of Ref.~\cite{Sext-RDT} (green) and from the OTM via Eq.~\eqref{OTM-6C}. The error function $\hbox{Err}$ is displayed too (blue, multiplied by 10), indicating an excellent applicability of the OTM formulas.}
\rule{0mm}{0mm}
\end{figure}

\begin{figure}[]
\rule{0mm}{0mm}  
\centerline{\includegraphics[width=8.3cm]{f2000_ORM_Re.eps}
\ \ \ \  \  \includegraphics[width=8.3cm]{f2000_ORM_Im.eps}}
  \caption{\label{fig_OTM2} (Color) Same plot of Fig.~\ref{fig_OTM1} but with a realistic error model of the old ESRF storage ring comprising distributed focusing errors and sources of betatron coupling. The latter corrupt the possibility of inferring accurately the focusing RDTs from the OTM, as suggested by the sizeable error function $\hbox{Err}$ (blue, multplied by 10).}
\rule{0mm}{0mm}
\end{figure}
\end{widetext}


\end{document}